# From gamma rays to radio waves: Dark Matter searches across the spectrum

Elena Pinetti

This dissertation is submitted for the degree of
**Philosophiae Doctor**

August 2021

Scuola di dottorato dell'Università di Torino

École doctorale Physique en Île-de-France

Supervisors: Prof. Nicolao Fornengo
Dr. Marco Cirelli

# Table of Contents

# Acknowledgements

The doctoral degree was an important milestone in my life. If you have met me 20 years ago, on my first day of school, you would have seen a little girl with pigtails and a big smile who, when asked "What will you do when you grow up?", always answered: "I'll be a crazy scientist!". And so here I am today, to fulfil the dream of that little girl. Over the past 20 years there have been many people who have been part of this journey and I would like to take this opportunity to thank those who have really made a positive impact in my life.

First, I'd like to thank both my supervisors, Nicolao Fornengo and Marco Cirelli, for being my trustworthy guides. You were the greatest supervisors I could hope for: supportive, amusing, always ready to joke and available to answer my questions. Thank you for encouraging me to be independent and develop critical thinking: two crucial skills for a scientist.

Thanks Stefano Camera, Filippo Sala, Alessandro Cuoco, Kallia Petraki, Bradley Kavanagh, Marco Regis and all my collaborators for expanding my horizons with useful discussions and for the lovely time we spent together in the office and during conferences.

Thanks to the PhD students of Turin, for all the fun we had while preparing the exams and participating to the summer schools together.

Thanks to the PhD students in Paris for the warm welcome when I arrived in the lab and for being so entertaining. I'll miss our lunches together with the Eiffel Tower framed in the window.

Thanks to the wonderful ladies of the Supernova Foundation: being part of this amazing community of women in physics inspires me everyday.

Thank you Jan-Albert, for showing me the treasures of South Africa. Thank you for all the beautiful sunsets: from the orange glow of the Atlantic ocean, to the illuminated plains of the savanna - surrounded by lions. I will always remember the adrenaline filled days of spotting leopards and being chased by the elephants.

Thank you, Aurelio, for supporting me in everything I do and for welcoming all my ideas, no matter how crazy they are. You are always there for me and I am so lucky to have you in my life.

Finally, I want to thank my parents, Sylvie and Mario, who have always encouraged me to study hard and aim high. Vi voglio un mondo di bene.

Looking back to the past three years, I am very happy and grateful to all of you for having filled my days with joy. You have taught me a very important lesson that I will always bring with me: the key to happiness is to be surrounded by positive people.

## An ode to dark matter

I'd like to share a poem that was submitted to a competition regarding physics outreach in Italy. My goal was to introduce the youth to the wonders of the Universe in a creative and original way, to inspire and motivate the next generation of scientists. This is an ode to dark matter:

Nell'Universo mi nascondo
Invisibile sono agli occhi del mondo
Mi piace poco interagire
Anche in mezzo agli amici preferisco non apparire
Mi dicono tutti che sono molto pesante
A confronto è snello un elefante
Sono di carattere neutrale
Forse persino un po' asociale
Mi muovo molto lentamente
Così risulto ancora meno appariscente
Sogno una vita con più colori
Nonostante non mi manchino gli ammiratori
Secondo alcuni sono fredda come il ghiaccio
In realtà avrei tanto bisogno di un abbraccio
Altri invece dicono che sono ardente
Ed è questo che mi rende molto attraente
Sono all'origine di tanti problemi
Perché mi piace vivere fuori dagli schemi
Il modello standard non fa al caso mio
E per secoli ho vissuto nell'oblio
Poi negli anni trenta uno scienziato ho incuriosito
Un ammasso di galassie lo aveva molto colpito
Tanta massa e poca luce
Le curve gravitazionali non riproduce
Negli anni settanta la conferma arrivò
Di quel giorno sempre mi ricorderò
Era donna, era scienziata,

Di Fisica era appassionata
Al mondo annunciò la mia esistenza
Da quel momento non fui più solo fantascienza
E’ partita una caccia al tesoro
Il perché di tanto trambusto davvero lo ignoro
La mia natura cercano di svelare
Ma io non ci penso proprio a farmi acchiappare
Alcuni credono che io sia particella
Altri dicono che di una nuova gravità sono sentinella
Per cercarmi usano grandi telescopi
Da lontano mi ricordano gli omerici ciclopi
Laboratori dal polo sud al sottosuolo
Interferometri per le onde gravitazionali da quadrupolo
Cercano un segnale da decadimento o annichilazione
Dicono segnerebbe l’inizio di una rivoluzione
Elettroni, neutrini, e positroni
Potrei produrne a milioni
Così come onde radio, raggi X e gamma
Manca solo che un poeta di me scriva l’epigramma
Continuate pure a cercarmi, cari umani,
Della Scienza siate sempre severi guardiani
Vi auguro buona fortuna per questa avventura
Alla ricerca della Materia Oscura.

# Abstract

One of the most fascinating open issues in modern physics is dark matter (DM), but so far the only evidence found of its existence is of gravitational nature. In fact, all the current probes are related to the gravitational impact of the presence of DM in the Universe: such as the rotation curves of spiral galaxies, the dynamics of galaxy clusters, the large-scale structure of the Universe, the cosmic microwave background (CMB), as well as weak and strong lensing effects. It is believed to be a new form of matter, however its intimate nature remains a puzzle. If DM is made up of unknown elementary particles or exotic macroscopic bodies, it is expected to produce characteristic signals of particle physics, revealing its true nature. The observational evidences of the presence of DM in the Universe are tackled in Chapter 1.

The subject of this thesis pertains to the indirect detection of DM particles, a method which relies on the idea that annihilating or decaying DM could produce a variety of astrophysical messengers. The messengers that we expect are charged cosmic rays, neutrinos and photons. Different messengers require different detection techniques and carry different kinds of information. For instance, neutrinos and photons are useful to trace the origin of the emitting sources, being neutral particles, while charged cosmic rays are deflected by the magnetic fields in the galaxies. However, charged cosmic rays are invaluable signals to study DM candidates, and antimatter is particularly interesting since it is associated with a relatively low astrophysical background on Earth. In addition, different messengers are usually associated to different DM candidates, for instance with different masses and types of interactions. Thus, a multi-messenger strategy is beneficial. Chapter 2 introduces various detection techniques, with a special focus on indirect detection with photons.

There are several ways to look for a signal indirectly produced by DM. In Part I we focus on a promising method called the cross-correlation technique. It aims to correlate two distinct features of DM: on one side, an electromagnetic signal, which is a manifestation of the particle nature of DM; on the other side, a gravitational tracer of the DM distribution in the Universe. If the cross-correlation between these two features yield a positive signal, it would indicate that the cause of the gravitational anomalies indeed consists of new elementary particles, which we call DM. We computed for the first time the cross-correlation signal between the Unresolved $\gamma$-Ray Background (UGRB) and the 21cm line of neutral hydrogen (HI). Specifically, we derived the contribution to the UGRB expected from annihilation events of DM particles, but also from the astrophysical background: notably BL Lac, flat-spectrum radio quasars, misaligned active galactic nuclei and star-forming galaxies. The HI distribution can be measured by using the Intensity Mapping technique. This is a cutting-edge method used to map the photons produced with a wavelength of 21cm, emitted by the spin-flip transition of HI – which may occur when the electron and proton of an HI atom have parallel spin. Since the HI mostly resides within DM halos, the 21cm line turns out to be a promising gravitational tracer of the matter distribution in the Universe. The precise redshift information provided by the intensity mapping technique can enable us to disentangle a subdominant DM contribution from the overwhelming astrophysical background. We considered FERMI-LAT as a benchmark experiment for the UGRB, while for the 21cm intensity mapping measurement we referred to the up-coming radio telescope Square Kilometre Array (SKA), currently under construction, as well as its precursor MEERKAT. We found that the combination of FERMI-LAT and SKA will be able to detect a signal of astrophysical origin with a signal-to-noise ratio above $5\sigma$. In addition, this combination of detectors will be able to probe Weakly Interacting Massive Particles (WIMPs) with thermal annihilation cross-section up to a DM mass of 150 GeV. Furthermore, we showed that a combination of two next-generation telescopes for both $\gamma$ rays and HI intensity mapping might be sensitive up to the TeV scale, allowing us to probe the entire mass window of WIMPs.

Alternative DM candidates which are getting increasing attention are sub-GeV particles, which are the subject of Part II. The astrophysical messengers that we expect in this case are electrons and positrons, neutrinos and photons. The solar magnetic field deflects sub-GeV charged particles and thus space telescopes in orbit around the Earth have no access to the electrons and positrons. Low-energy neutrinos from DM would be arduous to detect in the overwhelming neutrinos' background. Sub-GeV $\gamma$ rays would be be an interesting signal to consider, but unfortunately FERMI-LAT, which is the most powerful among the recent $\gamma$-ray detectors, is not sensitive in the energy range of interest. Indeed, one of the challenges in indirect detection of sub-GeV DM is the so-called "MeV gap": there is a scarcity of recent data with high sensitivity in the energy range between 1 MeV and hundreds of MeV. As a consequence, we need to find alternative ways to study DM candidates with mass in this energy window. In our work we proposed to probe sub-GeV DM by looking at photon energies much lower than the DM mass. In particular, the electrons and positrons produced by DM particles can do inverse Compton scattering on the low-energy radiation fields in the Galaxy (CMB, infrared light from dust, starlight) and give rise to X rays. This secondary emission falls in the energy range covered by the INTEGRAL data, which we used to determine conservative bounds on the DM annihilation cross-section. We considered three annihilation channels: electron, muon and pion. As a result, we derived competitive constraints for DM particles with a mass between 150 MeV and 1.5 GeV.

Lastly, clusters of galaxies represent an interesting target for DM searches. They are thought to be linked by extended filaments, which are the focus of Part III. These connective structures are very faint, thus extremely difficult to detect with our current telescopes. They are expected to include warm-hot intergalactic medium and magnetic fields, facilitating an acceleration of the cosmic rays and emission of synchrotron radiation. A popular method to study faint filament regions is image stacking. This technique consists of averaging over repeated samples of the same kind of astrophysical objects. The benefit is that the number of photons coming from the astrophysical sources ("true value") stays relatively constant, unlike the random noise whose average over multiple images is averaged out. As a result, the stacking technique allows us to increase the signal-to-noise ratio, when applied to approximately static astrophysical objects. Since the location of filaments is not known a priori, a proxy for galaxy clusters is needed. In this regard, luminous red galaxies (LRGs) are valuable tracers of the large-scale structure since they are usually located in the vicinity of the centres of clusters. The Sloan Digital Sky Survey has catalogued over a million of these galaxies. Thus, one can consider physical pairs of nearby LRGs as a tracer of nearby clusters, potentially connected by faint filaments. In our work we adopted radio maps from the GLEAM and OVRO-LWA surveys together with the stacking of LRGs pairs. We found the first-ever robust detection of the stacked radio emission from large filaments (1−15 Mpc), connecting pairs of nearby LRGs. The signal is compatible with synchrotron emission from filaments of the cosmic web, providing direct evidence of one of the cornerstones of our current understanding of the large-scale structure in the Universe. Notably, we showed that the excess detected in the radio signal can be interpreted as synchrotron radiation originated from DM particles. We found that candidates with a mass around 5−10 GeV decaying into electron-positron pairs, can produce a signal compatible with the observations. On the contrary, hadronic decays and annihilation events are disfavoured. The observed brightness temperatures associated to the filament regions can be used to constrain the lifetime of decaying DM. We obtained competitive bounds for masses in the range 3−10 GeV and magnetic fields above 130 nGauss.

Finally, Part IV draws the conclusions and gives a perspective of potential future developments.

# Sommario

Una delle questioni aperte più affascinanti della fisica moderna è la materia oscura (MO). Ad oggi, le prove della sua esistenza sono solo di natura gravitazionale, ossia sono legate all'impatto gravitazionale della presenza di MO nell'Universo. Ne sono un esempio le curve di rotazione delle galassie a spirale, la dinamica degli ammassi di galassie, la struttura su larga scala dell'Universo, il fondo cosmico a microonde (CMB), ma anche gli effetti di lente debole e forte. La possibilità più accreditata è che la MO sia una nuova forma di materia, ma la sua natura rimane un enigma. In particolare, se la MO è costituita da nuove particelle elementari o da altri oggetti esotici macroscopici, ci si aspetta che produca segnali caratteristici della fisica delle particelle che ne rivelino la natura. Le evidenze osservative della presenza di MO nell'Universo sono discusse nel capitolo 1.

L'argomento di questa tesi riguarda la rivelazione indiretta di particelle di MO, basata sull'idea che l'annichilazione o il decadimento della MO potrebbe produrre una grande varietà di messaggeri astrofisici. I messaggeri che ci aspettiamo sono raggi cosmici carichi, neutrini e fotoni. E' importante sottolineare che diversi messaggeri richiedono diverse tecniche di rivelazione e procurano diversi tipi di informazioni. Ad esempio, neutrini e fotoni conservano l'informazione sulla posizione delle sorgenti che li hanno emessi, essendo particelle neutre, invece i raggi cosmici carichi vengono deviati dai campi magnetici presenti nelle galassie. Tuttavia, i raggi cosmici carichi sono anch'essi segnali preziosi per studiare i candidati di MO e l'antimateria è particolarmente interessante poiché è associata a uno fondo astrofisico relativamente basso a Terra. Inoltre, diversi messaggeri ci permettono di studiare diversi candidati di MO, ossia particelle con un diverso intervallo di masse e tipi di interazioni. Di conseguenza è necessario adottare un approccio multimessaggero. A questo proposito, il capitolo 2 introduce varie tecniche di rivelazione, con particolare attenzione alla ricerca indiretta con i fotoni.

Esistono diversi modi per cercare un segnale indiretto da MO. La Parte I si concentra su un metodo molto promettente chiamato tecnica della correlazione incrociata. Tale tecnica ha lo scopo di correlare due caratteristiche distintive della MO: un segnale elettromagnetico, che è una manifestazione della natura particellare della MO, e un tracciatore gravitazionale della distribuzione di materia nell'Universo. Se la correlazione incrociata tra queste due osservabili producesse un segnale positivo, ciò sarebbe la prova che la materia invisibile di cui misuriamo gli effetti gravitazionali è effettivamente formata da nuove particelle elementari. Nel nostro lavoro abbiamo calcolato per la prima volta il segnale di correlazione incrociata tra il fondo non risolto di raggi $\gamma$ (UGRB) e la linea a 21 cm emessa dell'idrogeno neutro (HI). Nello specifico, abbiamo stimato il contributo dell'UGRB prodotto sia dagli eventi di annichilazione delle particelle di MO sia dalle sorgenti astrofisiche, in particolare BL Lacertae (comunement noti come BL Lac), radio quasar a spettro piatto, nuclei galattici attivi disallineati e galassie ad alta formazione stellare. Per quanto riguarda la distribuzione di HI nell'Universo, può essere misurata utilizzando la tecnica della mappatura di intensità. Si tratta di un metodo all'avanguardia utilizzato per mappare la temperatura di brillanza della riga a 21cm. Tali fotoni sono prodotti da una transizione energetica dell'HI che può verificarsi quando l'elettrone e il protone di un atomo di HI hanno entrambi spin parallelo. Poiché l'HI risiede principalmente all'interno di aloni di MO, la linea a 21 cm costituisce un promettente tracciatore gravitazionale della distribuzione di materia nell'Universo. Infatti, la precisa informazione sul redshift fornita dalla tecnica di mappatura dell'intensità può aiutarci a far emergere il contributo di MO dal fondo astrofisico dominante. Per quanto riguarda gli esperimenti di rifermento, abbiamo considerato il telescopio spaziale a raggi $\gamma$ FERMI-LAT per l'UGRB, mentre abbiamo fatto riferimento al radio telescope Square Kilometre Array (SKA) e al suo precursore MEERKAT per la mappatura dell'intensità dell'HI. La combinazione di FERMI-LAT e SKA ha le potenzialità per rivelare un segnale di origine astrofisica con un rapporto segnale-rumore superiore oltre $5\sigma$. Inoltre, questa combinazione di rivelatori sarà in grado di escludere le particelle massive a interazione debole (WIMP) con una sezione d'urto di annichilazione termica fino a una massa DM di 150 GeV, con un livello di confidenza del 95%. Inoltre, abbiamo mostrato che la combinazione di due telescopi di nuova generazione per i raggi $\gamma$ e la mappatura dell'intensità dell'HI sarebbe sensibile fino alla scala del TeV, consentendoci di studiare l'intera finestra di massa delle WIMP.

Molti altri candidati di MO stanno ricevendo una crescente attenzione. Tra questi spiccano le particelle sub-GeV che sono oggetto della Parte II. I messaggeri astrofisici che ci aspettiamo in questo caso sono elettroni e positroni, neutrini e fotoni. Il campo magnetico solare devia le particelle cariche con un'energia cinetica inferiore al GeV e quindi i telescopi spaziali in orbita attorno alla Terra non hanno accesso agli elettroni e ai positroni. I neutrini a bassa energia provenienti dalla MO sono di difficile rivelazione a causa del fondo di neutrini solari. I raggi $\gamma$ rappresentano un segnale interessante da considerare, ma sfortunatamente FERMI-LAT, che è il più potente tra i recenti rivelatori di raggi $\gamma$, non è sufficientemente sensibile nell'intervallo energetico di interesse. A questo proposito, una delle sfide nella rilevazione indiretta di MO con una massa dell'ordine del sub-GeV è il cosiddetto "gap MeV": c'è una scarsità di dati recenti e con una buona sensitività nell'intervallo di energia tra 1 MeV e centinaia di MeV. Di conseguenza, dobbiamo trovare modalità alternative per studiare i candidati di MO che hanno massa in questa finestra energetica. Nel nostro lavoro abbiamo proposto di studiare la MO con massa dell'ordine del sub-GeV osservando fotoni con energie molto inferiori rispetto alle masse delle particelle che desideriamo studiare. In particolare, gli elettroni e i positroni prodotti dalle particelle di

MO possono fare scattering Compton inverso sui campi di radiazione a bassa energia nella Galassia (CMB, luce infrarossa da polvere, luce stellare nella banda ottica) e dare origine a raggi X. Questa emissione secondaria rientra nell'intervallo di energia coperto dai dati INTEGRAL che abbiamo usato per determinare i limiti conservativi sulla sezione d'urto di annichilazione della MO. Abbiamo considerato tre canali di annichilazione (elettrone, muone e pione) e abbiamo derivato limiti competitivi per particelle di MO con una massa compresa tra 150 MeV e 1.5 GeV.

Infine, gli ammassi di galassie sono anche di particolare interesse per le ricerche di MO. Si pensa che siano collegati da filamenti diffusi, che rappresentano il fulcro della Parte III. Queste strutture connettive sono caratterizzate da un'emissione molto debole, quindi sono estremamente difficili da rivelare con i nostri attuali telescopi. Si pensa che i filamenti contengano del gas caldo e dei campi magnetici che inducono l'accelerazione dei raggi cosmici e l'emissione di radiazione di sincrotrone. Un metodo spesso adottato per studiare i filamenti deboli è lo "stacking" delle immagini. Questa tecnica consiste nel fare la media su numerose immagini ripetute degli stessi oggetti astrofisici. Il vantaggio è che il numero di fotoni provenienti dalle sorgenti astrofisiche ("valore vero") rimane relativamente costante, a differenza del rumore di fondo la cui media su più immagini converge a zero. Di conseguenza, la tecnica dello stacking ci consente di aumentare il rapporto segnale-rumore, se applicata a oggetti astrofisici approssimativamente statici. Poiché la posizione dei filamenti non è nota a priori, è necessario un tracciatore per individuare la posizione degli ammassi di galassie. A questo proposito, le galassie rosse luminose (LRG) sono preziosi tracciatori della struttura su larga scala poiché di solito si trovano in prossimità dei centri degli ammassi. Lo Sloan Digital Sky Survey ha catalogato oltre un milione di queste galassie. Pertanto, si possono considerare coppie fisiche di LRG vicine come tracciatori di cluster vicini, potenzialmente collegati da filamenti. Nel nostro lavoro abbiamo adottato le mappe radio dei cataloghi GLEAM e OVRO-LWA per effettuare lo stacking di coppie di LRG. Abbiamo trovato il primo segnale dell'emissione radio da stacking proveniente da filamenti di grande dimensione ($1-15$ Mpc), che collegano coppie di LRG vicine. Il segnale è compatibile con l'emissione di sincrotrone dalla rete cosmica, pertanto costituisce la prova diretta di uno dei capisaldi della nostra attuale comprensione della struttura su larga scala nell'Universo. In particolare, l'eccesso rivelato nel segnale radio può essere interpretato come radiazione di sincrotrone originata da particelle di MO. Abbiamo scoperto che i candidati con una massa di circa $5-10$ GeV che decadono in coppie elettrone-positrone possono produrre un segnale compatibile con le osservazioni. Al contrario, i decadimenti di tipo adronico e gli eventi di annichilazione sono sfavoriti. Le temperature di luminosità osservate associate ai filamenti possono essere utilizzate per ricavare i limiti sulla vita media della MO. In particolare, abbiamo ottenuto dei limiti competitivi per masse nell'intervallo $3-10$ GeV e per campi magnetici superiori a 130 nGauss.

Infine, la Parte IV trae le conclusioni e offre una panoramica di potenziali sviluppi futuri.

# Résumé

L'un des problèmes ouverts parmi les plus fascinants de la physique moderne est la matière noire (MN), mais à ce jour, la seule preuve trouvée de son existence est de nature gravitationnelle. En effet, toutes les sondes actuelles sont liées à l'impact gravitationnel de la présence de MN dans l'Univers : telles que les courbes de rotation des galaxies spirales, la dynamique des amas de galaxies, la structure à grande échelle de l'Univers, le fond diffus cosmologique (CMB), ainsi que des effets de lentille faibles et forts. On pense qu'il s'agit d'une nouvelle forme de matière, mais sa nature intime reste une énigme. Si la MN est composée de particules élémentaires inconnues ou de corps macroscopiques exotiques, elle devrait produire des signaux caractéristiques de la physique des particules, révélant sa vraie nature. Les preuves observationnelles de la présence de MN dans l'Univers sont abordées au chapitre 1.

Le sujet de cette thèse porte sur la détection indirecte des particules de MN, une méthode qui repose sur l'idée que l'annihilation ou la désintégration de la MN pourrait produire une variété de messagers astrophysiques. Les messagers que nous attendons sont des rayons cosmiques chargés, des neutrinos et des photons. Différents messagers nécessitent différentes techniques de détection et transportent différents types d'informations. Par exemple, les neutrinos et les photons sont utiles pour retracer l'origine des sources émettrices, étant des particules neutres, tandis que les rayons cosmiques chargés sont déviés par les champs magnétiques dans les galaxies. Cependant, les rayons cosmiques chargés sont des signaux précieux pour étudier les candidats de MN et l'antimatière est particulièrement intéressante car elle est associée à un fond astrophysique relativement faible sur Terre. De plus, différents messagers sont généralement associés à différents candidats de MN, par exemple avec différentes masses et types d'interactions. Ainsi, une stratégie multi-messagers est potentiellement fructueuse. Le chapitre 2 présente diverses techniques de détection, avec une attention particuliére sur la détection indirecte avec des photons.

Il existe plusieurs façons de rechercher un signal produit indirectement par la MN. Dans la partie I, nous nous concentrons sur une méthode prometteuse appelée technique de corrélation croisée. Elle vise à corréler deux caractéristiques distinctes de la MN : un signal électromagnétique, qui est une manifestation de la nature particulaire de la MN, et un traceur gravitationnel de la distribution de la MN dans l'Univers. Si la corrélation croisée entre ces deux caractéristiques donne un signal positif, cela indiquerait que la cause des anomalies gravitationnelles est bien constituée de nouvelles particules élémentaires, que nous appelons MN. Nous avons calculé pour la première fois le signal de corrélation croisée entre le fond $\gamma$ non résolu (UGRB) et la ligne de 21 cm d'hydrogène neutre (HI). Plus précisément, nous avons dérivé la contribution à l'UGRB attendue des événements d'annihilation des particules de MN, mais aussi du fond astrophysique: notamment BL Lacs, quasars radio à spectre plat, noyaux galactiques actifs désalignés et galaxies à formation d'étoiles. La distribution de HI peut être mesurée en utilisant la technique de la cartographie d'intensité. Il s'agit d'une méthode de pointe utilisée pour cartographier les photons d'une longueur d'onde de 21 cm, produits par une transition énergétique de l'HI - qui peut se produire lorsque l'électron et le proton d'un atome de HI ont les spins parallèles. Étant donné que le HI réside principalement dans les halos de MN, la ligne de 21 cm s'avère être un traceur gravitationnel prometteur de la distribution de la matière dans l'Univers. Les informations précises de décalage vers le rouge fournies par la technique de la cartographie d'intensité peuvent nous permettre de départager une contribution sous-dominante de la MN du fond astrophysique écrasant. Nous avons considéré le télescope spatial à rayon $\gamma$ FERMI-LAT comme expérience de référence pour l'UGRB. Pour la mesure de la cartographie d'intensité de 21 cm, nous nous sommes référés au futur radiotélescope Square Kilometre Array (SKA), actuellement en construction, ainsi qu'à son précurseur MEERKAT. Nous avons trouvé que la combinaison de FERMI-LAT et SKA sera capable de détecter un signal d'origine astrophysique avec un rapport signal sur bruit supérieur à $5\sigma$. De plus, cette combinaison de détecteurs sera capable de sonder les particules massives interagissent faiblement (WIMPs) avec une section efficace d'annihilation thermique jusqu'à une masse de MN de 150 GeV. De plus, nous avons montré qu'une combinaison de deux télescopes de nouvelle génération pour les rayons $\gamma$ et la cartographie d'intensité HI pourrait être sensible jusqu'à l'échelle du TeV, nous permettant de sonder toute la fenêtre de masse des WIMPs.

Des candidats de MN alternatifs qui attirent de plus en plus l'attention sont les particules sub-GeV, qui font l'objet de la partie II. Les messagers astrophysiques que nous attendons dans ce cas sont les électrons et les positons, les neutrinos et les photons. Le champ magnétique solaire dévie les particules chargées sous-GeV et donc les télescopes spatiaux en orbite autour de la Terre n'ont pas accès aux électrons et aux positons (à l'exception de VOYAGER1). Les neutrinos de basse énergie produits par la MN seraient difficiles à détecter à cause des neutrinos solaires. De plus, les rayons $\gamma$ sont certainement un signal intéressant à considérer, mais malheureusement FERMI-LAT, qui est le plus puissant parmi les détecteurs de rayons $\gamma$ récents, n'est pas suffisamment sensible dans la gamme d'énergie d'intérêt. En effet, l'un des défis de la détection indirecte de MN avec une masse de l'ordre du sub-GeV est ce que l'on appelle le "MeV gap" : il existe une pénurie de données récentes avec une haute sensibilité dans la gamme d'énergie comprise entre 1 MeV et des centaines de MeV. En conséquence, nous devons trouver des moyens alternatifs pour étudier les candidats de MN avec une masse dans cette fenêtre d'énergie. Dans notre travail, nous avons proposé de sonder ces particules de MN en regardant des énergies de photons bien inférieures aux masses de la MN. En particulier, les électrons et les positons produits par les particules de MN peuvent effectuer une diffusion Compton inverse sur le champs de rayonnement de faible énergie dans la Galaxie (CMB, lumière infrarouge de la poussière, lumière optique des étoiles) et peuvent donner naissance à des rayons X. Cette émission secondaire se situe dans la gamme d'énergie couverte par les données du télescope INTEGRAL, que nous avons utilisées pour déterminer des limites sur la section efficace d'annihilation de MN. Nous avons considéré trois canaux d'annihilation : l'électron, le muon et le pion. En conséquence, nous avons dérivé des limites competitives pour les particules de MN d'une masse comprise entre 150 MeV et 1.5 GeV.

Enfin, les amas de galaxies représentent une cible intéressante pour les recherches de MN. On pense qu'ils soient liés par des filaments diffus, qui sont au centre de la partie III. Ces structures connectives sont très faibles, donc extrêmement difficiles à détecter avec nos télescopes actuels. Ils devraient inclure un milieu intergalactique chaud et des champs magnétiques, facilitant une accélération des rayons cosmiques et l'émission de rayonnement synchrotron. Une méthode réputé pour étudier les filaments est la "superposition" d'images. Cette technique consiste à faire la moyenne sur des échantillons répétés des mêmes objets astrophysiques. L'avantage est que le nombre de photons provenant des sources astrophysiques ("valeur vraie") reste relativement constant, contrairement au bruit aléatoire dont la moyenne sur plusieurs images converge vers zéro. En conséquence, la technique de superposition nous permet d'augmenter le rapport signal sur bruit, lorsqu'elle est appliquée à des objets astrophysiques approximativement statiques. Comme l'emplacement des filaments n'est pas connu a priori, un traceur pour les amas de galaxies est nécessaire. À cet égard, les galaxies rouges lumineuses (LRG) sont de précieux traceurs de la structure à grande échelle puisqu'elles sont généralement situées à proximité des centres des amas. Le Sloan Digital Sky Survey a répertorié plus d'un million de ces galaxies. Ainsi, on peut considérer des paires physiques de LRGs proches comme un traceur d'amas proches, potentiellement connectés par de faibles filaments. Dans notre travail, nous avons adopté des cartes radio des relevés GLEAM et OVRO-LWA ainsi que la superposition de paires de LRGs. Nous avons obtenu la toute première détection robuste de l'émission radio du stackig de filaments de grandes dimensions (1−15 Mpc), qui connectent des paires de LRGs à proximité. Le signal est compatible avec l'émission synchrotron de la toile cosmique, fournissant une preuve directe de l'une des pierres angulaires de notre compréhension actuelle de la structure à grande échelle de l'Univers. Notamment, l'excès détecté dans le signal radio peut être interprété comme un rayonnement synchrotron provenant de particules de MN. Nous avons obtenu que les candidats avec une masse d'environ 5−10 GeV se désintégrant en paires électron-positron, peuvent produire un signal compatible avec les observations. Au contraire, les désintégrations hadroniques et les événements d'annihilation sont défavorisés. Les températures de brillance observées associées aux régions des filaments peuvent être utilisées pour poser des limites à la durée de vie des particules de MN. Nous avons obtenu des limites compétitives pour des masses comprises entre 3 et 10 GeV et des champs magnétiques supérieurs à 130 nGauss.

Enfin, la partie IV tire les conclusions et donne une perspective des développements futurs potentiels.

# Financial support

This research was sponsored by

- Department of Excellence grant 2018-2022, awarded by the Italian Ministry of Education, University and Research (MIUR)
- Research grant of the Italo-French University, under Bando Vinci 2020
- Cassini Junior grant, awarded by the French Embassy in Italy
- Prize "Giuliano Preparata" 2021 for young researchers, awarded by the Italian Physics Society
- Ministero degli Affari Esteri della Cooperazione Internazionale - Direzione Generale per la Promozione del Sistema Paese Progetto di Grande Rilevanza ZA18GR02
- Scholarship for talented graduate students, National Association of Railway Clubs
- Research grant 'The Anisotropic Dark Universe' No. CSTO161409, funded by Compagnia di Sanpaolo and University of Turin
- Research grant TAsP (Theoretical Astroparticle Physics) funded by Istituto Nazionale di Fisica Nucleare (INFN)
- Research grant 'The Dark Universe: A Synergic Multimessenger Approach' No. 2017X7X85K funded by Italian Ministry of Education, University and Research (MIUR)
- European Research Council (Erc) under the EU Seventh Framework Programme (FP7/2007-2013)/ Erc Starting Grant (agreement n. 278234 — 'NewDark' project)
- Cnrs 80 | Prime grant ('DaMeFer' project)

# Acronyms

| | |
|---|---|
| **2PCF** | Two-point correlation function |
| **AGN** | Active galactic nuclei |
| **APS** | Angular power spectrum |
| **BBN** | Big Bang Nucleosynthesis |
| **BL Lac** | BL Lacertae |
| **CMB** | Cosmic microwave background |
| **DM** | Dark matter |
| **dSph** | Dwarf spheroidal galaxy |
| **EBL** | Extragalactic background light |
| **EM** | Electromagnetic |
| **FSR** | Final-state radiation |
| **FSRQ** | Flat-spectrum radio quasars |
| **GC** | Galactic Centre |
| **GLF** | Gamma-ray luminosity function |
| **GP** | Galactic Plane |
| **HI** | Neutral hydrogen |
| **ICS** | Inverse Compton scattering |
| **IM** | intensity mapping |
| **IR** | Infrared light |
| **ISRF** | Interstellar radiation field |
| **LRG** | Luminous red galaxy |
| **LSP** | Lightest supersymmetric particle |
| **LSS** | Large-scale structure |
| **mAGN** | Misaligned active galactic nuclei |
| **MeerKAT** | MeerKaroo Array Telescope |
| **MSSM** | Minimal Supersymmetric Standard Model |
| **NFW** | Navarro-Frenk-White |
| **Opt** | Optical starlight |
| **PBH** | Primordial black holes |
| **PS** | Power spectrum |
| **PSF** | Point-spread function |
| **Rad** | Radiative decay emission |
| **SED** | Spectral energy distribution |
| **SFG** | Star-forming galaxies |
| **SKA** | Square Kilometre Array |
| **SM** | Standard Model |
| **SNR** | Signal-to-noise ratio |
| **UGRB** | Unresolved gamma-ray background |
| **WIMP** | Weakly Interacting Massive Particle |

# List of Publications

## This work is based on

## Articles submitted for publication

# Chapter 1

# Dark matter in the Universe

One of the most fascinating mysteries of our Universe is the so-called dark matter (DM) problem. The first hint of the existence of DM dates back to 1933, when the Swiss astronomer Fritz Zwicky was studying the content of matter in the Coma Cluster. According to his calculation, the luminous matter was two orders of magnitude less abundant than what was required for the dynamical stability of the system. His results attracted some attention for decades and were finally put under a new light when Professor Vera Rubin observed a similar discrepancy in the rotation curves of spiral galaxies. Analogous conclusions have been reached through the observation of the X-ray emission produced by the intracluster gas as well as via the gravitational lensing observations. Cosmological probes, such as the cosmic microwave background and the large-scale structure, lead to the same finding. An attractive solution is that DM is a new form of matter, possibly made up of new particles or other hypothetical astrophysical candidates, for example primordial black holes. In this chapter we summarise the best-known evidences of the existence of DM in the Universe (Sections 1.1–1.4) and we introduce the cosmological model adopted throughout this dissertation (Section 1.5). The thermodynamics of the Universe is discussed in Section 1.6. Finally, Section 1.7 presents examples of DM candidates and their main properties.

## 1.1 Rotation curves: evidence for dark matter in spiral galaxies

The American astronomer Edwin Powell Hubble observed unusual nebulae in the sky at the Mount Wilson Observatory, California. In 1924, he concluded that these astrophysical objects were far too distant to be part of our Milky Way: these nebulae were neighbouring galaxies. An important part of the legacy that Hubble left is a classification of different galaxies, among which he described the spiral galaxies in his book *The Realm of Nebulae* (1936). These objects are characterised by a central bulge, rich in stars, and a flat, rotating disk consisting of stars, gas and dust. The name "spiral galaxies" refers to the presence of bright spiral arms, characterised by young stars and star-forming regions. A common feature in many spiral galaxies is the presence of a bar. This is the case of the Milky Way and of the Andromeda Galaxy, which was one of the first galaxies identified by Hubble. In the 1970s Vera Rubin conducted pioneering work on the distribution of mass contained in several spiral galaxies. She made use of the fact that stars and gas in the disk rotate in approximately circular orbits and their dynamical stability is a consequence of the equilibrium between the gravitational potential and the rotational kinetic energy. Let us consider a galaxy with a spherical mass distribution: the mass enclosed within the radius $r$ can be written as $M(r)$. The rotation curve describes the circular velocity as a function of the distance from the galactic centre and therefore, it represents a useful tool to characterise the kinematics of the galaxy. If we consider a star with mass $m$ and rotational velocity $v_{\rm rot}(r)$ at distance $r$ from the centre of the galaxy, we can derive its rotational curve from the condition of dynamical equilibrium, that is to say by equating centripetal and gravitational forces:

$$m\,\frac{v_{\rm rot}^2(r)}{r} = G\,\frac{m\,M(r)}{r^2}\,. \tag{1.1}$$

Thus, the mass of the galaxies within radius $r$ can be derived from the rotation curve:

$$M(\leq r) = \frac{v_{\rm rot}^2(r)\,r}{G}\,. \tag{1.2}$$

In the outskirts of the galaxies, one expects the circular velocities to decrease with the square root of the distance as $v_{\rm rot} \propto r^{-1/2}$. This behaviour is called Keplerian, since it was first derived by Johannes Kepler to explain the planetary motions in the Solar System. For instance, these laws explain why Pluto and Neptune move slower around the Sun as compared to the Earth. The rotation curve of the galaxy can be obtained through optical measurements of emission lines, mostly originating from molecular hydrogen. However, the disk extends much further than ionised gas: the outer regions contain clouds of neutral hydrogen (HI). The spin-flip transition of HI atoms results in the emission of a characteristic line with wavelength equals to 21cm, which can be mapped using radio telescopes. If the gas is moving along the line of sight, the wavelength of this radiation is modified by the Doppler effect. By measuring this Doppler shift, we can infer the gas rotation curve and map the galaxies far beyond their optical limit. The surprising result observed by Vera Rubin [7] and confirmed in later studies [8] is that the rotational velocity remains constant even at a large distance from the galactic centre. An example is illustrated in Fig. 1.1. The green part highlights the data obtained from optical spectroscopy, while the blue region refers to the radio observations of the 21cm line emitted by neutral hydrogen atoms. The luminous disk has a Keplerian behaviour (denoted as "disk"), tapering off at large distance from the galactic centre, while the gas content (denoted as "gas") is not enough to account for the measured rotation curve. By looking at Eq. (1.2), we can remark that a flat rotation curve implies $M(r) \propto r$ : the total mass grows linearly with the distance from the galactic centre. This result can be formulated in terms of the mass-to-light ratio $M/L$ with respect to the solar one, $M_\odot/L_\odot$. It is found that $M/L$ is in the range (1–10) $M_\odot/L_\odot$ in the visible regions of the spiral galaxies, while it grows up to (10–20) $M_\odot/L_\odot$ in the outskirts. These observations can be explained by assuming that the spiral galaxies are enclosed in a DM halo, and the baryonic constituents (gas distribution, bulge, disk) represent a minor component of the total matter content. If we assume the presence of such a halo (dash-dotted line in the figure), the theoretical modelling (solid line) becomes consistent with the observations (data points).

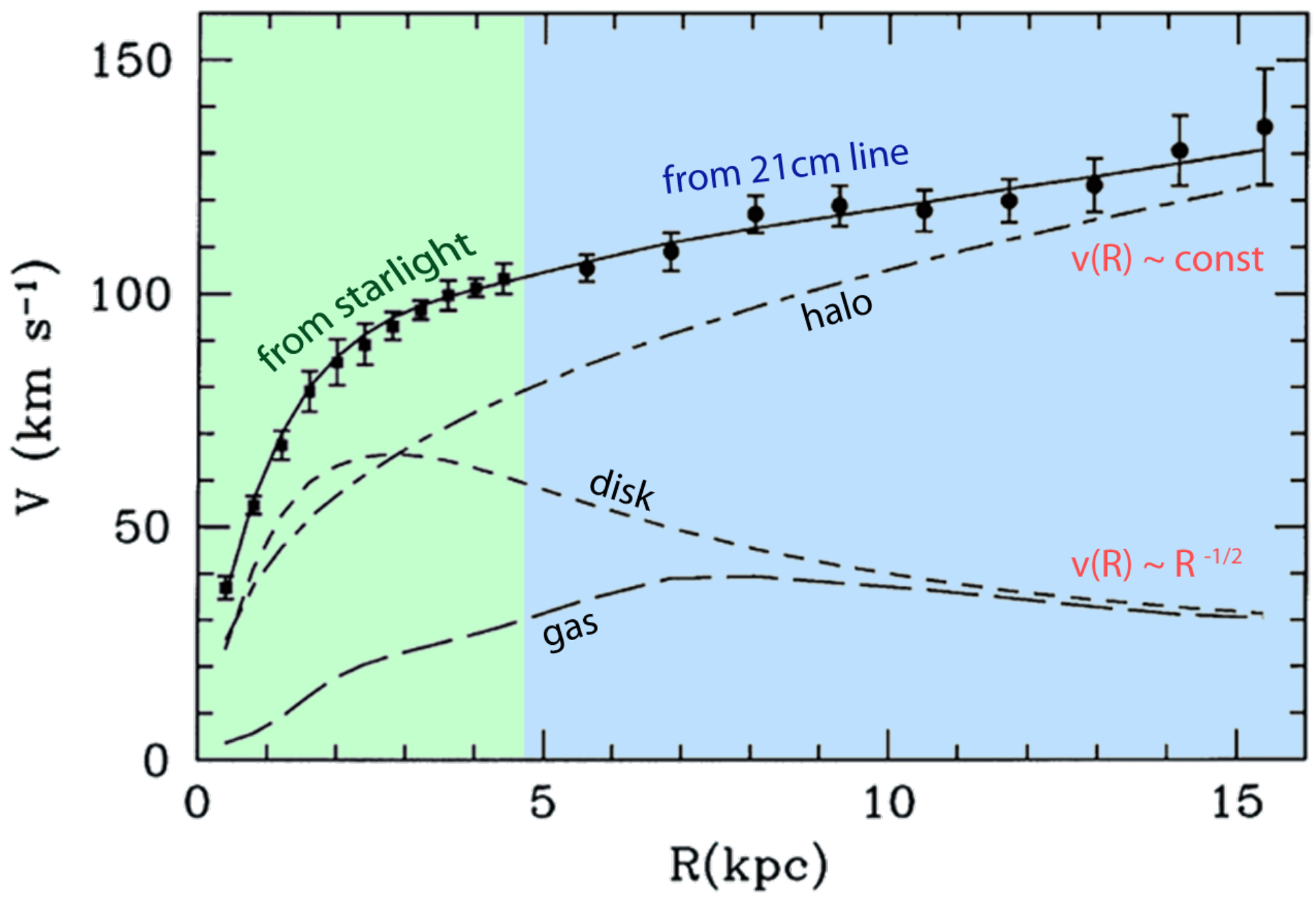

**Fig. 1.1.** Rotation curve of the galaxy M33. The dashed lines refer to baryonic matter (disk, gas), while the dash-dotted line represents the contribution from a DM halo. The green area highlights the data obtained from optical spectroscopy, while the blue region refers to the radio observations of the 21cm line emitted by neutral hydrogen atoms. Credit: adapted from [6].

## 1.2 Virial theorem: how to estimate the mass of a cluster

Let us consider a self-gravitating system of particles, which interact through the gravitational force. The $i$-th particle has mass $m_i$ and due to the interaction with the other particles, it undergoes an acceleration $\ddot{\boldsymbol{r}}_i$ given by

$$\ddot{\boldsymbol{r}}_i = \sum_{j \neq i} \frac{G\, m_j\, (\boldsymbol{r}_j - \boldsymbol{r}_i)}{|\,\boldsymbol{r}_i - \boldsymbol{r}_j\,|^3}\,. \tag{1.3}$$

Taking the scalar product with $m_i \boldsymbol{r}_i$, we get

$$m_i\,(\boldsymbol{r}_i \cdot \ddot{\boldsymbol{r}}_i) = \sum_{j \neq i} G\, m_i\, m_j\, \frac{\boldsymbol{r}_i \cdot (\boldsymbol{r}_j - \boldsymbol{r}_i)}{|\,\boldsymbol{r}_j - \boldsymbol{r}_i\,|^3}\,. \tag{1.4}$$

The time derivative of $\boldsymbol{r}_i \cdot \boldsymbol{r}_i$ is

$$\frac{\mathrm{d}}{\mathrm{d}t}(\boldsymbol{r}_i \cdot \boldsymbol{r}_i) = 2\,\dot{\boldsymbol{r}}_i \cdot \boldsymbol{r}_i\,. \tag{1.5}$$

By differentiating again with respect to the time coordinate, we get

$$\frac{1}{2}\frac{\mathrm{d}^2}{\mathrm{d}t^2}\boldsymbol{r}_i^2 = \frac{\mathrm{d}}{\mathrm{d}t}(\dot{\boldsymbol{r}}_i \cdot \boldsymbol{r}_i) = \ddot{\boldsymbol{r}}_i \cdot \boldsymbol{r}_i + \dot{\boldsymbol{r}}_i \cdot \dot{\boldsymbol{r}}_i = \ddot{\boldsymbol{r}}_i \cdot \boldsymbol{r}_i + \dot{\boldsymbol{r}}_i^2\,, \tag{1.6}$$

therefore

$$\frac{1}{2}\frac{\mathrm{d}^2}{\mathrm{d}t^2}\left(\boldsymbol{r}_i^2\right) - \dot{\boldsymbol{r}}_i^2 = \ddot{\boldsymbol{r}}_i \cdot \boldsymbol{r}_i\,. \tag{1.7}$$

By multiplying Eq. (1.7) by $m_i$ and summing over the particles we get

$$\frac{1}{2}\frac{\mathrm{d}^2}{\mathrm{d}t^2}\sum_i m_i \boldsymbol{r}_i^2 - \sum_i m_i \dot{\boldsymbol{r}}_i^2 = \sum_i m_i \ddot{\boldsymbol{r}}_i \cdot \boldsymbol{r}_i\,. \tag{1.8}$$

Then, using Eq. (1.4)

$$\frac{1}{2}\frac{\mathrm{d}^2}{\mathrm{d}t^2}\sum_i m_i \boldsymbol{r}_i^2 - \sum_i m_i\, \dot{\boldsymbol{r}}_i^2 = \sum_i \sum_{j\neq i} G\, m_i\, m_j \frac{\boldsymbol{r}_i \cdot (\boldsymbol{r}_j - \boldsymbol{r}_i)}{|\,\boldsymbol{r}_j - \boldsymbol{r}_i\,|^3}\,. \tag{1.9}$$

The double sum on the right side corresponds to summing over all the elements $ij$ and $ji$ of a matrix with a vanishing diagonal term. By summing over $ij$ and $ji$, we obtain

$$G\, m_i\, m_j \left[\frac{\boldsymbol{r}_i \cdot (\boldsymbol{r}_j - \boldsymbol{r}_i)}{|\,\boldsymbol{r}_i - \boldsymbol{r}_j\,|^3} + \frac{\boldsymbol{r}_j \cdot (\boldsymbol{r}_i - \boldsymbol{r}_j)}{|\,\boldsymbol{r}_j - \boldsymbol{r}_i\,|^3}\right] = -\frac{G\, m_i\, m_j}{|\,\boldsymbol{r}_i - \boldsymbol{r}_j\,|}\,. \tag{1.10}$$

Thus, Eq. (1.9) becomes

$$\frac{1}{2}\frac{\mathrm{d}^2}{\mathrm{d}t^2}\sum_i m_i \boldsymbol{r}_i^2 - \sum_i m_i \dot{\boldsymbol{r}}_i^2 = -\frac{1}{2}\sum_{j\neq i}\frac{G\, m_i\, m_j}{|\,\boldsymbol{r}_i - \boldsymbol{r}_j\,|}\,, \tag{1.11}$$

where the factor 1/2 is necessary to avoid double counting. Now, we recall that the kinetic energy of the system is

$$T = \frac{1}{2}\sum_i m_i\, \dot{\boldsymbol{r}}_i^2 \tag{1.12}$$

and the gravitational potential is given by

$$U = -\frac{1}{2}\sum_{j\neq i}\frac{G\, m_i\, m_j}{|\,\boldsymbol{r}_i - \boldsymbol{r}_j\,|}\,. \tag{1.13}$$

Therefore, Eq. (1.11) can be written as

$$\frac{1}{2}\frac{\mathrm{d}^2}{\mathrm{d}t^2}\sum_i m_i\, \boldsymbol{r}_i^2 = 2T - |\,U\,|\,. \tag{1.14}$$

This result is called the virial theorem and for stationary systems it holds

$$\frac{1}{2}\frac{\mathrm{d}^2}{\mathrm{d}t^2}\sum_i m_i\, \boldsymbol{r}_i^2 = 0\,, \tag{1.15}$$

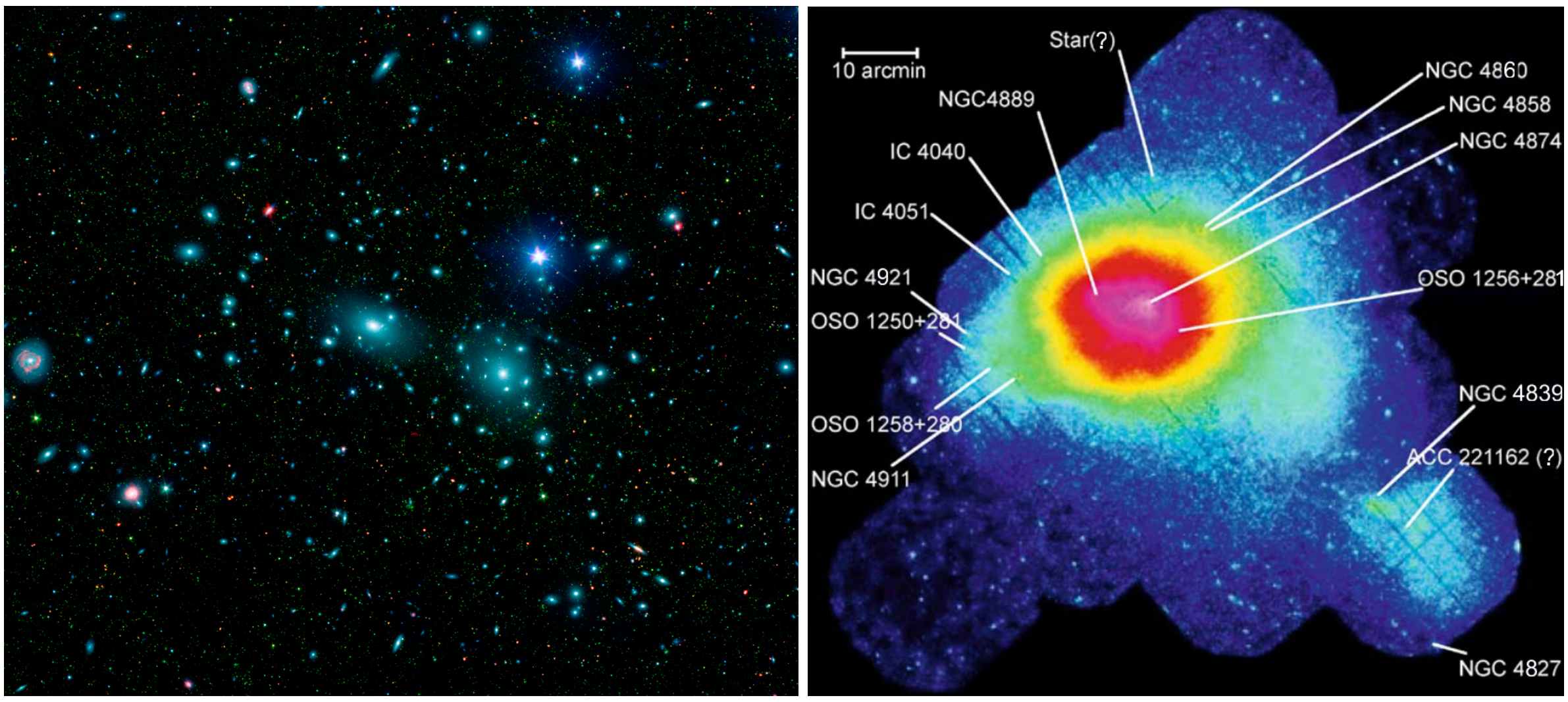

**Fig. 1.2.** *Left:* Central region of the Coma Cluster. Credit: Sloan Digital Sky Survey. *Right:* X-ray image of the Coma Cluster. Credit: XMM-Netwon Observatory.

which implies

$$T = \frac{1}{2\,|\,U\,|}\,. \tag{1.16}$$

The virial theorem was first derived by the German physicist Rudolf Clausius in 1870. It is noteworthy that this theorem holds independently of any assumption on the velocity distribution of the particles in the system, meaning that it is valid both in a cluster with random velocities or in the disk of spiral galaxies, characterised by rotational velocity about the centre. However, since the Doppler shifts of a spectral line can be used to measure only the radial component of the velocity, it is necessary to make some assumptions on the spatial and speed distributions of galaxies in a cluster, or likewise stars in a galaxy. On one hand, by assuming that the system is spherically symmetric, we can express the potential in terms of a characteristic separation $R_{\rm cl}$:

$$|\,U\,| = \frac{G M^2}{R_{\rm cl}}\,, \tag{1.17}$$

where $R_{\rm cl}$ can be derived from a measurement of the surface distribution of galaxies or stars. On the other hand, under the assumption that the velocity distribution is isotropic, we have that the velocity dispersion along the line of sight and in the two orthogonal directions are equal: $v_r^2 = v_\theta^2 = v_\phi^2$. Therefore, by denoting the radial velocity as $v_r^2$, the total velocity dispersion is $\langle v^2 \rangle = 3\,\langle v_r^2 \rangle$ and the kinetic energy can be written as $T = 3/2\,M\,\langle v_r^2 \rangle$. Therefore, from Eq. (1.16) we obtain a relation between the total mass of the system and its velocity dispersion:

$$M = \frac{3\,\langle v_r^2 \rangle\, R_{\rm cl}}{G}\,. \tag{1.18}$$

In 1933, Fritz Zwicky was the first to apply the virial theorem to estimate the mass of a cluster of galaxies called the Coma Cluster, also known as Abell 1656 (see Fig. 1.2). His physical intuition was that the stability of a cluster requires an equilibrium between the gravitational potential (which would induce the collapse of the system towards its centre) and the relative motions of the galaxies inside the cluster (which would induce the evaporation of the system). Zwicky measured the velocity dispersion $\langle v_r \rangle \sim 10^8$ cm s$^{-1}$ and obtained $R_{\rm Coma} \simeq 10^{24}$ cm from the observed angular diameter of the cluster. Regarding the theoretical predictions, he assumed that the Coma Cluster enclosed 800 galaxies with a mass about $10^9 M_\odot$, which implies $M_{\rm Coma} \simeq 8 \cdot 10^{11} M_\odot$, and by applying the virial theorem, he expected a velocity dispersion around $10^6$ cm s$^{-1}$. In order to have a good agreement between measurements and predictions, the Coma Cluster should have a mass about two orders of magnitude larger with respect to the typical luminous mass in a cluster expected at that time. In other words, Zwicky argued that the Coma Cluster should be dominated by a form of invisible matter, which he named *dunkle materie*, the German expression for dark matter. Nevertheless, the Coma Cluster never ceases to surprise. In 1996 Colless and Dunn [9] found a sub-cluster in the outskirt of Abel 1656, with NGC 4839 being the brightest member. The main cluster and the sub-cluster are clearly discernible in Fig. 1.2 (right). They also identified a subclustering in the central region of the main cluster, around NGC4889 and NGC 4884, which are on the verge of coalescing.

## 1.3 X-ray observations from hot gas in clusters of galaxies

The XMM-Newton observation of Abel 1656 in Fig. 1.2 (right) is a good example of how useful X-ray astronomy can be in the investigation of clusters. In the 1970s interesting measurements were made by the UHURU satellite, which provided evidence that clusters are indeed bright X-ray sources. The origin of this intense X-ray emission has been identified as thermal bremsstrahlung radiation emitted by intracluster hot gas. This intracluster medium is expected to be fully ionised, since typical temperatures are about $10^7 - 10^8$ K, corresponding to an energy of 1−10 keV for the emitted radiation. Let us assume that a cluster is a spherically symmetric system and the total mass $M(\leq r)$ within radius $r$ is the sum of luminous, gaseous and dark matter. The gravitational potential $U(r)$ and the density distribution $\rho_{\rm tot}(r)$ are related through the Poisson equation:

$$\nabla U(r) = 4\pi G \rho_{\rm tot}(r)\ . \tag{1.19}$$

Thus, considering the spherical symmetry,

$$\frac{\mathrm{d}\phi}{\mathrm{d}r} = \frac{4\pi G}{r^2}\int_0^r dx\, x^2 \rho_{\rm tot}(x) = \frac{G M(r)}{r^2}\ . \tag{1.20}$$

Under the assumption of gas in hydrostatic equilibrium, it holds

$$\frac{\mathrm{d}p}{\mathrm{d}r} = -\rho\frac{\mathrm{d}U}{\mathrm{d}r} = -\frac{GM(\leq r)\rho}{r^2}\ , \tag{1.21}$$

where the pressure $p$ and the density $\rho$ are position-dependent and they are related to the temperature $T$ through the ideal gas law

$$p = \frac{\rho k T}{\mu m_H}\ , \tag{1.22}$$

where $k$ is the Boltzmann constant and $m_H$ is the mass of an hydrogen atom. The mean molecular weight $\mu$ is about 0.6 for a fully ionised plasma [10]. From Eq. (1.21), we get

$$M = -\frac{r^2}{G\rho}\frac{\mathrm{d}p}{\mathrm{d}r}\,. \tag{1.23}$$

Then, inserting Eq. (1.22),

$$\begin{aligned} M &= -\frac{r^2}{G\rho}\frac{\mathrm{d}}{\mathrm{d}r}\left(\rho\frac{kT}{\mu}\right) \\ &= -\frac{k\,r}{\mu G}\frac{r}{\rho}\frac{\mathrm{d}}{\mathrm{d}r}(\rho T) \\ &= -\frac{k\,r}{\mu G}\frac{1}{\rho}\frac{\mathrm{d}}{\mathrm{d}\ln r}(\rho T) \\ &= -\frac{k\,r}{\mu G}\left[\frac{T}{\rho}\frac{\mathrm{d}\rho}{\mathrm{d}\ln r}+\frac{\mathrm{d}T}{\mathrm{d}\ln r}\right] \\ &= -\frac{k\,T}{\mu G}\,r\left[\frac{\mathrm{d}\ln\rho}{\mathrm{d}\ln r}+\frac{\mathrm{d}\ln T}{\mathrm{d}\ln r}\right]. \end{aligned} \tag{1.24}$$

Therefore, the total mass of a galaxy cluster can be inferred by measuring the distribution of the gas density and temperature. The physical quantity that we measure with our telescopes is the X-ray surface brightness, which is defined as the integral of the emissivity over the line of sight

$$S_\nu(a) = \frac{1}{2\pi}\int_a^\infty \mathrm{d}r\, k_\nu(r)\,\frac{r}{\sqrt{r^2-a^2}}\,, \tag{1.25}$$

where $a$ is the projected radius from the centre of the cluster. The thermal bremsstrahlung emissivity is

$$k_\nu = \left(\frac{\pi}{6}\right)^2 \frac{Z^2 e^6}{3\pi^2 m_e^2 c^3 \epsilon_0^2}\left(\frac{m_e}{k\,T}\right)^2 n_e\, n_i\, \exp\left(-\frac{m_e v^2}{k\,T}\right)\bar{g}_{\mathrm{ff}}\,, \tag{1.26}$$

where $-e$ and $+Ze$ represent the electric charge of the electrons and ions in the hot plasma, $m_e$ is the electron mass, $k$ stands for the Boltzmann constant, $\epsilon_0$ is the vacuum permittivity and $v$ represents the velocity of the particles. The velocity-averaged Gaunt factor $\bar{g}_{\mathrm{ff}}$ takes into account the quantum corrections and it is of order unity in the regime $h\nu \ll kT$. The derivation of Eq. (1.26) is included in Appendix A. Note that $k_\nu$ is roughly constant with $h\nu$ at low frequencies (flat spectrum), while it drops exponentially when $h\nu \sim k\,T$. The cut-off frequency is fixed to the value corresponding to the exponential term equal to 1/e and it depends only on the temperature. Therefore, by measuring the cut-off frequency, we can estimate the temperature of the plasma

$$\frac{h\nu}{k\,T} = 1 \quad\Longrightarrow\quad T = \frac{h\nu}{k}\,. \tag{1.27}$$

Once the temperature of the plasma is determined, it is possible to estimate the column density of the plasma, by measuring $S_\nu$. In this regard, it is noteworthy that the integral in Eq. (1.25) is an Abel integral [11], so it can be inverted to determine the total emissivity of the plasma as a function of radius

$$k_\nu(r) = \frac{4}{r}\frac{\mathrm{d}}{\mathrm{d}r}\int_r^\infty \mathrm{d}a\, S_\nu(a)\,\frac{a}{\sqrt{a^2 - r^2}}\,. \tag{1.28}$$

Thus, measuring the X-ray surface brightness is a powerful way to estimate the total mass of a galaxy cluster.

## 1.4 Gravitational lensing

The so-called gravitational lensing is a recent impressive method to derive the mass of an astrophysical object. It describes the deflection of light caused by the gravitational bending of space-time in the presence of massive objects (the lens). Depending on the distortion effects produced on the background images, gravitational lensing phenomena are commonly separated into two categories: strong and weak. We briefly introduce the former, then we discuss the implications of the latter. The formalism of strong lensing is quite informative to understand the effect of the gravitational potential on the path of a light ray and its implications in observational cosmology. The first attempt to derive the deflection of light dates back to 1804 when the German astronomer Johann Soldner was conducting his research at the Berlin Observatory. By applying Newtonian physics, he estimated a deflection angle $\Delta\phi$ = 0.84 arcsec [12]. More than one century later, Einstein calculated the deflection angle by applying general relativity, which takes into consideration the effect of a curved space-time on the photons' path. He found $\Delta\phi = 4GM_\odot/2R_\odot$ = 1.74 arcsec, a factor two larger with respect to the estimate of Soldner [13]. In 1919 Eddington measured the deflection of light during the solar eclipse: the observed value was in agreement with Einstein's prediction. After the correct explanation of the perihelion precession of Mercury, the deflection of light by the Sun is considered the second observational test of general relativity, which finally convinced the scientific community to accept general relativity as the new standard theory of gravitation. Five years later, in 1924, the Russian physicist Orest Danilovich Chwolson argued that if a star is located behind a massive astrophysical object, this would produce a circular image. The same effect was studied in 1916 by Einstein, who also claimed that if a light source, a gravitational lens and an observer are perfectly aligned, the background source will appear as a ring. Nowadays this effect has been observed numerous times and it is known as "Einstein ring" (see Fig. 1.3, right). On the other hand, if the two astrophysical objects are not aligned, the observer will see two magnified images, as shown in Fig. 1.4. S1 and S2 represent the two images of the background source and they are observed along the tangents to the real light path. It is worth noting that the typical size of a galaxy is about 50 kpc, while the size of a cluster is around 1 Mpc: both these values are much smaller than the lens-observer and lens-source distances, which are typically of order Gpc. Therefore, to a first approximation we can consider galaxies and clusters of galaxies as point-like lenses. This assumption is generally known as the

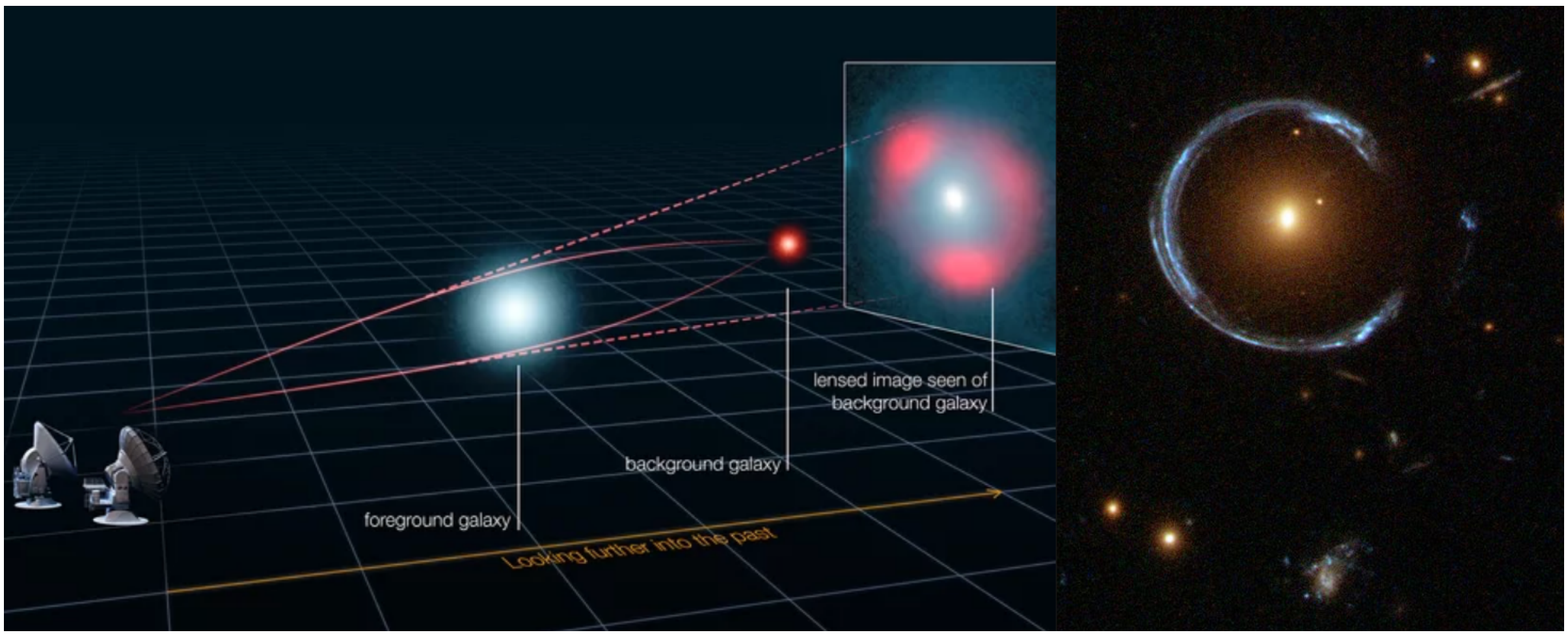

**Fig. 1.3.** *Left:* This image represents schematically how the gravitational lensing effect works: if there is a massive astrophysical object between the Earth and a background galaxy, we observe a distorted image with our telescopes. Credit: ALMA (ESO/NRAO/NAOJ), L. Calçada (ESO), Y. Hezaveh et al. *Right:* If the foreground and background galaxies are perfectly aligned, the lensed image has the shape of a circle, called Einstein ring. Credit: Hubble Space Telescope.

"thin lens approximation". The deflection angle due to a lens of mass $M$ is given by

$$\tilde{\alpha}(\xi) = \frac{G\,M(\xi)}{c^2\,\xi}\,, \tag{1.29}$$

where $\xi$ is called collision parameter and it almost coincides with the distance of closest approach between the lens and the light ray, since the angles involved in the gravitational lensing effect are small. The right panel of Fig. 1.4 illustrates the geometry of the problem. The distance between the observer and the lens is denoted by $D_{\rm L}$, while $D_{\rm LS}$ indicates the distance between the lens and the direction connecting the source to the two images. Then, $D_{\rm S} = D_{\rm L} + D_{\rm LS}$. In the case of cosmological distances, $D_{\rm S}$ (and thus also $D_{\rm L}$ and $D_{\rm LS}$) represents an angular diameter distance, defined as the ratio between the physical length of an object and its angular size (i.e. the subtended angle) as seen from the Earth. From Fig. 1.4 we can notice that

$$\theta D_S = \beta D_S + \tilde{\alpha} D_{LS}\,. \tag{1.30}$$

The reduced deflection angle is defined as

$$\alpha(\theta) = \frac{D_{LS}}{D_S}\,\tilde{\alpha}\,. \tag{1.31}$$

Inserting Eq. (1.31) in Eq. (1.4), we obtain the lens equation

$$\beta = \theta - \alpha(\theta)\,. \tag{1.32}$$

By inserting Eq. (1.29) into Eq. (1.4) and using the relation $\xi = D_L\,\theta$, we get

$$\beta = \theta - \frac{D_{LS}}{D_L\,D_S}\,\frac{4\pi G\,M}{c^2}\,\frac{1}{\theta}\,. \tag{1.33}$$

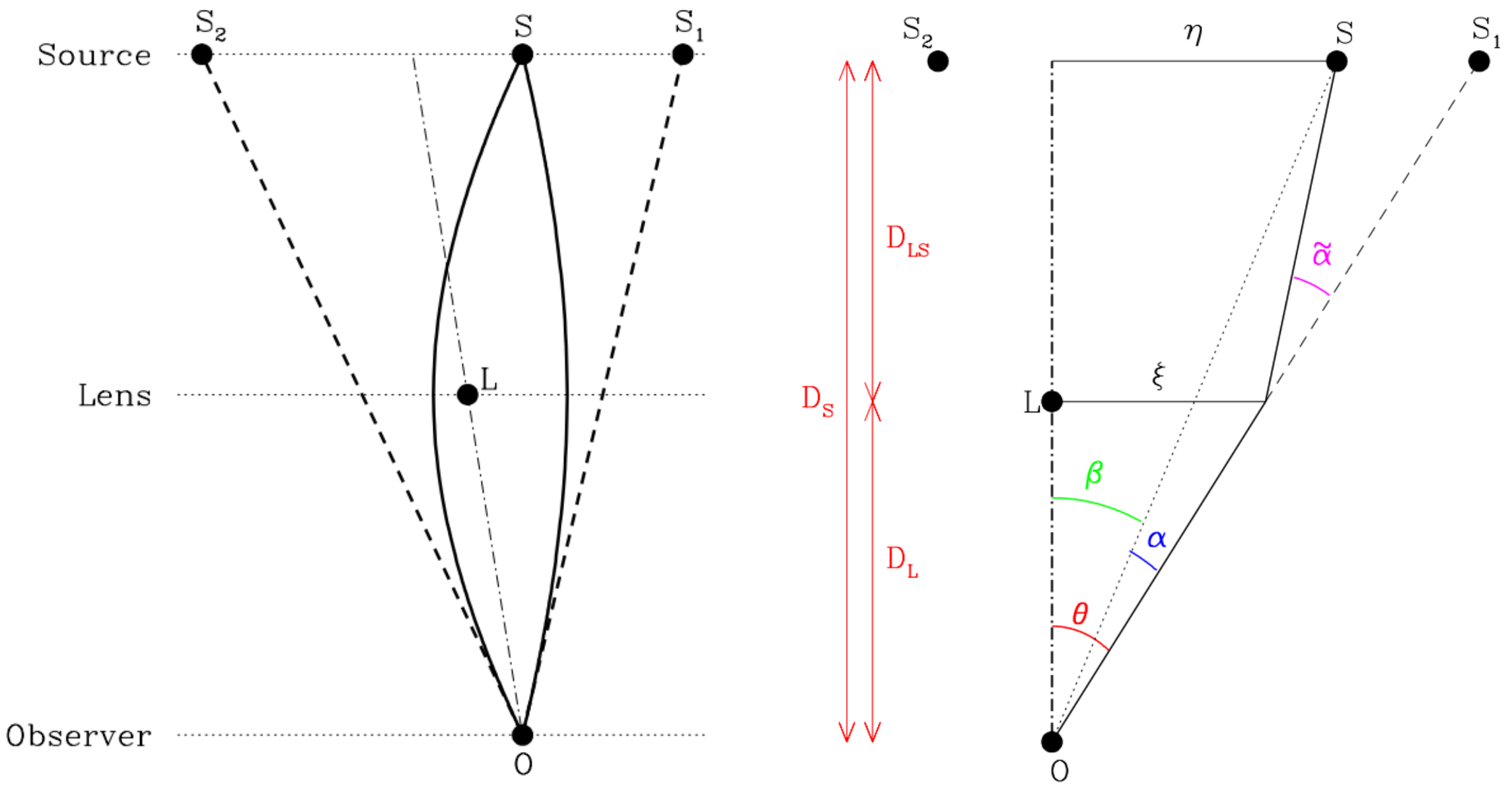

**Fig. 1.4.** *Left:* Schematic image of the gravitational lensing in the event that the observer O, the lens L and the background source S are not aligned. This would produce multiple images. *Right:* This figure includes the relevant angles and distances necessary to derive the lens equation and the Einstein radius. Credit: [14] .

If source and lens are perfectly aligned, then $\beta = 0$ and the observer will see the image of a ring with an angular radius

$$\theta_E = \sqrt{\frac{D_{LS}}{D_L D_S} \frac{4\pi G M}{c^2}} \,, \tag{1.34}$$

known as Einstein radius. By expressing the distance in Gpc and the mass in terms of the solar mass, one can write

$$\theta_E = \frac{3 \cdot 10^{-6}}{\sqrt{D_{\rm Gpc}}} \sqrt{\frac{M}{M_\odot}} \,. \tag{1.35}$$

So, a lens with a mass $M \sim 10^{15} M_\odot$ at cosmological distances produces an Einstein ring with angular radius of the order of tens of arcseconds. Eq. (1.33) can be expressed in terms of $\theta_E$

$$\beta = \theta - \frac{\theta_E}{\theta} \tag{1.36}$$

with two solutions

$$\theta = \pm \frac{\beta \pm \sqrt{\beta^2 + \theta_E^2}}{2} \tag{1.37}$$

corresponding to the two magnified images. One of the solutions is positive, while the other has a negative sign. The physical meaning of the negative sign is that the corresponding image is mirror-inverted with respect to the other image, if the object is extended. Also, the two images are located on opposite sides with respect to the position of the lens. Fig. 1.5 shows how a background source behind a lens can appear. From right to left: if the two astrophysical objects are aligned, the background source appears as an Einstein ring; if they are not perfectly aligned, an almost complete ring can appear, but also arcs, arclets and distinct magnified images, depending on the entity of the misalignment. The more they are misaligned, the more the resulting image will differ from a complete ring. Then, the mass of a cluster can be determined from the location of the arc. Provided the conservation of the total surface brightness, the entity of the magnification is determined as the ratio between the solid angle subtended by the image and the source. If we consider an infinitesimal angular portion of an arc of azimuthal angle $d\varphi$, the solid angle of the source from the observer's point of view is $d\varphi\,\beta\,d\beta$. By replacing $\beta \to \theta$, we obtain the corresponding solid angle for the lens. Thus, the magnification $\mu$ can be written as

$$\mu = \frac{\theta}{\beta}\frac{d\theta}{d\beta}\,. \tag{1.38}$$

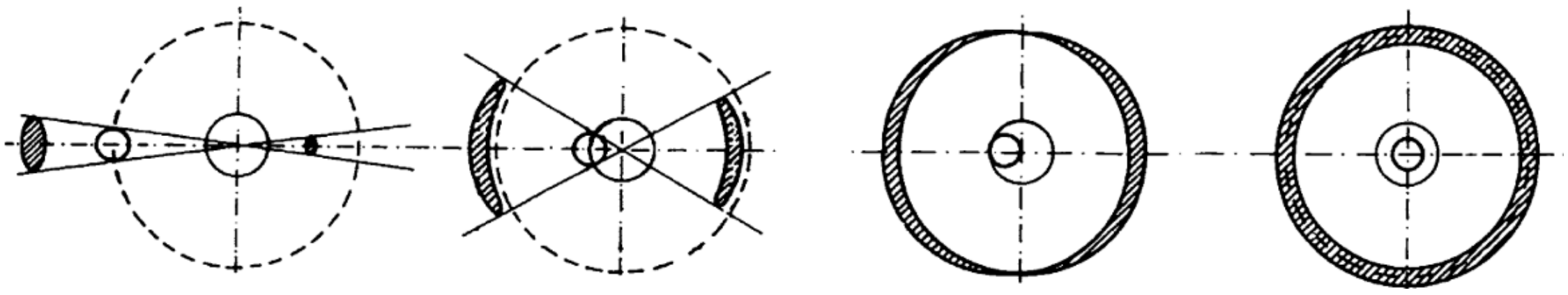

**Fig. 1.5.** Schematic figure with different possible lensed images, depending on the relative position between the lens and the background source. From right to left: if the two astrophysical objects are aligned, the background source appears as an Einstein ring; if they are almost perfectly aligned, the shape of a ring can still be identified; more prominent misalignment can result in arcs, arclets and distinct magnified images, depending on the entity of the misalignment. The more the foreground and background galaxies are misaligned, the more the resulting image will differ from a complete ring. Credit: [10] .

By using the lens equation, we have

$$\frac{1}{\mu} = \frac{\beta}{\theta}\frac{d\beta}{d\theta} = \left(1 - \frac{\theta_E^2}{\theta^2}\right)\left(1 + \frac{\theta_E^2}{\theta^2}\right) = 1 - \frac{\theta_E^4}{\theta^4}\,. \tag{1.39}$$

Thus,

$$\mu_{1,2} = \left(1 - \frac{\theta_E^4}{\theta_{1,2}^4}\right)^{-1} = \frac{1}{2} \pm \frac{u^2+2}{2u\sqrt{u^2+4}}\,, \tag{1.40}$$

where in the last equality we used the so-called angular impact parameter $u = \beta/\theta_E$, which represents the angular distance between the background source and the lens in units of the Einstein radius. As expected, the magnification of the image inside the Einstein radius is negative, meaning that it is mirror-inverted. Also, $\mu$ is divergent for $\beta \to 0$. Physically speaking, this means that the magnification of the Einstein ring for a point source is infinite. However, this is not a problem since astrophysical objects in the real world have a finite dimension, resulting in a finite magnification. The total magnification is

$$\mu = |\mu_1| + |\mu_2| = \frac{u^2+2}{u\sqrt{u^2+4}} \,. \tag{1.41}$$

It is worth mentioning that $\mu > 1$, while $\mu_1 + \mu_2 = 1$. Note that these formulas have been derived under the assumptions of spherical symmetry and thin lens. Reality is usually much more complicated, but this toy-model scenario is useful to have an intuitive understanding of the problem.

Gravitational lensing became a major tool in cosmology in 1979, with the first observation of two images associated with the so-called Twin Quasars, also known as QSO 0975+561 [15]. It is observed as two separate images because of the gravitational effect due to the presence of a galaxy between the Earth and the quasar. At present time, gravitational lensing is considered one of the most valuable tools for DM searches, since it directly traces the total matter distribution in the Universe. The values for the mass-to-luminosity ratio derived by looking at rings around galaxies and arcs inside clusters are in good agreement with the dynamical estimate obtained from X-ray emission produced by hot gas and from the rotational curves. Typical values for $M/L$ are about 10−20 for galaxies and 100−300 for clusters, confirming the presence of a dominant component of DM. However, the above-mentioned phenomena (Einstein rings, arcs and multiple images) are quite rare events and they represent examples of the so-called strong gravitational lensing. As you move away from the lens, the effect becomes too weak to produce multiple images or arcs. These outer regions can still be probed by looking at another phenomenon of gravitational origin, called weak lensing. It consists in looking for slight shape distortions in large samples of galaxies. This way, even if the effect on the individual galaxy is undetectable with our current instruments, we are still able to measure a cumulative lensing signal through a statistical analysis. To link the potential of the lens to the observables, we first recall that the deflection angle is the gradient of an effective potential $\psi$

$$\vec{\alpha} = \vec{\nabla}_\theta \, \psi \,. \tag{1.42}$$

This potential can be expressed in terms of the Newtonian potential associated to the lens through

$$\psi(\vec{\theta}) = \frac{D_{LS}}{D_L D_S} \frac{2}{c^2} \int \mathrm{d}z \, \Phi(\vec{r}) \,. \tag{1.43}$$

We can define the Jacobian matrix $A$

$$A = \frac{\partial \vec{\beta}}{\partial \vec{\theta}} = \delta_{ij} - \frac{\partial \vec{\alpha}_i}{\partial \theta_j} = \delta_{ij} - \frac{\partial^2 \psi}{\partial \theta_i \partial \theta_j} \,. \tag{1.44}$$

To ease the notation we can define

$$\psi_{ij} = \frac{\partial^2 \psi}{\partial \theta_i \partial \theta_j} \,. \tag{1.45}$$

Therefore, $A$ can be written as

$$A = \begin{pmatrix} 1-\psi_{11} & -\psi_{12} \\ -\psi_{21} & 1-\psi_{22} \end{pmatrix} \,. \tag{1.46}$$

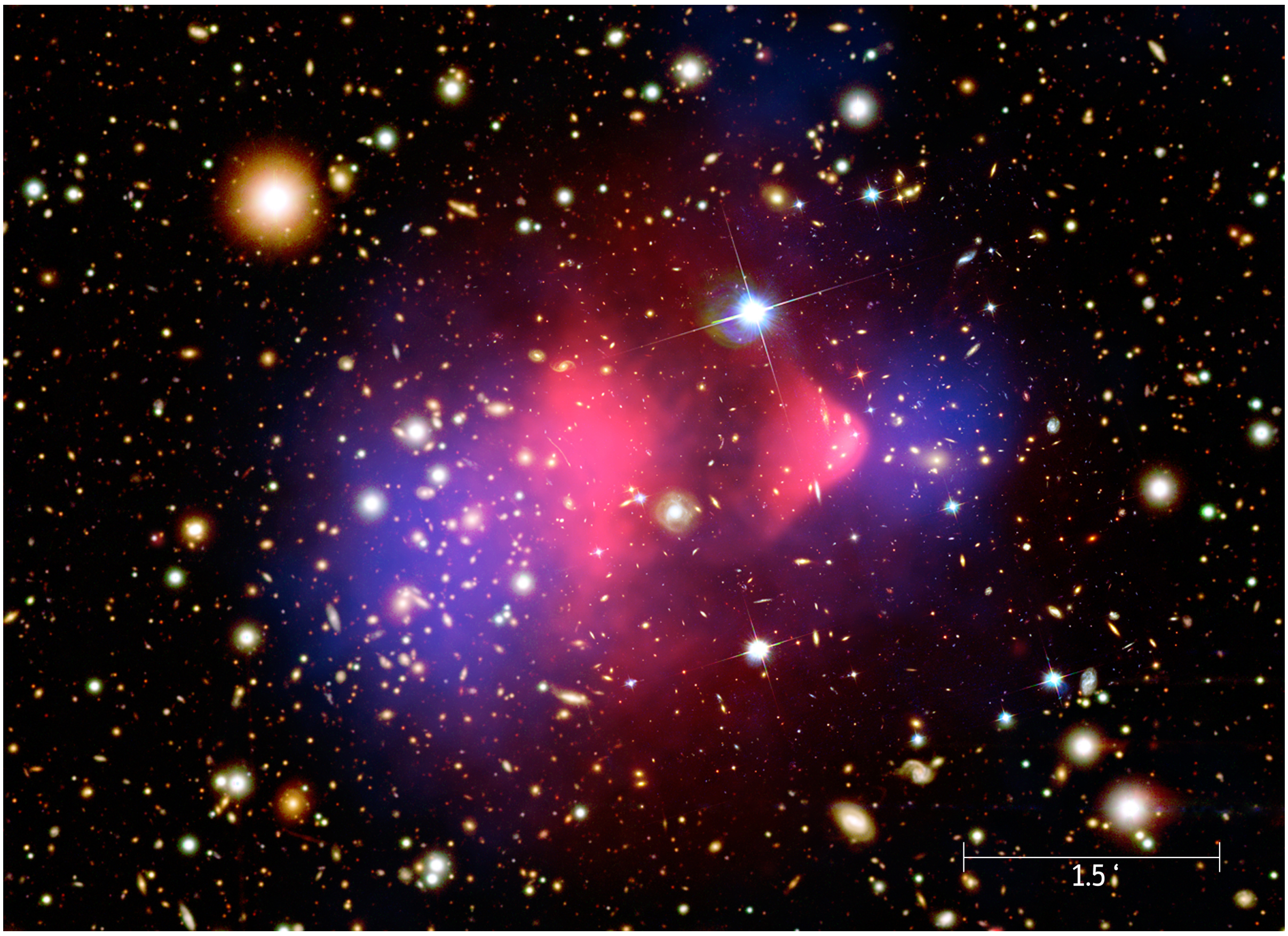

**Fig. 1.6.** Bullet Cluster as seen with different techniques: optical light identifying galaxies (orange, white and yellow spots), X rays produced by hot plasma (pink), matter distribution inferred with the gravitational lensing technique (blue). Credit: NASA/CXC/M. Weiss - Chandra X-Ray Observatory: 1E 0657-56.

The convergence $k$ is half the Laplacian of the effective potential. Thus,

$$k = \frac{1}{2}\nabla^2\psi = \frac{1}{2}\left(\psi_{11} + \psi_{22}\right). \tag{1.47}$$

Now, let us define the two shear components

$$\begin{cases} \gamma_1(\vec{\theta}) & = \frac{1}{2}\left(\psi_{11} - \psi_{22}\right) \\ \gamma_2(\vec{\theta}) & = \psi_{12} = \psi_{21}. \end{cases} \tag{1.48}$$

Plugging into the Jacobian matrix, we get

$$A = \begin{pmatrix} 1 - k - \gamma_1 & -\gamma_2 \\ -\gamma_2 & 1 - k + \gamma_1 \end{pmatrix}. \tag{1.49}$$

The determinant of $A$ is the inverse of the magnification $\mu$. Thus, we can write

$$\mu = \frac{1}{\det A} = \frac{1}{(1-k) - \gamma_1^2 - \gamma_2^2}. \tag{1.50}$$

The weak lensing effect is typically employed to estimate the mass of galaxy clusters and the DM distribution in the Universe. One of the most beautiful pieces of evidence for the presence of DM in the Universe associated with the application of gravitational lensing is the Bullet Cluster, also known as 1E 0657-56. This is a system of two colliding clusters and an example of how essential it is to investigate the same astrophysical object with different techniques. It has been studied both with optical and X-rays telescopes as well as with the gravitational lensing technique. These components are all shown in Fig. 1.6 with different colours. The yellow, orange and white spots are associated with the visible light emitted by galaxies and they have been observed with the Hubble Space Telescope and Magellan Telescope. Centrally located, there is the emission of X rays measured by the Chandra X-ray Observatory. It traces the presence of the hot gas and provides us with two key pieces of information: firstly, most of the baryonic matter is in form of hot plasma; secondly, the bullet-shaped area is the result of hot gas from the large cluster colliding with the gas in the smaller cluster. Finally, the DM component is determined by using the gravitational lensing of background galaxies. This gravitational effect is stronger in two distinct regions, shown in blue, located near the visible galaxies. The physical interpretation is that the two clusters were embedded in two DM halos and during the collision the DM component does not (or very little) interact with either the gas or the DM itself. Instead, the hot gas is collisional and produces a shock with the shape of a bullet. The observation of the Bullet Cluster is particularly relevant because it cannot be easily interpreted invoking Modified Newtonian dynamics[1]. Therefore, one of the reasons why the Bullet Cluster is a notorious evidence for DM is that it favours a DM interpretation in terms of elementary particles, which interact gravitationally and extremely weakly (if not at all) within themselves and with the baryonic matter.

## 1.5 Cosmology

The cosmological model that we adopt is known as ΛCDM. It is based on two assumptions:

- the equations of general relativity are properly able to describe the space-time in terms of the energy and matter content in the Universe,
- the Universe on the large scale is homogeneous and isotropic, which is the so-called cosmological principle.

According to General Relativity, gravity can be expressed as a distortion of space-time. The metric tensor $g_{\mu\nu}$ is the quantity that describes the geometry and it is useful to define the infinitesimal interval $\mathrm{d}\tau$ between two events in the space-time

$$\mathrm{d}\tau^2 = g_{\mu\nu}\,\mathrm{d}x^\mu\,\mathrm{d}x^\nu\,, \tag{1.51}$$

[1]Modified Newtonian dynamics (MOND) [16] invokes a modification of Newton's law to explain the dynamics of galaxies, as an alternative to assuming the existence of exotic physics (such as particle dark matter, primordial black holes, etc.).

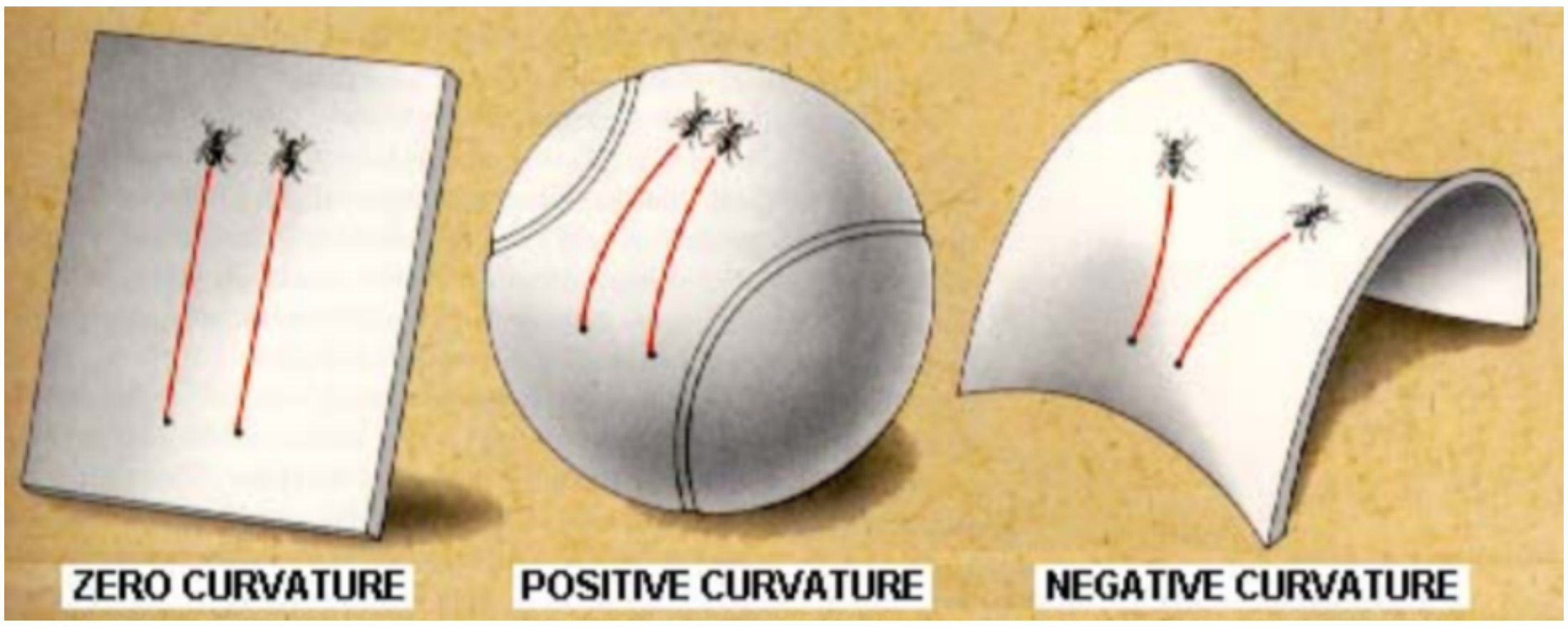

**Fig. 1.7.** Schematic representation of the curvature in two dimensions. Credit: Steve Phillipps.

where the indices $\mu, \nu$ run between 0 and 3, with 0 denoting the time coordinate and 1,2,3 representing the three spatial dimensions. The Einstein equation in natural units reads

$$R_{\mu\nu} - \frac{1}{2} g_{\mu\nu} R = 8\pi G T_{\mu\nu} + \Lambda g_{\mu\nu} \, . \tag{1.52}$$

This equation establishes a connection between the Riemann tensor $R^{\delta}{}_{\alpha\beta\gamma}$, associated to the space-time curvature, and the stress-energy tensor $T_{\mu\nu}$, carrying the information on the sources that generate the gravitational fields. $\Lambda$ represents the cosmological constant. This term was added afterwards and today it is usually associated to dark energy, a new hypothetical form of energy responsible for the accelerated expansion of our Universe. One can also redefine $T_{\mu\nu}$ as $T'_{\mu\nu} = T_{\mu\nu} + \frac{\Lambda}{8\pi G} g_{\mu\nu}$ and the right-hand term in Eq. (1.52) can be reduced to $8\pi G T'_{\mu\nu}$. The quantity $G$ is the gravitational constant, while the Ricci tensor is $R_{\mu\nu} = R^{\alpha}{}_{\mu\alpha\nu}$ and the scalar curvature is $R = R^{\mu}{}_{\mu}$. So, the left-hand side of Eq. (1.52) is related to the geometry of the space-time, while the right-hand side is associated to the content of matter and energy. We can adopt the spherical coordinates $x^i = (t, r, \theta, \phi)$ and define $\mathrm{d}\Omega = \mathrm{d}\theta^2 + \sin^2\theta \, \mathrm{d}\phi$. By defining the proper time as given in Eq. (1.51) in natural units, then the cosmological principle implies that

$$\mathrm{d}\tau^2 = \mathrm{d}t^2 - a^2(t) \left[ \frac{\mathrm{d}r^2}{1 - k r^2} + r^2 \mathrm{d}\Omega \right] . \tag{1.53}$$

This metric is known as the Robertson-Walker metric and it describes an homogeneous and isotropic space-time. All the information on the expansion of our Universe is encoded in the so-called scale factor $a(t)$, which is independent of the spatial coordinate, as a consequence of the homogeneity. The convention is that the scale factor at present time $a_0$ is unity. The constant $k$ represents the curvature, and without loss of generality, it can be redefined to take three discrete values: 0 (flat Universe), +1 (spherical), −1 (hyperbolic). Fig. 1.7 represents schematically different surfaces corresponding to the three possible values of the curvature. By pulling the metric given by Eq. (1.53) into Eq. (1.52) for $\mu = \nu = 0$, one obtains the first Friedmann equation

$$H^2 + \frac{k}{a^2} = \frac{8\pi G}{3} \rho \, , \tag{1.54}$$

where $H = \dot{a}/a$ is the Hubble parameter, which measures the expansion rate of our Universe. By considering $\mu = \nu = i$ in the Einstein equation, one gets the second Friedmann equation

$$2\frac{\ddot{a}}{a} = -\frac{8\pi G}{3}\left(\rho + 3p\right). \tag{1.55}$$

This equation expresses the acceleration $\ddot{a}$. In particular, if $\ddot{a} > 0$, the expansion is accelerated, while if $\ddot{a} < 0$ it is decelerated. It can be shown that the Universe can be described as a perfect fluid with density $\rho$ and pressure $p$, related by $p = W\rho$. A Universe ruled by non-relativistic matter will have $W = 0$, while if a relativistic component (like radiation) is the leading element $W = 1/3$ and in the event of a $\Lambda$-dominated space-time we will have $W = -1$. Thus, we can rewrite Eq. (1.55) as

$$2\frac{\ddot{a}}{a} = -\frac{8\pi G}{3}\rho(1+3W), \tag{1.56}$$

implying that the Universe is characterised by an accelerated expansion whenever $W < -1/3$. One can show that the Universe has seen three different phases. It was first dominated by radiation and then by matter. These first two phases were characterised by a decelerated expansion. The moment in which their densities are equal is called matter-radiation equality and corresponds to $a \sim 3\cdot 10^{-4}$. Due to the expansion of the Universe, the wavelength of the photons increases: in jargon it is redshifted. We can define the redshift $z$ as the quantity that links the emitted wavelength $\lambda_e$ of a photon with the observed one $\lambda_o$:

$$\frac{\lambda_o}{\lambda_e} = 1 + z. \tag{1.57}$$

The scale factor can be related to $z$ via

$$1 + z = \frac{a_0}{a(t)}. \tag{1.58}$$

Thus, $z = 0$ corresponds to today, while $z > 0$ represents a time in the past. Recently, at $a \simeq 0.72$ (or equivalently $z \simeq 0.39$) a new phase has started, where dark energy is the leading component and the expansion of the Universe is accelerated. Fig. 1.8 illustrates the density of the three components in the Universe (radiation, matter, dark energy) as a function of the scale factor. The density parameters are normalised to the critical density, whose value at present time is

$$\rho_c = \frac{3H_0^2}{8\pi G} \simeq 1.88\cdot 10^{26}\, h^2\, \mathrm{kg\ m^{-3}} \simeq 2.78\cdot 10^{11}\, h^2\, M_\odot\, \mathrm{Mpc}^{-3}, \tag{1.59}$$

assuming the Hubble constant today $H_0 = (67.4 \pm 0.5)$ km/s/Mpc, as estimated by the Planck collaboration using the CMB data [17]. The parameter $h$ is defined as

$$h = \frac{H_0}{100\,\mathrm{km/s/Mpc}}. \tag{1.60}$$

It has to be recalled that the measurements of the Cepheids made by the Hubble Space Telescope yields to $H_0 = (73.52 \pm 1.62)$ km/s/Mpc [18]. The two estimates are in disagreement and this represents one the most debated topic of modern cosmology. However, a discussion on this issue is far beyond the scope of this thesis. The critical density of the Universe is useful to define the cosmological parameters

$$\Omega_i = \frac{\rho_i}{\rho_c}, \tag{1.61}$$

where the index $i$ represents the different components: radiation, matter and dark energy. The total normalised density parameter is $\Omega_{\rm tot} = \Omega_r + \Omega_m + \Omega_\Lambda$. The matter term can be split into the sum of two distinguished components: baryonic and dark matter. Eq. (1.54) can be written as

$$\Omega_k = \Omega_{\rm tot} - 1\,, \tag{1.62}$$

where $\Omega_k$ is is the spatial curvature density:

$$\Omega_k = \frac{k}{a^2 H^2}\,. \tag{1.63}$$

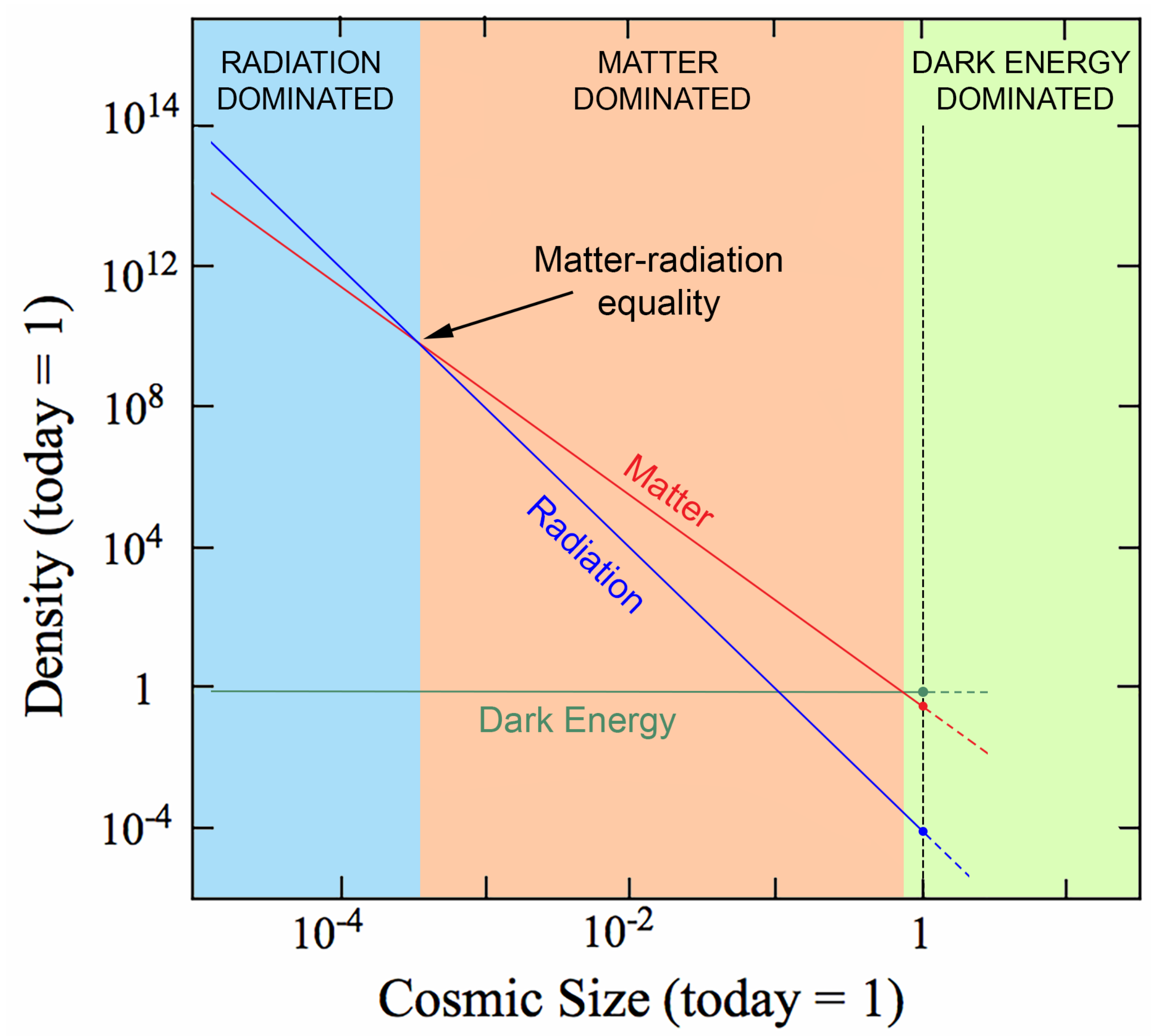

**Fig. 1.8.** Evolution of the density (normalised to the critical density today) as a function of the scale factor for the three components in the Universe: radiation (blue line), matter (red), dark energy (green). The black dashed curve outlines the scale factor today ($a_0 = 1$) and the coloured dots emphasise the value of the corresponding component at present time. The matter-radiation equality occurs at $a \simeq 3 \cdot 10^{-4}$ and marks the transition between a radiation-dominated and a matter-dominated era. Finally, in recent times ($a \simeq 0.72$), dark energy took over and became the major component. The lines transition to dashed at $a = 1$, indicating the future densities. Credit: adapted figure from Mark Whittle.

Therefore, a flat universe corresponds to $\Omega_{\rm tot} = 1$. It is worth noting that the measurements by WMAP [19] and Planck [17] appear to suggest that we live in a flat Universe. Thus, if the Universe has zero curvature, the evolution of $\Omega_i$ with redshift reads

$$\Omega_i(z) = \frac{\Omega_{i,0}\,(1+z)^{3(1+W)}}{\sum_i \Omega_{i,0}\,(1+z)^{3(1+W)}}\,, \tag{1.64}$$

while for the Hubble parameter, it holds

$$H(z) = H_0 \left[\sum_i \Omega_{i,0}\,(1+z)^{3(1+W)}\right]^{1/2}\,. \tag{1.65}$$

At the time of writing, the most recent results of Planck [17] suggest that the content of matter is $\Omega_m = 0.315 \pm 0.007$, of which dark matter represents the most abundant component $\Omega_c h^2 = 0.120 \pm 0.001$, as compared to the baryonic content $\Omega_b h^2 = 0.0224 \pm 0.0001$. However, the Universe today is ruled by dark energy, with $\Omega_\Lambda = 0.685 \pm 0.007$. This component is compatible with a cosmological constant, since $W_\Lambda = -1.03 \pm 0.03$. The radiation content is considerably smaller: $\Omega_r = 1 - \Omega_m - \Omega_\Lambda \sim 10^{-5}$.

## 1.6 Thermodynamics of the Universe

The early Universe was populated by numerous species of particles, which represent the degrees of freedom of the system. These species can be relativistic or non relativistic. Particularly, relativistic particles can either be massless or have a momentum much larger than their mass, $|p| \gg m$. On the other hand, for non-relativistic species it holds $|p| \ll m$. In the phase space, each species $a$ is associated to six coordinates $(x^\mu, p^\mu)$, with $\mu = 0, ..., 3$, and it is characterised by the statistical distribution $f_a(x^\mu, p^\mu)$. These species may interact, depending on their particle nature, and through these interactions they may reach local thermal equilibrium. In order to derive the condition of local thermal equilibrium for the primordial plasma, let us consider two particles $a$ and $b$ that interact with one another. The interaction rate $\Gamma$ depends on the magnitude of the interaction, associated to the cross-section $\sigma_{ab}$, and on the flux, which is given by the number density of targets $n_b$ times the relative speed $v_r$. Thus, we can write

$$\Gamma_a = \sum_b n_b\, v_r\, \sigma_{ab}\,. \tag{1.66}$$

The condition of thermal equilibrium is that the interaction rate has to be larger than the expansion rate of the Universe, which is related to the Hubble parameter. Therefore, if $\Gamma_a \gg H$, the species $a$ is in thermal equilibrium with the cosmic plasma. Now, recalling that for the cosmological principle the Universe is homogeneous and isotropic, the phase-space distribution $f_a$ of the species $a$ is independent on $x^i$ and $p^i$, with $i = 1,2,3$. It is possible to use the temperature of a species as the evolutionary parameter. Indeed, the number density can be written as

$$n_a(T) = \frac{g_a}{(2\pi)^3} \int \mathrm{d}^3 p\, f_a(p,T)\,, \tag{1.67}$$

where $(2\pi)^3$ is the volume of the smallest cell in natural units ($\hbar = 1$) according to quantum mechanics, while $g_a$ represents the internal degrees of freedom of the species (for example 2 for an electron since it can have spin $\pm 1/2$, or 3 for the colours of quarks). Likewise, the energy density is given by

$$\rho_a(T) = \frac{g_a}{(2\pi)^3} \int \mathrm{d}^3 p\, E\, f_a(p,T)\,. \tag{1.68}$$

The pressure associated to an isotropic gas is

$$P_a(T) = \frac{g_a}{(2\pi)^3} \int \mathrm{d}^3 p\, \frac{p^2}{3E} f_a(p,T)\,, \tag{1.69}$$

while the average energy is defined as

$$\langle E \rangle = \frac{\int \mathrm{d}^3 p\, E\, f_a(p,T)}{\int \mathrm{d}^3 p\, f_a(p,T)} = \frac{\rho_a(T)}{n_a(T)}\,. \tag{1.70}$$

The thermal equilibrium in primordial plasma is realised for instance through the following interactions:

$$e^\pm + \gamma \rightleftarrows e^\pm + \gamma\,, \tag{1.71}$$

while the chemical equilibrium is guaranteed by interactions of the type

$$e^\pm + \gamma \rightleftarrows e^\pm + \gamma + \gamma \tag{1.72}$$

$$e^+ + e^- \rightleftarrows 2\gamma \tag{1.73}$$

$$e^+ + e^- \rightleftarrows 3\gamma\,. \tag{1.74}$$

For species in thermal equilibrium, the phase-space distribution is

$$f_a(T) = \frac{1}{e^{\frac{E-\mu_a}{T}} \pm 1}\,, \tag{1.75}$$

where + is for fermions and − for bosons. The chemical potential $\mu$ is a measure of the asymmetry of a species. Indeed, Eq. (1.72) shows unambiguously that $\mu_\gamma = 0$ for photons, while Eqs. (1.73) and (1.74) imply that $\mu_{e^+} = -\mu_{e^-}$. In general, particles and antiparticles have the same potential with opposite sign. Let us consider a particle $\chi$ and its antiparticle $\overline{\chi}$. Plugging Eq. (1.75) into

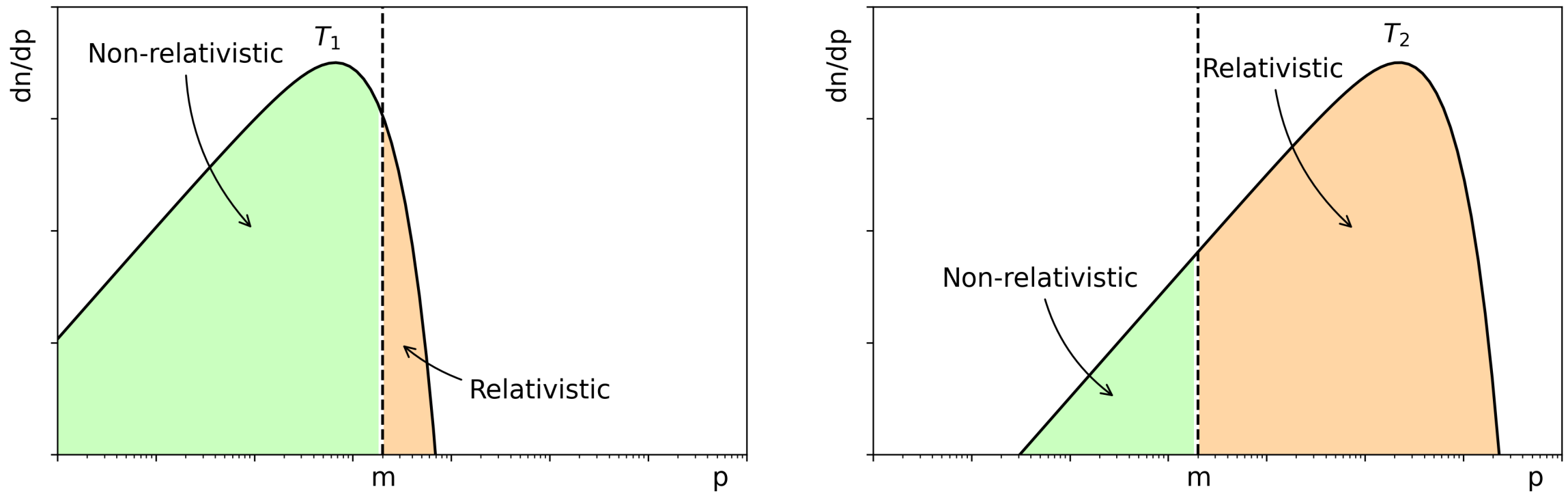

**Fig. 1.9.** The evolution of the function d$n$/d$p$ as a function of the impulse for a non-relativistic (left) and relativistic gas (right). The vertical dashed line denotes the mass of the particles under consideration. The green area corresponds to $p < m$, while the orange region concerns particles with $p > m$. If the peak temperature $T < m$, the gas is non-relativistic: this is the case of $T_1$ in the left panel. Conversely, if $T > m$ the gas is relativistic, as is the case for $T_2$ in the right panel.

Eq. (1.67), we learn two relevant pieces of information. Firstly, we see that at the equilibrium the species are asymmetric, $n^{\chi}_{eq} \neq n^{\bar{\chi}}_{eq}$, if their chemical potential is non-zero. In general the term $\mu_a/T$ in Eq. (1.75) is negligible for all species in the primordial plasma, due to the high temperatures involved. Secondly, we can derive that

$$\frac{\mathrm{d}n}{\mathrm{d}p} = \frac{4\pi g_a}{(2\pi)^3} p^2 \left[e^{\frac{p}{T}} \pm 1\right]^{-1} . \tag{1.76}$$

Fig. 1.9 qualitatively illustrates d$n$/d$p$ as a function of the momentum in two different situations. In both panels, it is clear that the term $p^2$ rules for low $p$, while the exponential cut-off becomes dominant at high $p$. The vertical line corresponds to the value of the mass which separates the relativistic region (orange) and the non-relativistic part (green). The peak of the curve d$n$/d$p$ is related to the temperature. In the left panel, most particles have $p \ll m$, thus we say that the gas is non relativistic. On the contrary, in the right panel the gas is considered relativistic. In other terms, the momentum is an individual property of each particle, while the temperature is a collective property of the gas. In this respect, a particle is relativistic when $p \gg m$, and likewise, a gas is relativistic when $T \gg m$. For higher temperature, the tail of d$n$/d$p$ moves toward the right, as we can observe in Fig. 1.9 with $T_2 > T_1$. Photons are always relativistic since they are massless particles. Instead, former relativistic gas can become non relativistic, as a result of the expansion and related cooling of the Universe. For a relativistic particle, it holds $p, T \gg m$ and $E \simeq p$, which brings to

$$n_a(T) = \frac{g_a}{\pi^2} T^3 \zeta(3) \begin{cases} 1 & \text{BE} \\ \frac{3}{4} & \text{FD} \end{cases} \tag{1.77}$$

$$\rho_a(T) = \frac{\pi^2}{30} g_a T^4 \begin{cases} 1 & \text{BE} \\ \frac{7}{8} & \text{FD} \end{cases} \tag{1.78}$$

$$P_a(T) = \frac{\rho_a(T)}{3} \tag{1.79}$$

$$\langle E \rangle = \begin{cases} 2.701\, T & \text{BE} \\ 3.151\, T & \text{FD} \end{cases} \tag{1.80}$$

where $\zeta$ is the Riemann zeta function, BE stands for the Bose-Einstein distribution and FD for the Fermi-Dirac species. Non-relativistic species have $p, T \ll m$ and $E \simeq m + \frac{p^2}{2m}$, which leads to

$$\begin{aligned} n_a(T) &= g_a \left[\frac{m\,T}{2\pi}\right]^{3/2} e^{-\frac{m}{T}} \\ \rho_a(T) &= m\, n_a(T) \\ P_a(T) &\simeq 0 \\ \langle E \rangle &\simeq m \end{aligned} \tag{1.81}$$

It is noteworthy that the energy density for non-relativistic species in Eq. (1.80) decreases exponentially. The total energy density is the sum over all the degrees of freedom in the system

$$\begin{aligned} \rho(T) &= \sum_a \rho_a(T) \\ &\simeq \sum_{a,\mathrm{rel}} \rho_a(T) \\ &= \frac{\pi^2}{30} g_*(T)\, T^4 , \end{aligned} \tag{1.82}$$

where in the second equality we used the fact that non-relativistic species are Boltzmann-suppressed in the radiation-dominated Universe because of the high temperature. The weight function $g_*$ is the weighted sum of $g_a$ over all the relativistic species involved

$$g_*(T) = \sum_{a=\mathrm{BE}} g_a + \frac{7}{8} \sum_{a=\mathrm{FD}} g_a \,. \tag{1.83}$$

However, different species may have different temperatures, thus the general expression of $g_*$ reads

$$g_*(T) = \sum_{a=\mathrm{BE}} g_a \left(\frac{T_a}{T}\right)^4 + \frac{7}{8} \sum_{a=\mathrm{FD}} g_a \left(\frac{T_a}{T}\right)^4 . \tag{1.84}$$

This function is constant in temperature until a species becomes non-relativistic, when $T < m$. Then, it drops as the temperature becomes lower than the mass of the species. To a first approximation, it behaves as a step function since the transition between relativistic and non-relativistic regime is rapid. If we consider bosons,

- for $s = 0$: $g = 1$ because there is only one possible spin
- for $s = 0$ and $m = 0$: $g = 2$ because there are two transverse polarizations
- for $s = 0$ and $m \neq 0$: $g = 3$ because there are two transverse and one longitudinal polarizations.

Instead, if we consider fermions,

- for $s = \frac{1}{2}$ and left-handed helicity: $g = 1$
- for $s = \frac{1}{2}$ and right-handed helicity: $g = 1$.

So, an electron will have $g_e = 2$ since it has both states of helicity, while neutrinos will have $g_\nu = 1$ since they are left-handed. Here we have considered only the particles of the standard model (SM), but in principle we could have additional possibilities if new particles will be discovered in the future and if we consider extensions of the SM. Since the particles are in plasma with finite temperature and density, their propagation is not free and one should consider the plasma effects. However, this is beyond the scope of this thesis.

An additional important quantity describing the thermodynamics of the system is the entropy $S$, which can be expressed as

$$S = s a^3 , \tag{1.85}$$

where $s$ is the entropy density and $a^3$ represents the fiducial volume. The entropy density can be related to the pressure, density and temperature of the species through

$$s = \frac{S}{V} = \frac{P + \rho}{T} . \tag{1.86}$$

Therefore, in the radiation-dominated era of the Universe, the entropy density of a species $a$ is

$$s_a(T) = \frac{2\pi^2}{45} g_a T_a^3 \begin{cases} 1 & \text{BE} \\ \frac{7}{8} & \text{FD} \end{cases} \tag{1.87}$$

and the total entropy density is given by

$$s(T) = \sum_a s_a(T) = \frac{2\pi^2}{45} g_{*s}(T)\, T^3 \tag{1.88}$$

with

$$g_{*s}(T) = \sum_{a=\text{BE}} g_a \left(\frac{T_a}{T}\right)^3 + \frac{7}{8} \sum_{a=\text{FD}} g_a \left(\frac{T_a}{T}\right)^3 . \tag{1.89}$$

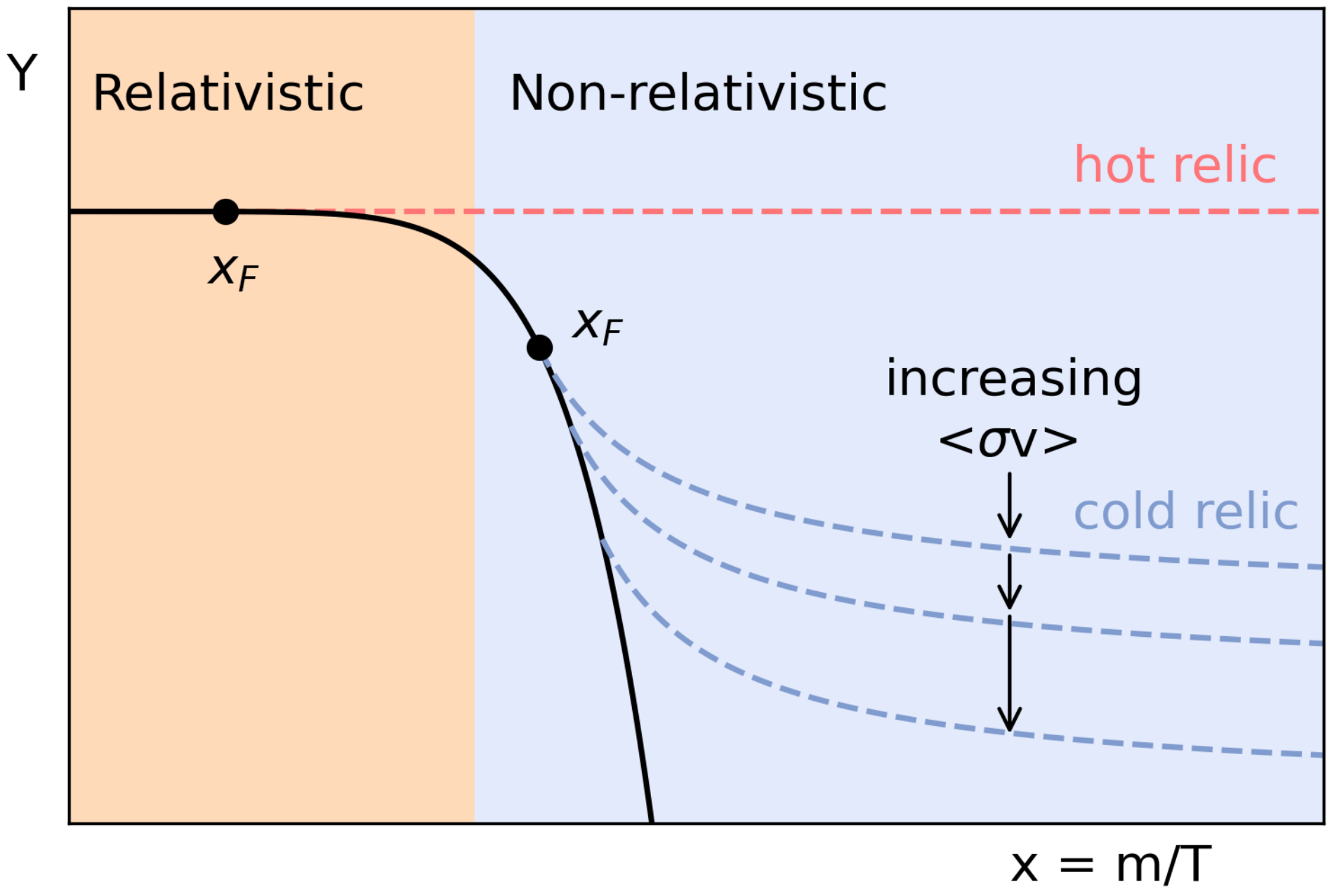

**Fig. 1.10.** The evolution of the abundance with the parameter $x = m/T$. If the decoupling occurs in the relativistic regime ($x_F < 1$), the particle becomes a hot relic, while if the freeze-out takes place in the non-relativistic regime ($x_F > 1$), it is a cold relic. Larger annihilation cross-sections lead to smaller relic abundances.

It is worthy to mention that the entropy scales as $T^3$, while the density as $T^4$. We can define the abundance of a species in thermodynamic equilibrium as

$$Y_a = \frac{n_a(T)}{s_a(T)} = \begin{cases} \dfrac{45\zeta(3)}{2\pi^4}\dfrac{g_a}{g_{*s}(T)} & \text{Relativistic species} \\[2ex] \dfrac{45\zeta(3)}{4\sqrt{2}\pi^{7/2}}\dfrac{g_a}{g_{*s}(T)}\left(\dfrac{m}{T}\right)^{3/2} e^{-\dfrac{m}{T}} & \text{Non-relativistic species} \end{cases} \tag{1.90}$$

The abundance is useful to determine the decoupling of a species. We define the decoupling temperature $T_d$ as the temperature beneath which the species passes from equilibrium to non-equilibrium. Practically, it is the temperature at which

$$\Gamma_a(T_d) = H(T_d)\,. \tag{1.91}$$

If the decoupling from the primordial plasma occurs during its relativistic regime ($m \gg T$), it produces a so-called hot relic. On the contrary, when the decoupling takes place in the non-relativistic regime, we have a cold relic. If the particle is unstable, its abundance decays exponentially after decoupling:

$$Y_a = Y_D\, e^{(t-t_D)/\tau_a}\,, \tag{1.92}$$

where $Y_D$ and $t_D$ are the abundance and the time at the decoupling, respectively. In order to have a significant abundance today, the mean lifetime $\tau_a$ of the particle has to be much larger than the

age of the Universe:

$$\tau_a \gg t_0\,. \tag{1.93}$$

Let us consider the case of annihilation of DM particles into SM particles,

$$\chi + \overline{\chi} \quad \rightleftarrows \quad \mathrm{SM} + \overline{\mathrm{SM}}\,. \tag{1.94}$$

The evolution of the number density of DM particles is determined by the Boltzmann equation

$$\frac{\mathrm{d}n}{\mathrm{d}t} = -3Hn - \langle\sigma v\rangle \left(n^2 - n_{\mathrm{eq}}^2\right), \tag{1.95}$$

where $n_{\mathrm{eq}}$ stands for the number density at the chemical equilibrium. The first term is related to the decrease of $n$ due to the expansion of the Universe, while the second and the third terms concern the annihilation and inverse annihilation of DM particles, respectively. If we define $x = m_\chi/T$, Eq. (1.95) can be expressed as

$$\frac{\mathrm{d}Y}{\mathrm{d}x} = -0.264 M_p \frac{g_{*s}(T)}{\sqrt{g_*(T)}} \frac{m_\chi}{x^2} \langle\sigma v\rangle \left(Y^2 - Y_{\mathrm{eq}}^2\right), \tag{1.96}$$

where $M_p = \sqrt{\hbar c^5/G}$ is the Planck mass. Fig. 1.10 illustrates the evolution of the abundance as a function of $x$. If we denote with $x_F$ the value of $x$ at the decoupling (known as freeze-out of the particle), hot relics correspond to particles with $x_F < 1$ (decoupling in the relativistic regime), while for cold relics we have $x_F > 1$ (freeze-out in the non-relativistic regime). From Fig. 1.10, we also learn that the abundance of hot relics is not strongly affected by the moment of decoupling. On the contrary, the abundance of cold relics changes significantly with $x$: in order to remain longer in equilibrium and decouple later, DM has to interact a lot, which leads to a lower abundance. For cold relics we have

$$Y_{\mathrm{eq}} = 0.145 \frac{g}{g_{*s}(x)} x^{3/2} e^{-x}, \tag{1.97}$$

and for $Y \gg Y_{\mathrm{eq}}$, the abundance at present time can be expressed as

$$\frac{1}{Y_0} = 0.264 M_p \frac{g_{*s}(T_F)}{\sqrt{g_*(T_F)}} \int_{T_0}^{T_F} \mathrm{d}T \, \langle\sigma v\rangle\,. \tag{1.98}$$

where $T_0$ is the present-day temperature. By integrating the velocity-averaged cross-section up to the freeze-out temperature $T_F$, we can define the quantity

$$\langle\sigma v\rangle_{\mathrm{int}} = \frac{1}{m_\chi} \int_0^{T_F} \mathrm{d}T \, \langle\sigma v\rangle\,. \tag{1.99}$$

Therefore

$$Y_0 \sim \frac{1}{\langle\sigma v\rangle_{\mathrm{int}}}\,. \tag{1.100}$$

Recalling that $\Omega_i = \rho_{0,i}/\rho_c$, where $\rho_{0,i}$ is the density today for the species $i$ and $\rho_c$ is the critical density of the Universe, after some non−trivial calculations we obtain

$$\Omega_\chi h^2 = 8.5 \cdot 10^{-11} \frac{\sqrt{g_*(T_F)}}{g_{*s}(T_F)} \frac{\mathrm{GeV}^{-2}}{\langle \sigma v \rangle_{\mathrm{int}}} . \tag{1.101}$$

We can expand the cross-section in polynomials

$$\langle \sigma v \rangle = a + \frac{b}{x} , \tag{1.102}$$

which brings to

$$\langle \sigma v \rangle_{\mathrm{int}} = \frac{1}{x_F} \left( a + \frac{1}{2} \frac{b}{x} \right) , \tag{1.103}$$

where the first addend in Eqs. (1.102) and (1.103) refers to the s-wave contribution, which is independent of speed, while the second addend is related to the p-wave component, which is speed dependent. The coefficients $a$ and $b$ are independent of the velocity. Let us consider the case in which $a \gg b x^{-1}$. As previously introduced, we can determine the value of $x_F$ by equating $\Gamma(x_F) = H(x_F)$, and we find that $x_F$ ranges between 15 and 25. Therefore, the freeze-out temperature for a cold DM relic is

$$T_F \simeq \frac{m_N}{20} . \tag{1.104}$$

For a weakly interacting particle whose annihilation is s-wave we can write

$$\langle \sigma v \rangle_{\mathrm{int}} = \frac{1}{x_F} G_F^2 m_\chi^2 . \tag{1.105}$$

We can set a limit on the mass of the DM particle by requiring that

$$\Omega_\chi h^2 \leq \left( \Omega_\chi h^2 \right)_{\mathrm{obs}} , \tag{1.106}$$

which implies

$$m_\chi \gtrsim 0.8\,\mathrm{GeV} , \tag{1.107}$$

where the limit $m_\chi$ is known as the Lee-Weinberg limit. Moreover, by equating $\Omega_\chi h^2$ in Eq. (1.101) to the observed value, we obtain

$$\langle \sigma v \rangle \simeq 3 \cdot 10^{-26}\,\mathrm{cm}^3\,\mathrm{s}^{-1} , \tag{1.108}$$

which is considered the natural scale of the so-called weakly interacting massive particles (WIMPs). Let us consider two DM particles which annihilate into two SM particles $a$ and $b$ via the weak interaction. Fig. 1.11 illustrates a possible diagram: $Z_0$ is one of the vector bosons of the weak interaction and $g$ represents the coupling constant. We can assume that the DM particles annihilate at rest. Thus, the centre-of-mass energy is $\sqrt{s} = 2m_\chi$ and the kinematic limit is

$$m_\chi > \frac{m_a + m_b}{2} . \tag{1.109}$$

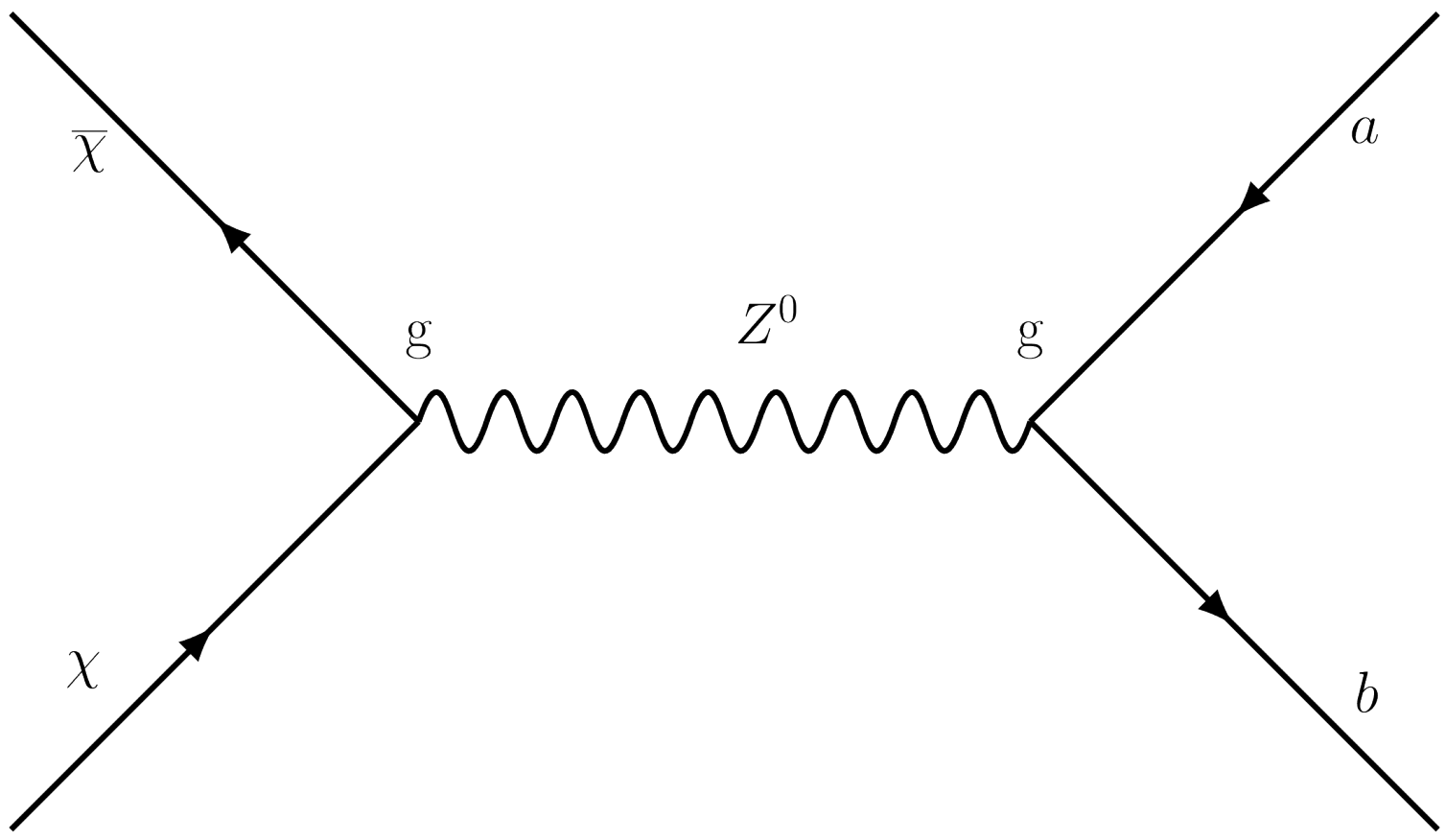

**Fig. 1.11.** Feynman diagram for two DM particles $\chi$ and $\overline{\chi}$ which annihilate into two standard model particles $a$ and $b$ via the weak interactions. The boson $Z^0$ is one of the mediators of the weak interaction and $g$ is the coupling constant.

The propagator includes a factor

$$\frac{1}{q^2 - m_Z^2} = \frac{1}{(2m_\chi)^2 - m_Z^2} . \tag{1.110}$$

We can identify three regimes:

- $m_\chi \ll m_Z$
- $(2m_\chi)^2 \ll m_Z^2$
- $m_\chi \gg m_Z$ .

By recalling that $\Omega_\chi h^2 \sim 1/\langle \sigma v \rangle$, we can estimate the behaviour of the relic abundance in the three regimes. When $m_\chi \ll m_Z$, it holds

$$\frac{1}{q^2 - m_Z^2} \sim \frac{1}{m_Z^2} \tag{1.111}$$

and $\langle \sigma v \rangle \sim G_F^2 m_\chi^2$. Thus, $\Omega_\chi h^2 \sim m_\chi^{-2}$. In the second regime, $m_\chi \sim m_Z/2$, the propagator is resonant and as a consequence, the cross-section is resonant as well. Therefore, the relic abundance drops significantly. Finally, in the third regime

$$\frac{1}{q^2 - m_Z^2} \sim \frac{1}{(2m_\chi)^2} . \tag{1.112}$$

Since $q^2 = (2m_\chi)^2 \gg m_Z$, we can write

$$\frac{1}{q^2} \sim \frac{1}{(2m_\chi)^2} . \tag{1.113}$$

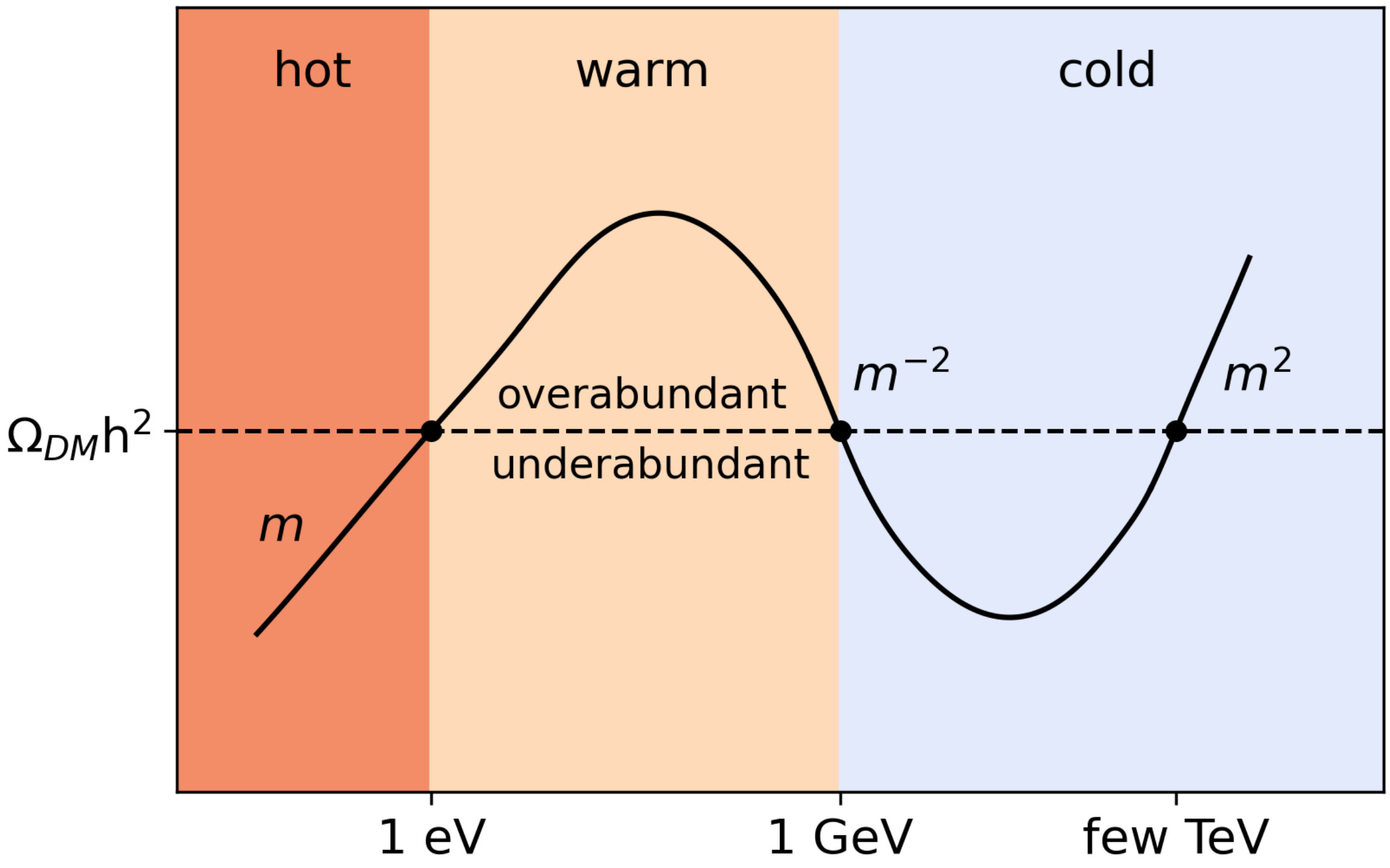

**Fig. 1.12.** Relic abundance of DM particles as a function of their mass. The red area refers to the hot regime, the orange region to the warm regime, the blue area refers to cold relics. The dashed line represents the DM density parameter, which sets the boundary between overabundant and underabundant species.

Hence,

$$\langle \sigma v \rangle \sim \left( \frac{g^2}{q^2} \right)^2 s \sim \frac{g^4}{m_\chi^4} m_\chi^2 \sim m_\chi^2, \tag{1.114}$$

which leads to $\Omega_\chi h^2 \sim m_\chi^2$. The WIMP paradigm or WIMP miracle refers to the fact that weakly interacting particles with a mass in the range between GeV and TeV manage to explain the entire relic abundance observed in the Universe. The WIMP particles are

- cold
- weakly interacting
- stable (or with a mean lifetime much larger than the age of the Universe)
- characterised by a relic abundance which is roughly the same at the freeze-out.

Fig. 1.12 illustrates the relic abundance as a function of the mass of the DM particle. In the range GeV−TeV of the WIMP, we can identify the three regimes mentioned above. The horizontal line corresponds to the observed value. Particles in the warm regime, that is to say with a mass in the keV−MeV range, are overabundant. Therefore, either they decay or they have to interact much more such that this curve drops. However, particles in this mass range can still account for DM in the Universe if they are produced with different mechanisms, rather than the thermalization in the primordial plasma. In other words, the figure shows that a thermal production in the early Universe is compatible with the observed relic abundances for DM particles with mass below 1 eV or in the WIMP range.

## 1.7 Dark matter candidates

The SM of particle physics is a theory that describes three out the four fundamental interactions (strong, weak and electromagnetic, while gravity is excluded) and how all the elementary particles known so far interact through them. It is based on the gauge group $SU(3)\otimes SU(2)\otimes U(1)$ where $SU(3)$ is associated to the strong interaction and $SU(2)\otimes U(1)$ describes the electroweak force, which is the unification of the electromagnetic and weak interactions. Fig. 1.13 illustrates the SM particle content. There are three generations of fermions, divided into quarks and leptons. The former include up, down, charm, strange, top and bottom quarks. Leptons include electrons, muons, taus and their respective neutrinos. Each of these particles has a corresponding antiparticle with the same mass and opposite quantum numbers. The interactions are mediated by bosons: gluons for the strong force, photons for the electromagnetic interaction and the three gauge bosons $W^{\pm}$, $Z^0$ for the weak force. The Higgs boson is a massive scalar particle, necessary to explain why particles acquire mass: the so-called Higgs mechanism. The SM represents one of the most successful theories of particle physics, being able to predict the existence of particles like the top quark and the tau neutrino (experimentally discovered at Fermilab in 1995 [20, 21] and 2000 [22], respectively), as well as the Higgs boson (detected at CERN in 2012 [23, 24]). However, none of the SM particles succeeds in explaining DM in the Universe and this reputable theory also fails in describing numerous observed phenomena (for instance the neutrinos' oscillation and how they acquire mass). As a consequence, theoretical physicists in the last decades had a great time in picturing extensions of the SM and new particles which could explain DM in the Universe and all the other unexplained phenomena. We already mentioned the most notorious candidate for dark matter: weakly interacting massive particles. Their fame is justified by the fact that they provide the correct DM abundance observed in the Universe and they naturally arise in some supersymmetric models. In the following, we will briefly mention a few particle models that have gained the attention of the community.

**Neutralino and sneutrino.** One example is the Minimal Supersymmetric Standard Model (MSSM), according to which for every particle of the SM there is a supersymmetric particle which has the same quantum numbers, except for the spin that differs by 1/2. The fact that none of these particles have been observed so far may be an indication that this symmetry is broken, but it remains unclear on which scale it might appear. In order to be compatible with the current limits to the proton lifetime, one can impose the so-called R-parity. It is a discrete symmetry,

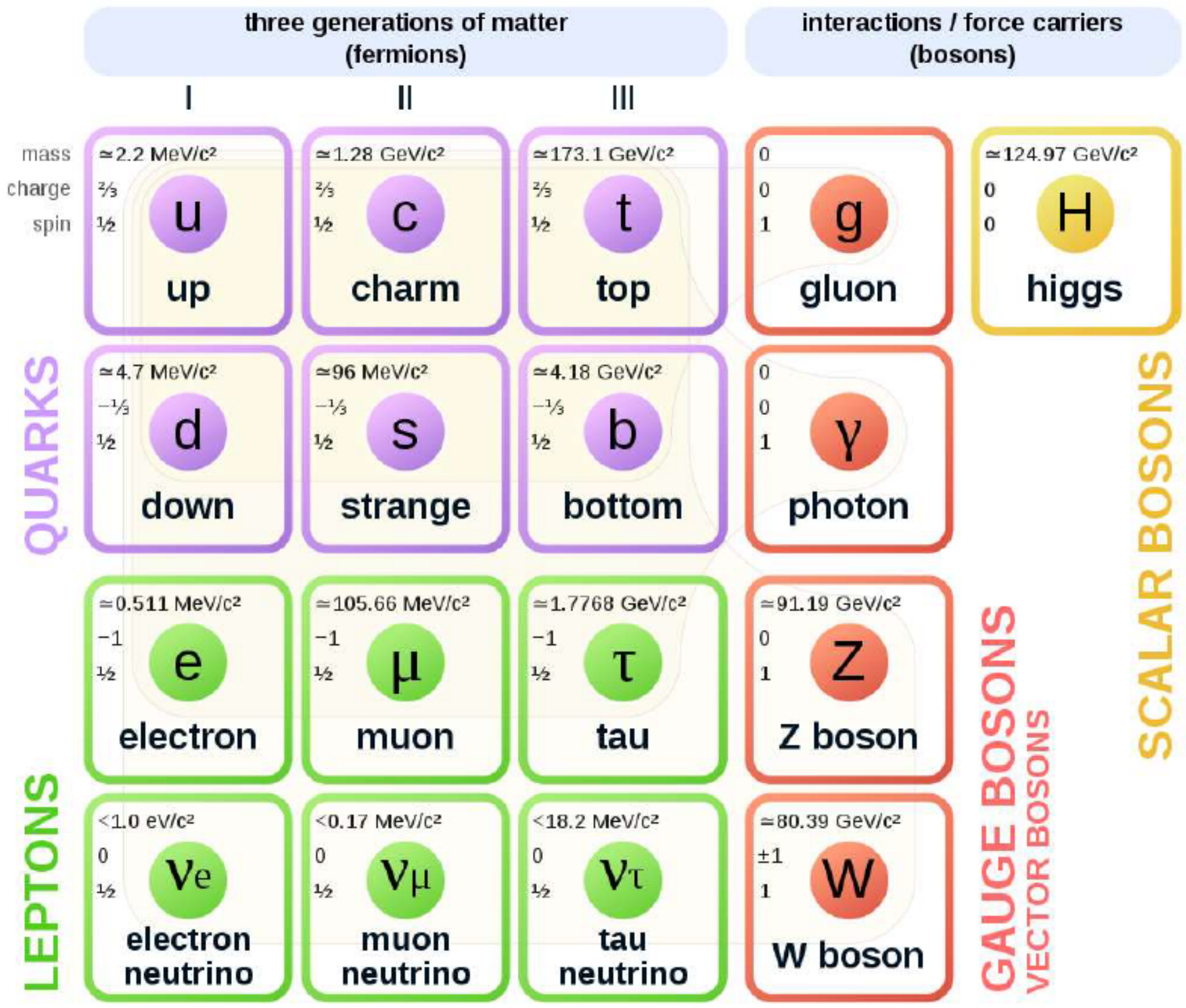

**Fig. 1.13.** Standard Model of particle physics. Credit: [25].

characterised by

$$R = (-1)^{3B+L+2s} \tag{1.115}$$

where $B$ is the baryon number, $L$ the lepton number and $s$ is the spin of the particle. All known SM particles have $R = 1$, while their supersymmetric partners have $R = -1$. Also, this model includes two Higgs doublets and four neutral fermions, called neutralinos. Here comes the interesting part: in the MSSM theory, if the R-parity is conserved, the interactions of SM particles can only produce an even number of supersymmetric particles and the decay products of the latter can only be lighter supersymmetric particles. As a consequence, the lightest supersymmetric particle (LSP) must be stable. One of the most appealing candidates is in fact the lightest neutralino since it is neutral and it is expected to have a mass in the range GeV−TeV. Therefore, it is considered a natural candidate for cold DM, being a very paradigmatic WIMP. Another option in supersymmetric models is the sneutrino, which represents the supersymmetric partner of neutrinos. The interested reader can find further information in Refs. [26, 27, 28, 29, 30, 31].

**Kaluka Klein model.** Other models propose the existence of extra dimensions which are compact. The simplest scenario considers a five-dimensional space-time, with an additional space dimension. In principle, for every particle of the SM there are infinite excitation modes $X^{(n)}$, where $n$ is the order of the excitation mode and the fundamental state $n = 0$ corresponds to the SM particle. These excitation states behave like particles with a higher mass with respect to the corresponding SM particle:

$$m^2_{X^{(n)}} = m^2_{X^0} + \left(\frac{n}{R}\right)^2 , \tag{1.116}$$

where $m_{X^{(n)}}$ is the mass of the particle of order $n$ and $R$ is the radius of the compactified dimension. The ensemble of these states is known as Kaluza-Klein tower. This model is an example of the Universal Extra Dimensions theory and it predicts the existence of a stable, neutral, weakly-interacting and heavy particle, called the Lightest Kaluza-Klein Particle, which constitutes a suitable candidate for cold DM. Further information can be found in Refs. [32, 33, 34].

**Minimal Dark Matter Model.** In the Minimal Dark Matter Model [35], instead, we consider only an additional quintuplet to the SM. The neutral particle of the quintuplet appears to be the lightest particle, as well. Its mass is expected to be of the order of a few TeV, making it a good candidate for cold DM. Both for the supersymmetric LSP and for this candidate of Minimal Dark Matter, one should take into account co-annihilation interactions, where the lightest particle annihilates not only with itself but also with the other particles in the multiplet. This effect is equivalent to having a larger "effective" cross-section. Therefore, co-annihilation processes lead to a lower relic abundance.

**Sterile neutrino.** Another interesting candidate is the sterile neutrino. The SM of particle physics predicts the existence of three generations of left-handed massless neutrinos. However, experiments involving atmospheric and solar neutrinos as well as accelerator and reactor neutrinos suggest that these particles are indeed massive. The value of their mass is still unknown, though the latest Planck results at the time of writing indicate an upper bound on the sum of the masses associated to the three eigenstates: $\sum m_\nu < 0.12$ eV [17]. Sterile neutrinos are proposed in some extensions of the SM to give mass to the active neutrinos. If they exist, sterile neutrinos can also represent a suitable candidate for DM, associated with a non-thermal mechanism of production. The results from the Large Electron-Positron accelerator on the Z boson are consistent with three generations of neutrinos [36]. Therefore, if the sterile neutrino exists, it is expected to have zero coupling to the vector bosons W and Z. It would be a right-handed particle with a mass in the keV range, produced through the oscillation of active neutrinos into sterile neutrinos. To delve deeper into this fascinating candidate, refer to Refs. [37, 38, 39].

**Axions.** The scientific community has focused on axions and axion-like particles as promising candidates. Their existence has been first proposed to solve the strong charge parity (CP) problem in quantum chromodynamics [40]. They are pseudoscalar particles and they can have a very low mass, order of $\mu$eV. These hypothetical candidates could be produced with a non-thermal mechanism. Let us consider a pseudoscalar $a$ which oscillates around its minimum. Its density and pressure can be expressed as

$$\begin{cases} \rho_a = \frac{1}{2}\dot{a}^2 + V(a) \\ p_a = \frac{1}{2}\dot{a}^2 - V(a)\,, \end{cases} \tag{1.117}$$

where $\dot{a}$ is the first derivative of the field and $V(a)$ is the potential. For a harmonic oscillator, the mean kinetic energy equals the average potential. Thus,

$$\begin{cases} \langle\rho_a\rangle = \langle K\rangle + \langle V\rangle = 2\langle V\rangle \\ \langle p_a\rangle = \langle K\rangle - \langle V\rangle = 0\,. \end{cases} \tag{1.118}$$

Therefore, a scalar field which oscillates towards its minimum behaves as non-relativistic matter and the axion can be a good candidate for cold DM. The interested reader can learn more by looking at Refs. [41, 42].

So far, we have mentioned heavy candidates and light candidates. There are also interesting candidates in the MeV range. For instance, light DM particles are characterised by a mass below 100 MeV and their spin can be either 0 or 1/2 [43]. These particles can annihilate into electron-positron pairs and the interaction is mediated by a new neutral light spin-1 gauge boson. It is beneficial to recall that MeV and keV particles are usually produced with a non-thermal mechanism, since a thermal production would result in an overabundant species with respect to the observed DM density (see Fig. 1.12).

**Primordial black holes.** The explanation behind the DM problem could also be in the form of compact objects, like primordial black holes (PBH). Unlike the stellar black holes which represent the final stage of very massive stars, these hypothetical candidates could originate from the density fluctuations that occurred in the first fractions of a second after the Big Bang. They are not comprised in the Big-Bang nucleosynthesis constraints that estimate a baryonic content around 5%, because they formed during the radiation era, well before the production of nuclei. Hence, they are classified as non-baryonic matter and their dynamical behaviour is consistent with cold DM. A nice overview on the status of PBH searches can be found in Ref. [44].

A nice and concise review of the numerous potential candidates and their production mechanisms can be found in Ref. [38]. In this doctoral thesis, a phenomenological approach is adopted, without a focus on specific candidates. The aim is to explore different regions of the DM parameter space with a multi-wavelength signal in a candidate-agnostic way, relying only on the particle relevant features for the DM studies (its mass and strength of interaction). The method of investigation is to look for multi-wavelength signals.

# Chapter 2

## Detection with a multi-wavelength strategy

Under the assumption that DM is made up of elementary particles which can interact with SM particles, a classical categorisation of DM searches takes into consideration three main directions: indirect detection, direct detection and production in colliders [45, 46]. In this regard, Section 2.1 focuses on indirect searches with photons, charged cosmic rays and neutrinos, with an excursus on the most promising astrophysical targets. The cosmological indirect searches based on the DM effects on the abundances of primordial light elements and on the cosmic microwave background are illustrated in Section 2.2. Finally, direct detection and DM production in colliders are briefly summarised in Section 2.3. This thesis focuses on indirect detection searches with a multi-wavelength strategy. Particular attention is placed on $\gamma$ rays, X rays and radio waves, which represent the three frequency bands at the core of this thesis.

## 2.1 Indirect detection

The indirect detection searches revolve around the idea that DM particles can annihilate or decay and produce a broad variety of SM particles. Typically the products of annihilation events consist of particle-antiparticle pairs. If these final states are unstable, they will decay into stable particles. The final stable products include photons, neutrinos, charged cosmic rays and they represent the DM signatures, which could be observed with our space and ground-based telescopes. Fig. 2.1 illustrates the production of these final states via some mechanisms, which will be explained in the following paragraphs. One main advantage of photons and neutrinos is that, being neutral particles, their propagation is not deflected by the ambient magnetic fields. Thus, their incoming trajectory traces back to the position of the source. On the contrary, cosmic rays are deflected by the galactic and extragalactic magnetic field, therefore we lose the information about their point of origin. Depending on the mass, different annihilation channels can be open. The energy spectrum depends on the annihilation channel and products (e.g. whether they hadronize, radiate or decay). In principle, DM particles can annihilate into different SM particles and the energy spectrum of each channel will be weighted by the corresponding branching ratio, which specifies the fraction of DM particles that annihilate through that specific channel. A general expression of the total energy spectrum is

$$\frac{\mathrm{d}N}{\mathrm{d}E} = \sum_i \mathrm{BR}_i \frac{\mathrm{d}N^i}{\mathrm{d}E}, \tag{2.1}$$

where $\mathrm{d}N^i/\mathrm{d}E$ represents the $i$-th annihilation channel and $\mathrm{BR}_i$ is the corresponding branching ratio. Different particle models will imply different branching ratios. Therefore, when adopting a phenomenological approach without focusing on any specific model, one usually assumes that all the DM particles annihilate (or decay) through a single channel. Physically speaking, this is equivalent to assume that there is a leading annihilation (or decay) channel. As a consequence of the conservation of energy and of the DM being non-relativistic, annihilation events of DM particles at rest cannot produce signals with an energy higher than the mass of the "parent" particles. In the case of DM decay, the product can not exceed half of the mass of the parent particle. This is due to the fact that the energy in the centre of mass is

$$\sqrt{s} = \begin{cases} 2m_\chi & \text{Annihilation} \\ m_\chi & \text{Decay}. \end{cases} \tag{2.2}$$

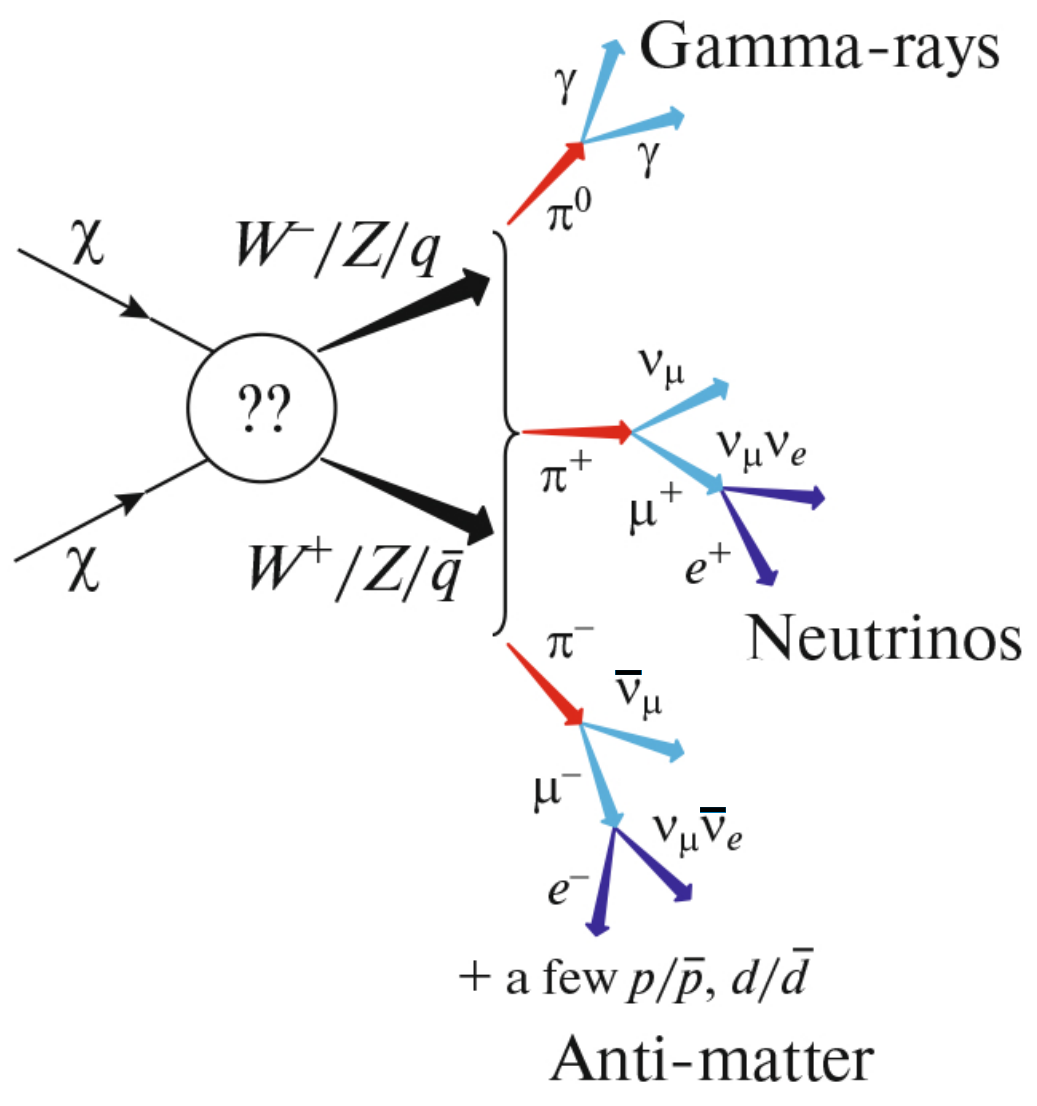

**Fig. 2.1.** Possible channels of DM annihilation with $\gamma$ rays, neutrinos and charged cosmic rays as final products. Credit: adapted figure from [47].

On one hand, indirect detection searches have the advantage that we look for DM in its natural environment. On the other hand, the astrophysical background is often poorly understood, leading to potential large and unknown systematic errors. In order to maximise the size of the signal over the astrophysical noise, we can focus on targets with a high mass-to-light ratio, where the amount of DM is expected to be much larger than baryonic matter, or we can look for distinctive features of a DM signal. Adopting a multi-messenger approach is also useful to get a better understanding of the astrophysical sources that represent our background. Different messengers are associated to different mechanisms of production, providing valuable insights on the properties of the emitters.

### 2.1.1 Photons

Concerning photon searches, we can identify two kinds of radiation: prompt and secondary. The former relates to an emission which is directly produced in DM annihilation/decay events. Let us consider for instance the annihilation of DM particles into quarks, gluons, Higgs and vector bosons. Quarks and gluons will hadronize and produce mostly pions, protons and antiprotons. Charged pions will rapidly decay, leading to electrons, positrons and neutrinos, while neutral pions will essentially produce gamma rays, as illustrated in Fig. 2.1. The muon and tau leptons will decay into electrons and neutrinos as well, and in the case of tau leptons there is also a production of charged and neutral pions, thus contributing to the primary photons. In the case of DM particles which annihilate into charged states, the final products will also include final state radiation. An additional example of prompt photons is DM decaying into pairs of photons or into a photon and a neutral state (such as Higgs or Z bosons, neutrinos or some new neutral particle). On one hand, this signal will have a characteristic spectral feature of a line, discernible from the astrophysical background. On the other hand, DM is neutral and this annihilation channel would be suppressed, making its detection quite challenging. In the case of prompt neutral signal, the differential flux produced by DM particles annihilating in a region of the sky with a solid angle $\Delta\Omega$, can be expressed as

$$\frac{\mathrm{d}^2\phi^{\mathrm{Ann}}}{\mathrm{d}E\,\mathrm{d}\Omega} = \frac{\eta}{4\pi}\frac{\langle\sigma v\rangle}{m_\chi^2}\frac{\mathrm{d}N}{\mathrm{d}E}J\,, \tag{2.3}$$

where the particle properties are included: $\langle\sigma v\rangle$ is the velocity-averaged annihilation cross-section, which encodes the information on the strength of the interaction, $m_\chi$ is mass of the DM particle, $\mathrm{d}N/\mathrm{d}E$ is the energy spectrum and the factor $\eta$ is 1/2 if DM is self-conjugate or 1/4 if not. The astrophysical information is encoded in the so-called J factor:

$$J = \int_{\mathrm{l.o.s.}} \mathrm{d}s\,\rho^2(s)\,, \tag{2.4}$$

where the integral is performed along the line of sight $s$. The DM density profile $\rho$ is squared in the case of annihilating DM because two particles are involved. This profile describes the spatial distribution of the DM particles as a function of the distance $r$ from the centre of the DM halo. More details on the halo model will be provided in Chapter 3, while in Part II we will discuss more thoroughly the differential photon flux and its application for Galactic DM searches. Regarding decaying DM, the differential flux for prompt neutral final states is given by

$$\frac{\mathrm{d}^2\phi_\gamma^{\mathrm{Dec}}}{\mathrm{d}E\,\mathrm{d}\Omega} = \frac{1}{4\pi}\frac{\Gamma}{m_\chi}\frac{\mathrm{d}N}{\mathrm{d}E}D\,, \tag{2.5}$$

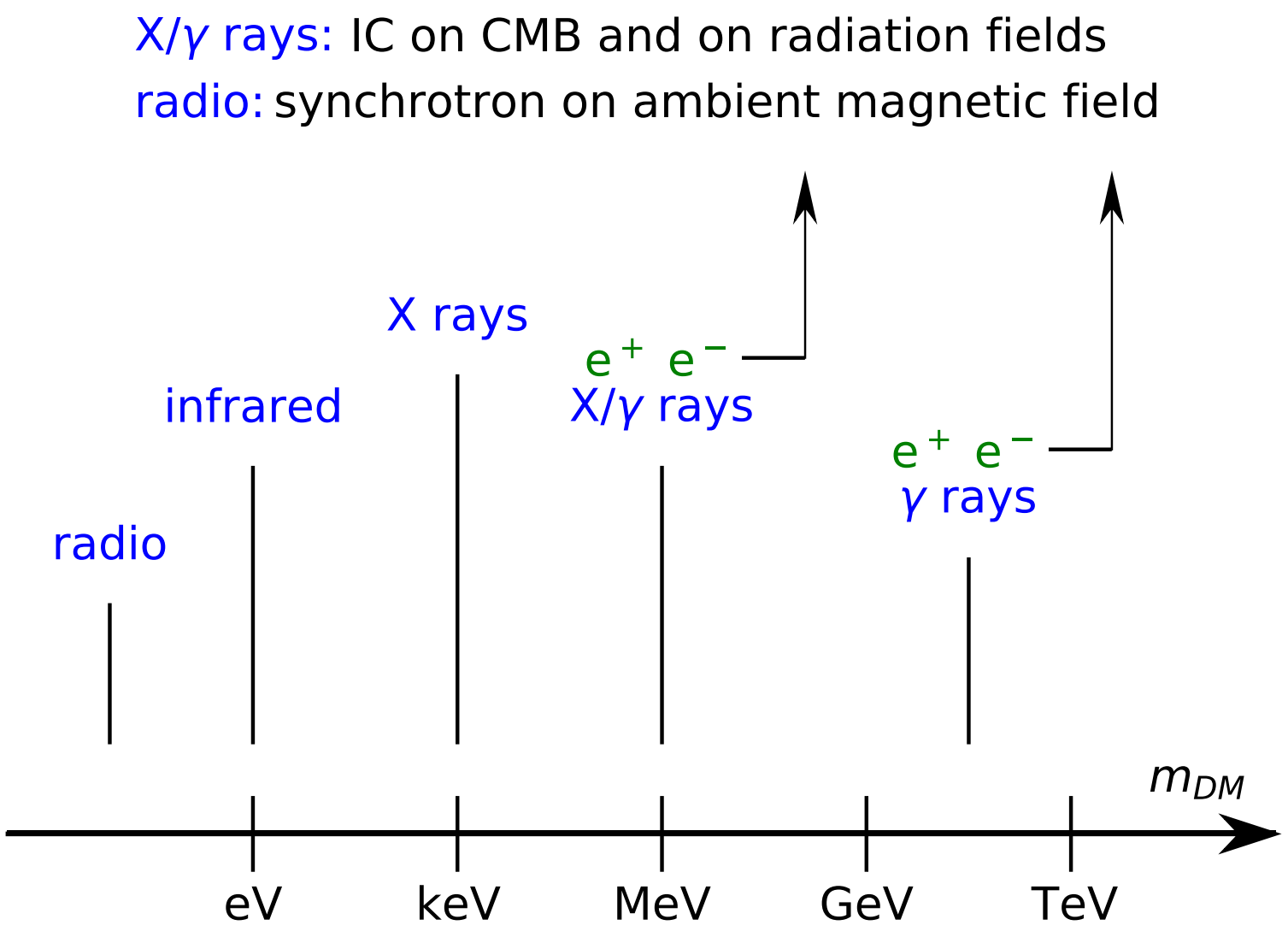

**Fig. 2.2.** This schematic figure shows which electromagnetic signal can be produced as a function of the DM particle's mass.

where Γ is the decay rate, d$N$/d$E$ is the decay energy spectrum and the D-factor reads

$$D = \int_{\text{l.o.s.}} \mathrm{d}s\, \rho(s)\,. \tag{2.6}$$

The decay rate is related to the decay lifetime of the DM particle through $\Gamma = 1/\tau$.

Regarding secondary photons, they can be produced through inverse Compton scattering, bremsstrahlung and synchrotron emission. DM particles could annihilate into primary $e^{\pm}$ pairs. These pairs can interact with the surrounding radiation fields, via inverse Compton scattering, or with the atomic nuclei, via bremsstrahlung. Inverse Compton scattering will be the protagonist of Part II, while bremsstrahlung has already been discussed in Chapter 1 and Appendix A. Finally, these $e^{\pm}$ can be decelerated in the presence of magnetic fields and emit synchrotron radiation, which will be the focus of Part III. Fig. 2.2 illustrates the various electromagnetic signals we might expect as a function of the DM mass. WIMP candidates or very light particles, such as axions at the scale of the $\mu$eV, could produce radio waves. At the electronvolt scale, we expect infrared emission, whereas particles with a mass of the order of keV (like sterile neutrinos in some models [37, 48]) could produce X rays. Finally, DM particles with a mass in the range MeV−TeV can produce $\gamma$ and X rays as primary photons, but they can also produce high-energy electron-positron pairs. As previously described, the latter can produce X rays and $\gamma$ rays through inverse Compton scattering on the interstellar radiation fields in the galaxy, but they can also interact with the ambient magnetic fields and emit synchrotron radiation in the radio band. We refer to diffuse emission to indicate the radiation that does not originate from point-like sources and to the cumulative radiation emitted by unresolved sources. We expect annihilating and decaying DM to contribute to this diffuse emission, though various alternative astrophysical processes can contribute as well, representing our astrophysical background. In the next paragraphs, the characteristic features of the different wavelengths will be discussed, bearing in mind that different photons not only are often produced by different mechanisms, but they also require different detection techniques. One main reason is the presence of gas in our atmosphere, which leads to the absorption of certain electromagnetic frequencies. Fig. 2.3 illustrates the atmospheric opacity as a function of the wavelength. The gas absorption is mainly associated with water vapour, carbon dioxide and ozone. Ultraviolet light is greatly absorbed by ozone, while carbon dioxide and water vapour mainly affect infrared radiation. Optical, near-infrared, millimetre and radio bands are the so-called atmospheric windows, namely regions of the electromagnetic spectrum that are visible from Earth. Astronomical observations of mid- and far-infrared, ultraviolet, X and $\gamma$ rays require space-based telescopes, though some detection techniques have been implemented to study ultra high-energy photons from Earth as well. The wavelengths relevant for this thesis are $\gamma$ rays, X rays and radio waves.

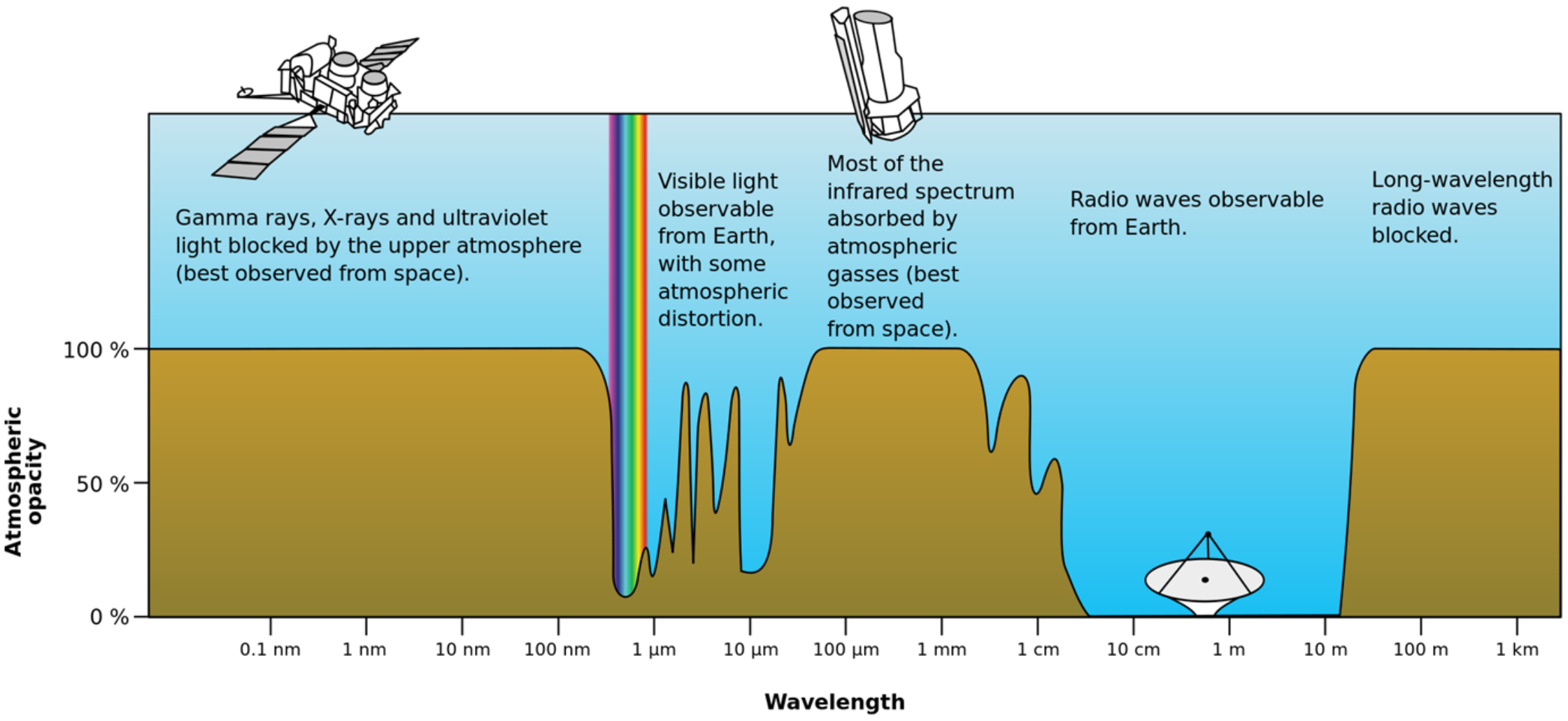

**Fig. 2.3.** Atmospheric opacity as a function of the wavelength. Our atmosphere is transparent to optical light as well as to near-infrared, millimetre and radio emission. Instead, $\gamma$ and X rays, ultraviolet radiation, long radio waves, mid- and far-infrared emissions need to be observed in space because of the gas absorption. Credit: NASA.

#### 2.1.1.1 $\gamma$ rays

Photons with energies above 100 keV are known as $\gamma$ rays. The detection of radiation above the ultraviolet wavelength is arduous due to the absorption by molecules in the atmosphere. Depending on their energy, we can classify these photons within three categories:

- low-energy (MeV regime) : 0.1 MeV < $E_\gamma$ < 30 MeV ,
- high-energy (GeV regime) : 30 MeV < $E_\gamma$ < 100 GeV ,

- very high-energy (TeV regime) : $E_\gamma > 100$ GeV .

In principle one could also take into account ultra-high-energy photons with $E_\gamma > 100$ TeV (PeV regime), originated from ultra-high-energy cosmic ray sources, but this category is beyond the scope of this thesis and will not be treated. This division is instrumental considering that different regimes are associated to different dominant mechanisms of production and distinct detection techniques. Low-energy $\gamma$ ray mostly originate from inverse Compton scattering and bremsstrahlung emission, while for high-energy and very high-energy photons the leading production mechanism is the decay of neutral pions. In the MeV range, Compton interaction and pair production are equally possible, while in the GeV region the pair production becomes the dominant detection mechanism and at above 1 TeV electromagnetic cascades take over. Thus, we can detect primary $\gamma$ rays and photons with an energy below the TeV scale with satellites, while very high-energy $\gamma$ rays can be probed from the cosmic-ray showers observed with Earth-based detectors. In other terms, space-based and ground-based experiments are complementary detectors, which make use of different physical processes to investigate $\gamma$ rays in different energy regimes.

*Satellites*. One main advantage of space observatories is that they are unaffected by the absorption and distortion effects of the electromagnetic waves caused by the atmosphere, and they are not subject to the light pollution provoked by the artificial light on Earth. Thus, they are ideal to study $\gamma$ and X rays, but also infrared and ultraviolet radiation which are also widely halted by the atmosphere and by light pollution. A drawback is that they are characterised by a small effective area ( $\sim 1\text{m}^2$ at most), due to the cost of space technology (mainly for the launch). This feature limits their sensitivity, especially at high energy. In the TeV range, the flux is typically too small to be measured with an effective surface of order $1\text{m}^2$. Thus, satellites are ideal to study low-energy and high-energy $\gamma$ rays, while very high-energy photons can be probed with ground-based telescopes. The state-of-the-art detector in the MeV region is the Imaging Compton Telescope (COMPTEL) [49] on board of the Compton Gamma Ray Observatory (CGRB). It covered the energy range 1–30 MeV and it exploits the Compton effect. The Compton scattering consists of an incoming photon colliding with an electron at rest and transferring part of its energy $E_\gamma$ and momentum $p_\gamma$ to the charged particle. The detector measures the recoil energy $E'_e$ of the electron and the energy $E'_\gamma$ of the outgoing photon. Recalling the relation for the Compton scattering and for the conservation of energy,

$$\begin{cases} E_\gamma + m_e c^2 = E'_\gamma + E'_e & \text{Conservation of energy} \\ \dfrac{1}{E'_\gamma} - \dfrac{1}{E_\gamma} = \dfrac{1-\cos\alpha}{m_e c^2} & \text{Compton scattering}\,, \end{cases} \tag{2.7}$$

one can derive the initial energy $E_\gamma$ of the incoming $\gamma$ ray and the scattering angle $\alpha$ between the initial and the final directions of the photon. This is the technique underlying Compton telescopes.

Now, considering that COMPTEL was launched in 1991 and operated until 2000, the MeV region lacks recent and sensitive data. This fact goes under the name of "MeV gap" and it represents one of the motivations behind the work illustrated in Part II.

As concerns the GeV region, photons are mainly detected through pair production. This mechanism relates to the creation of a particle-antiparticle pair from a neutral boson. In this case it refers to a photon converting into an electron-positron pair in proximity to a nucleus. This conversion can occur only in the field of a nucleus and not in free space. The reason is that in vacuum it is not possible to simultaneously conserve energy and momentum. A photon of energy $h\nu$ will produce an electron and a positron, each with momentum $p_e = \gamma m_e v$ and energy equal to $\gamma m_e c^2$, where $v$ represents their velocity and $\gamma = 1/\sqrt{1-v^2/c^2}$ is the Lorentz factor. For the conservation of energy

$$h\nu = 2\gamma m_e c^2 \,. \tag{2.8}$$

The conservation of momentum requires

$$\begin{cases} p_\gamma = p_{e^+,x} + p_{e^-,x} = 2\,p_e \cos\theta = 2\,\gamma\, m_e\, v\, \cos\theta \\ 0 = p_{e^+,y} + p_{e^-,y} \end{cases} \tag{2.9}$$

where $p_\gamma$ is the momentum of the photon, while $p_{e^\pm,x}$ and $p_{e^\pm,y}$ denote the momentum of the $e^\pm$ along the horizontal and vertical axes, respectively, and $\theta$ is the angle between the photon and electron directions. Eq. (2.9) implies that

$$\frac{h\nu}{c} = 2\gamma\, m_e\, v\, \cos\theta \,. \tag{2.10}$$

Thus, Eqs. (2.8) and (2.10) can be simultaneously verified only if $\cos\theta = 1$ and $v = c$. Since massive particles cannot travel at the speed of light in vacuum, this implies that the momentum and the energy of the initial photon cannot be simultaneously conserved in free space. Thus, the presence of a nucleus is necessary so that it can absorb a fraction of the initial photon's energy and momentum. For high-energy and very high-energy $\gamma$ rays, pair production is the leading kind of interaction with matter. An example of detector that utilises this physical process is the Energetic Gamma Ray Experiment Telescope (EGRET), aboard the CGRB. It collected data between 30 MeV and 30 GeV, and provided the first all-sky $\gamma$-ray map for high-energy photons. Its successor is the Fermi Gamma-Ray Space Telescope (usually simply called "FERMI") [50]. Launched in 2008, FERMI is the largest $\gamma$-ray space-based telescope ever built. The main detector on board is the Large Area Telescope[1] (LAT), which covers an energy range roughly between 20 MeV and 300 GeV. FERMI-LAT has an effective area of $1\text{m}^2$ and a large field of view of about 2.4 sr.

*Cherenkov telescopes.* Ground-based observatories are ideal to directly investigate the optical and radio windows, but they are also useful to indirectly probe very high-energy $\gamma$ rays through

[1] List of publications: https://www-glast.stanford.edu/cgi-bin/pubpub

Cherenkov radiation. When light propagates in a medium with refractive index $n$ greater than 1, its speed $v$ can be significantly lower than in vacuum, since it holds

$$v = \frac{c}{n}. \tag{2.11}$$

A particle that travels through such a material, faster than light, emits Cherenkov radiation. This effect can be exploited to study very energetic photons, which give rise to electromagnetic showers, through conversion into electron-positron pairs. Electromagnetic cascades usually start with a high-energy photon (or electron) which penetrates the atmosphere and produce electron-positron pairs in proximity of the nucleus of an atmospheric molecule. These secondary particles, in turn, produce photons via bremsstrahlung. A characteristic feature of these cascades is that they only include electrons, positrons and $\gamma$ rays. This aspect is convenient when it comes to distinguishing the arrival of a very high-energy photon from other particles such as pions, muons, etc. If the electrons and positrons of the electromagnetic shower travel faster than the speed of light in the air, they produce optical Cherenkov radiation, visible with our ground-based telescopes. Examples of Cherenkov detectors are the High Energy Stereoscopic System[2] (HESS) in Namibia, the Major Atmospheric Gamma Imaging Cherenkov telescope[3] (MAGIC) on the Canary Islands and the Very Energetic Radiation Imaging Telescope Array System[4] (VERITAS) in Arizona. We should keep in mind that these detectors measure the secondary particles in the cascades initiated by the primary $\gamma$ rays. Thus, they indirectly measure the very-high-energy radiation produced by DM particles or astrophysical sources. One drawback of these detectors is that they have a limited observational time, unlike satellites. The reason is that the signal produced from a very-high-energy $\gamma$ ray is faint (10 photons per square metre for 100 GeV primary photons), therefore data can be collected only when the sky is not bright (moonless time or moderate moonlight, no clouds). This requirement constrains the total observation time to 1000−1500 hours per year [51].

Concerning future missions covering the three energy ranges under consideration, the TeV region will be further explored by the Cherenkov Telescope Array[5] (CTA), currently under construction. It will be the most advanced Cherenkov detector ever built and it will allow us to study $\gamma$ rays from 20 GeV up to 300 TeV [53, 54, 55]. Developing a next-generation experiment in the GeV range with significantly better performances compared to Fermi is quite complicated. The main reason is that, with the current technologies, it would be extremely expensive to launch a new satellite of greater size on a space mission. Nevertheless, new satellites have been proposed, including the Gamma Astronomical Multifunctional Modular Apparatus[6] (GAMMA-400). This detector is designed to probe photons in the energy range between approximately 20 MeV and 1 TeV.

[2] List of publications: https://www.mpi-hd.mpg.de/hfm/HESS/pages/publications/
[3] List of publications: https://magic.mpp.mpg.de/backend/publications
[4] List of publications: https://veritas.sao.arizona.edu/the-science-of-veritas/publications
[5] List of publications: https://www.cta-observatory.org/science/library
[6] http://gamma400.lebedev.ru/public_e.html

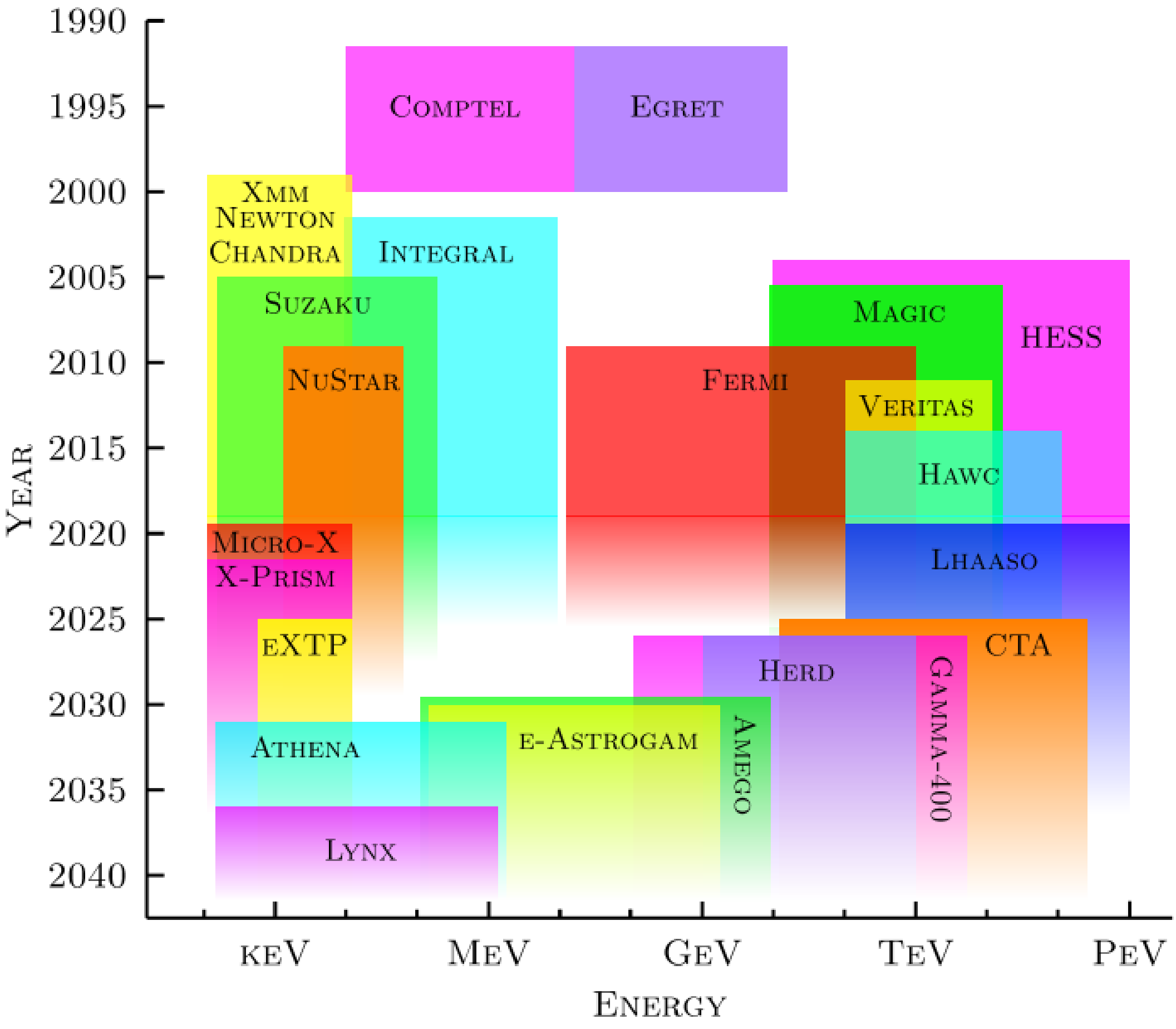

**Fig. 2.4.** A selection of $\gamma$-ray and X-ray telescopes distributed on a 2D plane, according to their mission dates (vertical axis) and approximate energy range covered by the detectors (horizontal axis). Different colours indicate different experiments. Credit: Rebecca K. Leane [52].

Regarding low-energy $\gamma$ rays, E-ASTROGAM [56] and AMEGO [57] are two proposed experiments, which would cover the MeV gap. A collection of $\gamma$-ray and X-ray detectors is displayed in Fig. 2.4. The vertical axis shows the mission dates, starting from 1990 up to 2040, and the horizontal axis indicates the approximate energy range probed by the different experiments.

After this discussion on the powerful detectors at our disposal, it is time to understand where to point them in the sky. The most promising targets for DM searches are:

*Galactic Centre*. The Galactic Centre (GC) is considered a promising target for DM searches for two main reasons: its proximity to Earth (approximately 8 kpc) and the fact that the DM density is expected to be higher towards the centre of the Milky Way. This line of argument is relevant especially if the DM profile is cusped. However, this region is also rich in astrophysical sources, making it challenging to disentangle the DM emission from the background. The situation gets even more complicated at the GeV scale since the interactions of cosmic rays with the dense gas component in the central part of the Milky Way result in a bright diffuse background. Refs. [58, 59] studied the morphology and energy spectrum of the $\gamma$-ray flux in a narrow region around the centre of the Milky Way, using the data of the FERMI-LAT telescope. They discovered a bump-like feature that peaks at energies around 2−4 GeV and interpreted this diffuse $\gamma$-ray excess in terms of emission from WIMP particles. Several groups have confirmed this observation and tried to give a conclusive explanation [60, 61, 62]. At the time of writing, the leading interpretation is the presence of an unresolved population of millisecond pulsars in the Galactic bulge [63, 61, 64, 65, 62]. Nature sends often ambiguous messages (and messengers in this case). After more than 10 years, the GC excess remains a puzzle and we will probably have to wait until the next-generation missions like CTA [66] to have a final say about this fascinating observation. The GC is considered an interesting target also because of the observation of an enigmatic 511 keV line, likely due to the annihilation of electron-positron pairs. This emission was first observed in the 1970s [67, 68, 69] and later confirmed by more recent measurements [70, 71]. It appears brighter in the bulge of our Galaxy and it has triggered many scientists in the attempt of identifying the source of the positrons which are at the origin of this peculiar line. Annihilation of light DM, with a mass order of MeV, has been suggested [72, 73, 74]. Alternative explanations include the merger of neutron stars [75], low-mass X-ray binaries [76] and the $\beta^+$ decay of elements produced in the nuclear reactions of stars [77, 78, 79]. At present time the origin of the 511 keV line remains a mystery to be solved.

*Dwarf spheroidal galaxies*. Dwarf spheroidals (dSphs) represent another interesting target. They are galaxies with a high mass-to-light ratio, thus they represent an ideal astrophysical system for DM searches. They are characterised by a scarcity of gas with no indication of recent star formation and approximately spherical shape. An additional advantage of dSphs is that they are satellite galaxies of the Milky Way, so they are relatively close to us. Moreover, since most of them are at high Galactic latitudes, the astrophysical background is much lower compared to the GC. The leading background for dSphs consists of the diffuse extragalactic emission. However, a drawback of this target is that they are extremely hard to find because of their low luminosity. Table 2.1 specifies some of the dSphs which are considered the best observational candidates for DM searches on the basis of their mass-to-light ratio, luminosity and distance from us. Stringent constraints on the DM cross-section has been derived using data on dSphs measured by the FERMI-LAT telescope [80]. However, they have been considerably weakened by a recent analysis [81] of systematics, associated with the uncertainties on the DM profile. Additional dSphs are expected to be detected by the Large Synoptic Survey Telescope (LSST) [82], which will likely improve the current DM constraints.

*Galaxy clusters*. As described in Chapter 1, galaxy clusters were the first system where the existence of a huge amount of invisible matter in the Universe was noticed. They represent the most massive bound DM structures, but they are quite distant from Earth. Examples of the closest galaxy clusters, which reside at 100−1000 Mpc from us, are the Coma Cluster, the Virgo, Fornax and Hercules clusters. The observations of $\gamma$ rays in these massive systems can be associated with DM particles or to a vast population of high-energy cosmic rays of astrophysical origin (e.g. active galactic nuclei and supernovae). Being able to discriminate between these two components is crucial. In this regard, a guideline principle is to look for differences in the astrophysical and DM gamma-ray spectrum and in their spatial distribution. In addition, numerical simulations and DM models predict the existence of substructures within the galaxy clusters [83, 84, 85, 86, 87], which could enhance the overall DM signal. The trade-off is a higher uncertainty in the results, since the size of the boost effect induced by the subhalos is still under debate [88, 89].

| dSph | $D_\odot$ (kpc) | $L$ ($10^3\ L_\odot$) | $M/L$ |
|---|---|---|---|
| Segue 1 | 23 | 0.3 | >1000 |
| UMa II | 32 | 2.8 | 1100 |
| Willman 1 | 38 | 0.9 | 700 |
| Coma Berenices | 44 | 2.6 | 450 |
| UMi | 66 | 290 | 580 |
| Sculptor | 79 | 2200 | 7 |
| Draco | 82 | 260 | 320 |
| Sextans | 86 | 500 | 90 |
| Carina | 101 | 430 | 40 |
| Fornax | 138 | 15500 | 10 |

**Table 2.1.** A list of dwarf spheroidal satellites of the Milky Way. The columns include, in order of appearance: name of the galaxy, distance from the Sun, luminosity, mass-to-light ratio [51].

*Diffuse emission*. The diffuse emission is of two origins: galactic and extragalactic. The leading contribution in the galactic component is imputable to the production and propagation of cosmic rays. These particles travel in the interstellar space and undergo inelastic collisions with the ambient gas. Neutral pions are formed and subsequently decay into $\gamma$ rays. Alternatively, electrons and positrons can up-scatter low-energy photons or produce bremsstrahlung radiation. The overall diffuse emission includes an extragalactic component, as well. This is often referred to as the unresolved gamma-ray background and it mainly consists of astrophysical emissions. Examples of extragalactic sources which contribute to the unresolved gamma-ray background are blazars and star-forming galaxies. Further information on this subject will be provided in Part I.

The relevant targets for DM indirect detection are displayed in Fig. 2.5. The left-hand panel highlights the different scales (red and black inscriptions in the middle) along with the targets (black inscriptions on the right) and with the astrophysical systems providing evidence for the existence of DM in the Universe (purple inscriptions on the left). The right-hand panel consists of a 2D plane with the advantages and disadvantages of each target, and it beautifully summarises the earlier discussion. The vertical axis refers to the likelihood of measuring a strong signal, whereas the horizontal axis emphasises the robustness of the constraints in case no signal is detected. The GC is in the top-left corner since it is a highly dense region, expected to be the brightest source of $\gamma$ rays from annihilating DM. However, it also exhibits large uncertainties due to an overwhelming astrophysical background as well as an inadequate knowledge of the DM density in the innermost region. Galaxy clusters are associated with the greatest uncertainty and a relatively strong signal since they are expected to host a large population of subhalos, which boost the overall signal from annihilating DM, but the contribution of the substructures is extremely uncertain to date. Dwarf galaxies provide the most robust constraints, being DM dominated, essentially without an astrophysical background and very close to us, nevertheless the signal is very low. The diffuse emission is separated into galactic and extragalactic contributions. The former benefits from the proximity to Earth, unlike the extragalactic component which represents the most difficult to determine and includes a largely unknown astrophysical background. One purpose of this thesis is to improve our understanding of the extragalactic unresolved background, which will be discussed in Part I.

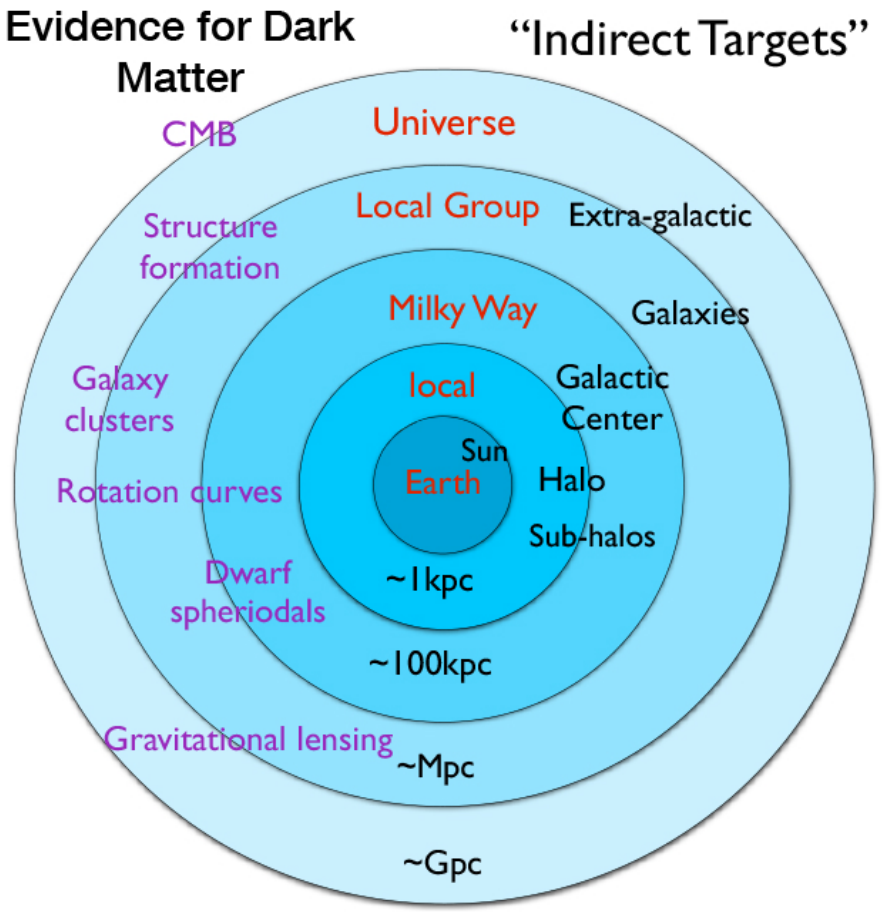

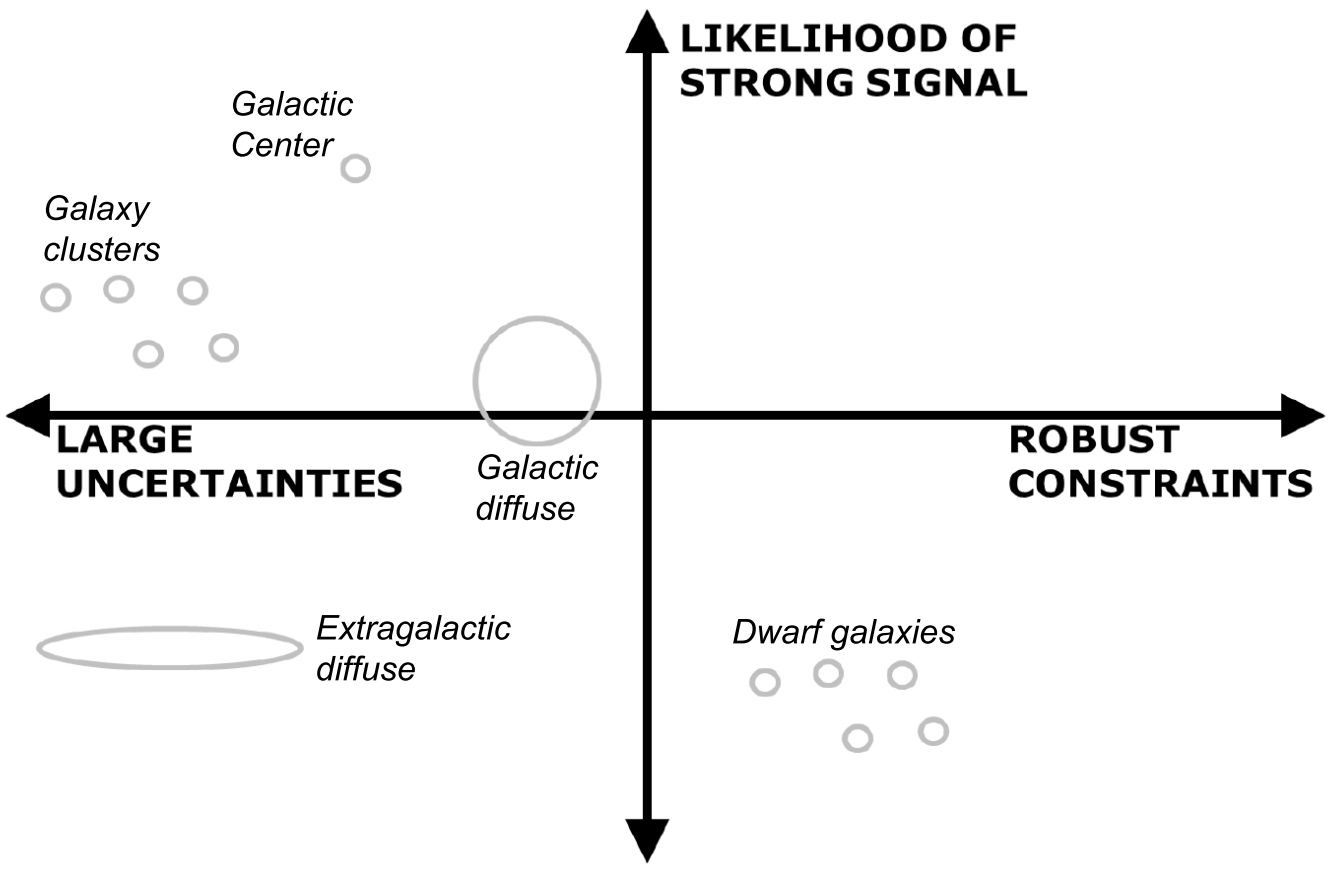

**Fig. 2.5.** *Right:* A schematic illustration showing the systems that provide evidence of the existence of DM (purple) and the indirect targets (black words on the right). The concentric circles highlight the different scales (red words at the top and black words at the bottom). Credit: adapted figure from [90]. *Left:* The targets for DM indirect detection are qualitatively displayed on a plane, according to the likelihood of detecting a strong signal (y-axis) and the robustness of the constraints (x-axis). Credit: adapted figure from [91].

#### 2.1.1.2 X rays

Photons with energy between 0.1 keV and 100 keV are known as X rays. In analogy with the $\gamma$-ray discussion, we can subdivide the X rays into two categories:

- soft X rays: 0.1 keV $< E_\gamma <$ 10 keV,
- hard X rays: 10 keV $< E_\gamma <$ 100 keV.

X rays are prevented from reaching the Earth's surface as a result of the photoelectric absorption caused by the atmosphere (see Fig. 2.3). During the photoelectric effect, a photon is absorbed by a bound electron and the energy is used to release the electron from the atomic potential. Other mechanisms of interaction with matter include the Compton scattering and the Rayleigh scattering. The former is the leading interaction process for hard X rays and it consists of an inelastic interaction between the photon and a bound electron, where the photon transfers part of its energy to the charged particle. Rayleigh scattering is an elastic interaction in which the incident photon provokes a resonant oscillation of the electron, which emits radiation in all directions at the same frequency of the incoming electromagnetic wave. This effect concerns more soft than hard X rays. However, photoelectric absorption represents the leading interaction mechanism for low-energy X rays. The dominant astrophysical sources in this band are active galactic nuclei, whose surrounding accretion disk is hot enough to emit bremsstrahlung radiation. Historically one of the most important experiments for X-ray searches was the UHURU satellite[7], which surveyed the sky in the 2−20 keV range. It located numerous X-ray sources, in particular binary systems containing a neutron star or a stellar black hole [92]. Following missions, such as the Roentgen Satellite[8] (ROSAT) in the 1990s, the ongoing X-ray Multi-Mirror Mission[9] (XMM-NEWTON) and Chandra X-ray Observatory, explored the soft X-ray spectrum. Measurements in the hard X-ray waveband were performed for instance by the Suzaku satellite[10], the Nuclear Spectroscopic Telescope Array[11] (NUSTAR) and the International Gamma-Ray Astrophysics Laboratory[12] (INTEGRAL). The latter is particularly relevant for this thesis. INTEGRAL is a mission of the European Space Agency which accommodates four instruments, including the spectrometer SPI. This experiment covers the energy range of hard X rays and soft $\gamma$ rays (27 keV – 8 MeV), and it will be employed to constrain MeV DM in Part II. Now, where do we want to point our telescopes? The promising targets for DM searches are essentially the same as $\gamma$ rays, since the guiding principle is always to look where the DM density is high. The relevant features that should be taken into account in multi-wavelength searches concern the astrophysical background, which may change for different messengers, as well as the production mechanisms of a DM signal, which depends for instance on the DM mass and types of interaction. Assuming that DM particles can annihilate into electron-positron pairs, these secondary charged particles can up-scatter the ambient radiation fields. Thus, in the case of MeV DM we can expect a contribution in the diffuse X-ray flux due to the inverse Compton scattering. This opportunity will be investigated in Part II. The main astrophysical background in this case will consist of the inverse Compton signal associated with electrons and positrons of astrophysical origin. Other relevant sources of X rays include the inner region of active galactic nuclei and clusters of galaxies. In these systems, the gravitational energy is converted into radiation through thermal bremsstrahlung, as described in Section 1.3 and Appendix A. It is worth mentioning that the X-ray waveband is also relevant to probe keV DM. An intriguing candidate in this mass range is the sterile neutrino since it can affect the X-ray spectra of galaxies and clusters. The most characteristic feature of sterile neutrino DM is its radiative decay emission [93, 94, 95, 96, 97]

$$N \longrightarrow \nu + \gamma \,. \tag{2.12}$$

[7] List of publications: https://heasarc.gsfc.nasa.gov/docs/uhuru/bib/uhuru_biblio.html
[8] List of publications: https://hera.gsfc.nasa.gov/docs/rosat/newsletters/biblio10.html
[9] List of publications: https://www.cosmos.esa.int/web/xmm-newton/publications-menu
[10] List of publications: http://www.astro.isas.jaxa.jp/suzaku/bibliography/
[11] List of publications: https://heasarc.gsfc.nasa.gov/docs/heasarc/biblio/pubs/nustar_atel.html
[12] List of publications: https://www.cosmos.esa.int/web/integral/scientific-publications

The final photon will have an energy equivalent to half of the neutrino mass. Thus, if DM is made up of sterile neutrinos, we should observe such a monochromatic emission in the sky, especially in DM-dominated systems. In this regard, an unidentified 3.5 keV line has been measured by XMM-NEWTON in the spectra of galaxy clusters [98], in the Milky Way [99] and Andromeda [100] galaxies. This emission line has been detected by other telescopes as well, such as Suzaku [101, 102]. A fascinating interpretation is that this line originates from the decay of DM particles with a mass of 7 keV [103]. Alternative explanations have been explored, including atomic transitions, instrumental effects or statistical fluctuations, but none of them is conclusive. Upcoming data from eROSITA[13] and future missions like the Advanced Telescope for High ENergy Astrophysics (ATHENA) will hopefully unveil the origin of this emission.

#### 2.1.1.3 Radio waves

The radio waveband refers to wavelengths above approximately 1cm (which corresponds to frequencies below 30 GHZ and energies below 0.1 meV). Fig. 2.3 shows that the atmosphere is transparent to a wide range of radio waves, allowing their detection from the ground, while very long wavelengths (above approximately 10m) tend to be absorbed. Thus, the data in this band come from ground-based telescopes and balloon-borne experiments. The prevailing contributions in the radio waveband are believed to come from active galactic nuclei, emission lines of neutral hydrogen atoms (further details about this line will be provided in Part I) and synchrotron radiation. In this regard, electrons and positrons produced by annihilating (or decaying) DM particles can emit synchrotron radiation. This type of emission can cover a broad region of the electromagnetic spectrum (from radio waves to hard X rays) and occurs whenever a relativistic charged particle is radially accelerated, for instance a fast electron moving in a magnetic field. For magnetic fields with a size of $\mu$G and GeV electrons, this radiation falls in the radio band.

[13] List of publications: https://www.mpe.mpg.de/455814/documents

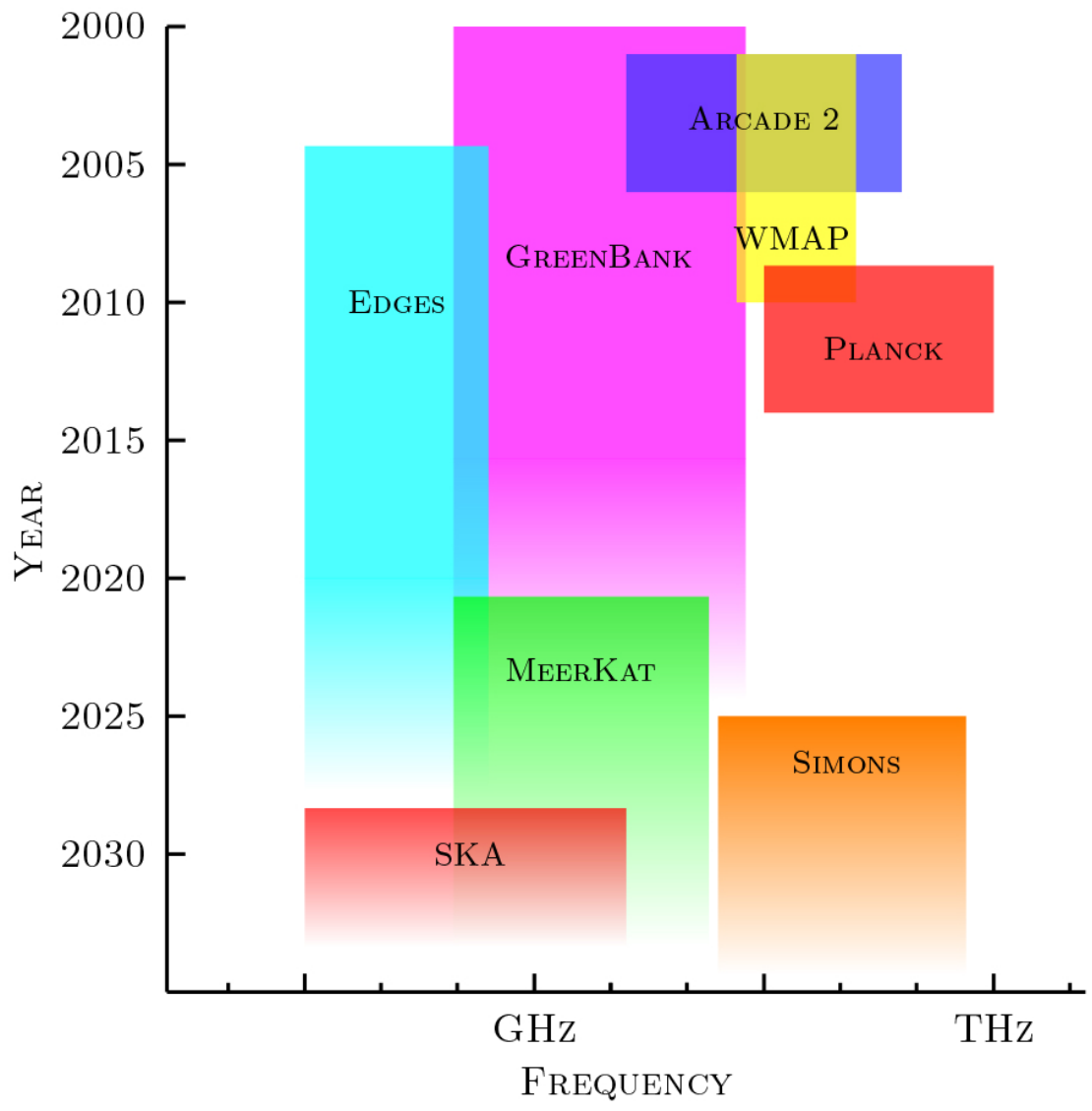

**Fig. 2.6.** A selection of radio and microwave telescopes distributed on a plane, according to their mission dates (vertical axis) and approximate frequency range covered by the detectors (horizontal axis). Different colours indicate different experiments. Credit: Rebecca K. Leane [52].

As discussed in Section 2.1.1.1, astrophysical systems with a high DM density represent desirable targets. In this regard, the GC has been the object of many investigations [104, 105, 106, 107, 108, 109], as well as the galactic halo and the substructures of the Milky Way [110, 111, 112]. It is worth mentioning that in the case of the GC, there is a huge uncertainty on the magnetic field profile. Nearby galaxies, including dwarfs spheroidals, offer a valuable proxy of a synchrotron signal from DM particles [113, 114, 115, 116, 117, 118, 119, 120, 121, 122, 123, 124, 125, 126, 127, 128], although they are affected by an uncertainty on the modelling of diffusion, which tends to be large in galaxies resulting in a lower radio flux. Instead, the diffusion effect is not significant in clusters of galaxies, making them an interesting target [129, 130, 131, 132, 133, 134]. However, nearby clusters are still very far from us compared to the typical distance of nearby galaxies, thus our telescopes will observe a small flux. This is life: you gain from one side and you lose from another. Hopefully, the increased sensitivity of next-generation telescopes will overcome this restraint. Extragalactic halos gained great interest after the detection of an isotropic radio excess by the Absolute Radiometer for Cosmology, Astrophysics and Diffuse Emission (ARCADE) experiment[14]. In particular, in 2009 the ARCADE 2 Collaboration found that the extragalactic radio background is much brighter than what is expected from a population of radio point sources [135] in the frequency range 3−90 GHz. A signal of astrophysical origin such as radio loud active galactic nuclei and star-forming galaxies, seems to be unlikely [136]. This emission appears to be compatible with DM synchrotron radiation from extragalactic halos [137, 138, 139, 140]. However, alternative explanations of the ARCADE 2 excess have also been proposed, such as merging of galaxy clusters and fast radio transients [141, 142]. At the time of writing, this signal remains a puzzle. The data from next-generation radio telescopes with high resolution, like the Square Kilometre Array, will likely shed some light over the origin of this interesting excess. In this regard, Fig. 2.6 shows a collection of radio (and microwave) experiments on a bidimensional plane, where the vertical axis refers to the missions dates, while the horizontal axis specifies the range of observing frequencies. Finally, an enticing possibility is to observe diffuse filaments linking clusters of galaxies. These connective structures are very faint, thus extremely difficult to detect with our current telescopes. They are expected to comprise warm-hot intergalactic medium and magnetic fields, yielding to an acceleration of the cosmic rays and emission of synchrotron radiation. Part III presents the first-ever robust detection of the stacked radio emission from large filaments (1−15 Mpc), which connect pairs of nearby luminous red galaxies. The signal is compatible with synchrotron emission from the cosmic web, providing direct evidence of one of the cornerstones of our current understanding of the large-scale structure in the Universe.

[14] List of publications: https://asd.gsfc.nasa.gov/archive/arcade/publications.html

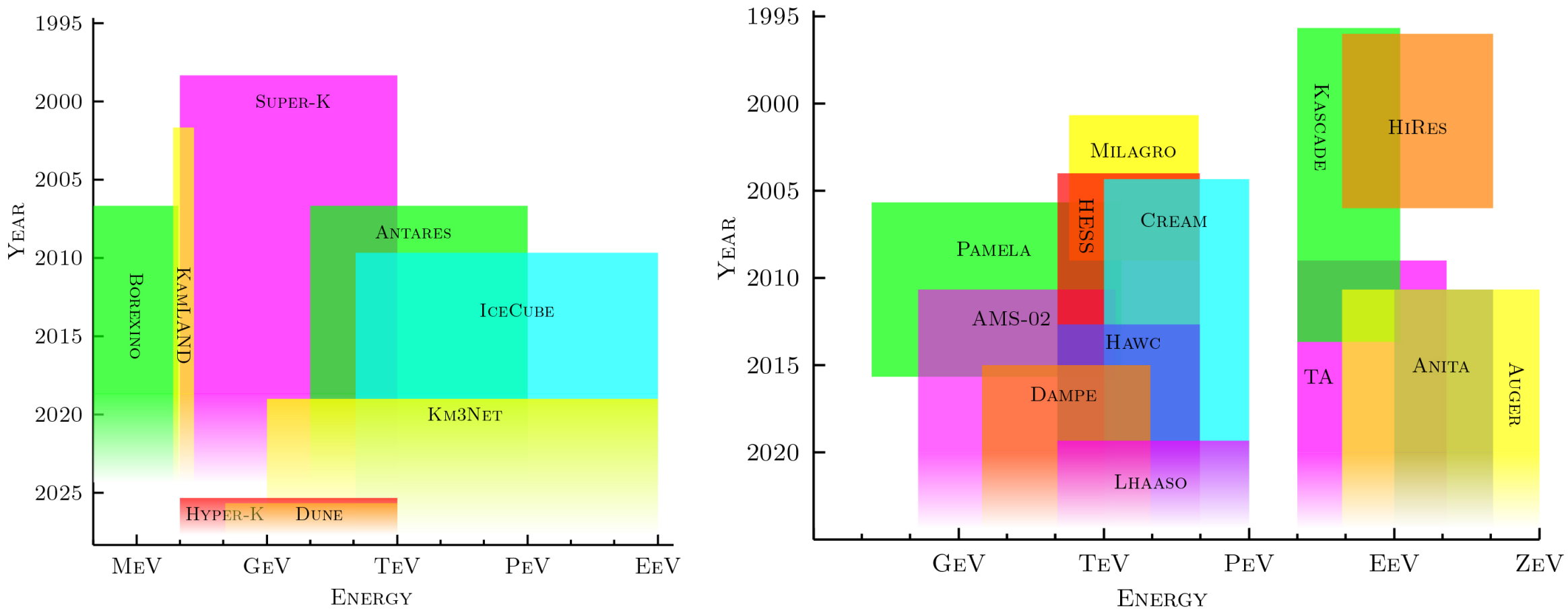

**Fig. 2.7.** A selection of neutrino (left panel) and cosmic-ray (right panel) detectors, distributed on a plane according to their mission dates (vertical axis) and approximate energy range covered by the experiments (horizontal axis). Different colours indicate different telescopes. Credit: Rebecca K. Leane [52].

### 2.1.2 Neutrinos

Like photons, neutrinos are not deviated by magnetic fields, therefore they trace back to their source. Neutrino detectors typically consist of a large volume of water or ice since they make use of the Cherenkov effect. When high-energy neutrinos interact with the detector medium they produce fast leptons: electrons in the case of $\nu_e$, muons in the case of $\nu_\mu$, tau particles in the case of $\nu_\tau$. Examples of neutrino experiments are the IceCube Neutrino Observatory[15] in the South Pole, involving approximately one cubic kilometre of ice, Astronomy with a Neutrino Telescope and Abyss environmental RESearch (ANTARES)[16] and Super-Kamiokande[17], which are water Cherenkov detectors in the Mediterranean sea and in Japan, respectively. Planned upgrades of these observatories leading to improved sensitivities are the Precision IceCube Next Generation Upgrade (PINGU), KM3NET[18] and Hyper-Kamiokande[19]. Fig. 2.7 on the left-hand side illustrates a collection of neutrino telescopes: the vertical axis delineates the mission dates, while the horizontal axis outlines the probed energy range.

Indirect detection with neutrinos can focus on the "traditional" targets used in photon-based searches such as the GC [143, 144], the Milky Way halo [145] and dwarf galaxies [146]. In this case, the differential flux has the same expression of Eq. (2.3), with the only caveat that here $\mathrm{d}N/\mathrm{d}E$ represents the energy spectrum into neutrinos. A relevant feature of neutrinos is that they are characterised by small interactions with matter. As a result, they can penetrate a larger volume of matter as compared to photons and charged cosmic rays. The positive scientist always looks on the bright side and takes the best out of every situation. In the case of neutrino searches, this translates into having two extra favourable targets for DM searches: the Sun and the Earth cores [147, 148, 149, 150, 151]. DM in the Milky Way halo can lose energy by interacting elastically with protons and nuclei in the Sun, thus becoming gravitationally bound to the Sun. When the capture rate and the annihilation rate reach equilibrium, the latter becomes independent of the self-annihilation cross-section and only depends on the DM-nuclei scattering cross-section (i.e. the capture cross-section). Thus, self-annihilating (or decaying) DM particles captured by the Sun can produce neutrinos, which can propagate and oscillate on their way to the Earth. The differential flux can be written as

$$\frac{\mathrm{d}\phi_\nu}{\mathrm{d}E} = \frac{\Gamma_A}{4\pi D^2}\frac{\mathrm{d}N_\nu}{\mathrm{d}E}\,, \tag{2.13}$$

where $\Gamma_A$ represents the DM annihilation rate, $D$ is the Sun-Earth distance and $\mathrm{d}N_\nu/\mathrm{d}E$ is the energy spectrum into neutrinos.

Similarly, DM particles in the GC can annihilate (or decay) into very high-energy neutrinos that can reach the Earth, interact with matter and some of them can be converted for instance into high-energy muons. The latter can produce Cherenkov light, measurable with our neutrino telescopes. One of the main advantages of using the Earth and Sun as targets is that they are essentially devoid of background, since they are not expected to be sources of high-energy neutrinos (except for the neutrino background originating from the interaction of cosmic rays with the solar atmosphere). In addition, the inner regions of these two targets are quite dense, so only neutrinos can easily propagate through them. As a result, the detection (or the missing) of an anomalous neutrino flux can be used to probe the DM-nucleon elastic cross-section [152]. Note that we can use this indirect probe in order to constrain DM-nucleon elastic cross-section, which is a typical observable of direct detection, discussed in Section 2.3.1.

[15] List of publications: https://icecube.wisc.edu/science/publications
[16] List of publications: https://antares.in2p3.fr/Publications/index.html
[17] List of publications: http://www-sk.icrr.u-tokyo.ac.jp/sk/pub/index.html
[18] List of publications: https://www.km3net.org/about-km3net/publications/pubblication/
[19] List of publications: https://www.hyperk.org/?page_id=59

IceCube has measured the TeV−PeV diffuse flux of neutrinos, whose origin is a puzzle [153, 154, 155, 156, 157, 158]. Numerous astrophysical sources have been proposed, such as blazars [159, 160, 161, 162, 163, 164, 165], supernovae [166, 167], galaxies [168], pulsar wind nebulae [169], $\gamma$-ray bursts [170, 171, 172, 173, 174, 175, 176, 177, 178, 179] and radio-bright active galactic nuclei [180, 181, 182], but none of these studies is able to explain this observation. An enticing alternative explanation is that the IceCube diffuse flux originates from DM particles [183, 184, 185, 186, 187, 188, 189, 190, 191, 192, 193, 194, 195, 196, 197]. Clearly neutrino experiments have still a lot of exciting science to uncover. Future data from next-generation detectors will hopefully be able to shed some light on this interesting observation and possibly open up new scientific horizons.

### 2.1.3 Charged cosmic rays

Cosmic rays usually refer to protons, antiprotons, electrons, positrons and light nuclei. Note that unbound neutrons are usually not considered since they are short-lived (they decay into protons after approximately 15 minutes). Cosmic rays can be detected with balloon experiments, satellites (especially at low energy), but they can also be studied indirectly through Cherenkov light using ground-based telescopes. Fig. 2.7 on the right-hand side illustrates a collection of cosmic-ray experiments. As in the previous figures, the vertical axis represents the mission dates and the horizontal line indicates the energy range covered by the detector. Considering that matter prevails over antimatter in the cosmic-ray spectrum and DM particles are usually expected to annihilate (or decay) into pairs of particle-antiparticle (thus, producing an equal quantity of matter and antimatter), DM searches with cosmic rays typically focus on antimatter. In this regard positrons, antiprotons and antideuterium have been the subject of numerous analysis [198, 199, 200, 201, 202, 203, 204, 205, 206, 207, 208, 209, 210, 211, 212, 213, 214, 215]. One of the main advantages of antimatter searches is that there is a relatively low background on Earth. However, indirect detection with charged cosmic rays have some significant disadvantages. Being charged particles, their direction is deflected by galactic and extragalactic magnetic fields, thus they do not point back to the source. Also, while propagating the cosmic rays can lose energy via several processes, depending on the type of cosmic ray under consideration. As a result, the energy spectrum observed with our telescopes may differ significantly from the one at source. Even with antimatter, it is crucial to take into account the astrophysical background. The Payload for Antimatter Matter Exploration and Light-nuclei Astrophysics[20] (PAMELA) detected an excess in the positron flux around 10 GeV [216]. This observation has been confirmed by other experiments, notably AMS-02 [217, 218], which also pointed out that this excess continues to the TeV scale with a cutoff approximately at 1 TeV. The most accepted explanation relates the positron excess to pulsar wind nebulae located near the Earth [219, 220, 221, 222, 223]. Alternative explanations in terms of DM particles have been put forward [224, 225, 226, 227, 228], but they are disfavoured. The main reason is that if DM was the cause of this excess, it should have also produced a $\gamma$-ray flux, which has not been detected by FERMI-LAT [229, 230]. Regarding antiprotons, they have been investigated as a possible signal to constrain the DM parameter space (see e.g. [231, 203, 204, 207, 232, 214]). The recent measurement of AMS-02 highlighted a potential excess in the $\overline{p}$ flux around 10 GeV, followed by several interpretations in terms of DM particles [233, 234, 235, 236, 237, 238, 239] In particular, a DM particle with a mass in the range 50−80 GeV annihilating into $b\overline{b}$ is able to explain both the antiprotons and GC excesses [233, 239]. A less alluring interpretation leads to cosmic-ray secondaries [240, 241]. Either way, this antiproton excess remains a quite debated topic because of the huge systematic uncertainties concerning notably the cosmic-ray propagation, the solar modulation and the antiproton production cross-section [237, 242, 239, 238, 243, 241]. Thus, a better comprehension of these systematics is required, with special consideration over the cosmic-ray propagation, in order to achieve a definitive theoretical explanation of the observed antiproton flux. Last but not least, antinuclei can be a promising probe, certainly worthy of investigation [244, 204, 245, 207, 246, 247]. In a nutshell, the main advantage is that antideuterium and antihelium at low kinetic energies are associated with a very low astrophysical background. A drawback is that the uncertainties on the production of antinuclei flux are quite large (both from DM and astrophysical sources). Future experiments, such as the General AntiParticle Spectrometer[21] (GAPS) balloon-borne experiment [248, 249, 250, 251, 252] and AMS-100 [253], will support us in the strenuous effort of searching for DM particles and identifying the origin of charged cosmic rays.

[20] List of publications: https://pamela-web.web.roma2.infn.it/?page_id=53

## 2.2 Cosmological indirect searches

In this section we discuss how DM can affect the abundance of primordial elements as well as the spectrum of the cosmic microwave background radiation. These are examples of cosmological indirect searches.

### 2.2.1 Big Bang Nucleosynthesis

According to the Big Bang model, during the first instants after the Big Bang the Universe can be described as a plasma of elementary particles. When the temperature drops below 150−200 MeV, around 1 $\mu$s after the Big Bang, quarks and gluons combine to form the first protons and neutrons. Therefore, we have the first nuclei of hydrogen and at this stage the Universe comprises a plasma mostly of photons, neutrinos, antineutrinos, electrons, positrons, protons and neutrons. The weak interaction is responsible for the chemical equilibrium between protons and neutrons. Typical processes that used to take place are

$$n \rightleftarrows p + e^- + \overline{\nu}_e \tag{2.14}$$

$$n + \nu_e \rightleftarrows p + e^- \tag{2.15}$$

$$p + \overline{\nu}_e \rightleftarrows n + e^+ \tag{2.16}$$

[21] List of publications: https://gaps.isas.jaxa.jp/publication.html

The temperature continues to decrease and when $T \sim 0.7$ MeV, the equilibrium between protons and neutrons ceases: the only significant process that remains is the decay of the neutrons, which increases the number of protons at the expense of the quantity of neutrons. At this stage the neutron-to-proton abundance is $n_n/n_p = 1/7$. Initially, the photodissociation prevents the formation of deuterium and consequently, of all the other light nuclei: this is the so-called "deuterium bottleneck". There is an abundance of photons compared to the number of protons ($n_\gamma/n_p \approx 10^9$) such that even when the average energy of photons is no longer higher than the deuterium binding energy ($B_D \simeq 2.2$ MeV), there are still some photons in the high-energy tail that can destroy the light nuclei. When the temperature drops below approximately 0.07 MeV, the Universe becomes cool enough for deuterium to survive and light nuclei can form. This phase is called Big Bang Nucleosynthesis (BBN). The first reaction to occur is

$$p + n \longrightarrow D + \gamma\,. \tag{2.17}$$

Since helium is more stable than deuterium, and among the light elements, $^4$He is associated with the highest binding energy per nucleon, we can safely assume that all the remaining neutrons form helium-4. The relevant nuclear reactions that occur are

$$\begin{aligned}
D + D &\longrightarrow {}^3\mathrm{He} + n \\
D + D &\longrightarrow {}^4\mathrm{He} + \gamma \\
D + p &\longrightarrow {}^3\mathrm{He} \\
D + n &\longrightarrow {}^3\mathrm{H} + \gamma \\
{}^3\mathrm{H} + p &\longrightarrow {}^3\mathrm{He} + n \\
{}^3\mathrm{H} + p &\longrightarrow {}^4\mathrm{He} + \gamma \\
{}^3\mathrm{H} + D &\longrightarrow {}^4\mathrm{He} + n \\
{}^3\mathrm{He} + n &\longrightarrow {}^4\mathrm{He} + \gamma \\
{}^3\mathrm{He} + D &\longrightarrow {}^4\mathrm{He} + p
\end{aligned} \tag{2.18}$$

Since neutrons are less abundant than protons, the number of neutrons determines the abundance of helium. In this regard, the abundance of light elements can be expressed in terms of the mass fraction

$$X_A = \frac{A\,n_A}{n_B}\,, \tag{2.19}$$

where $A$ represents the number of nucleons, $n_A$ the concentration of nuclei, $n_B$ the total number of baryons. The helium mass fraction, usually indicated with the symbol $Y_p$, is given by

$$Y_p = \frac{4n_4}{n_B} = \frac{4\frac{n_n}{2}}{n_p + n_n} = \frac{2\frac{n_n}{n_p}}{1+\frac{n_n}{n_p}} = \frac{1}{4}\,, \tag{2.20}$$

where in the second equality we used the fact that $^4$He = $2p+2n$ with $n_n/n_p = 1/7$ and the number of helium nuclei is half of the amount of neutrons. Thus, a good approximation is that 25% of baryons is made up of helium-4 and the remaining part consists of hydrogen nuclei. However, other relevant nuclear reactions that should be taken into account are

$$\begin{aligned} {}^4\text{He} + {}^3\text{H} &\longrightarrow {}^7\text{Li} + \gamma\,, \\ {}^4\text{He} + {}^3\text{He} &\longrightarrow {}^7\text{Be} + \gamma\,. \end{aligned} \tag{2.21}$$

Tritium and beryllium, being unstable isotopes, decay in $^3$He and $^7$Li, respectively. The formation of lithium-6 via the exothermic reaction

$${}^4\text{He} + D \longrightarrow {}^6\text{Li} + \gamma \tag{2.22}$$

can also occur, but it is suppressed relative to other processes, such as

$${}^4\text{He} + {}^3\text{H} \longrightarrow {}^7\text{Li} + \gamma\,. \tag{2.23}$$

Thus, BBN is responsible for the production of primordial deuterium, helium isotopes ($^3$He, $^4$He) and the stable lithium isotopes ($^6$Li and $^7$Li). Heavier elements are mostly produced in stars or supernovae. In the standard BBN scenario, the abundances of light nuclei depend solely on the baryon-to-photon number density ratio $\eta$, following the behaviour $\eta^{A-1}$. Thus, one can estimate $\eta$ by comparison of the theoretical prediction with the observations of the light nuclei abundances in metal-poor regions, where we expect low stellar nucleosynthesis (i.e. the abundance is expected to be closer to its primordial value). The range for $\eta$ inferred by $^4$He and $D$ is $5.8 \lesssim \eta/10^{10} \lesssim 6.6$. An alternative (and very precise) estimate of $\eta$ has been obtained by WMAP [254] and Planck [17]. The latest result from Planck [17] indicates $\eta = 6.09 \pm 0.06$, proving to be in good agreement with the standard BBN prediction. So far everything seems to be going the right way. However, the devil hides in the details. The prediction of $^7$Li abundance is a factor 3 higher compared to the value measured using absorption lines emitted by the photosphere of metal-poor stars. This is known as the lithium-7 problem [255]. An explanation in terms of systematic uncertainties seems very unlikely. This discrepancy may be the result of some stellar depletion mechanism, associated to the halo stars and to the low binding energy of lithium-7. A more intriguing explanation involves new physics. Indeed, DM can affect the mass fraction of the primordial elements by injecting the by-products of its decay or self-annihilation. These products can interact with light nuclei, mainly through fragmentation (spallation or photodissociation) and alter the final abundances of the light elements. It is convenient to distinguish between hadronic and electromagnetic channels. The former have impact for T $\lesssim$ 3 MeV and include mainly pions, nucleons and antinucleons, while the latter affect BBN when T $\lesssim$ 3 keV and consist of photons, electrons and positrons. Here are some examples of interactions due to the products of annihilating or decaying DM:

- High-energy nucleons can hit helium-4 via

$$N + {}^4\mathrm{He} \longrightarrow \begin{cases} N + 2p + 2n \\ N + D + D \\ D + {}^3\mathrm{He} \\ \dots \end{cases} \tag{2.24}$$

  The fragmentation of helium-4 can also occur via photodisintegration. These fragmentation processes result in a reduction of $^4$He as well as an increase in the abundances of deuterium and $^3$He.

- Charge pions can create extra neutrons in interactions like

$$\pi^- + p \longrightarrow \pi^0 + n\,, \tag{2.25}$$

  increasing the neutron-to-proton ratio and, accordingly, the helium mass fraction.

- Nuclei of helium-4 can be fragmented by very energetic protons and neutrons in spallation processes such as

$$\begin{aligned} n + {}^4\mathrm{He} &\longrightarrow D + p + 2n \\ n + {}^4\mathrm{He} &\longrightarrow {}^3\mathrm{H} + p + n\,, \end{aligned} \tag{2.26}$$

  which will raise the abundance of deuterium.

A distinctive feature of new physics is associated to an increase in lithium-6 abundance [256, 257, 258, 259, 260, 261], due to the injection of non-thermal $D$ and $^3$H in endothermic reactions such as

$$D + {}^4\mathrm{He} \longrightarrow {}^6\mathrm{Li} + \gamma \tag{2.27}$$

$${}^3\mathrm{H} + {}^4\mathrm{He} \longrightarrow {}^6\mathrm{Li} + n\,. \tag{2.28}$$

More importantly, the injection of secondary particles can enhance the conversion of beryllium-7 into lithium-7 via

$$n + {}^7\mathrm{Be} \longrightarrow p + {}^7\mathrm{Li}\,, \tag{2.29}$$

leading to the p-destruction of $^7$Li

$$p + {}^7\mathrm{Li} \longrightarrow {}^4\mathrm{He} + {}^4\mathrm{He} \tag{2.30}$$

which could solve the lithium-7 problem. However, this scenario also predicts that deuterium is much more abundant than we observe (measurements agree within 1% of standard BBN scenario). For this reason, the DM solution is disfavoured at present time and it is highly constrained by the comparison with the observed light abundances. The lithium-7 problem remains controversial and hardly explicable in standard BBN, despite this theory has a remarkable success in predicting the helium-4 and the deuterium abundances. The interested reader can find a comprehensive discussion of this fascinating topic in Refs. [262, 263, 264, 265].

### 2.2.2 Cosmic microwave background

Another important probe for cosmological indirect searches is the cosmic microwave background. At a certain point in the history of the Universe it was energetically favourable for electrons and positrons to combine and form neutral hydrogen. This moment, known as the epoch of recombination, occurred approximately 380 000 years after the Big Bang. These hydrogen atoms usually form with electrons in an excited state, due to the fact that the combination of protons and electrons into the lowest energy level (ground state) is not efficient. As a result, the electrons in the hydrogen atom will emit a photon in the energy transition to the ground state. Since neutral atoms do not interact with photons, the Universe becomes transparent to radiation and photons are free to travel without any significant interaction with matter: this is called photon decoupling. The resulting relic radiation goes by the name of cosmic microwave background (CMB) and it represents a unique probe of the early Universe. The set of points corresponding to the last interactions between photons and matter is known as the last scattering surface: it corresponds to the location where CMB originated at redshift z ∼ 1100. Thus, at the time of formation its wavelength was around 1100 times smaller and, accordingly, its energy was 1100 times higher. The temperature of the Universe was also higher: order of 3000 K [266, 267]. Before the decoupling, photons were kept in thermal equilibrium via the continuous interactions with electrons, that is why the CMB is characterised by a nearly perfect thermal black-body spectrum. The average temperature measured by the Cosmic Background Explorer[22] (COBE) satellite is (2.728 ± 0.004) K [268] and the peak of the spectrum at this temperature falls in the microwave band. Fig. 2.8 illustrates the intensity of the CMB as a function of the frequency. The black line refers to a thermal black-body spectrum with $T = 2.728$ K, while the data measured by the FIRAS instrument [269] on board of COBE are indicated by the points. The error bars have been multiplied by a factor of 500 to make them visible. This radiation is highly isotropic. However, precision measurements have pointed out the existence of small temperature fluctuations. These anisotropies are usually expressed by performing an expansion in spherical harmonics $Y_{\ell m}$ of

[22] List of publications: https://lambda.gsfc.nasa.gov/product/cobe/bibliography.cfm

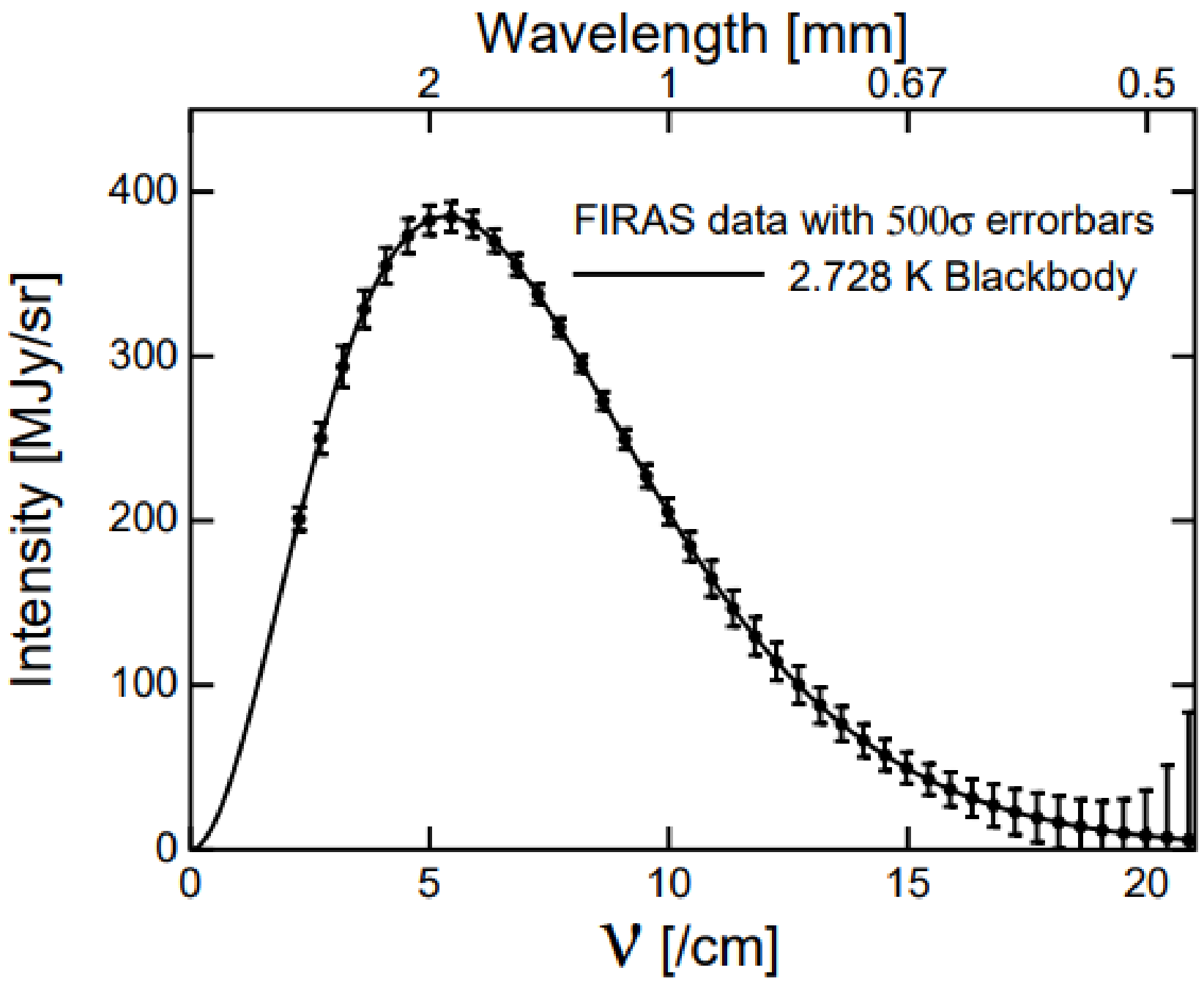

**Fig. 2.8.** Intensity of the CMB as a function of the frequency (lower x-axis) and wavelength (upper x-axis). The black line is a blackbody spectrum with $T$ = 2.728 K. The data points correspond to FIRAS data, with the error bars multiplied by 500. Credit: [270].

the CMB temperature spectrum:

$$T(\theta,\phi) = \sum_{\ell,m} a_{\ell m} Y_{\ell m}(\theta,\phi) \tag{2.31}$$

where $a_{\ell m}$ are the coefficients of the expansion. The largest anisotropy appears in the dipole (corresponding to the spherical harmonic $\ell = 1$) and its amplitude is approximately 0.1% of the mean temperature value [17]. The motion of an observer with velocity $\beta = v/c$ with respect to the rest frame of the CMB induces a Lorentz-boost in the temperature

$$T = T_0 \left[1 + \frac{v}{c}\cos\theta + \frac{1}{2}\left(\frac{v}{c}\right)^2 \cos 2\theta + O\left(v^3\right)\right], \tag{2.32}$$

where $T_0$ represents the unboosted temperature and $\theta$ is the angle between the line of sight and the direction of motion. The dipole pattern corresponds to the term proportional to $\cos\theta$ and it represents a frame-dependent term: it is associated with the Doppler effect due to our motion with respect to the frame in which the CMB appears isotropic. The term of order $v^2$ denotes the quadrupole anisotropy [271, 272]. It was first measured by the COBE instrument in 1992 [273], then by WMAP [254] and Planck [17], and it is about one part in $10^5$:

$$\frac{\Delta T}{T} \sim 10^{-5}\,. \tag{2.33}$$

In the Big Bang model the origin of the small thermal variations is associated with tiny quantum fluctuations of the matter density, which grew as a result of the gravitational attraction and led to the large-scale structure that we observe today. Thus, the anisotropies in the CMB represent a snapshot of the primordial density fluctuations which are responsible for the existence of clusters, galaxies, and so forth.

When the first stars formed, their burning processes produced ultraviolet radiation which re-ionised the hydrogen atoms. This phase is called reionisation era and it occurred in the redshift range $6 < z < 20$. Due to the expansion of the Universe, matter was quite far apart, making it more difficult for these photons to interact with the electrons. Thus, the Universe remained transparent. However, CMB photons may have scattered over the free electrons, leaving an imprint in the CMB spectrum. In particular, this effect could be quantified in terms of the Thomson scattering optical depth $\tau$, given by

$$\tau = \int_0^{z_i} \mathrm{d}z \, \frac{\mathrm{d}t}{\mathrm{d}z} \, \sigma_T \, n_e(z) \, , \tag{2.34}$$

where $z_i$ is the reionisation redshift, $\mathrm{d}t/\mathrm{d}z$ is fixed by the cosmology, $\sigma_T$ is the Thomson cross-section and $n_e$ represents the number density of free electrons. Moreover, the Thomson scattering experienced by an anisotropic radiation field induces a linear polarisation [274, 275, 276]. In this regard, the predicted polarisation in the CMB has been measured by the Degree Angular Scale Interferometer (DASI) [277].

The typical tool applied in the study of the CMB primary anisotropies is the angular power spectrum, defined as

$$C_\ell = \frac{1}{2\ell+1} \left\langle \sum_{m=-\ell}^{\ell} |a_{\ell m}|^2 \right\rangle \, , \tag{2.35}$$

where $\ell$ denotes the multipole and $a_{\ell m}$ are the coefficients of the expansion in spherical harmonics, as given in Eq. 2.31.

The presence of DM in the early Universe can affect the CMB spectrum. Annihilation or decay events of DM particles can give rise to relativistic $e^\pm$ which may heat the surrounding intergalactic gas as well as undergo inverse Compton scattering on the CMB photons which, in turn, can reionise the atoms of hydrogen. This reionisation will result into a higher optical depth [278, 279] as well as in distortions of the CMB temperature and polarisation spectra [280, 281, 282, 283]. The Sunyaev-Zeldovich effect also deserves some attention [284, 285]. This effect alters the CMB spectrum, as a consequence of the inverse Compton scattering induced by the hot electron gas in galaxy clusters. This effect can be adopted to constrain the DM parameter space: if DM injects high-energy electrons, this will contribute to the Sunyaev-Zeldovich effect and lead to spectral distortions in the CMB. Thus, it is possible to study DM particles by analysing the features of the CMB temperature and polarisation spectra. The interested reader can further investigate this fascinating subject by reading Refs. [286, 275, 287, 288, 289, 290]. A few additional comments about the CMB constraints on DM are also included in Part II, in comparison with the bounds on MeV DM derived by using the Galactic X-ray flux and the INTEGRAL data.

## 2.3 Other dark matter searches

### 2.3.1 Direct detection searches

DM particles in our Galaxy can interact with detectors and deposit some kinetic energy that could be measurable in our laboratories. Thus, the aim of direct detection searches is measuring the nuclear recoil energy due to the scattering of DM particles off target nuclei. Given the low interaction rate between DM particles and ordinary matter, the probability of multiple collisions within the detector is negligible. The differential recoil spectrum due to DM-nuclei interactions is given by

$$\frac{\mathrm{d}R}{\mathrm{d}E}(E,t) = \frac{\rho_0}{m_\chi m_N} \int d^3v\; v\, f(v,t)\, \frac{d\sigma}{dE}(E,v)\,. \tag{2.36}$$

It depends on the particle properties of the system: $m_\chi$, $m_N$ and $\mathrm{d}\sigma/\mathrm{d}E$, which represent the mass of the DM particles, the mass of the target nuclei and the differential cross-section, respectively. The astrophysical quantities involved are the local DM density $\rho_0$, the DM velocity $v$ and the DM velocity distribution in the detector reference frame $f(v,t)$, which is time-dependent as a consequence of the Earth's revolution around the Sun. In this regard, two peculiar features are predicted in a signal from DM particles in direct detection experiments [291]:

- an annual modulation of the recoil rate, related to the speed variation of the Earth moving along its orbit around the Sun,
- a daily modulation associated with the daily rotation of our planet.

Typically these detectors probe masses between few GeV and tens of TeV, thus they are particularly suitable for WIMP searches, and their sensitivity peaks for DM masses similar to the mass of the target nuclei. For instance, the elastic scattering of a WIMP particle with mass in the interval 10−100 GeV will produce a recoil in the energy range of 1−100 keV. The problem is that the expected DM event rate is subdominant with respect to the radiative background. In particular, the irreducible neutron background from environmental radioactivity can mimic a DM signal. The challenge is to discriminate between the background events and the signal from a new particle. In order to minimise as much as possible the background, experiments are built in underground laboratories, such as the Laboratori Nazionali del Gran Sasso (Italy). The material adopted in the detectors are typically Xenon, Germanium, Silicon, Iodine and Sodium. Examples of experiments are XENON [23] [292, 293, 294, 295, 296], the Large Underground Xenon experiment[24] (LUX) [297, 298, 299, 300], the DArk MAtter experiment[25] (DAMA) [301, 302, 303, 304, 305], the Cryogenic Rare Event Search with Superconducting Thermometers[26] (CRESST) [306, 307, 308, 309], the

[23] List of publications: http://xenon.astro.columbia.edu/XENON100_Experiment/Publications/
[24] List of publications: https://sites.brown.edu/luxdarkmatter/publications/
[25] List of publications: http://people.roma2.infn.it/ dama/web/publ.html
[26] List of publications: https://www.cresst.de/pubs.php

Cryogenic Dark Matter Search detector[27] (CDMS) [310, 311, 312, 313, 314] and Darkside[28] [315, 316, 317]. The DAMA Collaboration claims to have observed the seasonal variation in the measured event rate, compatible with a signal from WIMP particles. In particular, the flux from DM is predicted to reach its maximum around the 2nd of June and to become smaller in December, as a consequence of the orbital revolution around the Sun. The DAMA observation is puzzling in reason of the absence of any statistical significant excess above the background in other direct detection detectors [318, 295, 296, 299, 315, 319, 320, 321, 322].

### 2.3.2 Collider searches

Colliders are particle accelerators where two beams of particles are accelerated to very high energy in opposite directions and then coerced to collide. Numerous particles come out in the resultant inelastic scattering. An alluring prospect is that the impact of the two beams directly produces DM particles or generates some heavier particles beyond the SM that, in turn, decay into DM particles. The expected signature of a DM particle in a collider will be missing energy, especially in the case of WIMPs since they are neutral and weakly interacting (by analogy with neutrinos). Another possible signal would be missing energy plus a single photon or jet. However, any new particle produced in colliders has to be discovered in direct and/or indirect detection experiments as well, in order to be considered a suitable DM candidate. The reason is simple: even if we observe the signature of a particle with the properties expected for DM, we still need to find a matching signal coming from the astrophysical systems where we observe the DM in the Universe in order to prove that it is truly the particle we are looking for. Examples of colliders are the Large Hadron Collider (LHC) and the Double Annular Factory for Nice Experiments[29] (DAFNE). The former, located near Geneva (Switzerland), is currently the most powerful particle accelerator in the world and it mainly collides proton beams, while DAFNE is situated in the INFN Frascati National Laboratory (Italy) and represents an example of an electron-positron collider. Despite all the experimental efforts, there is no hint of any signature in current accelerators [323].

[27] List of publications: https://supercdms.slac.stanford.edu/publications
[28] List of publications: https://web.infn.it/darkside-bologna/index.php/en/publications
[29] List of publications: http://www.lnf.infn.it/acceleratori/dafne/pubs.html

# Part I

# Extragalatic hydrogen clouds rain $\gamma$ rays

CHAPTER 3

# Large-scale structure

Observing the large-scale structure (LSS) of the Universe can lead us to solve the mystery about the distribution and the nature of DM. N-body simulations and astronomical observations indicate that DM is hierarchical, and anisotropically distributed on small scales [324, 10, 325, 326, 327]. As a result, the electromagnetic (EM) signals produced by DM annihilation or decay events must show a certain degree of anisotropy since they trace the DM distribution. As discussed in Chapter 2, no signal unequivocally attributed to DM particles came out so far, even looking at promising targets where a strong DM signal and/or a low astrophysical background are expected. Thus, if DM particles annihilate or decay into $\gamma$ rays, this signal contributes to the unresolved $\gamma$-ray background (UGRB) [328]. This radiation background refers to the cumulative emission of the unresolved sources which are not sufficiently bright to be individually detected. It is expected to be mainly of extragalactic origin, including contributions from blazars, star-forming galaxies and misaligned active galactic nuclei, and it may also contain a subdominant DM signal. To a first approximation, the UGRB is isotropic, while at a deeper level anisotropies arise. Thus, we can look for an anisotropic EM signal from DM particles hidden among the anisotropies of the UGRB. This signal would be the evidence that DM is indeed a particle, annihilating (or decaying) into SM particles. Therefore, considerable efforts have been made in the past few years, in the hope of extracting such an elusive contribution. In order to discover a subdominant anisotropic DM signal in the UGRB, it is possible to work with statistical observables linked to the correlation of the spatial map of the signal, notably the two-point correlation function in real space (or its equivalent in Fourier space, the Fourier power spectrum) as well as the angular power spectrum in harmonic space. In the framework of indirect detection searches, there are three canonical ways to study signals due to anisotropic DM [329]: the auto-correlation of a single EM signal, the cross-correlation of two different EM signals, and finally the cross-correlation of an EM signal with a gravitational tracer of the DM distribution in the Universe. Among the three possibilities, likely the most promising option is cross-correlating an EM signal (an unambiguous indication of the particle nature of DM) with a gravitational tracer of the DM distribution, representing the evidence of the existence of DM in the Universe. A positive signal in this cross-correlation channel would be the evidence that DM is made up of elementary particles and it is not the exhibition of an alternative theory of gravity, as first proposed in Ref. [330]. Recently, many authors discussed and applied this technique to the cross-correlation of the UGRB anisotropy with numerous tracers of the LSS, such as cosmic shear [330, 331, 332, 333, 334, 335, 336], galaxy catalogues [337, 338, 339, 340, 341, 342, 343, 344, 345, 346], and CMB lensing [338, 347, 329]. In Chapter 4, we will propose a new promising gravitational tracer of DM: the neutral hydrogen distribution. Finally, in Chapter 5 we derive the forecasts for the cross-correlation signal between the brightness temperature of the 21-cm line emitted by neutral hydrogen atoms and the unresolved $\gamma$-ray background. The constraints on the DM annihilation cross-section are also obtained. This chapter is organised as follows: Section 3.1 introduces the two-point correlation function and the power spectrum as tools to probe the matter fluctuations; Section 3.2 discusses the linear power spectrum; the halo model framework and the non-linear power spectrum are illustrated in Sections 3.3 and 3.4, respectively.

## 3.1 Two-point correlation function and power spectrum

In accordance with the cosmological principle, we assume that all the cosmological random fields are statistically homogeneous and isotropic, namely the expectation values are invariant under global translations and global rotations. We also assume that the Universe has a smooth background with mean density $\overline{\rho}$. The fluctuations of the DM density field, at the position $\boldsymbol{x}$ and time $t$, relative to $\overline{\rho}$ are defined as

$$\delta(\boldsymbol{x}, t) = \frac{\rho(\boldsymbol{x}, t) - \overline{\rho}}{\overline{\rho}} , \tag{3.1}$$

known as the density contrast[1]. Thus, we can express $\rho(\boldsymbol{x}, t)$ in terms of the density contrast via

$$\rho(\boldsymbol{x}, t) = \overline{\rho}(t)\,[1 + \delta(\boldsymbol{x}, t)] . \tag{3.2}$$

The statistical tool employed to investigate the anisotropies in the matter density field is the $n$-point correlation function $\xi_n$, whose definition in real space is

$$\xi_n(\boldsymbol{x}_1, \ldots, \boldsymbol{x}_n) = \langle \delta(\boldsymbol{x}_1) \cdots \delta(\boldsymbol{x}_n) \rangle , \tag{3.3}$$

where $\langle \ldots \rangle$ denotes the ensemble average. We note that the one-point correlation function is $\xi_1 = \langle \delta \rangle$ and vanishes by definition of a random field: $\langle \delta \rangle = 0$. The two-point correlation function

[1] An alternative name in the literature for $\delta(\boldsymbol{x}, t)$ is density perturbation field.

(2PCF) $\xi_2$ will be one of the relevant observables in our analysis. In the following we will omit the subscript "2" to ease the notation. According to Eq. (3.3), the 2PCF is defined as the ensemble average of the density contrast at two different locations:

$$\xi(r) = \langle \delta(\boldsymbol{x}) \delta(\boldsymbol{x} + \boldsymbol{r}) \rangle \,. \tag{3.4}$$

Note that $\xi(r)$ depends only on the distance modulus $r = |\boldsymbol{r}|$, under the assumption of statistical homogeneity and isotropy. What is the physical interpretation of $\xi$ ? Let us consider a pair of objects enclosed in the volume elements $\mathrm{d}V_1$ and $\mathrm{d}V_2$, respectively, located at a distance $r$ from each other. The probability $\mathrm{d}P_{1,2}$ of finding two objects (for instance galaxies) at distance $r$ reads

$$\mathrm{d}P_{1,2} = \overline{n}^2 \,\mathrm{d}V_1 \,\mathrm{d}V_2 \,[1 + \xi(r)] \,, \tag{3.5}$$

where $\overline{n}$ denotes the mean number density of the objects under consideration. Eq. (3.5) shows that the 2PCF measures the excess over random probability to find these two objects at distance $r$, i.e. that these two objects are correlated. In other words, if two objects are randomly distributed (no correlation), $\xi = 0$ and the probability of finding a pair of objects depends only on the mean density squared $\overline{n}^2$, and not on the distance $r$ between them. A typical example is the clustering between galaxies. In this case, the correlation will asymptotically vanish at large distance, while two nearby galaxies usually tend to have a higher level of correlation, since they are likely gravitationally bound within the same cluster. Moreover, the 2PCF at $r = 0$ measures the variance of the field:

$$\xi(0) = \langle \delta(\boldsymbol{x})^2 \rangle = \sigma^2 \,. \tag{3.6}$$

The correlation function can be also analysed in Fourier space. For this purpose, we adopt the following conventions for the Fourier transform and the inverse Fourier transform:

$$\delta(\boldsymbol{k}) = \int \mathrm{d}^3 x \, \delta(\boldsymbol{x}) \, e^{-i\boldsymbol{k}\cdot\boldsymbol{x}} \,, \tag{3.7}$$

$$\delta(\boldsymbol{x}) = \int \frac{\mathrm{d}^3 k}{(2\pi)^3} \, \delta(\boldsymbol{k}) \, e^{i\boldsymbol{k}\cdot\boldsymbol{x}} \,. \tag{3.8}$$

Thus, the 2PCF in Fourier space reads

$$\langle \delta(\boldsymbol{k}) \, \delta^*(\boldsymbol{k}') \rangle = \int \mathrm{d}^3 x \int \mathrm{d}^3 x' \, \langle \delta(\boldsymbol{x}) \, \delta(\boldsymbol{x}') \rangle \, e^{-i\boldsymbol{k}\cdot\boldsymbol{x}} \, e^{i\boldsymbol{k}'\cdot\boldsymbol{x}'} \tag{3.9}$$

$$= \int \mathrm{d}^3 x \int \mathrm{d}^3 r \, \xi(r) e^{-i\boldsymbol{k}\cdot\boldsymbol{x}} \, e^{i\boldsymbol{k}'\cdot(\boldsymbol{x}+\boldsymbol{r})} \tag{3.10}$$

where $\boldsymbol{x}' = \boldsymbol{x} + \boldsymbol{r}$ and $\delta^*$ represents the complex conjugate of $\delta$. By recalling the definition of the Dirac delta[2]

$$\delta_D(\boldsymbol{k}_1, \ldots, \boldsymbol{k}_i) = \int \frac{\mathrm{d}^3 x}{(2\pi)^3} \, e^{-i\boldsymbol{x}\cdot(\boldsymbol{k}_1 + \cdots + \boldsymbol{k}_i)} \,, \tag{3.11}$$

[2]The subscript D is to avoid confusion with the density perturbation field $\delta(\boldsymbol{x})$.

Eq. (3.10) becomes

$$\langle\delta(\boldsymbol{k})\,\delta^\star(\boldsymbol{k}')\rangle = (2\pi)^3\,\delta_D(\boldsymbol{k}'-\boldsymbol{k})\int \mathrm{d}^3r\,\xi(r)\,e^{i\boldsymbol{k}'\cdot\boldsymbol{r}} \tag{3.12}$$

$$= (2\pi)^3\,\delta_D(\boldsymbol{k}'-\boldsymbol{k})P(k')\,, \tag{3.13}$$

where in the last equality we define the Fourier power spectrum (PS) as the Fourier transform of the 2PCF:

$$P(k) = \int \mathrm{d}^3r\,\xi(r)\,e^{i\boldsymbol{k}\cdot\boldsymbol{r}}\,. \tag{3.14}$$

According to the Wick theorem for Gaussian fields, all the correlations involving an odd number of density fluctuations vanish and the even-order correlations can be expressed in terms of the PS. Thus, the statistical properties of the matter fluctuations are completely determined by $P(k)$.

## 3.2 Linear power spectrum

Following Ref. [348], we can smooth the density perturbation field with a window function $W(\boldsymbol{x};R)$, characterised by a filter scale $R$:

$$\delta(\boldsymbol{x},R) = \int \mathrm{d}\boldsymbol{x}'\,\delta(\boldsymbol{x}')\,W(\boldsymbol{x}-\boldsymbol{x}',R)\,, \tag{3.15}$$

where the filter function is normalised through $\int \mathrm{d}^3x\,W(|\boldsymbol{x}|) = 1$. The convolution theorem states that a convolution in real space corresponds to a simple multiplication in Fourier space, implying

$$\delta(\boldsymbol{k},R) = \delta(\boldsymbol{k})\,\widetilde{W}(\boldsymbol{k},R)\,. \tag{3.16}$$

One of the most common filter functions is the top-hat window function

$$W(\boldsymbol{x},R) = \frac{3}{4\pi R^3}\begin{cases}1 & \text{for } |\boldsymbol{x}|\le R\\ 0 & \text{for } |\boldsymbol{x}|>R\end{cases} \tag{3.17}$$

whose Fourier transform reads

$$\widetilde{W}(kR) = \frac{3}{(kR)^3}\left[\sin(kR)-(kR)\cos(kR)\right]\,. \tag{3.18}$$

Given that $\delta(\boldsymbol{x},R)$ is defined as the convolution of the Gaussian field $\delta(\boldsymbol{x})$ with a top-hat filter, it follows that $\delta(\boldsymbol{x},R)$ is completely determined by its mean and variance. The former vanishes by definition of the density contrast, while the variance reads

$$\sigma^2(R) = \langle\,\delta^2(\boldsymbol{x},R)\,\rangle = \int_0^\infty \frac{\mathrm{d}k}{k}\,\frac{k^3\,P(k)}{2\pi^2}\,\left|\widetilde{W}(kR)\right|^2\,. \tag{3.19}$$

Inflationary models [349] typically predict a primordial PS generated by quantum fluctuations scaling as $P(k)\sim k^{n_s}$, where $n_s$ is the so-called spectral index. Thus, the initial PS at an initial time $t_i$ is usually parameterized as a scale-free, single power-law function. The parameter $n_s$

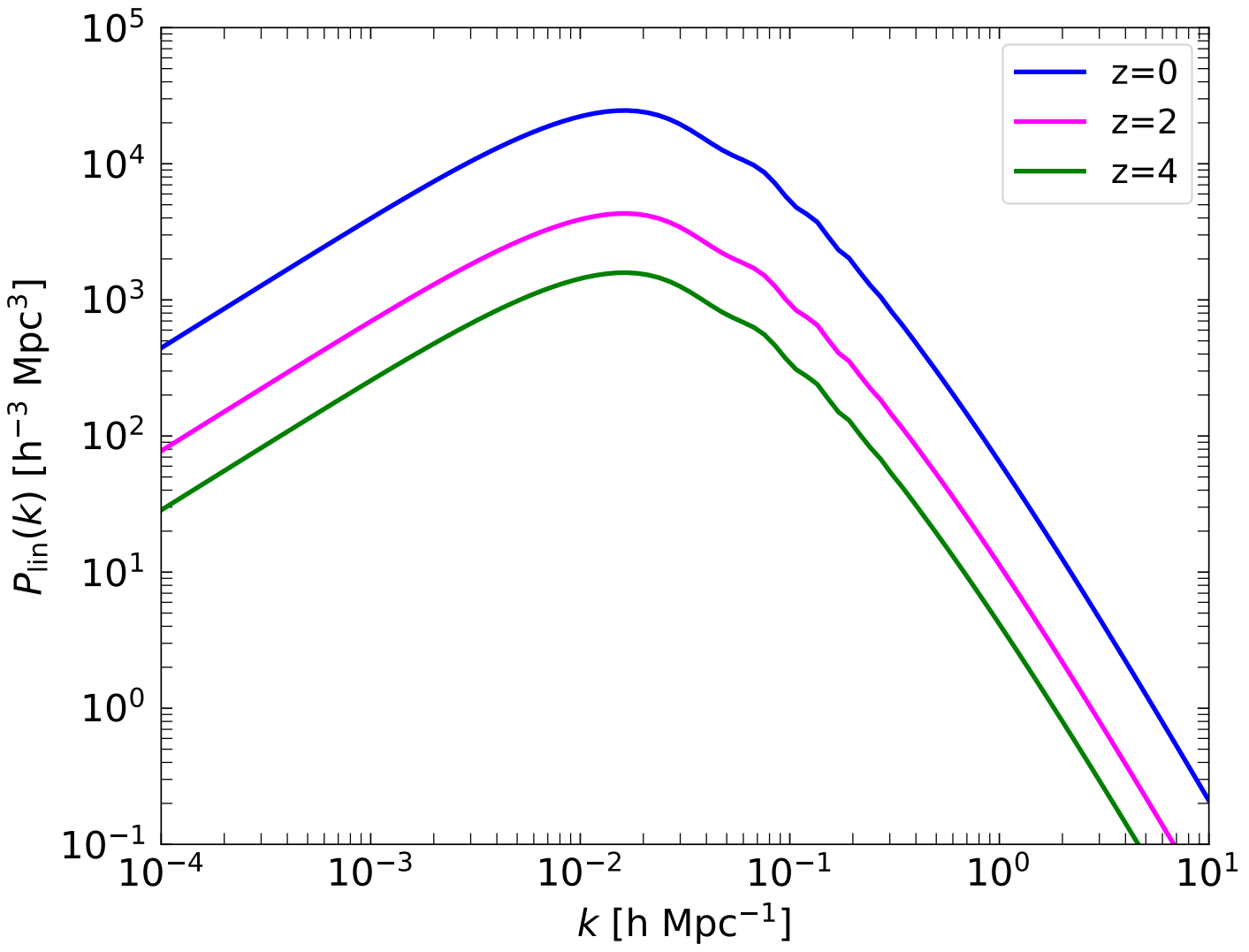

**Fig. 3.1.** Linear power spectrum as a function of the wavenumber for $z = 0$ (blue), $z = 2$ (magenta), $z = 4$ (green).

measures the departure from scale-invariance, namely

$$n_s(k) - 1 = \frac{\mathrm{d}\ln P(k)}{\mathrm{d}\ln k}\,. \tag{3.20}$$

Following the results of Ref. [350], we adopt $n_s = 0.96$. Fluctuations corresponding to different modes evolve independently, resulting in a modification of the matter PS which is encoded in the transfer function $T(k)$ [351] via

$$P(k) = A\,k^{n_s}\,T^2(k)\,, \tag{3.21}$$

where $A$ is the amplitude of the fluctuations. $T(k)$ is a decreasing function of $k$ and it outlines the evolution of fluctuations during the epoch of horizon crossing and the radiation-matter transition. It is normalised such that on large scales it is equal to unity: $T(k \to 0) = 1$.

In the linear regime $\delta \ll 1$, the linear PS is given by

$$\frac{k^3 P_{\rm lin}(k)}{2\pi^2} = \delta_H^2 \left(\frac{k}{H_0}\right)^{n_s+3} T^2(k)\,, \tag{3.22}$$

where $\delta_H = 4.2\cdot 10^{-5}$ is a normalisation constant [352]. We adopt the linear PS at $z = 0$ as given by the Code for Anisotropies in the Microwave Background (CAMB) [353]. Eq. (3.22) does not include the redshift evolution, which is encoded in the so-called growth factor [37]:

$$D_1(z) = \frac{5}{2}\Omega_m \frac{H(z)}{H_0}\int_z^\infty \mathrm{d}z\,(1+z)\frac{H_0}{H(z)}\,. \tag{3.23}$$

This function describes the growth of matter fluctuations and determines the evolution with $z$ of the linear PS

$$P_{\rm lin}(k,z) = P_{\rm lin}(k)\left(\frac{D_1(z)}{D_1(z=0)}\right)^2\,. \tag{3.24}$$

Fig. 3.1 illustrates the $P_{\rm lin}$ as a function of the wavenumber for three values of redshift: $z = 0$ (blue), $z = 2$ (magenta), $z = 4$ (green). The linear PS grows linearly with $k$ on large scales, then there is a turn-over around the scale corresponding to the matter-radiation equality $k_{\rm eq} \sim 0.01$ Mpc$^{-1}$, after which it declines as $k^{-3}$. Note that higher values of redshift are associated to a lower normalisation of the linear matter PS, because in the past there were fewer structures than today. It is noteworthy to mention that the linear regime applies to scales $k^{-1} > 10$ Mpc at $z = 0$, whereas on smaller scales we need to take into account the effects of non-linearities, as will be discussed in Section 3.4.

## 3.3 Halo model

The halo model is a useful formalism to study the spatial clustering of any object enclosed in a DM halo (galaxies, gas, etc.) and of DM itself [354, 355]. This model is instrumental to compute the PS in the non-linear regime, as will be explained in Section 3.4. In this section we will discuss the key idea behind this popular model and the major elements that characterise a halo.

Numerical simulations of structure formation within the ΛCDM paradigm show that initial small fluctuations grow linearly with the expansion of the Universe until they reach a critical density, after which they collapse and form a bound object, called halo [356]. Therefore, a halo is the outcome of an overdense region which undergoes a collapse as a consequence of gravitational instability. As cosmic time proceeds, these small halos continue to grow in mass and size by accreting nearby material or by merging with other halos, thus giving rise to increasingly larger halos which virialize[3]. Hereafter we assume that the halos form via spherical collapse [356, 357], according to which a structure collapses and virializes in a DM halo whenever the density contrast overcomes the threshold

$$\delta_{\rm sc}(z) \simeq 1.686(1+z)\,. \tag{3.25}$$

A schematic representation of this clustering process is illustrated in Fig. 3.2. The solid black line denotes the density contrast $\delta(\boldsymbol{x})$ at position $\boldsymbol{x}$ and the dashed horizontal line represents $\delta_{\rm sc}(z=0)$. Whenever the matter density fluctuations exceed $\delta_{\rm sc}$, a halo is formed, here represented by the red-coloured regions. A halo is associated to a characteristic size, known as the virial radius $R_{\rm vir}$. This quantity corresponds to the radius of a spherical volume with mean density $\rho_{\rm halo}$ at redshift $z$ given by

$$\rho_{\rm halo}(z) = \Delta_{\rm vir}(z)\,\rho_m(z)\,, \tag{3.26}$$

where $\Delta_{\rm vir}$ denotes the matter overdensity at virialization. For a flat universe with $\Omega_r = 0$, a convenient fitting function [358] for $\Delta_{\rm vir}$ is

$$\Delta_{\rm vir}(z) = \frac{18\pi^2 + 82x(z) - 39x^2(z)}{\Omega_m(z)}\,, \tag{3.27}$$

[3] The virialization corresponds to reaching virial equilibrium.

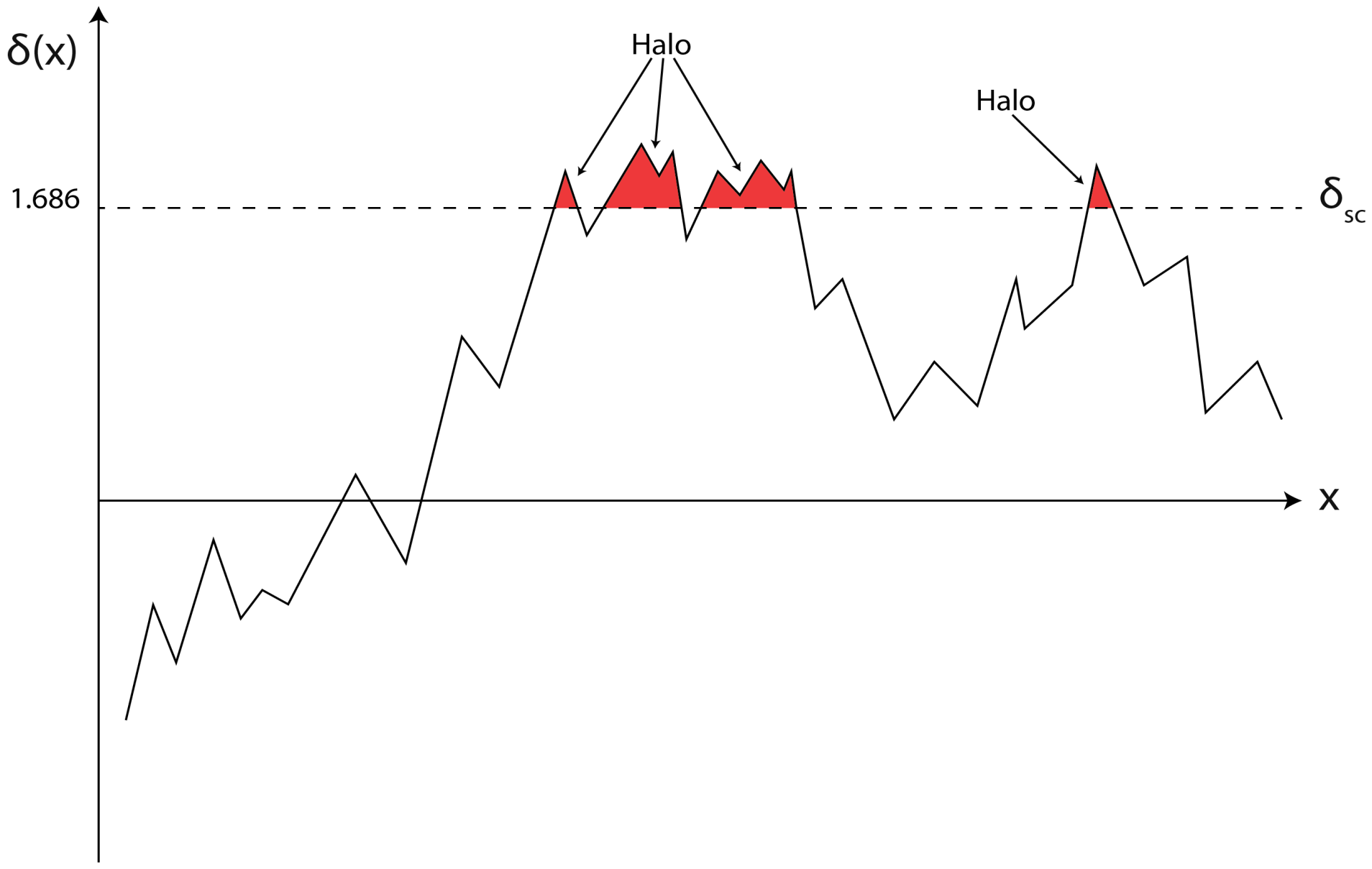

**Fig. 3.2.** Schematic representation of the halo formation. The solid black line represents the density fluctuation field $\delta(\boldsymbol{x})$ at the location $\boldsymbol{x}$, while the threshold overdensity $\delta_{\rm sc} = 1.686$ at $z = 0$ is displayed as a dashed curve. The red areas denote the overdensities which turn into halos.

where $x(z) = \Omega_m(z) - 1$. The mass enclosed within the virial radius, known as virial mass, reads

$$M_{\rm vir}(z) = \frac{4}{3}\pi R_{\rm vir}^3(z)\,\rho_{\rm halo}(z)\,. \tag{3.28}$$

In the following we indicate $M_{\rm vir}$ as $M$ to ease the notation. The halo model framework is based on the assumptions that all the mass in the Universe is partitioned into these distinct units and the distribution of mass can be divided into two levels: the mass distribution inside a single halo and the spatial distribution of the halos in the Universe. These two levels of the mass distribution translate into two contributions in the PS, as will be discussed in Section 3.4. To compute the non-linear PS, we need to introduce four main ingredients: the halo mass function, the halo concentration, the halo density profile and the halo bias. In the following, we will discuss each of these quantities.

### 3.3.1 Mass function

The halo mass function $\mathrm{d}n/\mathrm{d}M$ specifies the number density of halos of mass $M$ at redshift $z$. We adopt the model of Ref. [359], which expresses $\mathrm{d}n/\mathrm{d}M$ as follows:

$$\frac{\mathrm{d}n}{\mathrm{d}M}(M,z) = \frac{\overline{\rho}_0}{M}\,f(\nu)\,\frac{\mathrm{d}\nu}{\mathrm{d}M}\,, \tag{3.29}$$

where $\overline{\rho}_0$ denotes the background density at present time. The quantity $\nu$ reads

$$\nu(M,z) = \frac{\delta_{\rm sc}^2(z)}{\sigma^2(M,z)}\,, \tag{3.30}$$

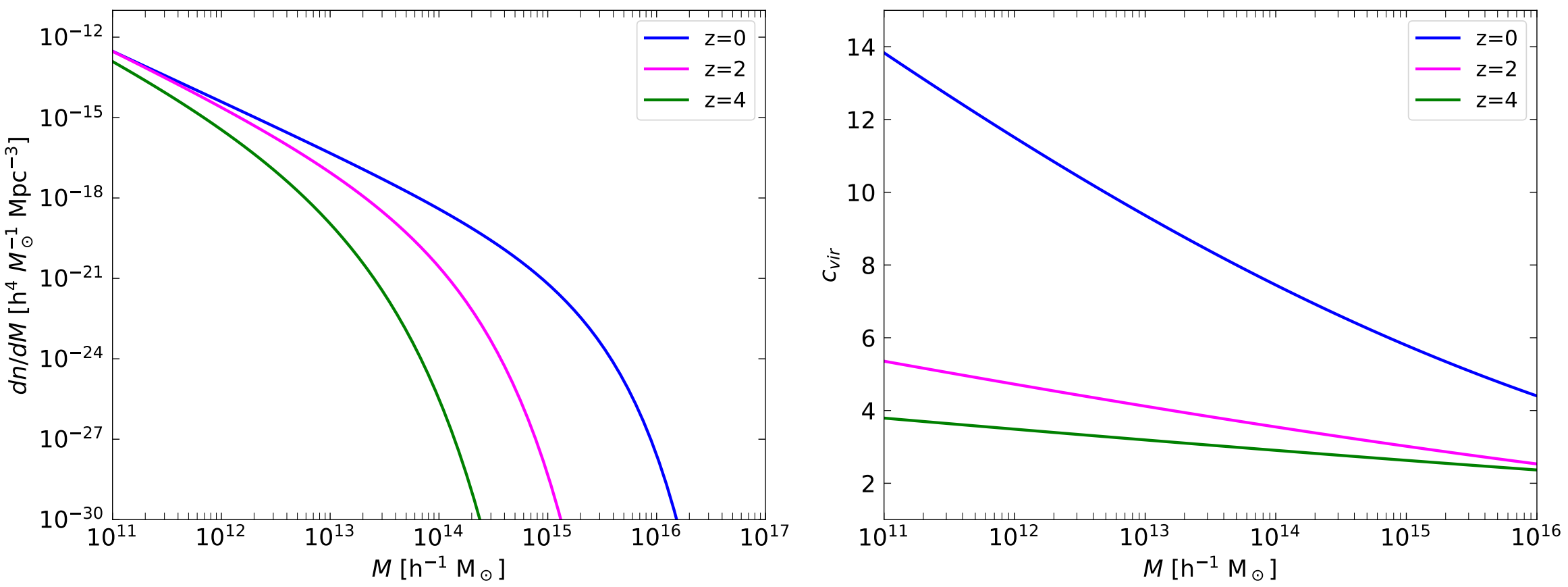

**Fig. 3.3.** Mass function (left) and concentration (right) as a function of the halo mass for $z = 0$ (blue), $z = 2$ (magenta), $z = 4$ (green).

where $\delta_{\rm sc}$ is the threshold for the formation of a halo, as given in Eq. (3.25) and $\sigma^2$ refers to the variance of the density perturbation, as given in Eq. (3.19). The so-called multiplicity function $f(\nu)$ specifies the fraction of mass within halos in units of $\mathrm{d}\ln\nu$ and is normalised to unity

$$\int_0^\infty \mathrm{d}\nu\, f(\nu) = 1\,, \tag{3.31}$$

such that the halo mass function is normalised to the background density of the Universe today:

$$\int \mathrm{d}M\, M \frac{\mathrm{d}n}{\mathrm{d}M} = \bar{\rho}_0\,. \tag{3.32}$$

We adopt the Sheth-Tormen fitting function [360], given by

$$\nu f_{ST}(\nu) = A(p)\,\left(1+(q\nu)^{-p}\right)\left(\frac{q\nu}{2\pi}\right)^{1/2}\exp\left(-\frac{q\nu}{2}\right) \tag{3.33}$$

with $p = 0.3$, $A(p) = \left[1+2^{-p}\,\Gamma(1/2-p)/\sqrt{\pi}\right]^{-1}$, $q = 0.75$ and $\Gamma$ denotes the gamma function. Fig. 3.3 (left) illustrates $\mathrm{d}n/\mathrm{d}M$ as a function of the halo mass for different redshift values: $z = 0$ (blue), $z = 2$ (magenta), $z = 4$ (green). We note that this function decreases with increasing mass, since massive halos are less abundant as a result of the hierarchical structure formation. Similarly, the halo mass function decreases with increasing redshift because there were fewer massive halos in the past.

### 3.3.2 Concentration

The halo concentration is necessary to compute the density profile, as will be explained in Section 3.3.3, and is defined as

$$c_{\rm vir} = \frac{R_{\rm vir}}{r_{-2}}\,, \tag{3.34}$$

here $r_{-2}$ denotes the radius where the density profile scales with logarithmic slope equal to −2 [361]. We adopt the semi-analytic model of Ref. [362]:

$$\log_{10} c_{200} = \alpha + \beta \log_{10}\left(\frac{M_{200}}{M_\odot}\right)\left[1+\gamma\left(\log_{10}\frac{M_{200}}{M_\odot}\right)^2\right], \tag{3.35}$$

where

$$\begin{cases} \alpha = 1.62774 - 0.2458\,(1+z) + 0.01716\,(1+z)^2 \\ \beta = 1.66079 + 0.00359\,(1+z) - 1.6901\,(1+z)^{0.00417} \\ \gamma = -0.02049 + 0.0253\,(1+z)^{-0.1044}\,. \end{cases} \tag{3.36}$$

This fitting function is valid for $0 \le z \le 4$, which is the redshift range we are interested in. We note that Eq. (3.35) provides

$$c_{200} = \frac{R_{200}}{r_{-2}}\,, \tag{3.37}$$

where $R_{200}$ represents the radius of a spherical volume with mean density 200 times larger than the background density[4]. In our analysis we work with virial quantities and therefore we need the corresponding concentration $c_{\rm vir}$, which is related to $c_{200}$ by [363]

$$c_{\rm vir} = a\, c_{200} + b\,, \tag{3.38}$$

where the coefficients $a$ and $b$ read

$$\begin{cases} a = -1.119\,\log_{10}\Delta_c + 3.537 \\ b = -0.967\,\log_{10}\Delta_c + 2.181 \end{cases} \tag{3.39}$$

and $\Delta_c(z) = \Delta_{\rm vir}(z)\,\Omega_m(z)$, as given by Eq. (3.27). However, the transition between $c_{200}$ and $c_{\rm vir}$ has an additional complication, since $c_{200}$ is a function of $M_{200}$, while $c_{\rm vir}$ depends on $M_{\rm vir}$. The two quantities are slightly different, as we can note from their definition

$$\begin{cases} M_{\rm vir} = \dfrac{4\pi}{3}\,\Delta_c(z)\,\rho_c(z) R_{\rm vir}^3 \\ M_{200} = \dfrac{4\pi}{3}\,200\,\rho_c(z) R_{200}^3 \end{cases} \tag{3.40}$$

where $\rho_c = 3H^2/8\pi G$. Thus, we can determine $M_{200}$ in terms of $M_{\rm vir}$ via

$$\frac{M_{200}}{M_{\rm vir}} = \frac{200}{\Delta_c(z)}\left(\frac{R_{200}}{R_{\rm vir}}\right) = \frac{200}{\Delta_c(z)}\left(\frac{c_{200}}{c_{\rm vir}}\right), \tag{3.41}$$

where in the last equation we have used the relation $R_{\rm vir}/R_{200} = c_{\rm vir}/c_{200}$ [363]. Fig. 3.3 (right) illustrates $c_{\rm vir}$ as a function of the halo mass for different values of redshift: $z = 0$ (blue), $z = 2$ (magenta), $z = 4$ (green). The concentration is a decreasing function of the halo mass. Indeed, if halos reflect the peaks in the initial fluctuation field, then massive halos have to be identified with higher peaks, which are less concentrated. As a result, massive halos are expected to have lower concentrations compared to low-mass halos. This could also be considered a corroboration of the hierarchical model of structure formation: more massive halos form later and so lower concentrations reflect the lower background density at the time of formation.

[4] It is useful to remind that two slightly different definitions of the overdensity with respect to the mean density background are typically used: $\Delta_c \simeq 178$ and $\Delta_{200} = 200$.

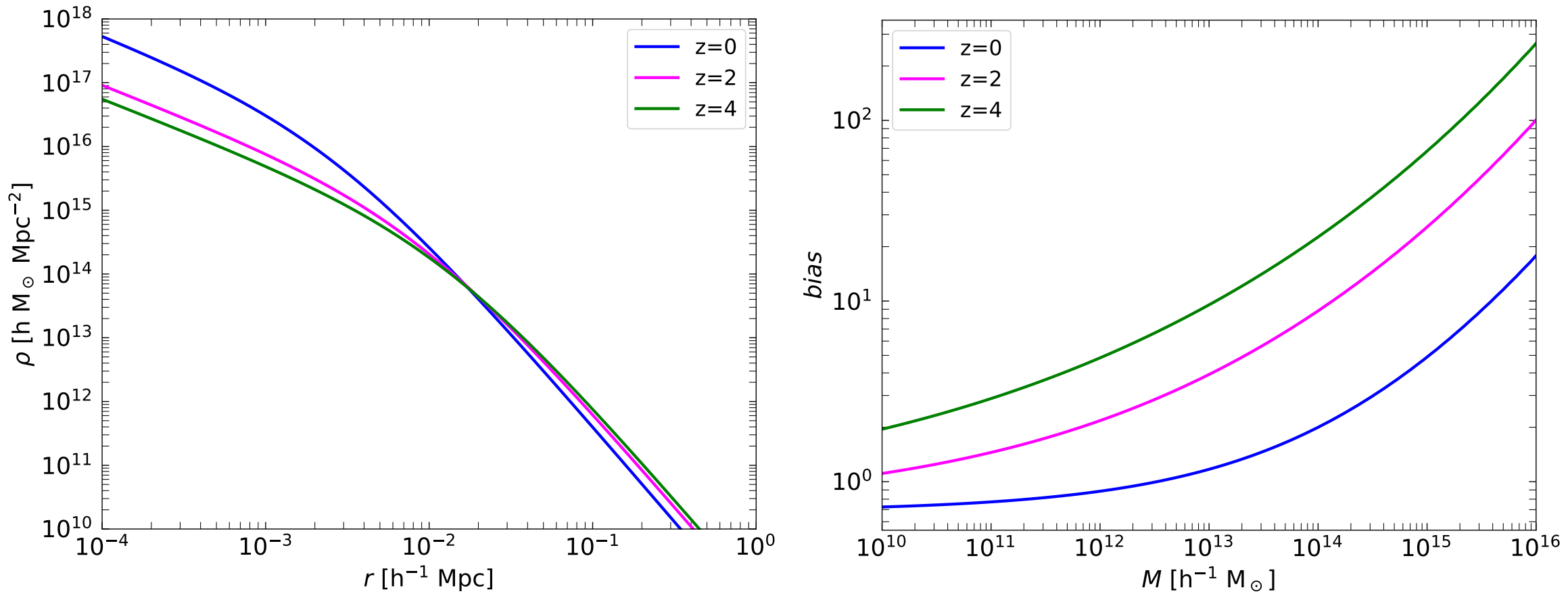

**Fig. 3.4.** *Left:* Navarro-Frenk-White density profile for $M = 10^{10} M_\odot$ as a function of the distance $r$ from the centre. *Right:* Halo bias as a function of the halo mass. Different colours denote different values of redshift: $z = 0$ (blue), $z = 2$ (magenta), $z = 4$ (green).

### 3.3.3 Density profile

The halo density profile $\rho$ describes the distribution of mass within a halo. To a first approximation, a DM halo can be approximated as a spherical object, with its mass distribution completely determined by $\rho(r)$, where $r$ denotes the distance from the halo centre. High-resolution N-body simulations suggest that the DM profile is steeper than $\rho \propto r^{-2}$ in the outer region of the halos and shallower in the inner part [364]. Such consideration implies that the circular velocity profile $v_c(r) = \sqrt{GM(<r)/r}$ exhibits a peak, which acts as a natural scale size $r_{\mathrm{v,\,max}}$. One of the most popular fitting functions determined from N-body simulations is the so-called Navarro-Frenk-White (NFW) profile [365, 366, 367]

$$\rho(r, M, z) = \rho_s \left(\frac{r_s}{r}\right)\left(1 + \frac{r}{r_s}\right)^{-2}, \tag{3.42}$$

where $r_s$ is the scale radius and $\rho_s$ is the DM density at the scale radius. The scale radius is linked to the peak circular velocity scale via $r_{\mathrm{v,\,max}} = 2.163\, r_s$. We note that in the case of a NFW profile it holds $r_s = r_{-2}$, since we recall that $r_{-2}$ corresponds to the radius where the density profile scales as $\rho \propto r^{-2}$. As a result, this density profile is completely determined by the two parameters $(M,\, c_{\rm vir})$, or equivalently $(r_s,\, \rho_s)$. The two pairs of parameters are related via

$$\begin{cases} r_s & = & \dfrac{R_{\rm vir}}{c_{\rm vir}} \\ \rho_s & = & \dfrac{M}{4\pi r_s^3}\left[\ln(1+c_{\rm vir}) - \dfrac{c_{\rm vir}}{1+c_{\rm vir}}\right]^{-1} . \end{cases} \tag{3.43}$$

The NFW profile evolves as $\rho \propto r^{-1}$ in the core, $\rho \propto r^{-2}$ at $r = r_s$ and $\rho \propto r^{-3}$ for large r. Fig. 3.4 (left) illustrates the NFW profile as a function of the radial distance for a fixed halo mass $M = 10^{10}\, M_\odot$ and different values of redshift: $z = 0$ (blue), $z = 2$ (magenta), $z = 4$ (green). The redshift dependence affects the concentration which, in turn, modifies the scale radius. At fixed halo mass, a higher redshift corresponds to higher values of the scale radius, which affects the radial coordinate at which the change of slope occurs.

Let us define the normalised Fourier transform of the DM profile truncated at the virial radius

$$\tilde{u}(k) = \frac{\displaystyle\int {\rm d}^3x\; \rho(\boldsymbol{x})\, e^{-i\boldsymbol{k}\cdot\boldsymbol{x}}}{\displaystyle\int {\rm d}^3x\; \rho(\boldsymbol{x})}\,, \tag{3.44}$$

where $\boldsymbol{x}$ is the position relative to the halo centre. For a spherically symmetric profile, Eq. (3.44) simplifies to

$$\tilde{u}(k) = \int_0^{R_{\rm vir}} {\rm d}r\, \frac{4\pi r^2}{M} \frac{\sin kr}{kr} \rho(r)\,. \tag{3.45}$$

A similar definition will be used in the case of the neutral hydrogen distribution, as will be explained in Chapter 4.2. In the case of annihilating DM, the signal scales with the density squared, thus we need to replace $\rho$ with $\rho^2$:

$$\tilde{v}(k) = \int_0^{R_{\rm vir}} {\rm d}r\, \frac{4\pi r^2}{M} \frac{\sin kr}{kr} \rho^2(r)\,. \tag{3.46}$$

Moreover, N-body simulations suggest that DM halos include several self-bound substructures, called subhalos. The physical motivation is that not all the smaller structures are destroyed during the processes of merging and accretion, which produce more massive halos in the hierarchical $\Lambda$CDM paradigm. As mentioned in Chapter 2.1, the contribution of subhalos may play a key role in the field of DM indirect research since they are expected to enhance a signal from DM annihilation events and the closest among these substructures may also serve themselves as suitable targets for DM indirect detection [368]. The presence of substructures within the main halos can be taken into account by replacing $\rho^2$ in Eq. (3.46) with $(1+B)\rho^2$, where $B$ is the so-called boost function. We adopt the following model provided by Ref. [369]:

$$\log B(M, z=0) = \sum_{i=0}^{5} d_i \left[\log\left(\frac{M}{M_\odot}\right)\right]^i , \tag{3.47}$$

where $d_0 = -0.186$, $d_1 = 0.144$, $d_2 = -8.8\cdot 10^{-3}$, $d_3 = 1.13\cdot 10^{-3}$, $d_4 = -3.7\cdot 10^{-5}$, $d_5 = -2\cdot 10^{-7}$. Eq. (3.47) is valid at present day, whereas the boost factor evolves like

$$B(M,\, z) = \frac{B(M,\, z=0)}{1+z}\,. \tag{3.48}$$

### 3.3.4 Bias

As illustrated in Fig. 3.2, the DM halos do not trace the entire distribution of matter since they correspond to the peaks in the density fluctuations. Therefore, they are biased tracers. In this regard, the halo bias $b_h$ represents the link between the halos and the underlying matter distribution in the Universe. Let us define the halo density contrast as

$$\delta_h = \frac{n_M(\boldsymbol{x})}{\overline{n}_M} - 1\,, \tag{3.49}$$

where $\overline{n}_M$ denotes the average number density of DM halos with a given mass $M$ and $n_M$ represents the halo number density at position $\boldsymbol{x}$. Assuming that on large scales $\delta_h$ is linearly proportional to the matter overdensity, the halo bias is defined through the following relation:

$$\delta_h(\boldsymbol{x}) = b_h(M)\,\delta(\boldsymbol{x}) = b_h(M)\left(\frac{\rho_m(\boldsymbol{x})}{\overline{\rho}_m} - 1\right). \tag{3.50}$$

We adopt the model of Ref. [356]:

$$b_h(M,z) = 1 + \frac{q\nu - 1}{\delta_{\rm sc}(z)} + \frac{2p}{\delta_{\rm sc}(z)[1+(q\nu)^p]}\,, \tag{3.51}$$

where the parameter $q$, $p$ and $\nu$ are the same of Eq. (3.33). Fig. 3.4 (right) illustrates that the bias is an increasing function of the halo mass: more massive halos form later and are more biased. The reason is that the most massive halos are also more clustered than low-mass halos [356, 370, 371].

In this section we have introduced the mass function, the density profile and its Fourier transform, the boost factor accounting for the substructures, the concentration parameter and the halo bias. They represent the key elements which are necessary to evaluate the non-linear PS for the DM halos, as will be discussed in the next section.

## 3.4 Fourier power spectrum

The Fourier PS measures the correlation between two observables in $k$-space. The two points that we correlate may be located within the same DM halo or in two separate collapsed structures[5]. Fig. 3.5 provides a schematic representation of this situation. The orange spheres represent our observables, connected by lines which denote the correlation between them. White lines account for the correlation of points inside the same halo, while black lines denote the correlation for objects located in different virialized structures. The total correlation signal will include both contributions from the single and the two halos. Indeed, it can be proved that the totals PS (and likewise the 2PCF) can be split into two terms: one-halo and two-halo. The latter refers to the correlation among objects in different collapsed structures, while the one-halo contribution concerns the correlation among points within the same halo. The generic expression of the non-linear Fourier PS reads

$$P_{ij}(k,z) = P^{1h}_{ij}(k,z) + P^{2h}_{ij}(k,z)\,, \tag{3.52}$$

where $i$ and $j$ refer to the two observables that we want to cross-correlate. The auto-correlation scenario corresponds to $i = j$. The one-halo and two-halo terms are given by

$$P^{1h}_{ij}(k,z) = \int_{M_{\rm min}}^{M_{\rm max}} {\rm d}M \frac{{\rm d}n}{{\rm d}M} \widetilde{f}_i^* \widetilde{f}_j \tag{3.53}$$

$$P^{2h}_{ij}(k,z) = \left[\int_{M_{\rm min}}^{M_{\rm max}} {\rm d}M_1 \frac{{\rm d}n}{{\rm d}M_1} b_i \widetilde{f}_i^*\right] \left[\int_{M_{\rm min}}^{M_{\rm max}} {\rm d}M_2 \frac{{\rm d}n}{{\rm d}M_2} b_j \widetilde{f}_j\right] P_{\rm lin}(k,z)\,, \tag{3.54}$$

where $b$ is the bias and $\widetilde{f}_i$ stands for the Fourier transform of the fluctuation field

$$f_i = \frac{g_i}{\langle g \rangle} - 1\,. \tag{3.55}$$

[5] Here we use the terms "halo" and "collapsed (or virialized) structure" as synonyms.

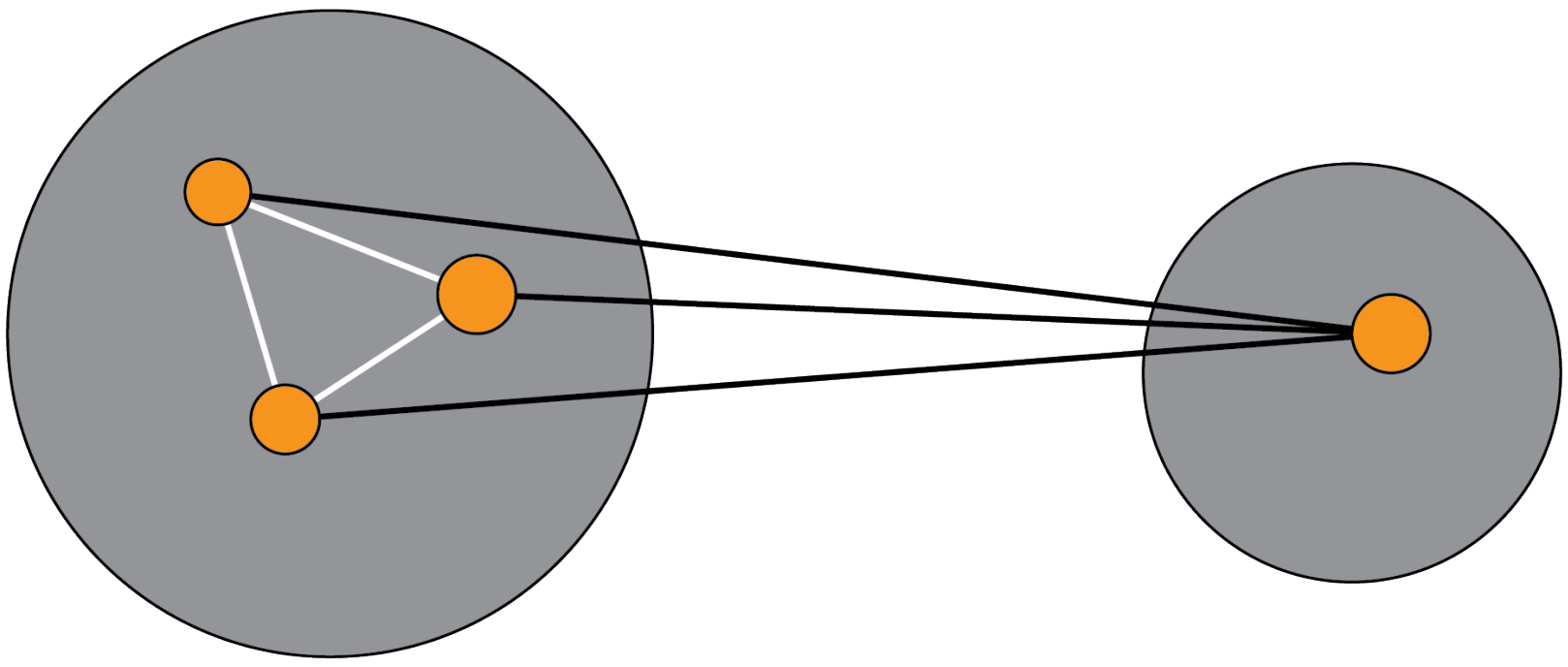

**Fig. 3.5.** Schematic representation of the correlation between observables located in the dark matter halos. The white lines connect objects within the same single halo, while the black lines link points in two distinct halos.

The variable $g_i$ represents the density field of the source $i$ and the average $\langle g \rangle$ can be expressed as

$$\overline{g} = \langle g \rangle = \int_{M_{\rm min}}^{M_{\rm max}} \frac{{\rm d}n}{{\rm d}M} \int {\rm d}^3 x\, g(\boldsymbol{x})\,. \tag{3.56}$$

In Eqs. (3.53) and (3.54), the symbol $*$ denotes the complex conjugate. Note that the one-halo term includes the two contributions of the density fluctuations from the same halo, while the two-halo component comprises the signals from two distinct halos, corresponding to the two terms in square brackets. The integrals are performed over the halo mass. The values of the upper and lower mass limits are model-dependent. The upper mass boundary is set to $M_{\rm max} = 10^{18}\,M_\odot$ and the results are very mildly dependent on $M_{\rm max}$. The lower limit $M_{\rm min}$ remains a very uncertain element because N-body simulations are still far from reaching the resolution necessary to probe clustering at small halo masses [372, 373, 374, 375, 376, 377]. However, in the early Universe the kinetic decoupling of DM particles[6] set a cut-off limit on the minimum mass allowed to form a protohalo [378, 379, 380, 381, 382, 383, 384]. In particular, the value of $M_{\rm min}$ is expected to lie in the mass range $10^{-12}\,M_\odot - 10^{-4}\,M_\odot$ [383]. A canonical value often cited in the literature and adopted in this work is $M_{\rm min} = 10^{-6}\,M_\odot$, corresponding generically to the WIMP free-streaming mass [329, 369]. The derivation of Eqs. (3.52)–(3.54) is performed in Appendix B.

[6] This line of argument typically refers to WIMP candidates.

In the next section, we will apply these general expressions for $P^{\rm 1h}_{ij}$ and $P^{\rm 2h}_{ij}$ to the specific case of the auto-correlation of two signals produced by DM particles. Section 3.4.2 introduces the modelling of the astrophysical sources and derives the auto-correlation PS for two signals of astrophysical origin. The cross-correlation PS between the $\gamma$-ray emitters and the neutral hydrogen distribution will be the focus of Chapter 4.

### 3.4.1 Dark matter

Recall that decaying DM produces an EM signal, that scales linearly with the DM density, while annihilation events of DM particles are expected to produce a flux scaling with $\rho^2$. Such a dependence on the DM density directly reflects on the expressions of the PS. In particular, in the case of two observables that depend linearly on the density, e.g. the auto-correlation of a decaying DM signal or the cross correlation of decaying DM with the cosmic shear, the one-halo and two-halo terms read

$$P^{\rm 1h}_{\delta\delta}(k,z) = \frac{1}{\overline{\rho}^2}\int_{M_{\rm min}}^{M_{\rm max}} {\rm d}M\, M^2\, \frac{{\rm d}n}{{\rm d}M}(M,z)\,|\tilde{u}(k,M,z)|^2 \tag{3.57}$$

$$P^{2h}_{\delta\delta}(k,z) = \left[\frac{1}{\overline{\rho}}\int_{M_{\rm min}}^{M_{\rm max}} {\rm d}M\, M\, \frac{{\rm d}n}{{\rm d}M}(M,z)\, b_h(M,z)\,\tilde{u}(k,M,z)\right]^2 P_{\rm lin}(k,z)\,. \tag{3.58}$$

where $\overline{\rho}$ is the background density at present day and the factor $M^2$ in due to the fact that $\tilde{u}$ in Eq. (3.45) is normalised to the halo mass. Note the Eqs. (3.57) and (3.58) apply to any signal which scales with the matter density contrast, produced by a biased gravitational tracer. In particular, we will employ a similar expression in Chapter 4, while discussing the PS of neutral hydrogen intensity mapping. In the case of the auto-correlation of two observables depending on the square of the DM density, e.g. the auto-correlation signal of annihilating DM, the one-halo and two-halo contributions can be expressed as

$$P^{\rm 1h}_{\delta^2\delta^2}(k,z) = \int_{M_{\rm min}}^{M_{\rm max}} {\rm d}M\, \frac{{\rm d}n}{{\rm d}M}(M,z)\left(\frac{\tilde{v}(k,M,z)}{\Delta^2(z)}\right)^2 \tag{3.59}$$

$$P^{\rm 2h}_{\delta^2\delta^2}(k,z) = \left[\int_{M_{\rm min}}^{M_{\rm max}} {\rm d}M\, \frac{{\rm d}n}{{\rm d}M}(M,z)\, b_h(M,z)\left(\frac{\tilde{v}(k,M,z)}{\Delta^2}\right)\right]^2 P_{\rm lin}(k,z)\,, \tag{3.60}$$

where $\Delta^2(z)$ is the so-called clumping factor. This function takes into account that $\tilde{v}$ is the Fourier transform of $\rho^2$, while Eqs. (3.57) and (3.58) involve $\langle\rho\rangle^2 \neq \langle\rho^2\rangle$. Thus, the clumping factor is defined as

$$\Delta^2(z) \equiv \frac{\langle\rho^2\rangle}{\overline{\rho}^2} = \int {\rm d}M\, \frac{{\rm d}n}{{\rm d}M}(M,z)\int {\rm d}^3x\, \frac{\rho^2(\boldsymbol{x}|M)}{\overline{\rho}^2}\,. \tag{3.61}$$

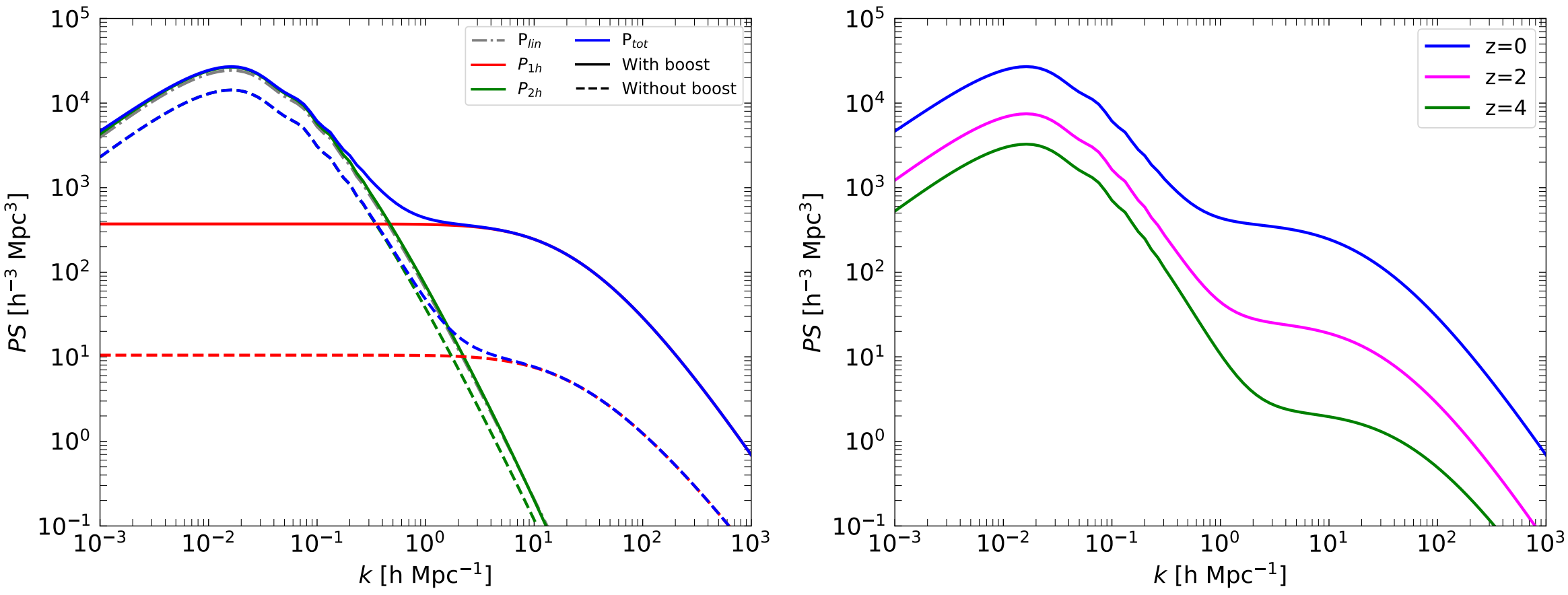

**Fig. 3.6.** *Left:* Power spectrum as a function of the wavenumber for annihilating DM at $z = 0$: linear (dash-dotted grey), one-halo (red), two-halo (green), total (blue). The case with the substructures is displayed with solid lines, while in dashed there is the scenario without subhalos. *Right:* Total power spectrum with the contribution of substructures included, for different values of redshift: $z = 0$ (blue), $z = 2$ (magenta), $z = 4$ (green).

In order to consider the additional contribution from the substructures within the main halos, we replace

$$\rho^2(\boldsymbol{x}|M,z) \Longrightarrow [1+B(M,z)]\,\rho^2(\boldsymbol{x}|M,z)\,, \tag{3.62}$$

where $B(M,z)$ is the boost factor, as given by Eq. (3.48). Similarly, we can define the power spectrum for the $\delta \times \delta^2$ channel, as will be discussed in Chapter 5 for the HI$\times\gamma$ rays signal, where we will make explicit the relevant expression. Fig. 3.6 (left) illustrates the PS for annihilating DM as a function of the wavenumber at redshift $z = 0$. The red lines denote the one-halo contribution, the green curves represent the two-halo term and the blue lines refer to the total PS. It is apparent that on large scales (small $k$), $P_{\rm tot}$ follows the same behaviour of the linear PS (dash-dotted grey line), since we are in the linear regime. The peak of the PS is located around the wavenumber corresponding to the matter-radiation equality and the slope after the maximum is ruled by the transfer function. At some point $k > 0.1\,h$ Mpc$^{-1}$, the one- and two-halo contributions become comparable[7]. This point occurs where the density perturbations become non-linear ($\delta \sim 1$) and $P_{\rm tot}$ departs from the linear PS, corresponding to the hump around 10 $h$ Mpc$^{-1}$. On small scales, $P_{\rm tot}$ declines because these scales correspond to regions where there is less clustering and the one-halo contribution becomes the leading term in the total PS. This implies that on small scales $P_{\rm tot}$ is mainly determined by the internal structure of the halo. The solid lines refer to considering the contribution of the substructures, while the dashed curves relate to the case without the boost factor. The effect of including the substructures is to boost the one-halo term by more than an order of magnitude and $P_{\rm 1h}$ becomes dominant earlier (i.e. on larger scales). In other words, if we consider the additional component of the substructures, the non-linearities take over earlier and the internal structure becomes more relevant. Viceversa, neglecting the boost factor results in underestimating the contribution of the internal structure to the total PS. Regarding the redshift dependence in the PS, it resides in the growth factor given by Eq. (3.23). Fig. 3.6 (right) illustrates the evolution of $P_{\rm tot}$ with $k$ for different values of redshifts: $z = 0$ (blue), $z = 2$ (magenta), $z = 4$ (green). All the curves also include the subhalos' contribution. The amplitude of the PS is inversely proportional to the redshift, because higher values of $z$ refer to epochs in the past when matter was less clustered in the halos.

[7] The precise range in which $P_{\rm 1h}$ and $P_{\rm 2h}$ are of the same order of magnitude depends on whether we take into account a boost from the substructures and also on the value of redshift under consideration.

### 3.4.2 Astrophysical sources

Astrophysical sources are expected to contribute to the UGRB. In this work, we take into consideration star-forming galaxies and active galactic nuclei, further subdivided into BL Lacertae objects (BL Lac), flat-spectrum radio quasars (FSRQ) and misaligned active galactic nuclei. Active galactic nuclei (AGN) represent the central region of a galaxy, characterised by a high excess of non-stellar emission (called "active galaxy") at all wavelengths of the electromagnetic spectrum. A supermassive black hole is expected to be situated at the centre of these active galaxies and the accretion of matter leads to two relativistic jets. The orientation of these jets determines whether an AGN is a misaligned active galactic nucleus (mAGN) or a blazar [385]. According to the unified model of AGN, a jet misalignment of 14° with respect to the line of sight distinguishes the two categories. In particular, blazars are characterised by a relativistic jet pointing towards the Earth and they are usually subdivided into two classes: BL Lac and FSRQ. A main difference between the two classes is that FSRQ are associated to strong and broad optical emission lines and a high radio luminosity, unlike BL Lac [386, 387]. There is also a distinction in their energy distributions with BL Lac having a harder energy spectrum in the MeV−GeV range [388]. AGN with a jet misalignment with respect to the line of sight above 14° are classified as misaligned AGN [385, 388]. Finally, star-forming galaxies (SFG) are characterised by a tremendous stellar activity. The interactions of cosmic-ray protons accelerated by supernova remnants with the interstellar medium produce pions, which subsequently decay and produce $\gamma$ rays. However, the detection of the $\gamma$-ray flux from SFG is challenging because their luminosity is typically much smaller compared to AGN [389].

The emission from the astrophysical sources is typically modelled by specifying the $\gamma$-ray luminosity function and the spectral energy distribution. The $\gamma$-ray luminosity function (GLF) provides the number of sources per unit of luminosity $L$, co-moving volume $V_c$ and spectral index $\Gamma$:

$$\phi_i(L, z, \Gamma) = \frac{\mathrm{d}N_i}{\mathrm{d}L\,\mathrm{d}V_c\,\mathrm{d}\Gamma} \tag{3.63}$$

with $i$ = {BL Lac, FSRQ, mAGN, SFG}. The explicit expression adopted for each class is illustrated in Appendix C. The energy spectrum $\mathcal{G}(E)$ represents the number of $\gamma$ rays in the energy range $(E, E+\mathrm{d}E)$ per unit time, in absence of any absorption from the extragalactic background light. We adopt a simple power-law

$$\mathcal{G}(E) = \left(\frac{E}{\rm GeV}\right)^{-\Gamma}, \tag{3.64}$$

where $E$ is the photon energy observed by the detector and the value of $\Gamma$ for each source is shown in the second column of Table 3.1. The spectral energy distribution (SED) reads

$$\frac{\mathrm{d}\mathcal{N}}{\mathrm{d}E} = k\,\mathcal{G}(E)\,, \tag{3.65}$$

where $k$ is a normalisation constant. The second most energetic diffuse background, after the CMB, is represented by the stellar emission which escapes the stars and reaches the intergalactic medium. This radiation is known as extragalactic background light (EBL) and covers the entire electromagnetic spectrum. The EBL is relevant in $\gamma$-ray astronomy because it is responsible for the opacity of the Universe due to photon absorption of the high-energy photons via pair-production on the low-energy radiation fields in the intergalactic medium. In particular, the survival probability at Earth for a $\gamma$-ray photon originated at redshift $z$ is given by $\exp[-\tau_{\gamma\gamma}(E_r, z)]$, where $E_r$ is the rest-frame energy at the source and $\tau_{\gamma\gamma}$ is the so-called optical depth. As a result, the observed flux $F_{\rm obs}$ is attenuated by this exponential factor:

$$F_{\rm obs} = \int \mathrm{d}E\, k\,\mathcal{G}(E)\,\exp(-\tau_{\gamma\gamma}[E(1+z), z])\,, \tag{3.66}$$

where we adopt the parameterisation of $\tau_{\gamma\gamma}$ as given in Ref. [394], which takes into account the opacity of the Universe in the range 1 GeV − 80 TeV, and we recall that the observed energy $E$ is related to $E_r$ via $E_r = E(1+z)$.

| | $\Gamma$ | $L_{\rm min}$ [erg/s] | $L_{\rm max}$ [erg/s] | Ref. |
|---|---|---|---|---|
| BL Lac | 2.11 | $7\times10^{43}$ | $10^{52}$ | [390] |
| FSRQ | 2.44 | $10^{44}$ | $10^{52}$ | [391] |
| mAGN | 2.37 | $10^{40}$ | $10^{50}$ | [392, 388] |
| SFG | 2.7 | $10^{37}$ | $10^{42}$ | [393] |

**Table 3.1.** Spectral index $\Gamma$, minimum luminosity $L_{\rm min}$ and maximum luminosity $L_{\rm max}$, reference for the gamma-ray luminosity function for the classes of astrophysical sources adopted in this analysis.

In this work we consider astrophysical sources as point sources. Also, the astrophysical sources are better characterised by their luminosity rather than mass of the hosting halo, thus it is convenient to introduce the luminosity $L$ in the energy range (0.1, 100) GeV in the rest frame[8] of the astrophysical source, which is defined as

$$L = \int_{0.1\ \rm GeV}^{100\ \rm GeV} \mathrm{d}E_r\, \frac{\mathrm{d}L}{\mathrm{d}E_r}(E_r)\,. \tag{3.67}$$

[8] Since we define the luminosity in the rest frame of the source, we do not have to include the absorption.

The differential luminosity $\mathrm{d}L/\mathrm{d}E_r$ and the SED at redshift $z$ are linked via

$$\frac{\mathrm{d}L}{\mathrm{d}E_r}(E_r) = \frac{4\pi d_L^2(z)}{1+z} E \frac{\mathrm{d}\mathcal{N}}{\mathrm{d}E}(E), \tag{3.68}$$

where $d_L = (1+z)\chi(z)$ is the luminosity distance and the comoving distance $\chi$ is defined as

$$\chi(z) = \int_0^z \frac{c\,\mathrm{d}z'}{H(z')}. \tag{3.69}$$

Thus, Eq. (3.67) in the observed reference frame becomes

$$L = 4\pi d_L^2(z) \int_{\frac{0.1}{1+z}}^{\frac{100}{1+z}} \mathrm{d}E\, E\, \frac{\mathrm{d}\mathcal{N}}{\mathrm{d}E}(E). \tag{3.70}$$

In the expressions of the power spectra for the astrophysical sources we need to replace the halo mass $M$ with the luminosity $L$ and the halo mass function with the GLF. In particular, the auto-correlation one-halo and two-halo terms for the astrophysical components read

$$P_{SS}^{1h}(k,z) = \int_{L_{\min}}^{L_{\max}} \mathrm{d}L\, \phi(L,z) \left(\frac{L}{\langle g_s \rangle}\right)^2 \tag{3.71}$$

$$P_{SS}^{2h}(k,z) = \left[\int_{L_{\min}}^{L_{\max}} \mathrm{d}L\, \phi(L,z)\, b(M(L),z)\, \frac{L}{\langle g_s \rangle}\right]^2 P_{\mathrm{lin}}(k,z), \tag{3.72}$$

where the mean luminosity produced by unresolved sources at redshift $z$ is given by

$$\langle g_s \rangle (z) = \int_{L_{\min}}^{L_{\max}} \mathrm{d}L\, L\, \phi(L,z). \tag{3.73}$$

Given the luminosity of a certain class of astrophysical objects, we can determine the mass of the host DM halo, via the $M(L)$ relations specified in Appendix C. The bias of the astrophysical sources $b_S$ is defined as:

$$b_S(z) = \int_{L_{\min}}^{L_{\max}} \mathrm{d}L\, \phi(L,z)\, b(M(L), z)\, \frac{L}{\langle g_S \rangle}, \tag{3.74}$$

where the minimum and maximum luminosities $L_{\min}$ and $L_{\max}$ in the integral depend on the intrinsic properties of the source class. However, $L_{\max}$ cannot be larger than the luminosity threshold $L_{\mathrm{sens}}$ above which a source class can be resolved by the detector, in which case the upper bound is floored to $L_{\mathrm{sens}}$. We determine $L_{\mathrm{sens}}$ for each class in accordance with the detector flux sensitivity, for which we assume $F_{\mathrm{sens}} = 10^{-10}$ cm$^2$ s$^{-1}$ for photons in the energy interval $(1-100)$ GeV, well compatible with the sensitivity of FERMI-LAT in 8 years of data taking [395]. Following Eq. (3.70), the luminosity threshold in the observed reference frame can be expressed as

$$L_{\mathrm{sens}} = 4\pi d_L^2 \int_{0.1/(1+z)}^{100/(1+z)} \mathrm{d}E\, E\, k_{\mathrm{sens}}\, \mathcal{G}(E), \tag{3.75}$$

where the constant $k_{\mathrm{sens}}$ can be determined from the flux threshold of the detector via

$$F_{\mathrm{sens}} = \int_{1\ \mathrm{GeV}}^{100\ \mathrm{GeV}} \mathrm{d}E\, k_{\mathrm{sens}}\, \mathcal{G}(E)\, \exp(-\tau_{\gamma\gamma}[E(1+z),z]). \tag{3.76}$$

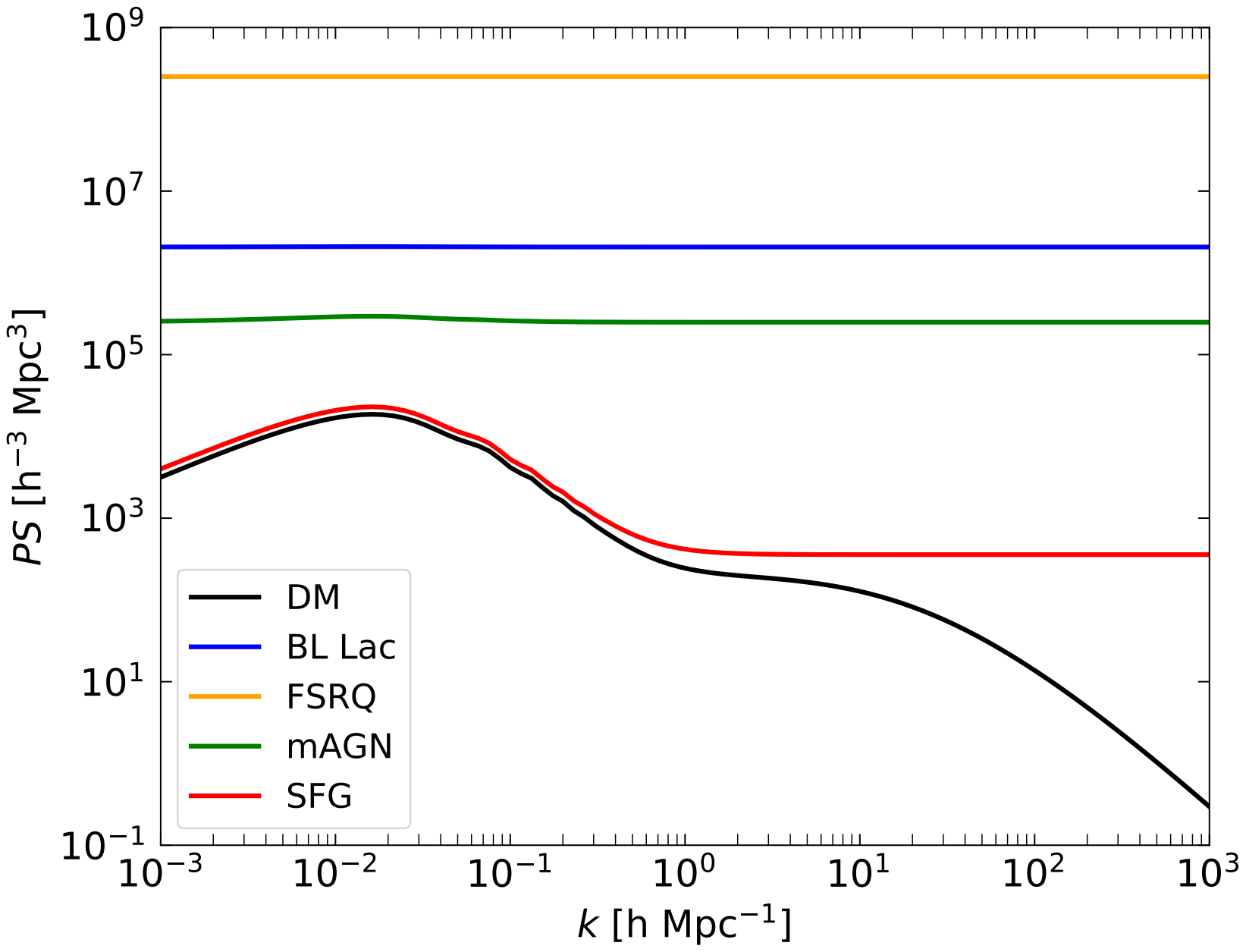

**Fig. 3.7.** Total auto-correlation power spectrum as a function of the wavenumber at $z = 0.5$ for annihilating DM (black), BL Lac (blue), FSRQ (orange), mAGN (green), SFG (red).

Table 3.1 illustrates the spectral indices (second column), the minimum and maximum luminosities (third and four columns, respectively) and the references of the GLF for each astrophysical source under consideration. Fig. 3.7 illustrates the auto-correlation power spectra for annihilating DM (black), BL Lac (blue), FSRQ (orange), mAGN (green) and SFG (red) at $z = 0.5$. It is noteworthy that the one-halo PS and the term in square-bracket in the two-halo contribution for the astrophysical sources are independent on $k$, because of the point-source approximation. In the case of SFG, on large scales the total PS is dominated by $P_{2h}$, thus it follows the $P_{lin}$, while on small scales $P_{1h}$ becomes the dominant contribution. Instead, the total PS of the other three astrophysical sources is always ruled by the one-halo term, therefore it is constant in $k$ along the entire range of wavenumbers under consideration. Note that the auto-correlation PS of DM has a peculiar behaviour in $k$, which is discernible from the astrophysical sources. However, $P_{\delta^2\delta^2}$ passes well below the astrophysical background. In the next chapter we will introduce the intensity mapping of neutral hydrogen, which we employ as a gravitational tracer of the matter distribution in the Universe. We will show that the cross-correlation PS between neutral hydrogen and annihilating DM is the same order of magnitude of the correlation between neutral hydrogen and the astrophysical sources, making this cross-correlation channel a promising tool to unveil a subdominant DM contribution in the UGRB.

# CHAPTER 4

## INTENSITY MAPPING OF NEUTRAL HYDROGEN

A typical method to probe the DM distribution in the Universe is by using cosmological galaxy surveys, which samples a large number of galaxies, however they have a limitation: they only detect discrete bright objects whose radiation is above the flux threshold of the detector. A promising novel technique capable of overcoming this drawback is the so-called intensity mapping (IM). This method consists of measuring the integrated emission of a spectral line originating both from galaxies and from the diffuse intergalactic medium. IM detects all sources radiating the emission line of interest and is optimal to probe faint or extended sources. This technique does not resolve individual galaxies and therefore does not require high angular resolution, making it more effective than traditional galaxy surveys regarding the trade off between sky area and observational time. The IM technique can be applied to a vast assortment of spectral lines and enables us to probe redshift ranges which are typically inaccessible by traditional galaxy surveys [396]. Examples of emissions that have gained the attention of the community are: the 21cm line emitted by neutral hydrogen atoms, the rotational transition of carbon-monoxide (CO) [397, 398, 399, 400, 401, 402], the atomic CII fine-structure line [403, 404] and the Lyman-alpha emission line [405, 406, 407, 408]. The growing interest for this emerging field is also driven by the numerous current and upcoming telescopes devoted to this technique. Several IM experiments observing different line emissions are shown in Fig. 4.1, illustrated by the different colours. The sky area and the redshift range probed by the different detectors are indicated on the vertical and horizontal axes, respectively. The figure distinguishes among running (solid), funded (dashed) and proposed (dotted) experiments, supporting the fact that IM is a fertile and emerging field. In this thesis we focused on IM of the 21cm line emitted by atoms of neutral hydrogen, and on the opportunities offered by the next-generation of radio telescopes: the Square Kilometre Array (SKA), as well as its precursor MeerKaroo Array Telescope (MEERKAT). The following chapter is organised as follows: Section 4.1 briefly introduces the origin of the 21cm line, Section 4.2 presents the halo model of neutral hydrogen, focusing on the key elements required to obtain the IM signal. Finally, Section 4.3 illustrates the cross-correlation power spectra between the neutral hydrogen IM and the $\gamma$-ray sources.

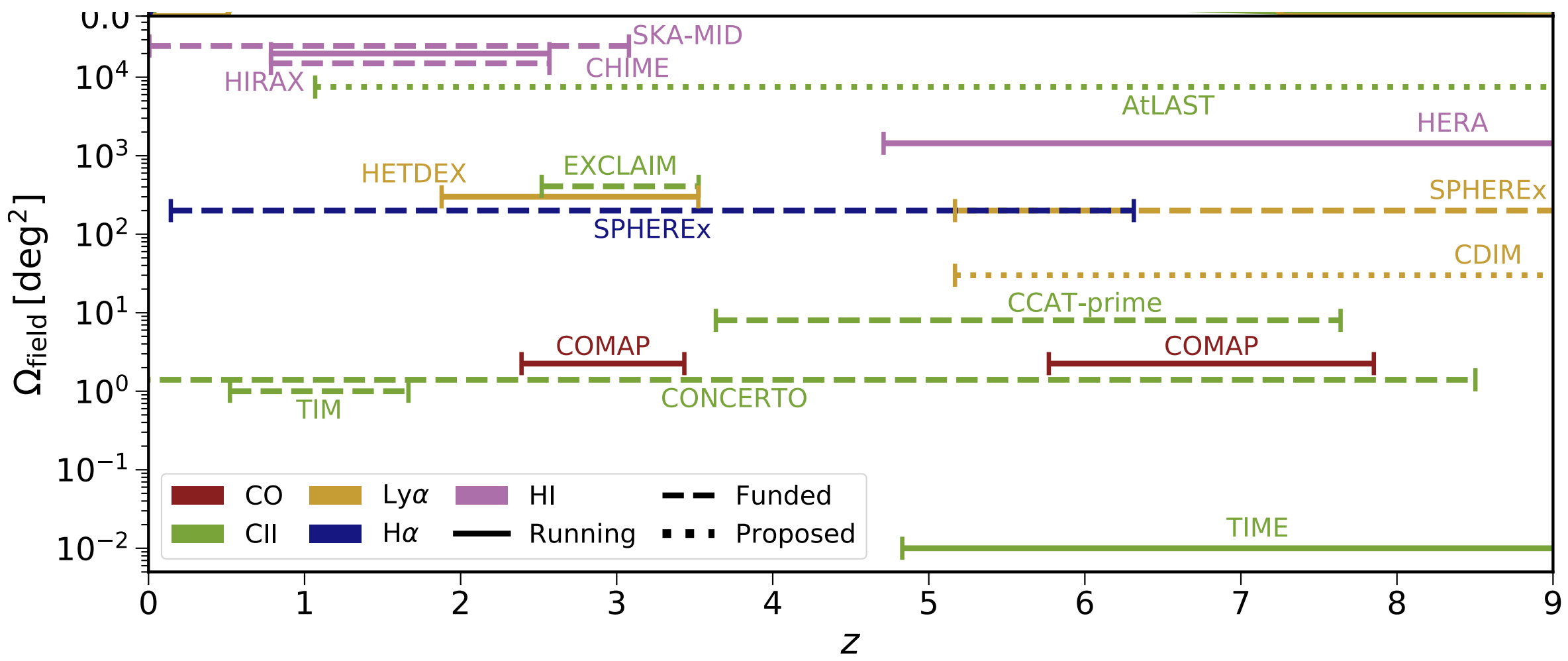

**Fig. 4.1.** A representative list of running (solid), funded (dashed) and proposed (dotted) experiments on line-intensity mapping. The vertical axis indicates the sky area of the detectors, while the horizontal axis illustrates the redshift range probed. Different colours refer to different emission lines: carbon-monoxide (red), [CII] fine-structure line (green), Lyman-alpha (yellow), $H_\alpha$ (blue), 21cm line (pink). Credit: J. L. Bernal Mera.

## 4.1 The 21cm line

The first prediction of the existence of a line from neutral hydrogen (HI) dates back to 1940 when Hendrik van de Hulst [409], following the suggestion of Jan Oort, postulated the existence of a 21cm emission related to the hyperfine structure of HI. The prediction was confirmed in 1951 and the following year the first maps of the HI distribution in our Galaxy were made, revealing the spiral arms of the Milky Way for the first time [410]. We recall that HI represents the most abundant element in our Universe, comprising approximately 74% of the total baryonic matter. Its ground state consists of an electron bound to a proton with two possible spin configurations: parallel or antiparallel. The former configuration is associated with a higher energy state and the two levels are separated by $\Delta E \approx 5.9 \cdot 10^{-6}$ eV. When the spin-flip transition from the higher to the lower energy level occurs, a spectral line is radiated at a wavelength $\lambda_e = 21$ cm, corresponding to a frequency $\nu_e \simeq 1420$ MHz, as illustrated in Fig. 4.2. This transition is very rare and it is expected to happen once every $10^7$ s. However, the hydrogen is so abundant that this line is frequently observed. Most neutral hydrogen resides inside galaxies post reionization era, which form part of DM halos. The contribution to the HI power spectrum originating from outside the halos is negligible, as shown in Ref. [411]. Therefore, the HI gas is an excellent gravitational tracer of the DM distribution in the Universe. The long lifetime of the spin-flip transition translates into a very narrow width of the line. As a result, we can extract a precise redshift information by using the simple relation between the observed frequency $\nu_o$ and the frequency of emission $\nu_e$:

$$\nu_o = \frac{\nu_e}{1+z}\,. \tag{4.1}$$

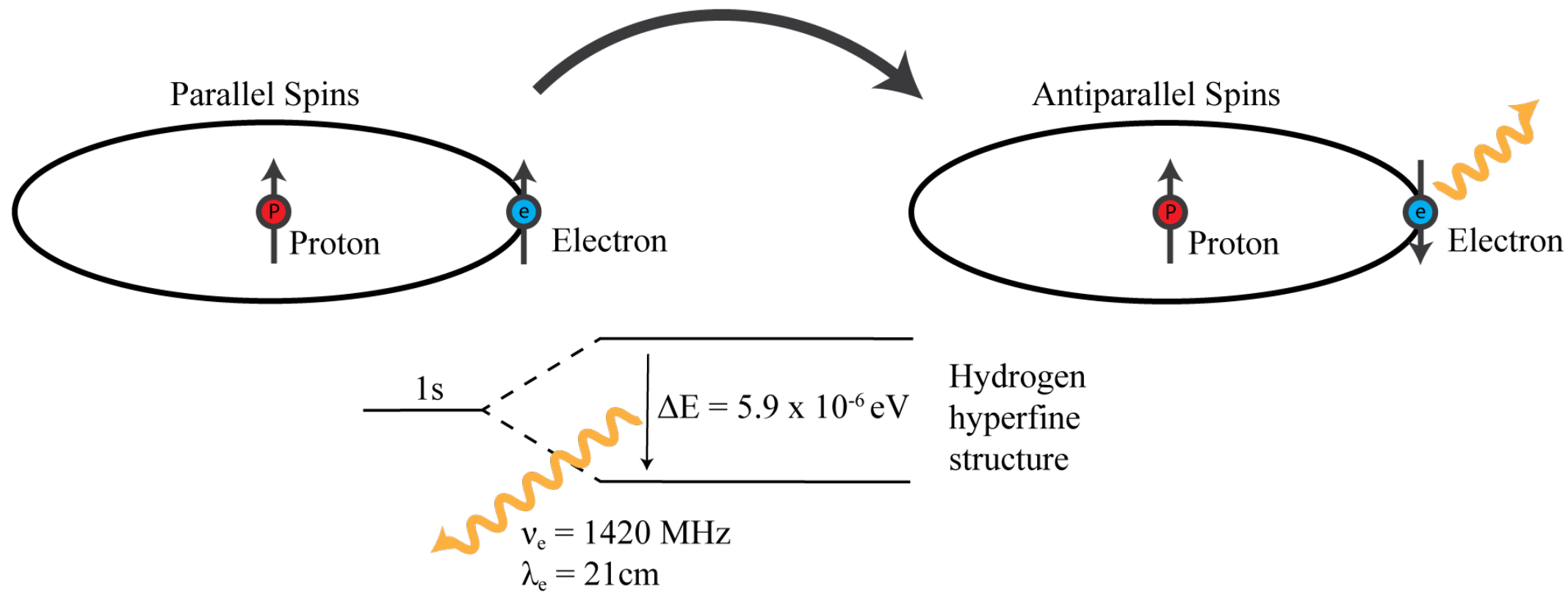

**Fig. 4.2.** Spin-flip transition of neutral hydrogen, leading to the emission of a line with wavelength $\lambda_e = 21$ cm and frequency $\nu_e = 1420$ MHz.

The study of the HI line has already provided valuable insights on the galactic dynamics by measuring the rotation curves, and now the next-generation radio telescopes offer the exciting prospect of using the 21cm line as a novel probe of cosmology. Our ambition is to utilise the precise spectroscopic redshift information provided by the sensitivity offered by SKA's HI intensity mapping surveys, in order to discriminate a DM contribution from the astrophysical background.

## 4.2 The neutral hydrogen halo model

The distribution and redshift evolution of HI atoms have been determined by Refs. [412, 413]. They provide a function to measure the average mass $M_{\rm HI}$ of HI enclosed within a DM halo of mass $M$ at redshift $z$, known as the HI-halo mass relation:

$$M_{\rm HI}(M) = \alpha f_{{\rm H},c}\left(\frac{M}{10^{11}h^{-1}M_\odot}\right)^\beta \exp\left[-\left(\frac{v_{c,0}}{v_c(M)}\right)^3\right]. \tag{4.2}$$

The cosmic hydrogen fraction is given by

$$f_{{\rm H},c} = (1-Y_p)\frac{\Omega_b}{\Omega_m}\,, \tag{4.3}$$

where $Y_p = 0.24$ denotes the cosmic helium fraction by mass. Eq. 4.2 depends on three parameters: the neutral fraction of HI in the halo $\alpha = 0.176$, the logarithmic slope $\beta = -0.69$ and the circular velocity $v_c$. We recall that

$$v_c = \sqrt{\frac{GM}{R_{\rm vir}}}\,, \tag{4.4}$$

where the minimum virial velocity of a DM halo which contains HI is found to be $v_{c,0} = 40.7$ km/s [412]. Given the HI-halo mass relation and the halo mass function, we can obtain the mean HI density at redshift $z$, given by

$$\bar{\rho}_{\rm HI}(z) = \int_{M_{\rm min}}^{M_{\rm max}} {\rm d}M\, \frac{{\rm d}n}{{\rm d}M}(M,z)\, M_{\rm HI}(M,z)\,, \tag{4.5}$$

which is necessary to compute the HI density fraction parameter

$$\Omega_{\rm HI}(z) = \frac{\bar{\rho}_{HI}(z)}{(1+z)^3\,\rho_{c,0}}\,. \tag{4.6}$$

The HI bias, $b_{\rm HI}(z)$, is linked to the clustering bias $b(z,M)$ through the relation

$$b_{\rm HI}(z) = \frac{1}{\bar{\rho}_{\rm HI}(z)} \int_{M_{\rm min}}^{M_{\rm max}} {\rm d}M\, \frac{{\rm d}n}{{\rm d}M}(M,z)\, M_{\rm HI}(M,z)\, b_h(M,z)\,. \tag{4.7}$$

Fig. 4.3 (left) illustrates the bias as a function of redshift for the different components under consideration: HI (purple), BL Lac (blue), FSRQ (orange), mAGN (green), SFG (red). The DM component is shown for two different halo masses: $10^6\,{\rm M}_\odot$ (solid) and $10^{10}\,{\rm M}_\odot$ (dashed). It is useful to recall that the bias parameter measures how well a certain class of objects traces the total matter distribution, as discussed in Chapter 3.3. DM halos and HI gas are more extended than the astrophysical objects, thus their distributions are a better proxy of the matter distribution (i.e. lower bias) with respect to the astrophysical sources. Also, the HI and the $\gamma$-ray emission from DM are extended distributions: a noticeable difference to take into account is that HI traces the total matter density, while gamma rays from DM annihilation trace the DM density squared, since each annihilation process involves two DM particles.

After the formation of a DM halo, hydrodynamical simulations [414] suggest that the hot gas follows an altered NFW profile [415] with a thermal core at $r \simeq 3/4\, r_s$. The HI radial distribution reads

$$\rho_{\rm HI}\,(r) = \frac{\rho_0\, r_s^3}{(r + 3/4 r_s)\,(r+r_s)^2}\,, \tag{4.8}$$

where $r$ denotes the distance from the centre of the DM halo and $r_s = R_{\rm vir}/c_{\rm HI}$ is the HI scale radius. The normalisation $\rho_0$ in Eq. (4.8) is determined by requiring

$$M_{\rm HI}(M,z) = \int_0^{R_{\rm vir}} {\rm d}r\; 4\pi\, r^2\, \rho_{\rm HI}(r,M,z)\,. \tag{4.9}$$

For the HI concentration parameter, we adopt the fitting function of Ref. [412]:

$$c_{\rm HI} = \frac{4\, c_{\rm HI,0}}{(1+z)^\gamma} \left(\frac{M}{10^{11}\, M_\odot}\right)^{-0.109}\,, \tag{4.10}$$

where $c_{\rm HI,\,0} = 139$ and $\gamma = 0.13$. Finally, the Fourier transform of Eq. (4.8) truncated to $R_{\rm vir}$ is defined as

$$\tilde{u}_{\rm HI}(k) = \frac{4\pi}{M_{\rm HI}} \int_0^{R_{\rm vir}} {\rm d}r\; r^2 \rho_{\rm HI}(r)\, \frac{\sin kr}{kr}\,. \tag{4.11}$$

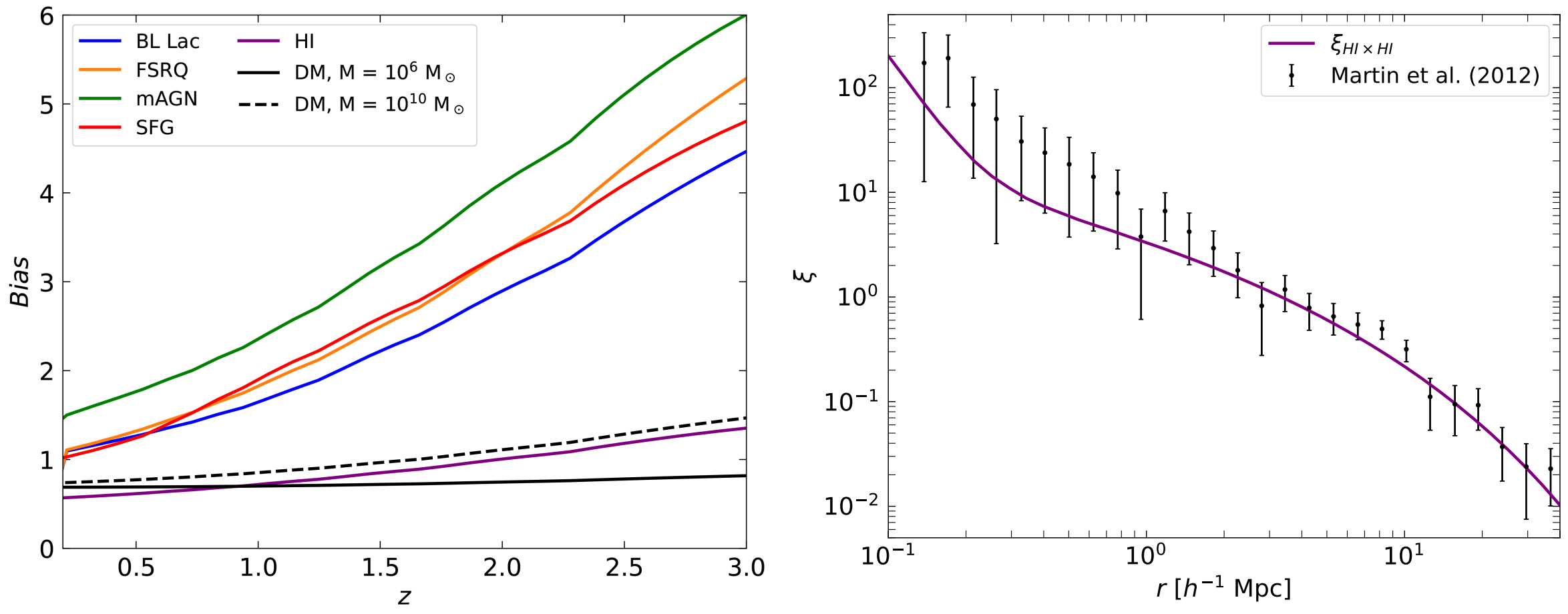

**Fig. 4.3.** *Left:* Bias as a function of redshift for: BL Lac (blue), FSRQ (orange), mAGN (green), SFG (red), HI (purple). The figure includes two DM scenarios (black), corresponding to two different halo masses: $10^6\,\mathrm{M}_\odot$ (solid) and $10^{10}\,\mathrm{M}_\odot$ (dashed). *Right:* The prediction of the real-space two-point correlation function (purple) compared to the data of the ALFALFA survey catalogue (black).

## 4.3 Neutral hydrogen clustering and power spectrum

We recall that in the halo model framework the total PS can be expressed as the sum of two contributions:

$$P_{\mathrm{HI}\times\mathrm{HI}} = P^{1\mathrm{h}}_{\mathrm{HI}\times\mathrm{HI}} + P^{2\mathrm{h}}_{\mathrm{HI}\times\mathrm{HI}}\,, \tag{4.12}$$

which can be written as

$$P^{1\mathrm{h}}_{\mathrm{HI}\times\mathrm{HI}}(k) = \frac{1}{\bar{\rho}_{\mathrm{HI}}} \int_{M_{\min}}^{M_{\max}} \mathrm{d}M\, M^2_{\mathrm{HI}}\, \tilde{u}^2_{\mathrm{HI}}(k)\,, \tag{4.13}$$

$$P^{2\mathrm{h}}_{\mathrm{HI}\times\mathrm{HI}}(k) = \left[\int_{M_{\min}}^{M_{\max}} \mathrm{d}M\, \frac{\mathrm{d}n}{\mathrm{d}M}\, \tilde{u}_{\mathrm{HI}}(k)\, \frac{M_{\mathrm{HI}}}{\bar{\rho}_{\mathrm{HI}}}\, b_h\right] P_{\mathrm{lin}}(k)\,. \tag{4.14}$$

In order to validate the model that we adopt for HI and its power spectrum, we calculate the real-space correlation function

$$\xi_{\mathrm{HI}}(r) = \frac{1}{2\pi^2} \int \frac{\mathrm{d}k}{k}\, k^3\, P_{\mathrm{HI}\times\mathrm{HI}}\, \frac{\sin kr}{kr} \tag{4.15}$$

and we compare the result with the measurement obtained with the Arecibo Legacy Fast ALFA (ALFALFA) survey catalogue [416]. The comparison between theory and data in the local Universe is illustrated in Fig. 4.3 (right). The measurements are displayed in black, while the purple curve refers to our theoretical prediction. We notice that our model is in good agreement with the data, especially considering large scales, which are the most relevant for our analysis of the cross-correlation signal. On small scales the model, while being in fair agreement with the data, nevertheless appears to slightly underestimate the experimental results for scales below 1 $h^{-1}$

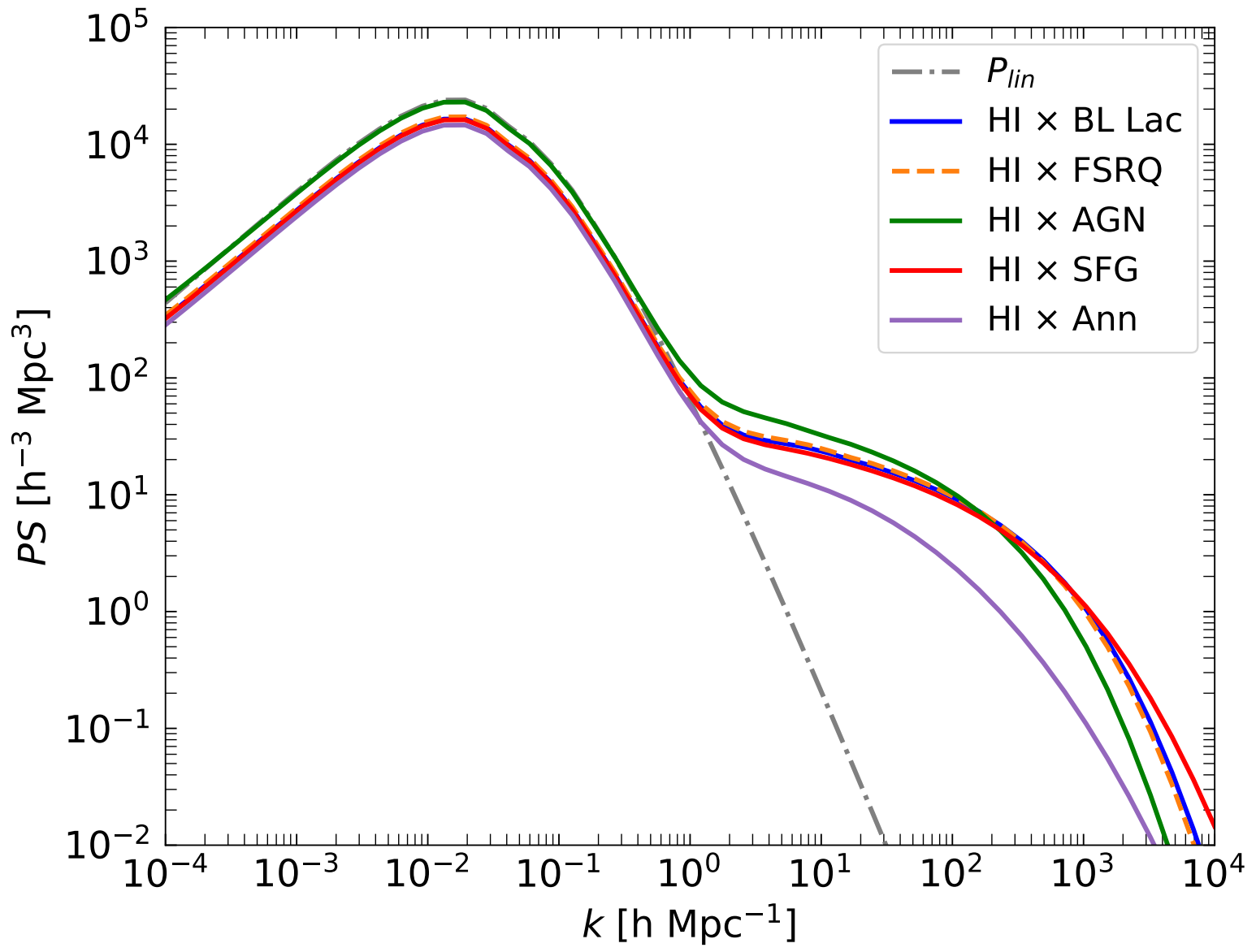

**Fig. 4.4.** Cross-correlation power spectra between the 21cm line and the $\gamma$-ray signal emitted by annihilating DM (black line), BL Lac (blue), FSRQ (orange), mAGN (green), SFG (red), calculated at $z = 0.5$.

Mpc. The origin of this discrepancy is likely due to the imperfect knowledge of the inner parts of the HI density profile on small scales[1].

Let us now turn to the discussion of the cross-correlation signal between the 21cm line and the $\gamma$-ray flux. Following Eqs. (3.53) and (3.54), the relevant expressions for the cross-correlation between the HI distribution and the DM emission are

$$P^{1h}_{HI\times DM}(k) = \int_{M_{min}}^{M_{max}} dM \, \frac{dn}{dM} \, \frac{\tilde{v}_{DM}(k)}{\Delta^2} \, \tilde{u}_{HI}(k) \, \frac{M_{HI}}{\bar{\rho}_{HI}} \, , \tag{4.16}$$

$$P^{2h}_{HI\times DM}(k) = \left[\int_{M_{min}}^{M_{max}} dM \, \frac{dn}{dM} \, \frac{\tilde{v}_{DM}(k)}{\Delta^2} \, b_h\right] \left[\int_{M_{min}}^{M_{max}} dM \, \frac{dn}{dM} \, \tilde{u}_{HI}(k) \, \frac{M_{HI}}{\bar{\rho}_{HI}} \, b_h\right] P_{lin}(k) \, . \tag{4.17}$$

In the case of the cross-correlation between the HI line and the astrophysical sources, the one-halo and two-halo terms read:

$$P^{1h}_{HI\times S}(k) = \int_{L_{min}}^{L_{max}} dL \, \frac{L}{\langle g_S \rangle} \, \tilde{u}_{HI}(M(L)) \, \phi(L) \, \frac{M_{HI}(M(L))}{\bar{\rho}_{HI}} , \tag{4.18}$$

$$P^{2h}_{HI\times S}(k,z) = \left[\int_{M_{min}}^{M_{max}} dM \, \frac{dn}{dM} \, \tilde{u}_{HI}(k) \, \frac{M_{HI}(M,z)}{\bar{\rho}_{HI}} \, b_h\right] \left[\int_{L_{min}}^{L_{max}} dL \, \frac{L}{\langle g_S \rangle} \, \phi(L) \, b_h(M(L))\right] P_{lin}(k) \, .$$

[1] See also Ref. [417] for an analysis of the impact of incorrect assumptions of HI clustering on cosmological parameter estimation.

The power spectra for all contributions at $z = 0.5$ are shown in Fig. 4.4. At large physical scales (small $k$) the power spectra are dominated by the two-halo term, which closely follows the linear matter power spectrum. Instead, at physical scales smaller than 1 $h^{-1}$ Mpc (large $k$) the one-halo term becomes dominant. We notice that on large scales, the cross-correlations involving the astrophysical sources are the same order of magnitude as the signal from annihilating DM. On small scales, $P_{\mathrm{HI}\times\mathrm{DM}}$ is below the other curves, but only by one order of magnitude. This situation is very different from the auto-correlation scenario illustrated in Fig. 3.7, where the astrophysical components are several orders of magnitude above the DM contribution, both on large and small scales. This is yet another confirmation that the cross-correlation between an EM signal and a gravitational tracer is a powerful technique to reveal a subdominant DM contribution. Also, the 21cm line proves to be a compelling gravitational proxy for the matter distribution.

# CHAPTER 5

## CROSS-CORRELATION SIGNAL AND DARK MATTER CONSTRAINTS

Cross-correlations between gravitational tracers of the large-scale distribution of matter in the Universe and the electromagnetic cosmic backgrounds promise to be a powerful tool to explore the origin of the unresolved components of these radiation fields, a place where an elusive dark matter particle signal might hide [330]. In fact, DM is expected to produce annihilation or decay signals, a most notable example being the $\gamma$-ray emission produced by essentially any kind of WIMPs. However, those signals are remarkably faint and immersed in an overwhelming astrophysical background. By including the three-dimensional spatial information of the gravitational tracer via the cross-correlation technique adds relevant details that can potentially assist in disentangling a DM signal from the other astrophysical emissions [331, 329, 337]. The cross-correlation technique has been employed to investigate the faint end of the UGRB by using galaxies [337, 338, 339, 340, 341, 342, 343, 344, 345, 346], clusters of galaxies [418, 419, 420, 421], lensing of the cosmic microwave background [347, 422] and the weak lensing effect of cosmic shear [330, 331, 332, 333, 334, 335, 336]. In this work we consider a novel possibility that will become available in the near future with the thorough investigation of the 21cm emission from cosmic HI, explored through the intensity mapping technique and made possible with the coming generation of radio telescopes, most notably the Square Kilometre Array (SKA). The 21cm emission works as a probe of the underlying matter field, and due to the large-scale structure of the Universe, it is intrinsically anisotropic. If DM consists of a new kind of elementary particle, able to produce a faint radiation by means of its self-annihilation, then this radiation traces the same DM structure probed by HI atoms and shares a statistically common pattern of fluctuations[1]. Emission from unresolved astrophysical sources also contributes to the cosmic radiation with a pattern that is necessarily correlated with the gravitational tracer, here represented by the 21cm line, as both are hosted by the same DM halos. For this reason, a level of cross-correlation between the HI brightness temperature with those cosmic radiation fields is expected. We study this cross-correlation signal by focusing on the high-energy tail of the cosmic radiation, namely on the broad $\gamma$-ray band, which is relevant both for the astrophysical sources and for DM particles. Specifically in terms of WIMPs, which have the ability to produce $\gamma$ rays through their annihilation products, like final state radiation of produced leptons or decays of hadrons generated in the annihilation process. The tool to investigate the cross-correlation signal is the angular power spectrum of the correlation between the fluctuations of the two fields, namely the 21cm line and the UGRB. We quantify the size of the signal and derive forecasts on the ability to detect this signal by adopting a full-sky, large field-of view $\gamma$-ray telescope like the Fermi Large-Area Telescope (FERMI-LAT), combined with the intensity mapping observations of SKA [423, 424] and, on a shorter timescale, its precursor MEERKAT [425]. We show that indeed the cross-correlation signal is potentially detectable with the SKA, with a possible hint attainable already with MEERKAT. We derive the prospects to set bounds on (or detect a signal for) the relevant DM particle physics properties, namely its mass and annihilation cross-section. Finally, we quantify what prospects can be achieved in the long term by the Phase 2 of SKA combined with a future generation $\gamma$-ray telescope with larger exposure and improved angular resolution. The chapter is organised as follows: Section 5.1 reviews the formalism of the angular power spectrum, Section 5.2 illustrates the window functions for the HI brightness temperature as well as $\gamma$ rays produced by DM and astrophysical sources. The experimental features of the detectors under consideration are outlined in Section 5.3. The cross-correlation angular power spectrum are displayed in Section 5.4 and the signal-to-noise ratio for a signal purely of astrophysical origin is illustrated in Section 5.5. Finally, the bounds on the DM parameter space are discussed in Section 5.6 and the main points of the analysis are summarised in Section 5.7.

[1] Note that DM decay is also a viable scenario, provided the decay is sufficiently suppressed in order to allow the particle to be long-lived on cosmological timescales. Nevertheless, in this work we focus on annihilating DM.

## 5.1 Formalism of the angular power spectrum

Let us define the intensity field $I_g$ in direction $\boldsymbol{n}$ as

$$I_g(\boldsymbol{n}) = \int \mathrm{d}\chi\, g(\chi,\boldsymbol{n})\,\widetilde{W}(\chi)\,, \tag{5.1}$$

where $\chi(z)$ is the radial comoving distance, $g(\chi,\boldsymbol{n})$ denotes the source density field, and $\boldsymbol{n}\cdot\boldsymbol{n} = 1$. The window function $\widetilde{W}(\chi)$ provides the redshift dependence of the observable under consideration. The normalised window function reads

$$W(\chi) = \langle g\rangle\,\widetilde{W}(\chi)\,, \tag{5.2}$$

with $\langle g\rangle$ denoting that the source density field is averaged over $\boldsymbol{n}$, and is related to the average intensity via

$$\langle I_g\rangle = \int \mathrm{d}\chi\, W(\chi)\,. \tag{5.3}$$

The intensity fluctuations are defined as

$$\delta I_g(\boldsymbol{n}) \equiv I_g(\boldsymbol{n}) - \langle I_g \rangle \,. \tag{5.4}$$

By expanding Eq. (5.4) in spherical harmonics $Y_{\ell m}$, we get

$$\delta I_g(\boldsymbol{n}) = \langle I_g \rangle \sum_{\ell m} a_{\ell m} Y_{\ell m}(\boldsymbol{n}). \tag{5.5}$$

Let the solid angle in direction $\boldsymbol{n}$ be denoted by $\Omega_{\boldsymbol{n}}$, the dimensionless coefficients $a_{\ell m}$ can be expressed as

$$a_{\ell m} = \frac{1}{\langle I_g \rangle} \int \mathrm{d}\Omega_{\boldsymbol{n}}\, \delta I_g(\boldsymbol{n}) Y^*_{\ell m}(\boldsymbol{n}) \tag{5.6}$$

$$= \frac{1}{\langle I_g \rangle} \int \mathrm{d}\Omega_{\boldsymbol{n}} \int \mathrm{d}\chi\, f_g(\chi, \boldsymbol{n})\, W(\chi)\, Y^*_{\ell m}(\boldsymbol{n}) \,, \tag{5.7}$$

where we recall that the fluctuation field is

$$f_g = \frac{g}{\langle g \rangle} - 1 \,. \tag{5.8}$$

The angular power spectrum (APS) measures the amplitude of the fluctuations and is defined as

$$C_\ell^{ij} = \frac{1}{2\ell + 1} \left\langle \sum_{m=-\ell}^{\ell} a^{(i)}_{\ell m} a^{*(j)}_{\ell m} \right\rangle , \tag{5.9}$$

where $i, j$ denote the two signals that we want to cross-correlate and the auto-correlation scenario corresponds to the case $i = j$. After some machinery, the APS can be expressed in terms of the Fourier PS:

$$C_\ell^{ij} = \frac{1}{\langle I_i \rangle \langle I_j \rangle} \int \frac{\mathrm{d}\chi}{\chi^2} W_i(\chi)\, W_j(\chi)\, P_{ij}\left(k = \frac{\ell}{\chi}, \chi\right) , \tag{5.10}$$

which is known as the dimensionless fluctuation APS. The detailed derivation of the above equation is explicitly discussed in Appendix D. In our analysis we employed the intensity APS, which corresponds to the fluctuation APS multiplied by the two mean intensity fields associated to the two observables. Thus, the unit of measurement of the intensity APS is $[C_\ell^{ij}] = [I_i] \times [I_j] \times [\mathrm{sr}]$[2].

## 5.2 Window functions

The window function is a weighting function which provides the redshift information of the observable and its shape strongly depends on the signal under consideration. In the following, we introduce the window functions for HI intensity mapping, annihilating DM and the four classes of astrophysical sources considered in our analysis.

[2]The steradian represents the unit of solid angles.

**Neutral hydrogen.** Following Ref. [426], we adopt a top-hat window function

$$W_{\rm HI}(z) = W_0(z)\, T_{\rm obs}(z)\,. \tag{5.11}$$

The normalisation reads

$$W_0(z) = \frac{\theta(z - z_{\rm min})\,\theta(z_{\rm max} - z)}{z_{\rm max} - z_{\rm min}}\,, \tag{5.12}$$

where $z_{\rm min}$ and $z_{\rm max}$ respectively represent the lower and upper edges of the redshift bin under consideration, and $\theta$ is the Heaviside step function. The mean observed brightness temperature can be expressed as:

$$T_{\rm obs}(z) = 44\,\mu K \left(\frac{\Omega_{\rm HI}(z)\,h}{2.45\cdot 10^{-4}}\right)\frac{(1+z)^2}{E(z)}\,,$$

where $E(z) = H(z)/H_0$. We recall that the dimensionless density is defined as

$$\Omega_{\rm HI} = \frac{8\pi G}{3H^2}\overline{\rho}_{\rm HI}\,, \tag{5.13}$$

with the mean HI density given by

$$\overline{\rho}_{\rm HI}(z) = \int_{M_{\rm min}}^{M_{\rm max}} {\rm d}M\, \frac{{\rm d}n}{{\rm d}M}(M,z)\, M_{\rm HI}(M,z)\,. \tag{5.14}$$

However, in the following we will use the redshift-independent value $\Omega_{\rm HI} = 2.45\cdot 10^{-4}$, as determined by Refs.[427, 426][3].

**Astrophysical sources.** The window function for astrophysical sources is obtained from the spectral energy distribution (SED) $\mathrm{d}\mathcal{N}/\mathrm{d}E$ weighted by the $\gamma$-ray luminosity function as follows:

$$W_\star(E,z) = \left(\frac{d_L(z)}{1+z}\right)^2 \int_{L_{\rm min}}^{L_{\rm max}} {\rm d}L\, \frac{{\rm d}\mathcal{N}}{{\rm d}E}(E,L,z)\,\phi(z)\,{\rm e}^{-\tau_{\gamma\gamma}[(1+z)E,z]}\,, \tag{5.15}$$

where $d_L$ is the luminosity distance and $L$ is the luminosity in the energy interval $(0.1-100)$ GeV. The SED, defined in Eq. (3.65), can be expressed in terms of $L$ through Eq. (3.70). We assume a power-law for the energy spectrum, which leads to a simple analytical expression for the differential flux

$$\frac{{\rm d}\mathcal{N}}{{\rm d}E} = \frac{L}{4\pi d_L^2}\left\{\frac{1}{2-\Gamma}\left[\left(\frac{100}{1+z}\right)^{2-\Gamma} - \left(\frac{0.1}{1+z}\right)^{2-\Gamma}\right]\right\}^{-1}\left(\frac{E}{\rm GeV}\right)^{-\Gamma}\,, \tag{5.16}$$

where we recall that Table 3.1 reports the values of the spectral index $\Gamma$ as well as the minimum and maximum luminosities $L_{\rm min}$ and $L_{\rm max}$ for each astrophysical source, while Appendix C illustrates the models adopted for the GLFs.

**Dark matter.** The window function for $\gamma$ rays from annihilating DM reads [430]

$$W_{\rm DM}(E,z) = \frac{1}{4\pi}\frac{\langle\sigma v\rangle}{2}\Delta^2(z)\left(\frac{\Omega_{\rm DM}\,\rho_{\rm c,0}}{m_\chi}\right)^2(1+z)^3\,\frac{{\rm d}N}{{\rm d}E}[(1+z)E]\,{\rm e}^{-\tau_{\gamma\gamma}[(1+z)E,z]}\,, \tag{5.17}$$

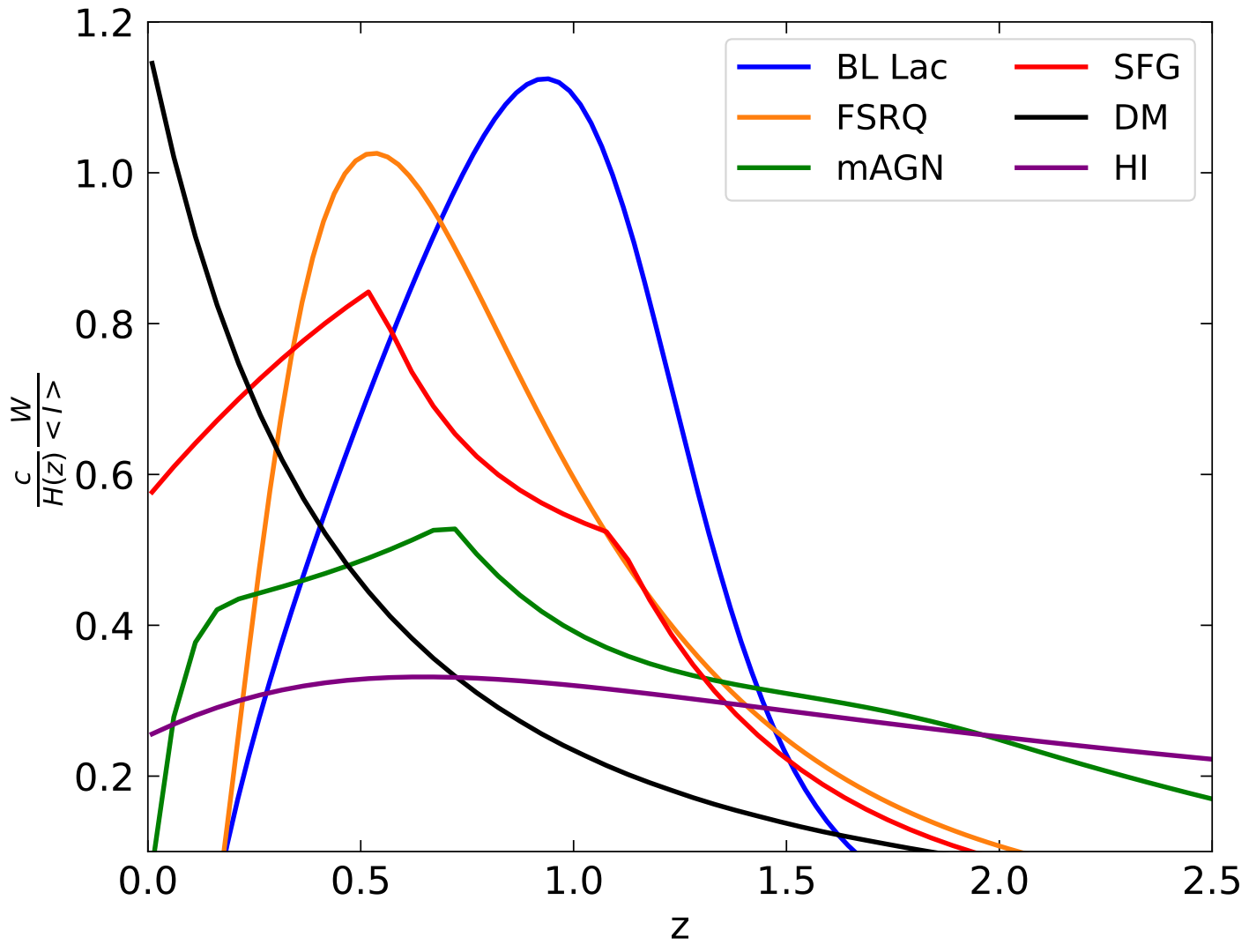

**Fig. 5.1.** Window functions as a function of redshift for the $\gamma$-ray emitters considered in our analysis, namely annihilating DM (black), BL Lac (blue), FSRQ (orange), mAGN (green), SFG (red), and HI (purple). The photon energy in the figure is set to $E = 5$ GeV.

where $dN/dE$ denotes the number of photons produced per annihilation event, for which we employed Pythia[4]. The derivation of Eq. (5.17) is illustrated in Appendix E.

The window functions, normalised to the Hubble parameter and to the average intensity, for the HI brightness temperature and the $\gamma$-ray flux are displayed in Fig. 5.1. We note that the $\gamma$-ray window functions for the unresolved astrophysical sources exhibit a peak at redshift in the range between 0.5 and 1. In the case of DM, whose emission is totally unresolved, the peak is prominent at low redshift and quickly decays as the redshift increases. For the 21cm brightness temperature, the window function is broad and almost featureless. However, the excellent frequency resolution of a radio telescope can be exploited to set apart specific redshift intervals: since the APS depends on the overlap of the two window functions associated with the two observables, redshift tomography can be used to outline the redshift range where the DM signal or the astrophysical emission is more prominent.

We validate the $\gamma$-ray model by comparing the average UGRB intensity $\langle I_\gamma \rangle$ defined in Eq. (5.3) (integrated over the energy bins of Table 5.1) with the measurements. Fig. 5.2 (left) illustrates the total intensity reported by the FERMI-LAT Collaboration [431] (blue-shaded region) and the results obtained in Ref. [345] (pink-shaded region) together with our theoretical estimate for the cumulative unresolved signal from the astrophysical sources (black line). We find that our nominal model is in good agreement with the observations. The total intensity shown in black refers to the sum of the astrophysical components only, demonstrating that the model is able to reproduce the UGRB intensity without DM across the majority of the energy range in the figure. The individual contributions from the astrophysical sources are also displayed: BL Lac (blue), FSRQ (orange), mAGN (green), SFG (red). For comparison, three DM cases are shown, referring to DM particles with $m_\chi = 10, 100, 1000$ GeV and thermal annihilation cross-section $\langle\sigma v\rangle = 3\times10^{-26}$ cm$^3$ s$^{-1}$. In order to further corroborate the models with respect to the anisotropies, we calculate the auto-correlation APS and compare them with the measurements of the anisotropies of the UGRB obtained in Ref. [395]. Fig. 5.2 (right) illustrates $C_\ell^{\gamma\gamma}$ as a function of the photon energy.

[3] Refs. [428, 429] show that there is at most a factor 2 of difference along the redshift range of interest.

[4] The Python code that produces the DM energy spectra modelled with Pythia is a private communication of Prof. Nicolao Fornengo.

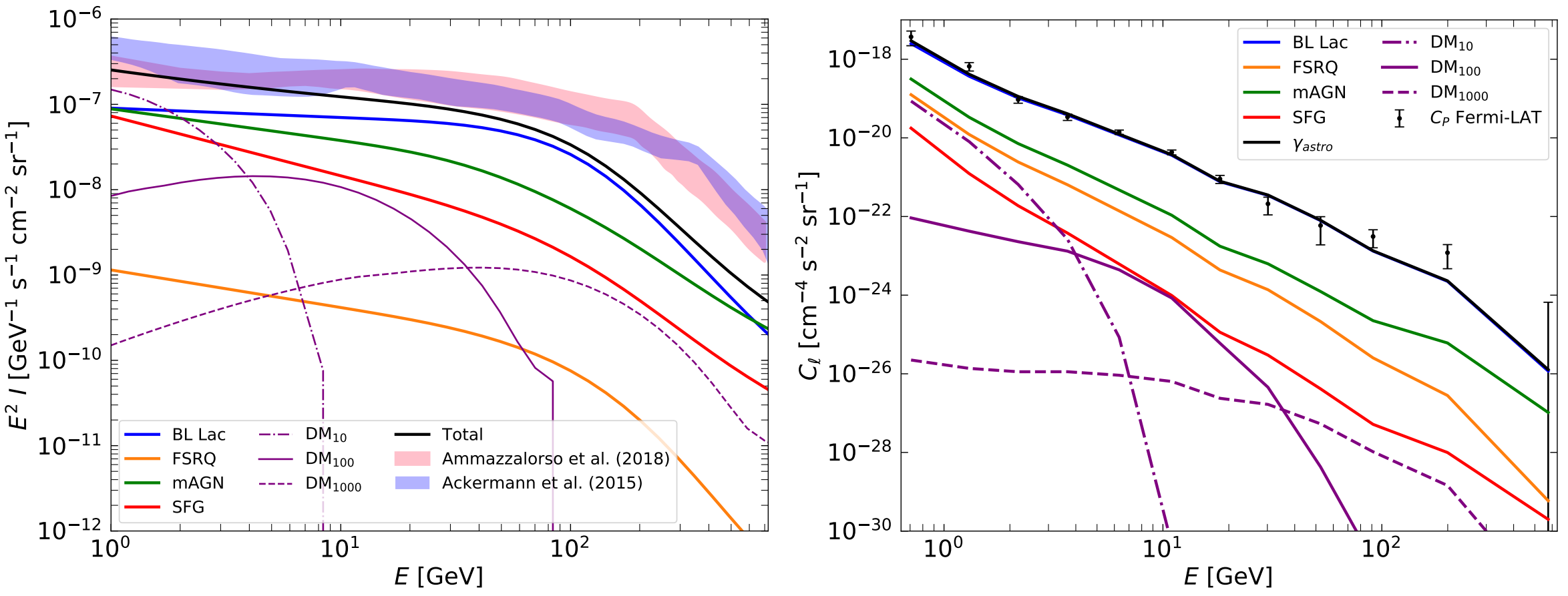

**Fig. 5.2.** *Left:* Measured total astrophysical flux (pink [345] and blue [431] shaded bands) together with the theoretical predictions for the contribution from BL Lac (blue), FSRQ (orange), mAGN (green) and SFG (red) as a function of the photon energy. The upper black line denotes the sum of all astrophysical components. The purple dash-dotted, solid and dashed lines show three predictions for a DM signal produced by particles with thermal cross-section and mass $m_\chi = 10, 100, 1000$ GeV, respectively. *Right:* Comparison between the theoretical estimation of the total auto-correlation APS of astrophysical origin, $C_\ell^{\gamma\gamma}$ (black), and the latest measurement of the FERMI-LAT detector, $C_\mathrm{P}$ [395] (dots with error bars), as a function of the energy bins. The individual contributions from the astrophysical sources and the predictions for annihilating DM are also shown with the same colours and line styles of the left panel.

The total theoretical signal of purely astrophysical origin (black line) is compared to the FERMI-LAT measurements (black crosses). It is apparent that there is an excellent agreement between the data and the theoretical model, dominated by the BL Lac component. It is also clear that the DM contribution to both the UGRB mean intensity and APS are expected to be subdominant. However, the different dependence on $E$ of the astrophysical sources and the DM emissions can be exploited in the cross-correlation analysis, in attempt to disentangle the two types of contributions and emphasise the DM component.

| Bin | $E_{\rm min}$ (GeV) | $E_{\rm max}$ (GeV) | $C_N$ ($\rm cm^{-4}\ s^{-2}\ sr^{-1}$) | $f_{\rm sky}$ | $\sigma_0^{\rm Fermi}$ (deg) | $E_{\rm b}$ (GeV) |
|---|---|---|---|---|---|---|
| 1 | 0.5 | 1.0 | $1.056 \times 10^{-17}$ | 0.134 | 0.87 | 0.71 |
| 2 | 1.0 | 1.7 | $3.548 \times 10^{-18}$ | 0.184 | 0.50 | 1.30 |
| 3 | 1.7 | 2.8 | $1.375 \times 10^{-18}$ | 0.398 | 0.33 | 2.18 |
| 4 | 2.8 | 4.8 | $8.324 \times 10^{-19}$ | 0.482 | 0.22 | 3.67 |
| 5 | 4.8 | 8.3 | $3.904 \times 10^{-19}$ | 0.549 | 0.15 | 6.31 |
| 6 | 8.3 | 14.5 | $1.768 \times 10^{-19}$ | 0.574 | 0.11 | 11.0 |
| 7 | 14.5 | 22.9 | $6.899 \times 10^{-20}$ | 0.574 | 0.09 | 18.2 |
| 8 | 22.9 | 39.8 | $3.895 \times 10^{-20}$ | 0.574 | 0.07 | 30.2 |
| 9 | 39.8 | 69.2 | $1.576 \times 10^{-20}$ | 0.574 | 0.07 | 52.5 |
| 10 | 69.2 | 120.2 | $6.205 \times 10^{-21}$ | 0.574 | 0.06 | 91.2 |
| 11 | 120.2 | 331.1 | $3.287 \times 10^{-21}$ | 0.597 | 0.06 | 199.5 |
| 12 | 331.1 | 1000. | $5.094 \times 10^{-22}$ | 0.597 | 0.06 | 575.4 |

**Table 5.1.** Technical specifications of the FERMI-LAT detector and $\gamma$-ray analysis. In order of appearance: lower and upper edges of the energy bin, noise $N^\gamma$, fraction of the sky $f_{\rm sky}$ outside the combined Galactic and point-source masks, 68° containment angle $\sigma_0^{\rm Fermi}$ of the FERMI-LAT point-spread function. The containment angles refer to the geometric centre of the energy bin $E_{\rm b} = \sqrt{E_{\rm min}\, E_{\rm max}}$ (last column).

## 5.3 Experiments

The technical specifications of the detectors, together with the auto-correlation signals, determine the variance $\Delta C_\ell^{ij}$ on the predicted cross-correlation APS. Under the hypothesis of gaussianity, it holds

$$\left(\Delta C_\ell^{ij}\right)^2 = \frac{1}{(2\ell+1) f_{\rm sky}} \left\{ \left(C_\ell^{ij}\right)^2 + \left[C_\ell^{ii} + \frac{C_N^i}{\left(B_\ell^i\right)^2}\right] \left[C_\ell^{jj} + \frac{C_N^j}{\left(B_\ell^j\right)^2}\right] \right\} \tag{5.18}$$

where $C_\ell^{ii}$ and $C_\ell^{jj}$ denote the auto-correlation APS of the two observables, $C_N$ is the noise of the experiment, $f_{\rm sky}$ represents the observed fraction of the sky and $B_\ell$ is the beam window function of the detector. In this section, we briefly introduce the characteristic features both of $\gamma$-ray and radio telescopes. Also, we discuss one by one all the factors which are necessary to estimate the error bars in the forecast of our cross-correlation signal.

**$\gamma$-ray detector.** The Fermi Gamma-Ray Space Telescope (usually abbreviated as FERMI) is a space observatory devoted to $\gamma$-ray astronomy in the energy range 100 MeV $\lesssim E \lesssim$ 1 TeV. The main instrument aboard is the Large Area Telescope (LAT), a pair-conversion detector with a broad field of view [50]. We adopt the measurements of the anisotropies, as given in Ref. [395], which are based on 8 years of FERMI-LAT observations (specifically Pass 8 data), and a selection of events with optimal angular resolution and background rejection. In our analysis we employed twelve energy bins, covering the interval from 0.5 GeV up to 1 TeV. We adopted the photon noise $C_N$ as derived in Ref. [395]. For future convenience, we also note that the photon noise of the

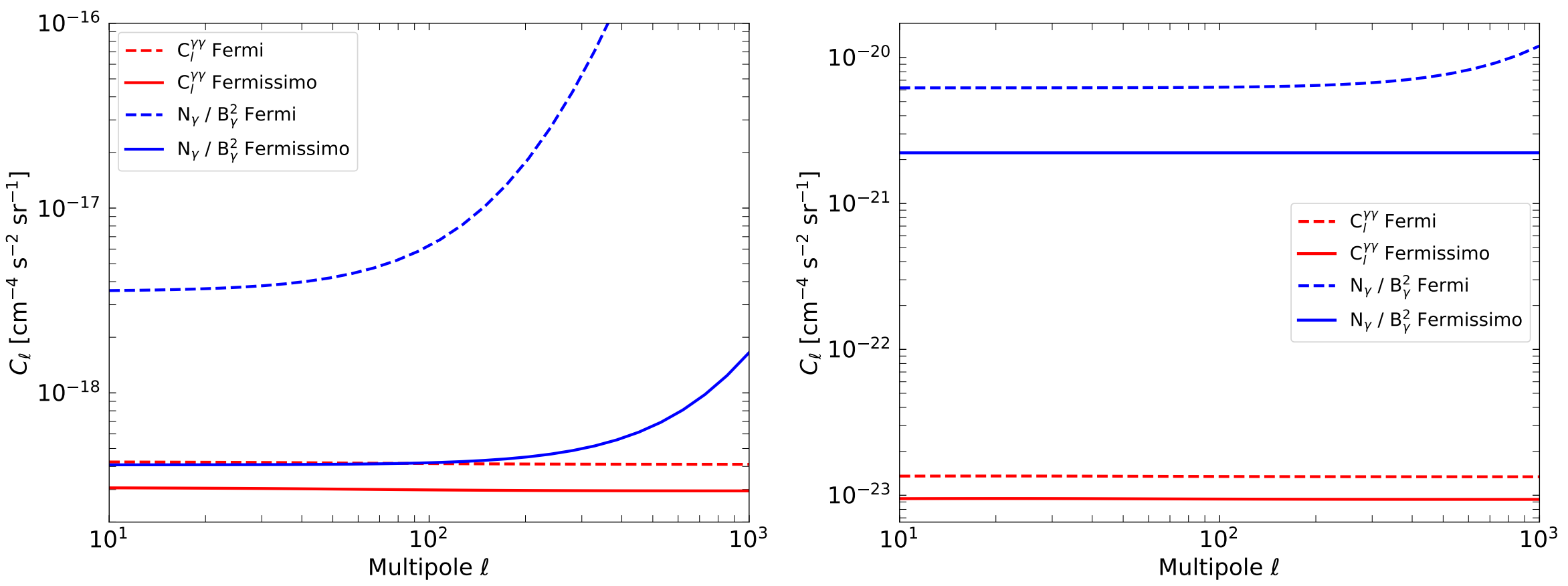

**Fig. 5.3.** Terms contributing to the Gaussian estimate of the variance in the cross-correlation signal and arising from the $\gamma$-ray auto-correlation signals $C_\ell^{\gamma\gamma}$ and its noise $N^\gamma$. The left panel refers to the energy bin number 2, while the right panel to the energy bin number 10, as reported in Table 5.1. The different lines refer to the two contributions for different $\gamma$-ray telescope specifications, as stated in the inset box.

detector can be approximated as [432]

$$C_N = \frac{4\pi f_{\rm sky} \langle I \rangle^2}{N_\gamma}\,, \tag{5.19}$$

where $N_\gamma$ is the total number of observed events outside the mask and $f_{\rm sky}$ denotes the fraction of sky observed by the detector. Note that the photon noise is related to the Poisson fluctuations between pixels in the FERMI maps and is caused by a finite event statistics, thus it will vanish in the utopian situation of infinite statistics. Instead, the fluctuations in the UGRB are related to the distribution of sources. As a result, the UGRB fluctuations could be observed above the $C_N$ if a large statistics is available. The angular resolution of the experiment is encoded in the point-spread function (PSF), which quantifies the degree of spreading (blurring) of a point source. The beam window function $B_\ell$ in multipole space can be defined in terms of the PSF

$$B_\ell = \int \mathrm{d}\Theta\, \mathrm{PSF}(\Theta)\, \sin\Theta\, P_\ell(\cos\Theta)\,, \tag{5.20}$$

where $P_\ell(\cos\Theta)$ denotes the Legendre polynomials of degree $\ell$. The full FERMI-LAT beam window function depends on the photon event class[5] as well as on the photon energy and is available through the FERMI tools [433]. We found that a good analytic approximation for $B_\ell^\gamma(E)$ is a modified Gaussian in multipole-space

$$B_\ell^\gamma(E) = \exp\left[-\frac{\sigma_{\rm b}(\ell,E)^2\,\ell^2}{2}\right], \tag{5.21}$$

[5]Each event is associated with a certain probability to be indeed a real photon event. The photon samples are divided into several event classes, which may differ for the contamination of background events. Thus, different categories are optimised for the study of different sources and phenomena.

where the dispersion evolves with $\ell$ for large multipoles

$$\sigma_{\rm b}(\ell,E) = \sigma_0^{\rm Fermi}(E)\left[1+0.25\,\sigma_0^{\rm Fermi}(E)\,\ell\right]^{-1}. \tag{5.22}$$

The normalisation $\sigma_0(E)$ is the 68% containment angle of the FERMI-LAT detector at energy $E$. The energy evolution of the $\sigma_0(E)$ parameter is well reproduced by

$$\sigma_0^{\rm Fermi}(E) = \sigma_0^{\rm Fermi}(E_{\rm ref}) \times \left(\frac{E}{E_{\rm ref}}\right)^{-0.95} + 0.05\,{\rm deg}, \tag{5.23}$$

with $E_{\rm ref} = 0.5$ GeV and $\sigma_0^{\rm Fermi}(E_{\rm ref}) = 1.20$ deg. The relation follows a power-law behaviour and flattens out at around 0.05 deg for high energies[6]. These empirical relations reproduce well the beam function adopted in Ref. [395] and derived from the FERMI-LAT data. The energy bins employed in the analysis and the technical specifications of the detectors are illustrated in Table 5.1, which shows: the lower and upper edges of each energy bin (second and third columns, respectively), the measured noise $C_N$ (fourth), the observed sky fraction $f_{\rm sky}$ and the angular resolution $\sigma_0^{\rm Fermi}$ (fifth) at the geometric centre of each bin $E = \sqrt{E_{\rm min}\,E_{\rm max}}$ (last column), as determined by the FERMI Collaboration [395].

In order to determine the potential of the cross-correlation signal with respect to investigating the particle nature of DM, we also examine the capabilities of a future $\gamma$-ray detector set-up, which we called FERMISSIMO[7]. The hypothetical experiment was attributed the following characteristics: First, we assume the exposure $A$ of the detector to be twice the size of the current FERMI-LAT specification adopted here. This implies that the limiting sensitivity to the unresolved sources $L_{\rm sens}$ is scaled down by the square-root of the increase in the exposure, i.e. by a factor $\sqrt{2}$, and the signal coming from the unresolved astrophysical sources gets slightly diminished (while the DM signal remains unchanged). We recall that the noise can be determined using Eq. (5.19), where $N_\gamma = 4\pi f_{\rm sky}\langle I\rangle A$, thus $C_N$ scales with the inverse of $A$. Second, we assume that the detector PSF can be improved, and we adopt the same behaviour of the beam function expressed in Eqs. (5.21) and (5.22) but with a better angular resolution [434], viz.

$$\sigma_0(E) = \alpha_\sigma \times \sigma_0(E_{\rm ref}) \times \left(\frac{E}{E_{\rm ref}}\right)^{-0.95} + 0.001\,, \tag{5.24}$$

for which we assume $\alpha_\sigma = 0.2$. Finally, a better angular resolution translates into a smaller mask, therefore we adopt a larger sky-fraction coverage of $f_{\rm sky} = 0.8$, which allows to slightly reduce the impact of noise. For definiteness, we work with the same energy bins of Table 5.1: this allows us to directly rescale the noise estimate in terms of the adopted changes in exposure and sky coverage. Fig. 5.3 details the contribution to the variance for a low (bin 1, left panel) and high (bin 9, right panel) energy bin, as a function of the multipole. The red lines refer to the auto-correlation signal $C_\ell^{\gamma\gamma}$, while the blue curves denote $C_N/B_\ell^2$. The dashed lines indicate the FERMI-LAT configuration, while the solid curves refer to the FERMISSIMO set-up. The exponential increase of the noise term for the low-energy bin is due to the detector beam function, which is suppressed for multipoles larger than 100 at low energies for FERMI-LAT. For larger energies the suppression becomes less relevant for the multipole range adopted here. We note the FERMI-LAT configuration is always noise-dominated, while the $N_\gamma/B_\ell^2$ for FERMISSIMO in the low-energy bins are the same order of magnitude of the signal. This will impact the error bars of the APS, as will be discussed in Sections 5.4 and 5.6 .

[6] We adhered to the specifications of the PSF2 event type response function of Ref. [395].

[7] This set-up has similar specifications to the ones assumed in Ref. [331], when forecasting the cross-correlation signal with cosmic shear.

| | $S$ (deg$^2$) | $t$ ($10^3$ hr) | $N_{\rm dish}$ | $D_{\rm dish}$ (m) | $D_{\rm interf}$ (km) | $N_{\rm b}$ | $f_{\rm sky}$ | $[z_{\rm min}, z_{\rm max}]$ | |
|---|---|---|---|---|---|---|---|---|---|
| MEERKAT | 4,000 | 4 | 64 | 13.5 | 1 | 1 | 0.097 | $[0.4, 1.45]$<br>$[0.0, 0.58]$ | UHF-band<br>L-band |
| SKA1 | 25,000 | 10 | 133+64 | 14.5 | 3 | 1 | 0.61 | $[0.35, 3]$<br>$[0.0, 0.5]$ | Band 1<br>Band 2 |
| SKA2 | 30,000 | 10 | 2,000 | 14.5 | 10 | 36 | 0.72 | $[0.35, 3]$<br>$[0.0, 0.5]$ | Band 1<br>Band 2 |

**Table 5.2.** Technical specifications for MEERKAT, SKA1 and SKA2, as used in our analysis: $S$ denotes the area covered by the survey, $t$ is the total observation time, $N_{\rm dish}$ indicates the number of dishes, $D_{\rm dish}$ is the dish diameter, $D_{\rm interf}$ represents the interferometer mean baseline, $N_{\rm b}$ is the number of beams, $f_{\rm sky}$ is the fraction of sky covered by the detector. The two last columns show the redshift bands adopted in the analysis and their assigned names.

**Intensity mapping radio telescope.** Intensity mapping is one of the main science drivers for cosmology with the next generation of radio telescopes currently under construction, notably the Square Kilometre Array [423, 435]. As of now, some of the SKA precursors are dedicating observing time to IM studies, most notably of all the MeerKaroo Array Telescope (MEERKAT) [425]. It is worth mentioning also purpose-built instruments such as: the Canadian Hydrogen Intensity Mapping Experiment (CHIME) [436], BAO In Neutral Gas Observations (BINGO) [437], TianLai [438]; or the Five hundred metre Aperture Spherical Telescope (FAST) [439]. IM surveys can be divided into two kinds, depending on whether they will operate the system in single-dish mode or interferometry [440]. The first set-up employs the auto-correlation signal from one or more dishes and is best suited to study correlations at large angular separations. Conversely, the interferometer set-up employs the cross-correlation signal from different array elements and it estimates the Fourier modes on the sky with high angular resolution. The most promising new generation radio telescope is certainly the SKA, which will be the world's largest radio telescope. Phase 1 of the SKA (SKA1) will cover a frequency range from 50 MHz to 14 GHz and will be arranged in two independent sub-arrays known as SKA1-LOW (the low-frequency instrument, in Australia), and SKA1-MID (operating at mid-frequencies, in the Karoo desert in South Africa). In our analysis we focus on the latter, which will observe frequencies higher than 350 MHz, corresponding to the late Universe at redshift below 3. SKA1-MID can be used both in single-dish and interferometer modes, which we consider in our analysis. Following Ref. [426], the detector beam window function of the radio telescope can be parameterised as

$$B_\ell^{\mathrm{HI}} = \exp\left[-\frac{\ell^2}{2}\left(\frac{1.22}{\sqrt{8\ln 2}}\frac{\lambda_o}{D}\right)^2\right], \tag{5.25}$$

where $\lambda_o$ is the observed wavelength of the HI line, related to the wavelength of emission $\lambda_e$ via $\lambda_o = \lambda_e\,(1+z) = 0.21\,(1+z)$ m. The reference length $D$ corresponds to the diameter of a dish $D_{\mathrm{dish}}$ for single-dish surveys, while for interferometers we will consider the length of the core baseline $D_{\mathrm{interf}}$, which in the case of SKA1 contains approximately 75% of the total number of dishes.

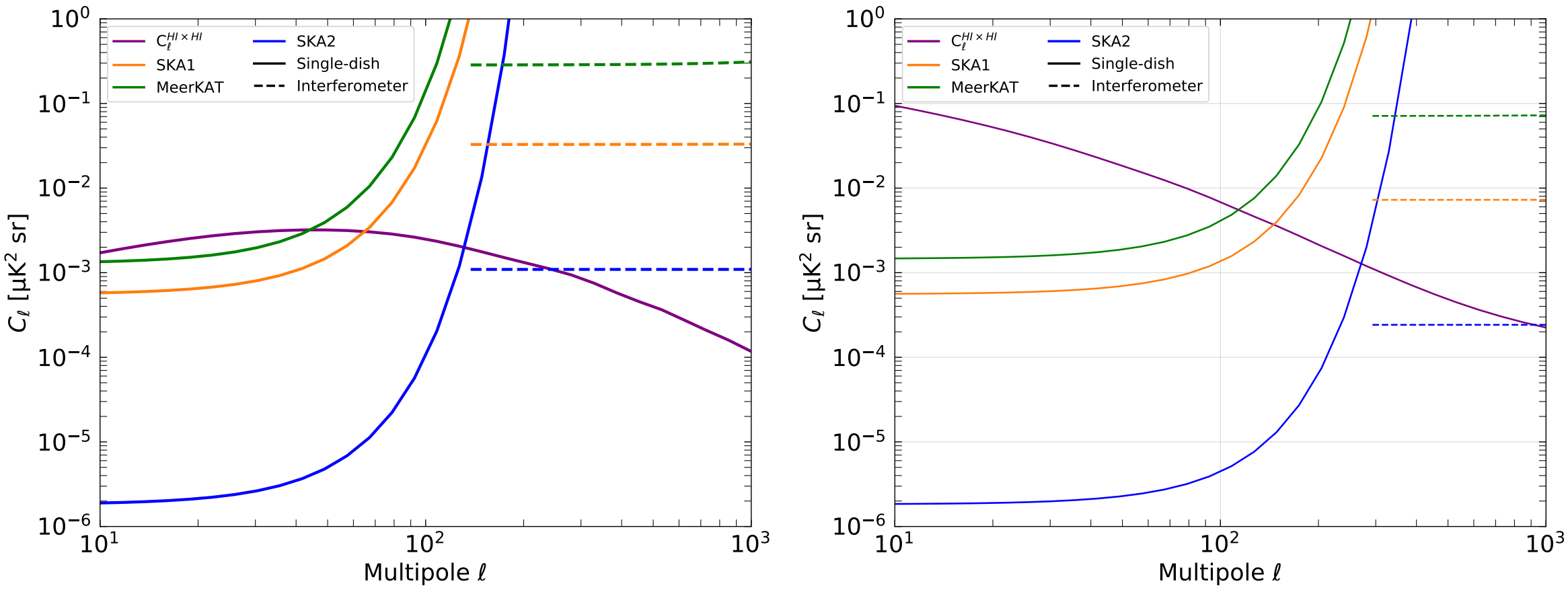

**Fig. 5.4.** Terms contributing to the Gaussian estimate of the variance in the cross-correlation signal and arising from the HI: the auto-correlation signals $C_\ell^{\mathrm{HI}\times\mathrm{HI}}$ and its noise $C_N^{\mathrm{HI}}$. The left panel refers to SKA Band 1, while the right panel to SKA Band 2, as reported in Table 5.2. Different colours represent different detectors: MEERKAT (green), SKA1 (orange), SKA2 (blue). The purple line denotes the signal $C_\ell^{\mathrm{HI}\times\mathrm{HI}}$. The solid curves indicate the single-dish configuration, while the interferometric set-up is displayed with dashed lines.

Regarding the noise of the single-dish survey, we follow Ref. [440] and adopt

$$C_{N,\mathrm{dish}} = \frac{T_{\mathrm{sys}}^2\, S}{N_{\mathrm{dish}}\, t\, \Delta\nu\, N_{\mathrm{b}}\, N_{\mathrm{pol}}\, \eta^2}, \tag{5.26}$$

where $T_{\mathrm{sys}}$ is the total system temperature, for which we take $T_{\mathrm{sys}}$ = 30 K + 60 (300 MHz/$\nu$)$^{2.55}$ K in all configurations, $S$ is the survey area, $N_{\mathrm{dish}}$ denotes the number of dishes, $t$ is the observation time, $\Delta\nu$ indicates the frequency band corresponding to the redshift bin considered, $N_{\mathrm{pol}} = 2$ denotes the number of polarisation states, and $\eta$ is the efficiency (assumed to be unity). Finally, $N_{\mathrm{b}}$ denotes the number of simultaneous beams that will be different from 1 when considering the use of phased array feeds[8] for SKA2 [444].

[8] Phased array feeds are multi-pixel and wide field-of-view receivers, where each receiving element acts as small antenna [441, 442]. This state-of-the-art technology is currently employed by the Australian SKA Pathfinder (ASKAP) [443, 441, 444].

For an interferometer survey, the noise can be written as [444]

$$C_{N,\mathrm{interf}} = \frac{T_{\mathrm{sys}}^2 S \,\mathrm{FoV}}{n(u)\, t\, \Delta\nu N_{\mathrm{b}} N_{\mathrm{pol}}\, \eta^2}\,, \tag{5.27}$$

where the average number density of baselines[9] is taken to be $n(u) = 0.005$ for SKA and a factor of 10 smaller for MEERKAT [444, 445]. The field of view is $\mathrm{FoV} \simeq \lambda_{\mathrm{o}}^2/D_{\mathrm{dish}}^2$. Eq. (5.27) is valid only for angular scales smaller than $\lambda_{\mathrm{o}}/D_{\mathrm{short}}$, where the shortest baseline $D_{\mathrm{short}}$ of the array is typically a few times $D_{\mathrm{dish}}$. For definiteness, we will adopt $D_{\mathrm{short}} = 2D_{\mathrm{dish}}$, which reflects in a minimal multipole $\ell_{\mathrm{cut}} = \pi D_{\mathrm{short}}/(1.22\lambda_{\mathrm{o}})$. Thus, for the interferometric case we will consider only $\ell \geq \ell_{\mathrm{cut}}$. Also, we investigate the potential of an upgraded version of SKA1 with enhanced capabilities, notably a longer baseline and a higher number of dishes. We name this set-up SKA Phase 2 (SKA2)[10]. Table 5.2 illustrates the instrumental specifications for the three IM radio telescopes adopted in our analysis: MEERKAT, SKA1 and SKA2. Fig. 5.4 displays $C_N/B_\ell^2$ for MEERKAT (green), SKA1 (orange), SKA2 (blue) against the auto-correlation signal $C_\ell^{\mathrm{HI}\times\mathrm{HI}}$ (purple). The left panel refers to $0.35 < z < 3.0$, while the right panel applies to $0.0 < z < 0.5$ (Band 2). The noise of the single-dish configuration (solid lines) blows up for multipoles of the order of 80 to 100, similar to FERMI-LAT at low photon energies, with FERMI-LAT resolution at high energies being better than the one obtained for the radio telescope in single-dish configuration. In this case, the range of multipoles which contribute information to the cross-correlation signal is limited by the single-dish resolution. However, in the interferometric configuration (dashed curves), the excellent angular resolution provides a large gain over the single-dish configuration for multipoles larger than $\ell_{\mathrm{cut}}$, i.e. 120 for Band 1 and 250 for Band 2. Therefore, we derive our results for two different configurations of the radio telescopes: single-dish and a combination of single-dish and interferometer, which takes into account the best between the two noise terms. For SKA2, Fig. 5.4 also shows that the error budget in Band 2 is (almost) always dominated by the HI auto-correlation term up to large multipoles, thus not limited by the noise term.

The discussion of the different sources of error in the $\gamma$-ray and IM detectors is instrumental to determine the maximal value of the multipole over which we focus our analysis: we adopt $\ell_{\mathrm{max}} = 1000$, which is also consistent with the analysis of other types of correlations performed with the FERMI-LAT data [345, 395]. We ensure that the Limber approximation [446, 447, 448] is valid by selecting the minimum multipole $\ell_{\mathrm{min}} = 10$.

## 5.4 Cross-correlation angular power spectrum

Now we have all the elements necessary to compute the cross-correlation APS and its error. Fig. 5.5 illustrates $\ell(\ell+1)C_\ell$ as a function of the multipole, for the combination FERMI-LAT × MEERKAT, while in Figs. 5.6 and 5.7 the radio telescopes are SKA1 and SKA2, respectively. In each figure, the left panel refers to the signal integrated in the higher redshift band (UHF-band for MEERKAT and Band 1 for SKA) and the right panel to the lower redshift band (L-band for MEERKAT and Band 2 for SKA), as reported in Table 5.2. The curves refer to the sum of the signal in the twelve $\gamma$-ray energy bins of Table 5.1. The dashed lines show the signal originated by astrophysical sources (as indicated in the inset boxes of the figures) and the purple solid line stands for the signal produced by annihilation events of DM particles with mass $m_\chi = 100$ GeV and thermal cross-section $\langle\sigma v\rangle = 3\times10^{-26}$ cm$^3$ s$^{-1}$, annihilating into a $b\bar{b}$ quark pair, representative of an hadronic final state. The DM signal, being directly proportional to the annihilation cross-section, can be increased or decreased by acting on $\langle\sigma v\rangle$, while a change in the mass values implies a different energy spectrum, as can be seen in Fig. 5.2 for some representative cases. The DM clustering properties can also boost the size of the APS of annihilating DM by up to a factor of a few [449] or suppress it by up to a factor of 10 [368].

[9]Two receivers provides one baseline. Similarly, N receivers correspond to N(N-1)/2 baselines.
[10]This set-up is based on the specifications indicated in Ref. [444].

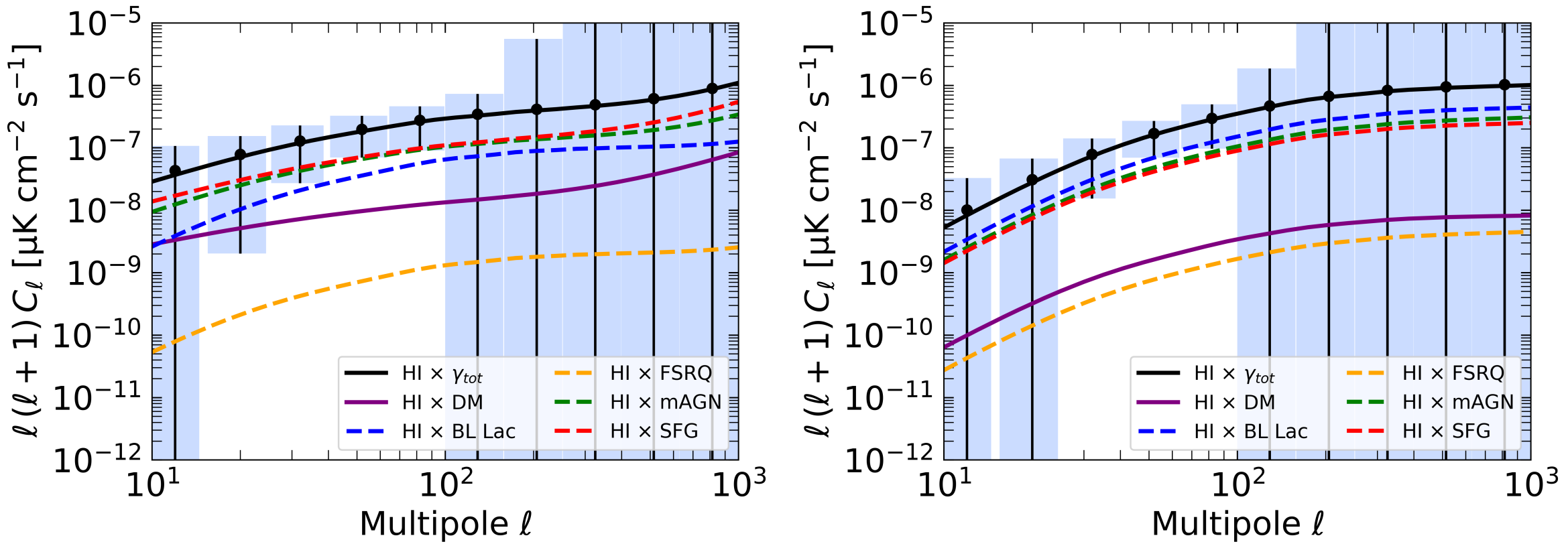

**Fig. 5.5.** Angular power spectrum of the cross-correlation between HI intensity mapping and $\gamma$ rays. The different lines refer to the theoretical predictions of the signal originated by the different $\gamma$ rays sources, as indicated in the inset box. The purple solid line, which refers to the signal due to annihilating DM, is obtained for a DM mass $m_\chi = 100$ GeV and a thermal annihilation rate $\langle\sigma v\rangle = 3\times10^{-26}$ cm$^3$ s$^{-1}$. The solid black line is the sum of all components. The error bars are obtained as the Gaussian estimate of the variance of the signal and refer to the combined (dish + interferometer) configuration. The results refer to the sum of the contributions in all FERMI-LAT energy bins of Table 5.1. The radio telescope configuration is the higher redshift MEERKAT UHF-band in the left panel and the lower redshift L-band in the right panel.

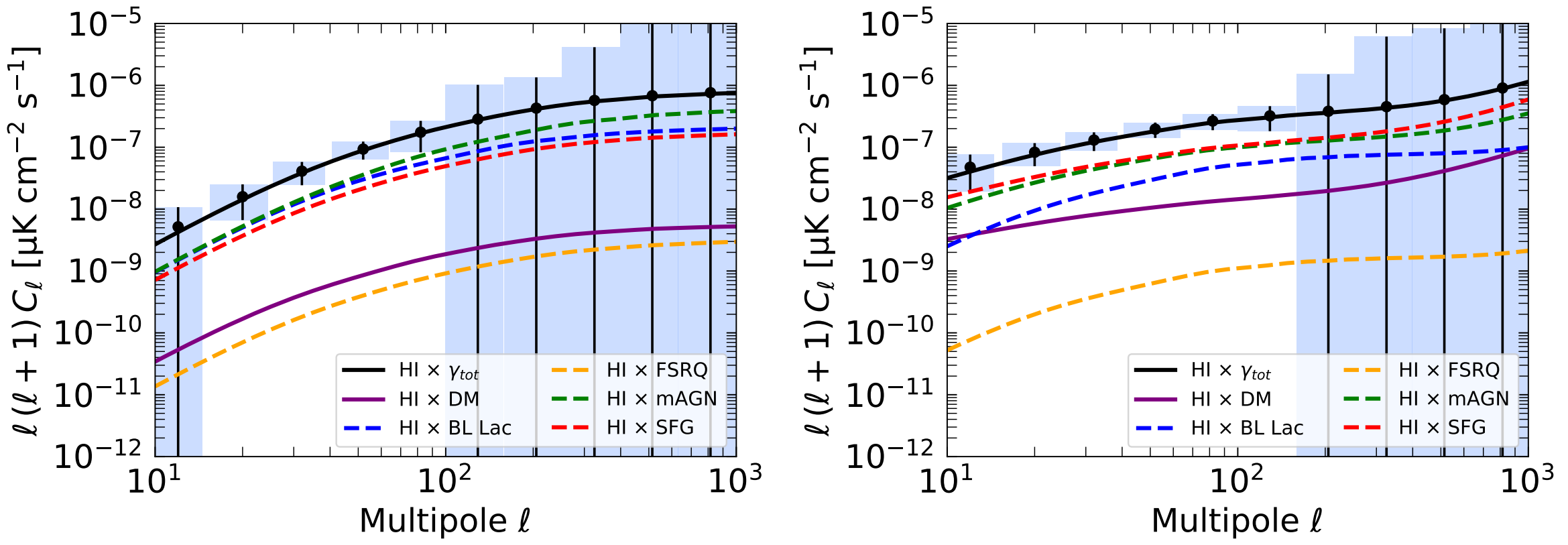

**Fig. 5.6.** The same as in Fig. 5.5, for the combination SKA1 × FERMI-LAT. The left panel refers to the higher redshift band of SKA, Band 1, while the right panel to the lower redshift Band 2.

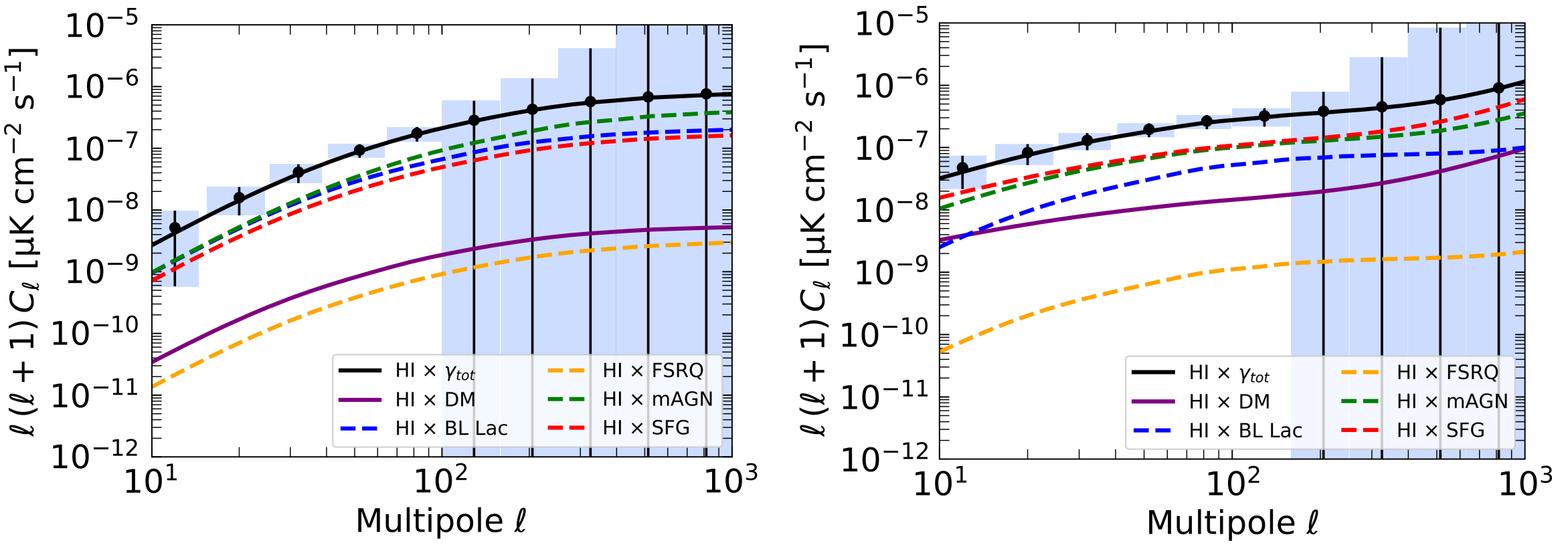

**Fig. 5.7.** The same as in Fig. 5.5, for the experimental configuration SKA2 × FERMI-LAT. The left panel refers to the higher redshift Band 1, the right panel to the lower redshift Band 2.

The relative size between the astrophysical sources and the DM signals is dictated by their different energy spectra, as shown in Fig. 5.2, and to the different redshift evolution of the corresponding window functions, as displayed in Fig. 5.1. A breakdown of the signal produced in different energy bins is shown in Fig. 5.8 for the lower redshift band (Band 2), illustrating the relative contributions at different energies. The four panels refer to the energy bins denoted as 1, 2, 3 and 9 in Table 5.1. Concerning the redshift dependence, the comparison of the left and right panels of Figs. 5.5, 5.6 and 5.7 shows that the signal coming from higher redshift (Band 1) is dominated by the two-halo term, while for the lower redshift (Band 2) the one-halo term emerges at multipoles larger than approximately 500. The cross-correlation signal due to the astrophysical sources is dominated by mAGN, SFG and BL Lac, with the latter being less important for the low-redshift emission of Band 2. This fact arises from an interplay between the redshift dependence of the window functions and the size of the $\gamma$-ray emission of each individual class of sources, as shown in Fig. 5.2 and Fig. 5.1. For instance, the dominance of SFG and mAGN for the signal coming from $z < 0.5$ is traced to the fact that these sources contribute more than BL Lac at low redshift.

For the nominal DM case shown here, the DM signal is subdominant by a factor of 10 to 50 for Band 1 and improves to become only of a factor of 3 to 5 smaller than the signal from the dominant classes of astrophysical sources for Band 2. This finding is due to the fact that the DM unresolved emission is peaked at very low redshift, contrarily to the emission from unresolved astrophysical sources, as discussed in Section 5.2. Therefore, the comparison between the left and right panels in Figs. 5.5, 5.6, and 5.7 suggests that higher redshifts (i.e. lower detected radio frequencies) are better suited to pinpoint the contribution from astrophysical sources, while the low redshift investigation can be focused on the search of a DM signal. In other words, detected frequencies in the range between 950 MHz and the rest-frame 21cm line frequency are the most promising for a DM search.

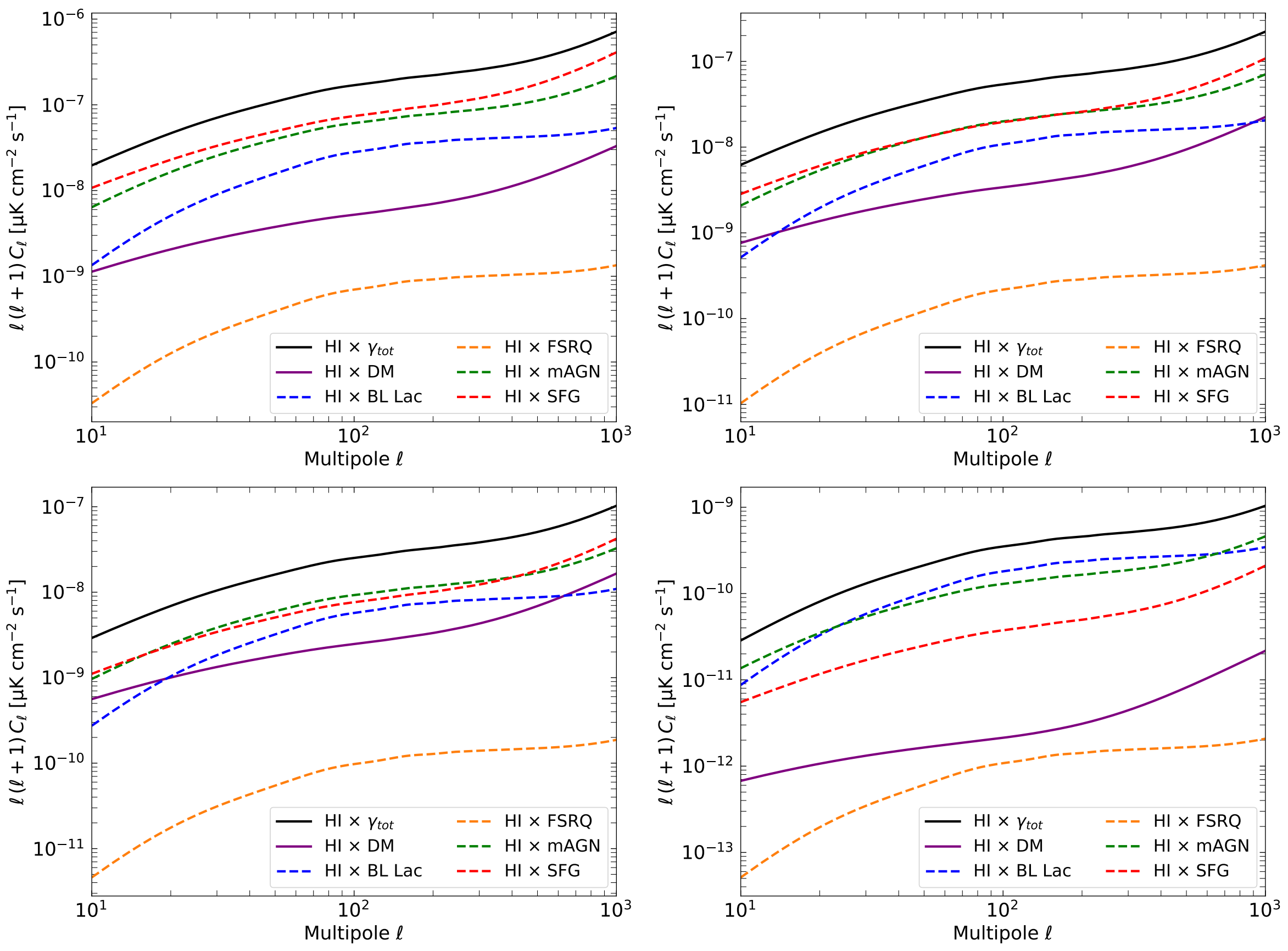

**Fig. 5.8.** Angular power spectrum of the cross-correlation between HI intensity mapping and $\gamma$ rays. The different lines refer to the theoretical predictions of the signal originated by the different $\gamma$ rays sources, as indicated in the inset box. The purple solid line refers to the signal due to annihilating DM and is obtained for a DM mass $m_\chi = 100$ GeV and a thermal annihilation rate $\langle\sigma v\rangle = 3\times10^{-26}$ cm$^3$ s$^{-1}$. The solid black line is the sum of all components. The results refer to the FERMI-LAT energy bins number 1, 2, 3 and 9 of Table 5.1. The radio telescope configuration is the lower redshift Band 2 of SKA1, as reported in Table 5.2.

Together with the signal predictions, Figs. 5.5, 5.6, and 5.7 also show the expected uncertainty on the signal. The error estimates are determined according to Eq. (5.18). In the error budget, the cross-correlation signal is always largely dominated by the two auto-correlation terms. Also, the contribution due to the $\gamma$-ray auto-correlation is largely dominated by the noise term, as discussed in Section 5.3. The error on the cross-correlation can be thus approximated by

$$\left(\Delta C_\ell^{\mathrm{HI}\times\gamma}\right)^2 \simeq \frac{1}{(2\ell+1)f_{\mathrm{sky}}}\left[\frac{C_N^\gamma}{\left(B_\ell^\gamma\right)^2}\times\left(C_\ell^{\mathrm{HI}\times\mathrm{HI}}+\frac{C_N^{\mathrm{HI}}}{\left(B_\ell^{\mathrm{HI}}\right)^2}\right)\right]. \tag{5.28}$$

The discussion in Section 5.3 on the properties of the different terms entering the error determination helps in understanding the behaviour of the error bars of the cross-correlation signal shown in Figs. 5.5, 5.6 and 5.7, which refer to the combined dish+interferometer case. At low multipoles, the error is large due to the low number of modes available in the measurement of the APS, while for high multipoles the error increases due to the size of the FERMI-LAT and MEERKAT or SKA beams. However, for all configurations there are windows in multipole where the total signal is potentially measurable. In particular, the comparison among Figs. 5.5, 5.6 and 5.7 emphasises the evolution and the improvement that can be obtained by progressing from MEERKAT to SKA1 to SKA2.

## 5.5 Signal-to-noise ratio for the astrophysical sources

To determine whether a cross-correlation signal is detectable, we adopt a signal-to-noise ratio (SNR) defined as

$$\mathrm{SNR}^2 = \sum_{\ell,a,r} \left( \frac{C_{\ell,a,r}^{\mathrm{HI}\times S}}{\Delta C_{\ell,a,r}^{\mathrm{HI}\times S}} \right)^2 , \tag{5.29}$$

where $a$ denotes the energy bin and $r$ the redshift bin. The sum extends over the total number of bins $N = N_{\mathrm{multipole}} \times N_{\mathrm{energy}} \times N_{\mathrm{redshift}}$, where the number of multipoles is $N_{\mathrm{multipole}} = \ell_{\max} - \ell_{\min}$, for which we adopt $\ell_{\min} = 10$ and $\ell_{\max} = 1000$, while the number of energy bins is $N_{\mathrm{energy}} = 12$, as reported in Table 5.1. Regarding the number of redshift bins, our benchmark case is $N_{\mathrm{redshift}} = 1$ in the bands reported in Table 5.2. We have investigated also a tomographic redshift binning in each band, with $N_{\mathrm{redshift}} = 5$ and 10, resulting in a marginal improvement over the results reported here for the single redshift bin. We perform the analysis exclusively on the astrophysical signal, in order to assess the potential of the cross-correlation technique to probe the UGRB independently on any assumption on the presence and size of a DM contribution. The results are shown in Table 5.3. We note that a hint for the presence of the cross-correlation signal between the 21cm brightness temperature and the UGRB is already possible with MEERKAT combined with a statistics of FERMI-LAT data comparable to the one already available: a SNR of 3.6/3.7 is in fact predicted for both the single-dish and combined (dish+interferometer) configurations, in both redshift bands. With SKA1, a SNR in excess of 5 can be obtained for both configurations in Band 2. Moreover, a clear identification of the signal is allowed with SKA 2 using both redshift bands, with a SNR ranging from 6.7 to 8.2.

## 5.6 Dark matter constraints

Now that we have assessed that a signal of astrophysical origin is potentially identifiable, we investigate what kind of bounds on the DM properties this cross-correlation technique can lead to. We determine whether a DM signal can be observed above the astrophysical signal by performing a test on a null hypothesis (presence of the astrophysical signal only) vs the alternative hypothesis where the astrophysical sources and the DM $\gamma$-ray emission are both present. We adopt the statistics

$$\Delta\chi^2 = \sum_{\ell,a,r} \left( \frac{C_{\ell,a,r}^{\mathrm{HI}\times\gamma}}{\Delta C_{\ell,a,r}^{\mathrm{HI}\times\gamma}} \right)^2 - \sum_{\ell,a,r} \left( \frac{C_{\ell,a,r}^{\mathrm{HI}\times S}}{\Delta C_{\ell,a,r}^{\mathrm{HI}\times S}} \right)^2 , \tag{5.30}$$

where $\gamma$ refers to the signal coming from both astrophysical sources and DM, while $S$ represents the astrophysical signal only. We perform a raster scan of the DM parameter space over the DM mass: in this case, for each DM mass the free parameter is the annihilation cross-section. In this way, the adopted statistics is distributed as a $\chi^2$ with one degree of freedom. We determine the level at which the cross-correlation technique can set a bound on the DM annihilation cross-section at $2\sigma$ level, i.e. we determine the values of $\langle\sigma v\rangle$ where $\Delta\chi^2 = 4$.

| | | Single-dish | Dish+Interferometer |
|---|---|---|---|
| MEERKAT | L−band | 3.6 | 3.6 |
| | UHF−band | 3.7 | 3.7 |
| SKA1 | Band 1 | 4.5 | 4.6 |
| | Band 2 | 5.7 | 5.7 |
| SKA2 | Band 1 | 7.1 | 8.2 |
| | Band 2 | 6.7 | 7.0 |

**Table 5.3.** Forecast of the signal-to-noise ratio expected for the cross-correlation between HI intensity mapping and the $\gamma$-ray emission from astrophysical sources, for different radio telescope configurations combined with FERMI-LAT and different redshift bands.

The results are shown in Fig. 5.9 for the combination of FERMI-LAT with MEERKAT (red thick curve), FERMI-LAT with SKA1 (blue), FERMI-LAT with SKA2 (green). For all cases, we have considered the combined dish+interferometer configuration and they all refer to the lower-redshift band (L-band for MEERKAT and Band 2 for SKA), which is the more promising redshift range for investigating DM, since its window function is prominently peaked at low redshift. The bounds attainable with Band 1 are a factor of 5 to 10 less constraining, therefore they are not reported here. The best obtainable constraints for MEERKAT are deeper in the parameter space as compared to most of the bounds already achieved by cross-correlating $\gamma$ rays with galaxies [337, 340, 341, 343, 345], clusters of galaxies [421] and cosmic shear [332, 333, 334, 335, 336]. SKA1 can improve the constraints by an additional factor of 4 as compared to MEERKAT, and test a DM particle with thermal annihilation cross-section for masses up to 130 GeV. SKA2 can further explore the thermal DM particle up to masses of 200 GeV.

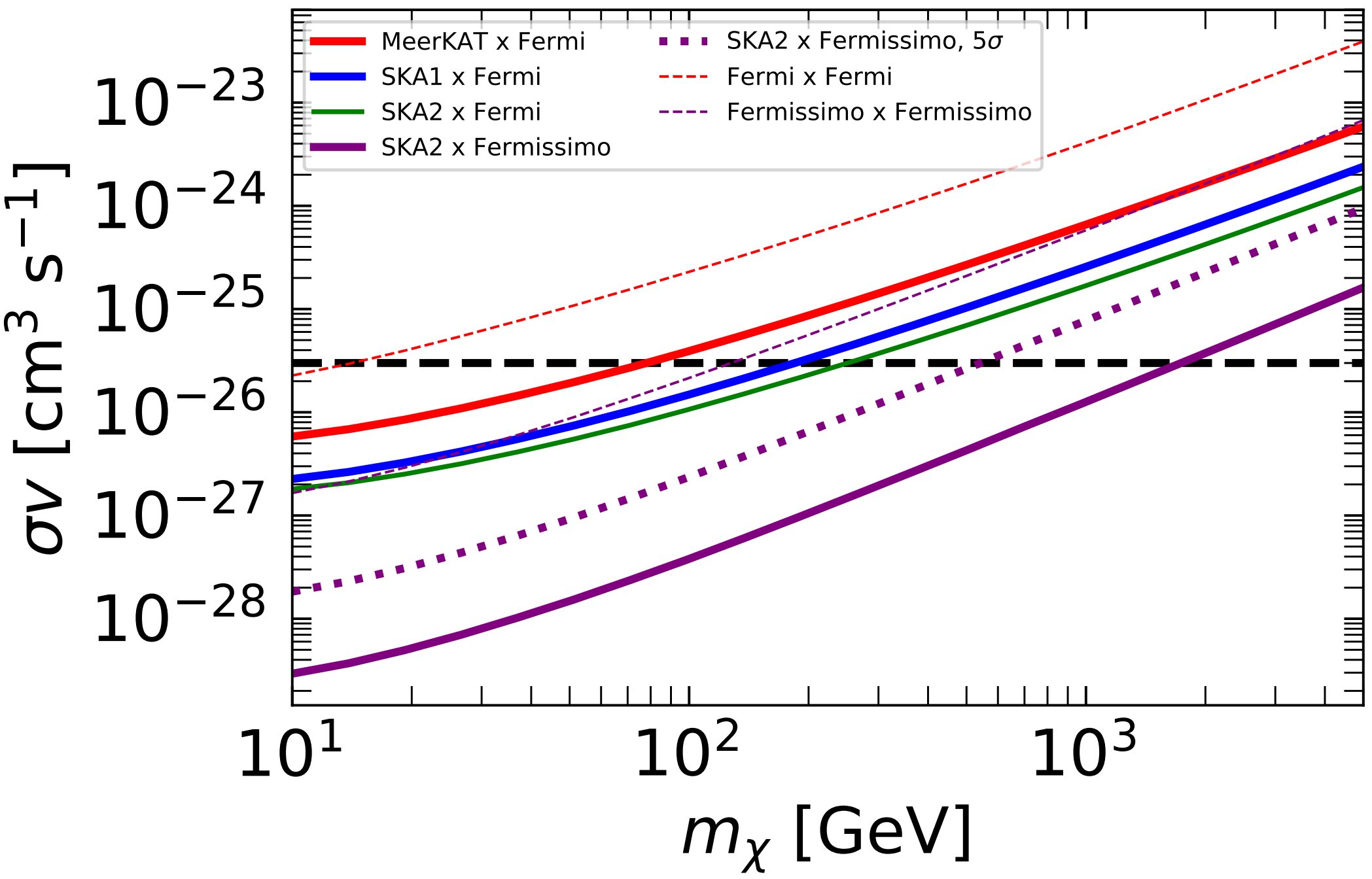

**Fig. 5.9.** Forecast for the bounds on the DM particle properties (mass $m_\chi$ and annihilation rate $\langle\sigma v\rangle$) attainable from the study of the cross-correlation between HI intensity mapping and the unresolved component of the $\gamma$-ray background, for different observational set-ups. Solid lines refer to the 95% C.L. expected upper bounds on $\langle\sigma v\rangle$ for each mass, while the dotted purple line refers to the region of the DM parameter space where a cross-correlation signal can be detected at $5\sigma$ C.L. in the configuration SKA2 × FERMISSIMO. The horizontal dashed line outlines the value of the thermal cross-section $\langle\sigma v\rangle = 3\times10^{-26}$ cm$^3$ s$^{-1}$.

We have focused our analysis on 8 years of data taking forFERMI-LAT. However, it is foreseeable that by the time SKA1 will provide intensity mapping data, FERMI-LAT will have provided about 50% more data. This would allow for slight improvement of the predicted bounds shown in Fig. 5.9. However, a leap in the exploration of the DM parameter space would require a new generation of $\gamma$-ray detectors. For this reason, we perform the analysis for the combination of SKA2 with FERMISSIMO. Fig. 5.10 illustrates the evolution of $\ell(\ell+1)C_\ell$ with multipoles: the impact of the improved $\gamma$-ray angular resolution is clearly visible, as compared to Fig. 5.7. There is a significant reduction of the error bars for multipoles beyond 300 and this combination of detectors allows to extend the analysis to larger multipoles, for which we set $\ell_{\max} = 2000$. The ensuing implications on the DM investigation are shown by the solid purple curve in Fig. 5.9. The 95% C.L. bound is shifted down by a factor between 10 and 60, as compared to the constraint arising from SKA2 in combination with FERMI-LAT. The whole mass range of a thermal DM particle up to the TeV scale can be tested. For comparison, the constraints from auto-correlation channels with FERMI-LAT (dashed red) and FERMISSIMO are displayed, showing that cross-correlating the UGRB with the HI brightness temperature provides stronger results.

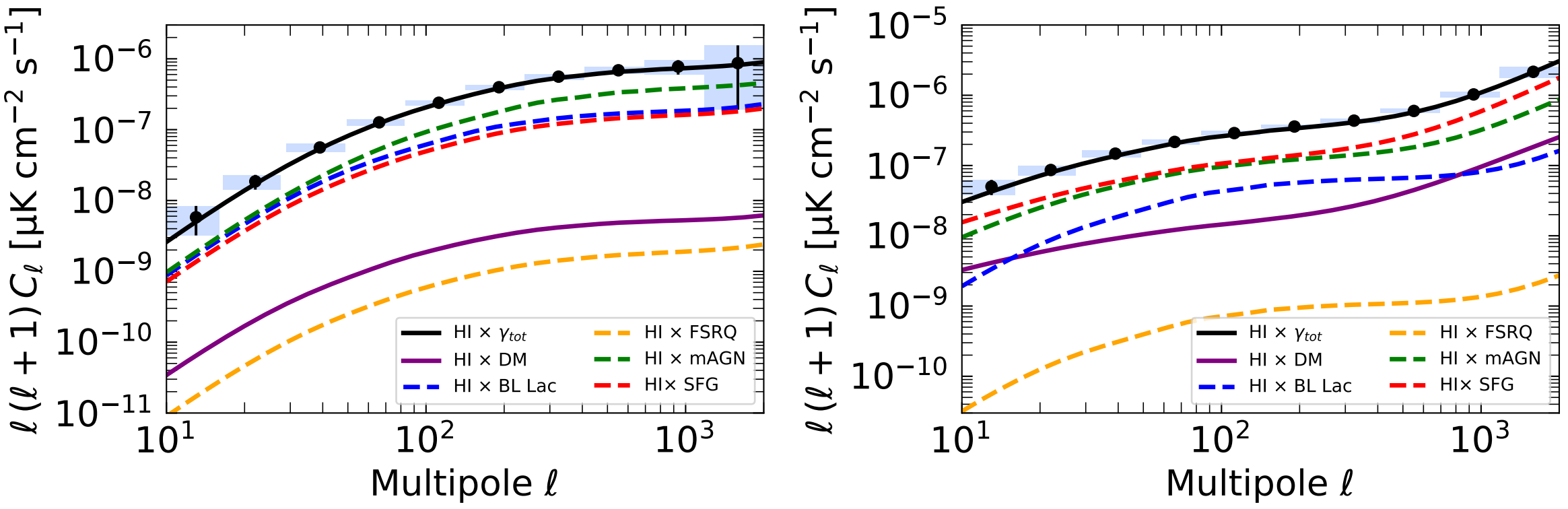

**Fig. 5.10.** The same as in Fig. 5.5 (except that the multipole scale is here extended up to $\ell = 2000$), for SKA2 × FERMISSIMO. The left panel refers to the higher redshift Band 1, the right panel to the lower redshift Band 2.

Finally, on the same figure we also show the $5\sigma$ detection reach (dotted purple line), which allows the detection of a DM particle with thermal cross-section up to masses of 400 GeV.

## 5.7 Summary

We have explored the idea to use the 21cm line of neutral hydrogen as a gravitational tracer of the matter distribution in the Universe to investigate the nature of the unresolved $\gamma$-ray background, through the adoption of the cross-correlation technique. Since the cosmological $\gamma$-ray emission predominantly occurs in the same cosmic structures hosting neutral hydrogen, a positive correlation is expected. We have quantified the size of this effect by investigating the small fluctuations due to the inhomogeneous distribution of matter in the late Universe. The large-scale structure distribution of matter in the Universe, from one side induces fluctuations in the 21cm brightness temperature emission of neutral hydrogen, on the other side produces fluctuations in the unresolved component of the $\gamma$-ray background. These fluctuations are due to either astrophysical sources hosted by those cosmic structures, or to DM in the form of particles which annihilate and produce $\gamma$ rays through their annihilation products.

We have studied the angular power spectrum of the cross-correlation between these two kinds of fluctuations and found that data from future campaigns of neutral hydrogen intensity mapping measurements combined with the current sensitivity of the FERMI-LAT $\gamma$-ray telescope, have the capability to detect the cross-correlation signal. We obtained that the combination of MEERKAT with the current FERMI-LAT statistics can provide a first hint of the cross-correlation signal due to astrophysical sources, with a signal-to-noise ratio of 3.7. We have then performed forecasts for SKA Phase 1 and Phase 2, again combined with the current sensitivity of FERMI-LAT: in these cases, the signal-to-noise ratio is predicted to increase to 5.7 and 8.2, respectively for SKA1 and SKA2.

Having found potential detectability of the signal originated by astrophysical sources, we have investigated the capabilities of this technique to probe particle DM signatures. We predict that the attainable bounds on the DM properties are quite competitive with those obtained from other techniques able to explore the unresolved side of the $\gamma$-ray background, notably the cross-correlation of $\gamma$ radiation with galaxy [337, 340, 341, 343, 345], galaxy cluster-catalogues [421], CMB [422], and the cosmic shear [332, 333, 334, 335, 336]. The enhanced capabilities of SKA Phase 2, combined with a future generation $\gamma$-ray telescope with improved specifications will instead allow to investigate the whole mass window for a thermal WIMP up to the TeV scale, with a $5\sigma$ detection possible for DM masses up to 400 GeV. In order to obtain this enhanced sensitivity, the main requirement for a future $\gamma$-ray telescope is an improved angular resolution, which would allow to better exploit the excellent angular resolution of the interferometric configuration of SKA2 in the determination of the cross-correlation signal. On the other hand, an exposure similar or slightly larger than the one currently attained by FERMI-LAT would be adequate.

# Part II

# Our Galaxy observed with X-ray vision

CHAPTER 6

# PHOTON FLUX FROM SUB-GEV DARK MATTER

Chapters 1 and 2 illustrate that DM in cosmic structures is expected to produce signals originating from its particle physics nature, among which the electromagnetic emission represents a relevant opportunity. One of the major candidates for DM are weak-scale particles, however no convincing signal from them has been observed so far. For this reason, alternative candidates are getting increasing attention, notably sub-GeV particles, which are the subject of this chapter. The challenge in indirect detection of sub-GeV DM is that there is a scarcity of competitive experiments in the energy range between 1 MeV and hundreds of MeV. Fig. 6.1 shows the sensitivity of numerous X-ray and $\gamma$-ray experiments as a function of the observed photon energy. It is apparent that there is a huge variety of experiments probing the GeV and TeV scales, while the MeV range is much less explored. In particular, above a few MeV up to approximately a hundred MeV, there are only the measurements of the COMPTEL telescope. This mission was operational between 1991 and 2000, so the measurements are not current and their sensitivity is quite low. This scarcity of measurements in the MeV range is referred to as the "MeV gap". As a consequence, we need to find alternative ways to study DM candidates with mass in this energy window. In this chapter we look at energies much lower than the mass of the sub-GeV DM particles by including the contribution from inverse Compton scattering in the total flux. In particular, the electrons and positrons produced by DM particles give rise to X rays by up-scattering the low-energy photons in the Galaxy. These X rays fall in the energy range covered by the INTEGRAL data (black line on the left in Fig. 6.1), which we use to constrain the DM annihilation cross-section [2]. This chapter is organised as follows: Section 6.1 introduces the coordinate system and the geometry of the problem, which are instrumental to compute the total flux produced by DM particles. A discussion on the annihilation channels and mass range of interest is also included. The prompt components to the total flux are presented in Section 6.2.

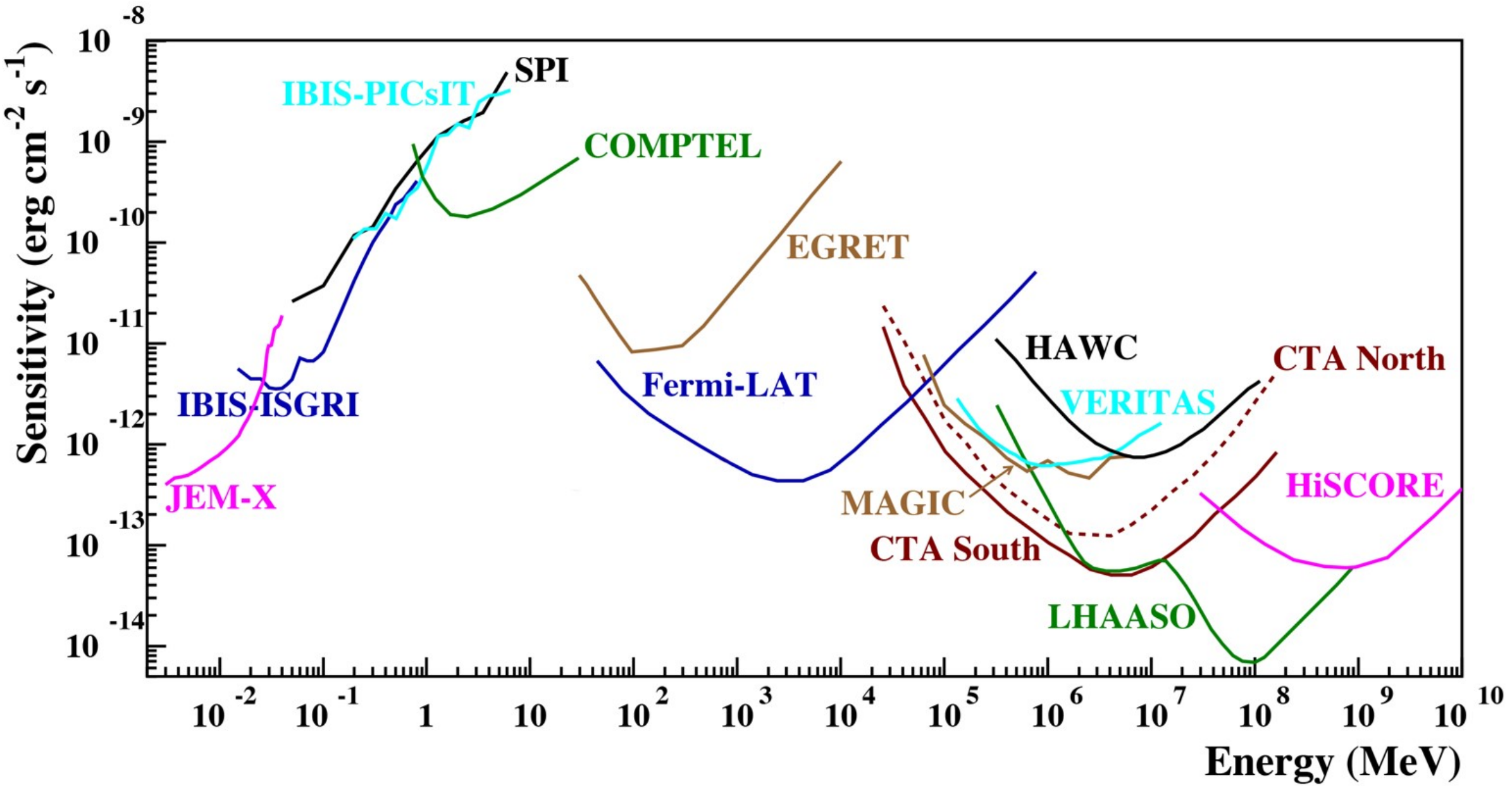

**Fig. 6.1.** Differential sensitivity of various X-ray and $\gamma$-ray experiments as a function of the energy. The space telescope relevant for this analysis is INTEGRAL/SPI (black line on the left). Credit: adapted figure from Ref. [450].

Finally, the inverse Compton scattering signal is thoroughly discussed in Section 6.3, where each element necessary to compute the corresponding flux is examined.

## 6.1 Introduction

Let us consider a direction of observation from the solar system that is identified by the vector $\vec{s}$, as illustrated in Fig. 6.2. The vector $\vec{r}$ represents the galactocentric distance, namely the physical separation between the GC and the generic point $P$, while the angle $\theta$ denotes the aperture between the direction of observation and the axis connecting the Sun to the GC. It holds that

$$\vec{r} = \vec{s} - \vec{R}_\odot \,, \tag{6.1}$$

where $R_\odot = 8.33$ kpc is the distance between the Sun and the GC. This relation implies

$$\vec{r}^2 = \left(\vec{s} - \vec{R}_\odot\right) \cdot \left(\vec{s} - \vec{R}_\odot\right) = s^2 + R_\odot^2 - 2\vec{s} \cdot \vec{R}_\odot = s^2 + R_\odot^2 - 2sR_\odot \cos\theta \,. \tag{6.2}$$

Equivalently, we can adopt the galactic coordinates, latitude $b$ and longitude $\ell$. The projection of $\vec{s}$ on the galactic plane is $s_{xy} = s\cos b$ and its orthogonal projections along the cartesian axes read

$$\begin{cases} s_x &= s\cos b \cos\ell \\ s_y &= s\cos b \sin\ell \\ s_z &= s\sin b \,. \end{cases} \tag{6.3}$$

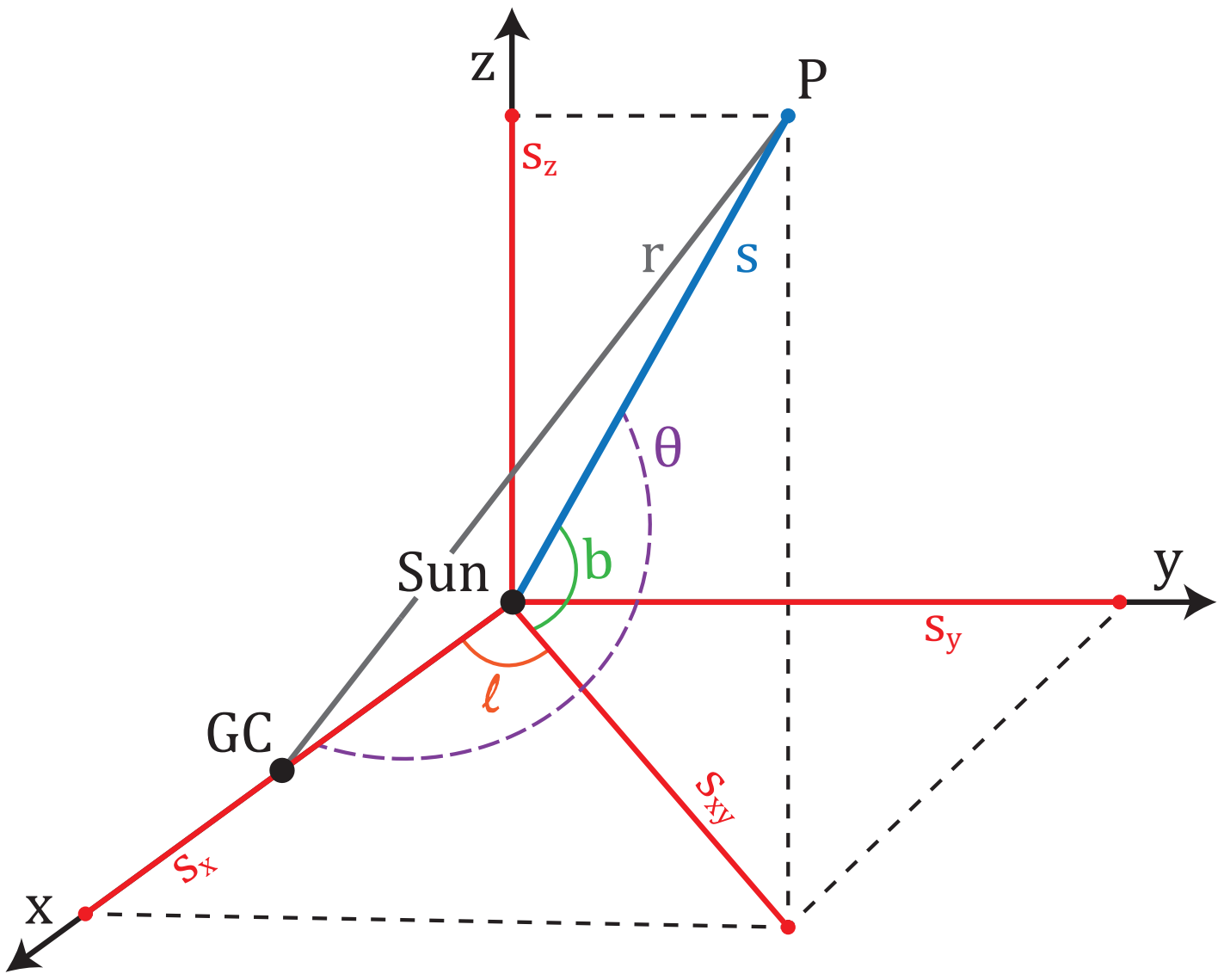

**Fig. 6.2.** Coordinate system, illustrating a generic source P at distance $\vec{s}$ from the Sun and $\vec{r}$ from the Galactic Centre.

In order to derive the relation between $\theta$, $b$, and $\ell$, we recall that

$$\vec{s} \cdot \vec{R}_\odot = \begin{cases} s R_\odot \cos\theta \\ R_\odot s_x = R_\odot s \cos b \cos \ell \end{cases} \tag{6.4}$$

which implies

$$\cos\theta = \cos b \cos \ell \,. \tag{6.5}$$

Note that $s = 0$ corresponds to the position of the Sun and $\theta = b$ when $\ell = 0$. At each point along the direction of observation, DM particles annihilate, thus contributing to the photon signal. We collect the overall signal by integrating all the contributions along the line of sight.

We focus on DM particles with a mass in the range 1 MeV $\leqslant \mathrm{m}_\chi \leqslant$ 5 GeV and we take into consideration three annihilation channels:

$$\chi\chi \to e^+ e^- \tag{6.6}$$

$$\chi\chi \to \mu^+ \mu^- \tag{6.7}$$

$$\chi\chi \to \pi^+ \pi^- \,. \tag{6.8}$$

The above annihilation channels are kinematically open whenever $m_\chi > m_i$, where $i = e, \mu, \pi$. The electron and muon channels are representative of leptonic DM channels, while the pion channel is representative of a hadronic DM channel. We do not consider the annihilation into a pair of neutral pions, since in this case the $\gamma$ rays (boosted to the DM frame) do not reach down to the energies covered by the INTEGRAL data. For each channel, the total photon flux is given by the sum of two main components: the emission from the charged particles in the final state and the photons produced via inverse Compton scattering by very energetic $e^{\pm}$, originated from the channels above. The former component includes the final state radiation and other radiative decays for the muon and the pion channels. An additional contribution to the total photon flux is given by the in-flight annihilation of DM-produced $e^+$ with ambient $e^-$ from the gas [451]. This process is however subdominant outside of the gas-dense region of the GC that we do not consider in our analysis, as will be discussed in Chapter 7.

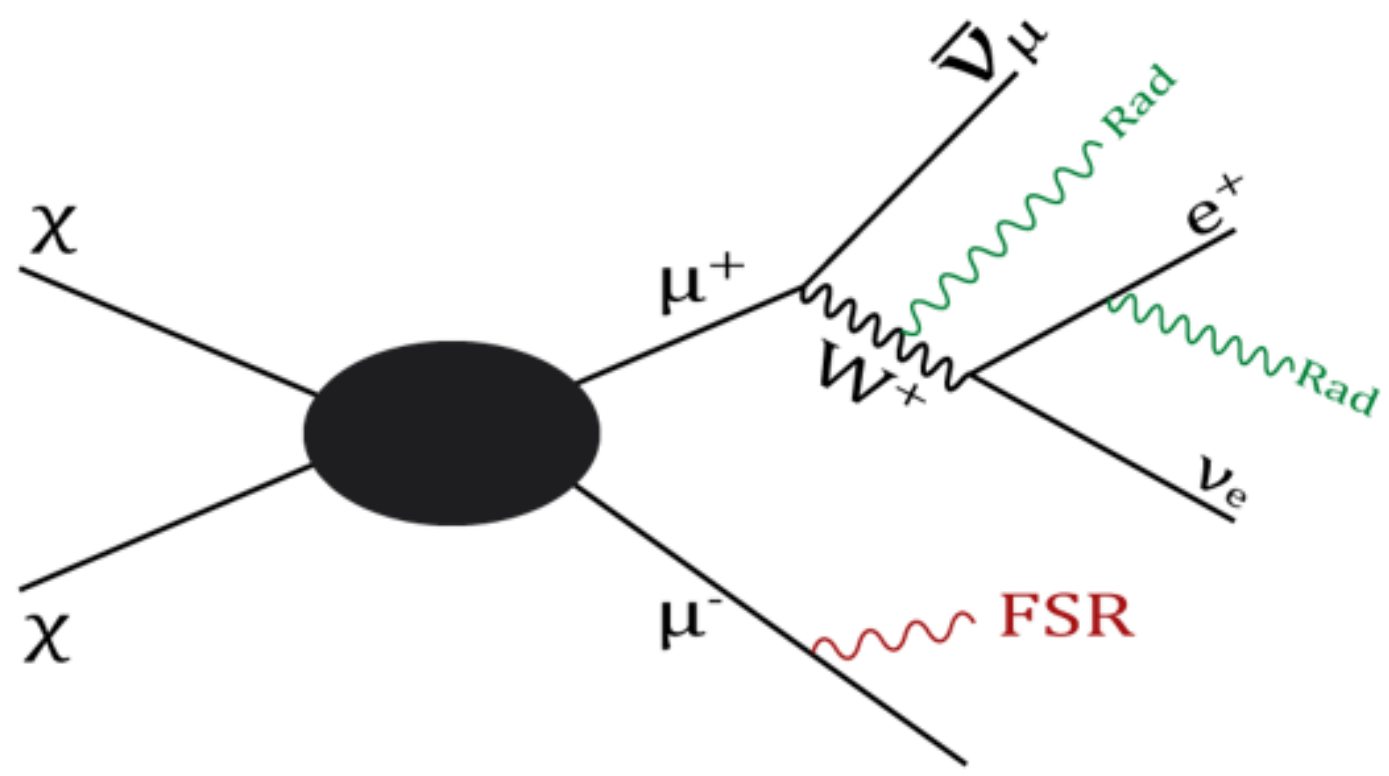

**Fig. 6.3.** Prompt photons from final state radiation and radiative decay emission in the case of dark matter particles annihilating into muons.

## 6.2 Prompt photon flux

Let us consider two DM particles which annihilate into muons. These secondary particles can radiate a photon and this is the so-called final state radiation (FSR). Alternatively the secondary particles can decay, and for instance, the W boson or the positron can radiate a photon as well: this is the radiative decay emission (Rad). Fig. 6.3 illustrates this scenario. The differential flux of the prompt photons produced in the annihilation events of DM particles is computed via the standard expression (see e.g. Ref. [452]):

$$\frac{\mathrm{d}\Phi_{\mathrm{prompt},\gamma}}{\mathrm{d}E_\gamma\,\mathrm{d}\Omega} = \frac{1}{2}\frac{R_\odot}{4\pi}\left(\frac{\rho_\odot}{m_\chi}\right)^2 J(\theta)\,\langle\sigma v\rangle_f\,\frac{\mathrm{d}N_\gamma^f}{\mathrm{d}E_\gamma}\,, \qquad J(\theta) = \int_{\mathrm{l.o.s.}} \frac{\mathrm{d}s}{R_\odot}\left(\frac{\rho\,(r\,(s,\theta))}{\rho_\odot}\right)^2, \tag{6.9}$$

where $E_\gamma$ denotes the photon energy and $\mathrm{d}\Omega$ is the solid angle. The normalised $J$ factor corresponds to the integration along the line of sight of the square of the galactic DM density profile $\rho(r)$, for which we adopt a standard NFW, as given in Eq. (3.42). The two parameters $r_s = 24.42$ kpc and $\rho_s = 0.184$ GeV/cm$^3$ are fixed such that the total mass of the Milky Way is $M_{\mathrm{MW}} = 4.7\cdot 10^{11}$ M$_\odot$ and the DM density at the location of the Sun is $\rho(R_\odot) = \rho_\odot = 0.3$ GeV/cm$^3$. The annihilation cross-section in each of the final states $f$ of Eqs. (6.6)–(6.8) is denoted as $\langle\sigma v\rangle_f$. Regarding the spectrum of FSR photons, we adopt the following expressions which we derive

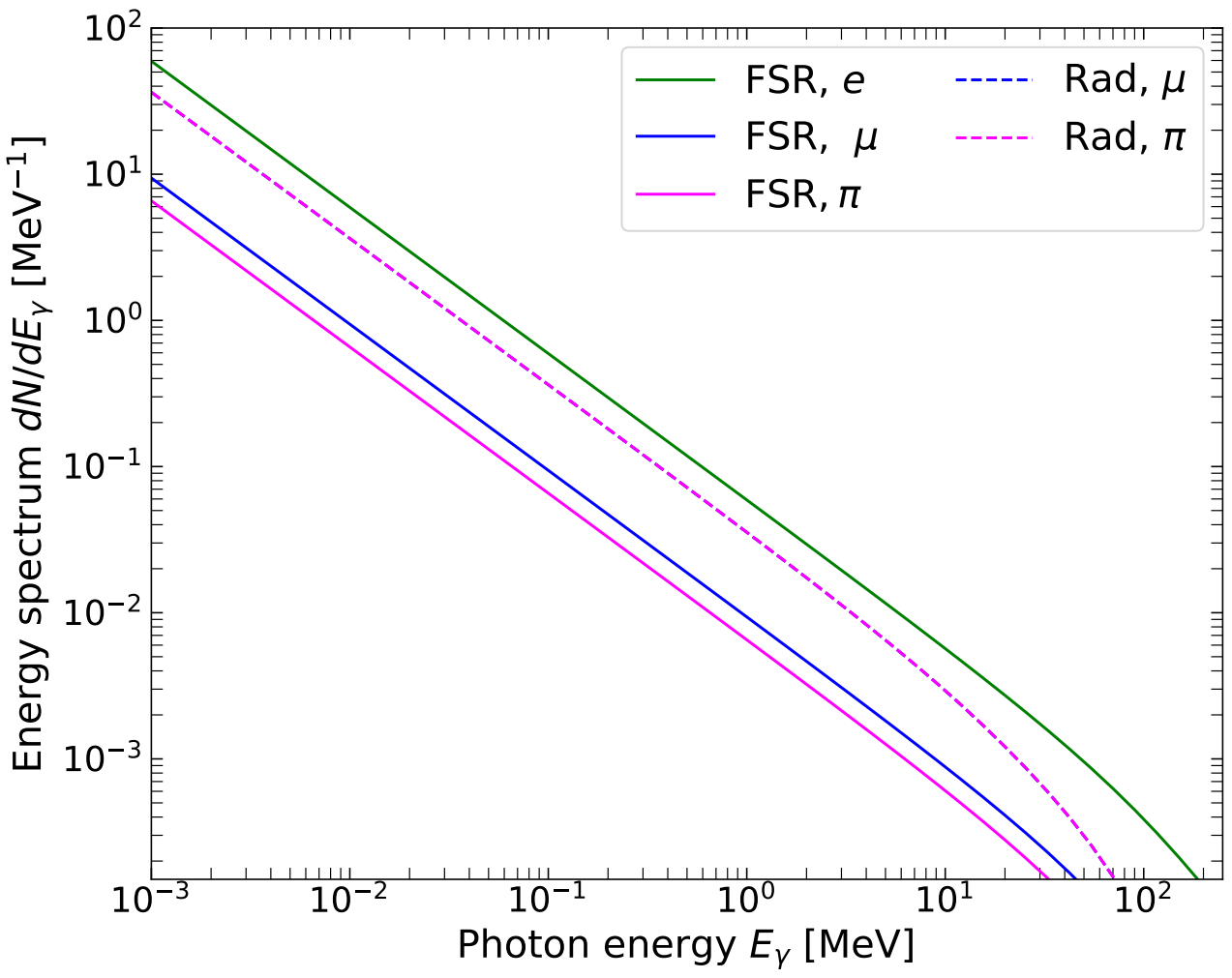

**Fig. 6.4.** The energy spectra for the final state radiation (FSR, solid line) and radiative decay (Rad, dashed) components from DM particles with a mass of 250 MeV, annihilating into $e^+e^-$ (green), $\mu^+\mu^-$ (blue), $\pi^+\pi^-$ (magenta).

from the analysis of Ref. [453] :

$$\frac{\mathrm{d}N_{\mathrm{FSR}\,\gamma}}{l^+l^-}\mathrm{d}E_\gamma = \frac{\alpha}{\pi\beta(3-\beta^2)m_\chi}\left[\mathcal{A}\ln\frac{1+R(\nu)}{1-R(\nu)} - 2\mathcal{B}R(\nu)\right], \tag{6.10}$$

where

$$\mathcal{A} = \left[\frac{\left(1+\beta^2\right)\left(3-\beta^2\right)}{\nu} - 2\left(3-\beta^2\right) + 2\nu\right], \tag{6.11}$$

$$\mathcal{B} = \left[\frac{3-\beta^2}{\nu}(1-\nu)+\nu\right], \tag{6.12}$$

where $l = e, \mu$ and we have defined: $\nu = E_\gamma/m_\chi$, $\beta^2 = 1-4\xi^2$ with $\xi = m_l/(2m_\chi)$ and $R(\nu) = \sqrt{1-4\xi^2/(1-\nu)}$. For the pion, we use

$$\frac{\mathrm{d}N^{\pi^+\pi^-}_{\mathrm{FSR}\,\gamma}}{\mathrm{d}E_\gamma} = \frac{2\alpha}{\pi\beta m_\chi}\left[\left(\frac{\nu}{\beta^2} - \frac{1-\nu}{\nu}\right)R(\nu) + \left(\frac{1+\beta^2}{2\nu} - 1\right)\ln\frac{1+R(\nu)}{1-R(\nu)}\right], \tag{6.13}$$

with the same definitions as above with $m_l \to m_\pi$.

The muon can undergo the radiative decay

$$\mu^- \to e^-\,\bar{\nu}_e\,\nu_\mu\,\gamma \tag{6.14}$$

$$\mu^+ \to e^+\,\nu_e\,\bar{\nu}_\mu\,\gamma \tag{6.15}$$

and this channel can be a relevant source of soft photons. For the photon spectrum we adopt the parameterization of Refs. [454, 455] which in turn are derived from Ref. [456]. Thus, in the muon rest frame we use the following expression

$$\left.\frac{dN^{\mu}_{\text{Rad}\,\gamma}}{dE_\gamma}\right|_{E_\mu=m_\mu} = \frac{\alpha\,(1-x)}{36\pi E_\gamma}\left[12\left(3-2x(1-x)^2\right)\log\left(\frac{1-x}{r}\right)+x\,(1-x)\,(46-55x)-102\right], \qquad (6.16)$$

where $x = 2E_\gamma/m_\mu$, $r = \left(m_e/m_\mu\right)^2$ and the maximal photon energy is $E^{\text{max}}_\gamma = m_\mu(1-r)/2 \simeq 52.8$ MeV. For muons in flight, Eq. (6.16) is boosted to the frame where the muon has energy $E_\mu = m_\chi$. For the process $\chi\chi \to \mu^+\mu^-$, a multiplicity factor of 2 needs to be applied, since a pair of muons is produced for each annihilation event.

Charged pions can also produce photons radiatively through the processes

$$\pi^- \to \ell^-\,\bar{\nu}_\ell\,\gamma \qquad (6.17)$$

$$\pi^+ \to \ell^+\,\nu_\ell\,\gamma \qquad (6.18)$$

with $\ell = e,\mu$. Also in this case we adopt the parameterization of Ref. [455], which has been derived from Ref. [457]. In the pion rest frame the expression we use is then

$$\left.\frac{dN^{\pi}_{\text{Rad}\,\gamma}}{dE_\gamma}\right|_{E_\pi=m_\pi} = \frac{\alpha\,[f(x)+g(x)]}{24\pi\,m_\pi\,f_\pi^2\,(r-1)^2\,(x-1)^2\,r\,x}, \qquad (6.19)$$

where $x = 2E_\gamma/m_\pi$, $r = (m_\ell/m_\pi)^2$, $f_\pi = 92.2$ MeV is the pion decay constant and

$$\begin{aligned} f(x) &= (r+x-1)\left[m_\pi^2 x^4\left(F_A^2+F_V^2\right)\left(r^2-rx+r-2(x-1)^2\right)\right.\\ &\quad -12\sqrt{2}\,f_\pi\,m_\pi\,r(x-1)x^2\left(F_A(r-2x+1)+xF_V\right)\\ &\quad \left.-24 f_\pi^2\,r(x-1)\left(4r(x-1)+(x-2)^2\right)\right], \\ g(x) &= 12\sqrt{2}\,f_\pi\,r(x-1)^2\log\left(\frac{r}{1-x}\right)\left[m_\pi x^2(F_A\,(x-2r)-x\,F_V)\right.\\ &\quad \left.+\sqrt{2}\,f_\pi(2r^2-2rx-x^2+2x-2)\right], \end{aligned} \qquad (6.20)$$

where the axial and vector form factors are $F_A = 0.0119$ and $F_V\left(q^2\right) = F_V(0)\left(1+aq^2\right)$ with $F_V(0) = 0.0254$, $a = 0.10$, $q^2 = (1-x)$ [455, 290]. When the pion decays into an on-shell muon, the radiative decay of the muon is again a source of low energy photons. Following Refs. [455], the total radiative charged pion spectrum in the pion rest frame can be phrased as

$$\left.\frac{dN^{\pi}_{\text{Rad Tot}\,\gamma}}{dE_\gamma}\right|_{E_\pi=m_\pi} = \sum_{\ell=e,\mu}\text{BR}(\pi\to\ell\nu_\ell)\left.\frac{dN^{\pi}_{\text{Rad}\,\gamma}}{dE_\gamma}\right|_{E_\pi=m_\pi} + \text{BR}(\pi\to\mu\nu_\mu)\left.\frac{dN^{\mu}_{\text{Rad}\,\gamma}}{dE_\gamma}\right|_{E_\mu=E_\star}, \qquad (6.21)$$

where $E_\star = (m_\pi^2+m_\mu^2)/(2m_\pi)$ is the muon energy in the pion rest frame. For the process $\chi\chi \to \pi^+\pi^-$, where the pion is in flight, Eq. (6.21) is boosted to the frame where $E_\pi = m_\chi$ and a multiplicity factor of 2 is applied, since also in this case a pair of pions is produced for each annihilation event. The energy spectra of prompt radiation from DM particles with a mass of 250 MeV are illustrated in Fig. 6.4. The solid lines denote the FSR and the dashed curves refer to the Rad component, for the three annihilation channels ($e^+e^-$ in green, $\mu^+\mu^-$ in blue, $\pi^+\pi^-$ in magenta). As pointed out in Ref. [455], the muon radiative decay provides a quite relevant soft-photon production channel. Both in the muon and in the pion annihilation channels, the Rad photon flux arising from the muon radiative decay is dominant over the FSR emission for DM masses up to a few GeV, as shown in Fig. 6.4. This will be explicitly discussed when we will derive the constraints on the DM annihilation cross-section in Chapter 7.

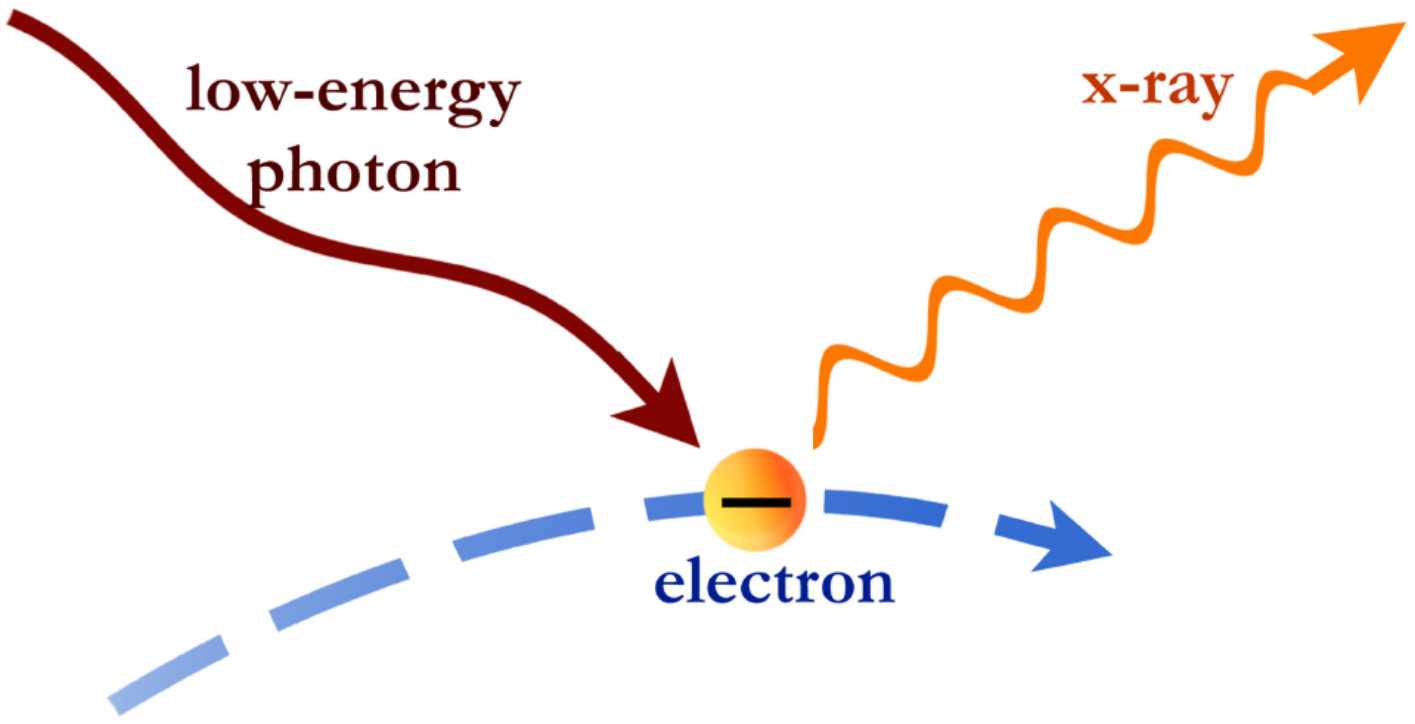

**Fig. 6.5.** Diagram illustrating the inverse Compton scattering of a low-energy photon (red line), colliding with a relativistic electron (blue). During the process, the electromagnetic wave gains energy, becoming an X ray (orange). Credit: [458].

## 6.3 Inverse Compton scattering

Let us recall the physics behind the inverse Compton scattering (ICS) signal. The DM particles in the Galactic halo can annihilate and produce pairs of electrons and positrons. These secondary particles propagate and may up-scatter the low-energy radiation fields in the Galaxy to the X-ray and $\gamma$-ray frequencies. Fig. 6.5 illustrates the ICS process. The propagation of the electrons and positrons can be described via the transport equation

$$\underbrace{-\frac{1}{r^2}\frac{\mathrm{d}}{\mathrm{d}r}\left[r^2 D\frac{\mathrm{d}f}{\mathrm{d}r}\right]}_{\text{diffusion}} + \underbrace{v_a\frac{\mathrm{d}f}{\mathrm{d}r}}_{\text{advection}} - \underbrace{\frac{1}{3r^2\frac{\mathrm{d}}{\mathrm{d}r}\left(r^2 v_c\right)p\frac{\mathrm{d}p}{\mathrm{d}p}}}_{\text{convection}} + \underbrace{\frac{1}{p^2}\frac{\mathrm{d}}{\mathrm{d}p}\left(\dot{p}\,p^2 f\right)}_{\text{radiative losses}} = \underbrace{q(r,p)}_{\text{source}} \tag{6.22}$$

where $f(r,p)$ is the phase-space density at the radius $r$ and momentum $p$. The function $f$ is related to the electron number density $n_e$ in the energy interval $(E, E+\mathrm{d}E)$ through

$$n_e(r,E)\,\mathrm{d}E = 4\pi p^2 f(r,p)\,\mathrm{d}p\,. \tag{6.23}$$

The first term in Eq. (6.22) describes the spatial diffusion, with $D(r,p)$ being the diffusion coefficient. The second term stands for the advective transport associated to a flow of velocity $v_a$ towards the black hole located in the GC. Instead, the third term relates to the convective transport with outflow away from the galactic plane with velocity $v_c$. The last term on the left outlines the radiative losses, with $\dot{p}(r,p) = \sum_i \mathrm{d}p_e/\mathrm{d}t \equiv b_{\mathrm{tot}}$ being the sum over the momentum loss rates due to the radiative processes $i$. In this context, the energy losses include bremsstrahlung, ionization, synchrotron radiation and ICS. Finally, the term on the right $q(r,p)$ represents the source term and it is related to the injection function of $e^{\pm}$ from DM through the relation

$$Q_e(r,E)\,\mathrm{d}E = 4\pi p^2 q(r,p)\,\mathrm{d}p\,. \tag{6.24}$$

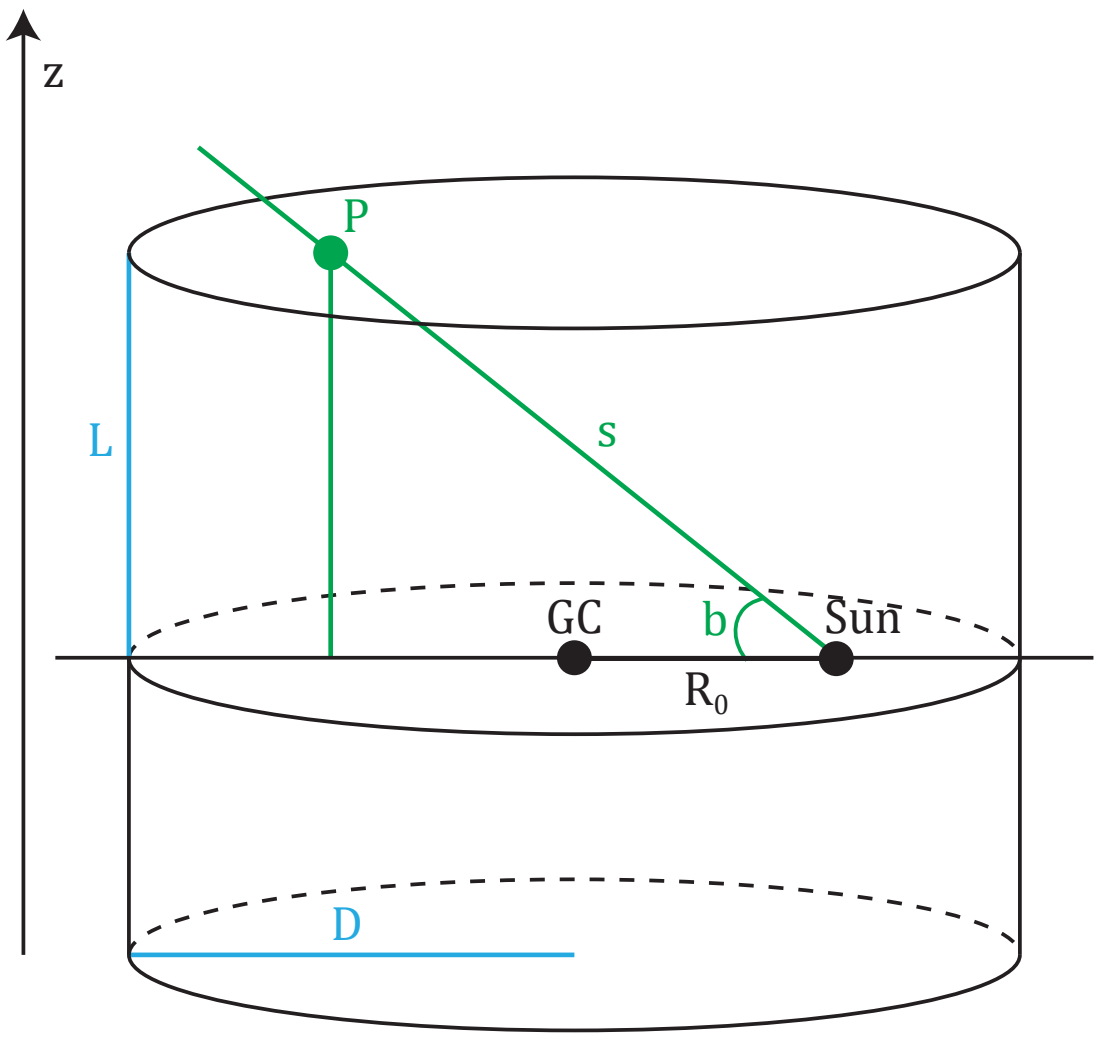

**Fig. 6.6.** The geometry of the galactic magnetic halo is assumed to be a cylinder with half-height $L$ = 4 kpc and radius $D$ = 20 kpc, outside of which the $e^{\pm}$ are no longer confined.

We consider regions of the galactic halo far from the central black hole, thus we can safely neglect the advection and convection terms. Also, we adopt a simplified treatment by neglecting diffusion. Physically speaking this is equivalent to assume that the electrons and positrons scatter off the ambient photons in the same location where these $e^{\pm}$ were produced by annihilation events of DM particles. This assumption is called on-the-spot approximation and in Chapter 7 we motivate its use in the regime of interest. Finally, we model the galactic magnetic halo as a cylinder of height $L$ and radius $D$, centred in the GC, as illustrated in Fig. 6.6. For convenience, we adopt the cylindrical coordinates $\vec{x} = (R,z)$, where $R$ represents the galactocentric radius on the galactic plane and $z$ measures the distance from the galactic plane, as illustrated in Fig. 6.2. The change of coordinates between spherical and cylindrical systems is simply given by

$$R^2 = r^2 - z^2\,, \tag{6.25}$$

or equivalently

$$\begin{cases} R^2 = (s_x - R_\odot)^2 + s_y^2 \\ z \;\; = s_z\,. \end{cases} \tag{6.26}$$

The location of the solar system in cylindrical coordinates is $(R_\odot, z_\odot) = (8.33\,\mathrm{kpc}, 0\,\mathrm{kpc})$. We assume that the electrons and positrons are confined inside this galactic magnetic halo, with boundary conditions such that the cosmic-ray density vanishes at the halo borders, or equivalently:

$$\begin{cases} R \leq D \\ z \leq L\,. \end{cases} \tag{6.27}$$

The maximum distance from the solar system, along the line of sight, in which a particle is considered trapped in the galactic magnetic field corresponds to $s_{\max} = \sqrt{L^2 + (D + R_\odot)^2}$, where the radius $D$ = 20 kpc. We adopt the "MED" configuration of Ref. [204]. Among the set of propagation parameters, this configuration assumes half-height $L$ = 4 kpc. In the following, we will illustrate the two key elements that are necessary to compute the ICS flux: the electron number density and the power radiated as photons. For the sake of completeness, the physical quantities on which they depend, such as the energy losses, are also discussed.

### 6.3.1 Electron number density

Under the above assumptions on the geometry of the galactic magnetic halo and the $e^\pm$ propagation, Eq. (6.22) simplifies to

$$-\frac{\partial}{\partial E_e}\left[b_{\rm tot}(E_e,\vec{x})\, n_e(E_e,\vec{x})\right] = Q_e(E_e,\vec{x})\,, \tag{6.28}$$

leading to the following convenient expression for the $e^\pm$ spectral number density:

$$n_e(E_e,\vec{x}) = \frac{1}{b_{\rm tot}(E_e,\vec{x})}\int_{E_e}^{m_\chi} \mathrm{d}\tilde{E}_e\, Q_e\left(\tilde{E}_e,\vec{x}\right)\,. \tag{6.29}$$

The injection term of $e^\pm$ from annihilation events of DM particles reads

$$Q_e(\tilde{E}_e,\vec{x}) = \frac{\langle\sigma v\rangle}{2}\left(\frac{\rho(\vec{x})}{m_\chi}\right)^2 \frac{\mathrm{d}N_e}{\mathrm{d}\tilde{E}_e}\,, \tag{6.30}$$

where $\mathrm{d}N_e/\mathrm{d}\tilde{E}_e$ is the $e^\pm$ energy spectrum. The integral of the injection term provides essentially the number of electrons at the location $(R,z)$ with energy higher than $E_e$. The impact of varying the DM profile will be discussed in Chapter 7.

**Energy spectrum.** Regarding the $e^\pm$ energy spectrum from DM annihilations, for the $e^+e^-$ channel it consists of a monochromatic line with $E_e = m_\chi$. For the $\mu^+\mu^-$ case, we take the electron spectrum from the muon decay obtained in the muon rest frame and we boost it to the DM annihilation frame, where the muon has energy $E_\mu = m_\chi$. This leads to [459]:

$$\frac{\mathrm{d}N_e^{\mu\to e\nu\bar{\nu}}}{\mathrm{d}E_e} = \frac{4\sqrt{\xi^2 - 4\varrho^2}}{m_\mu}\left[\xi(3-2\xi) + \varrho^2(3\xi - 4)\right]\,, \tag{6.31}$$

where $\varrho = m_e/m_\mu$, $\xi = 2E_e/m_\mu$ and the maximal electron energy is $E_e^{\max} = (m_\mu^2 + m_e^2)/(2m_\mu)$. For the $\pi^+\pi^-$ case, as a result of the decay chain $\pi \to \mu \to e$, we boost the electron spectrum from the muon decay of Eq. (6.31) to the rest frame of the pion, where the muon has energy $E_\mu = (m_\pi^2 + m_\mu^2)/(2m_\pi)$. Afterwards, we boost the ensuing distribution to the DM annihilation frame, where the pion has energy $E_\pi = m_\chi$. The expression of the boost from a frame where the particle $a$, produced by its parent $A$, has energy $E'$ and momentum $p'$ to a frame where the parent particle has energy $E_A$ [204] reads

$$\frac{\mathrm{d}N}{\mathrm{d}E} = \frac{1}{2\beta\gamma}\int_{E'_{\rm min}}^{E'_{\rm max}} \frac{1}{p'}\frac{\mathrm{d}N}{\mathrm{d}E'}\,, \tag{6.32}$$

where $\gamma = E_A/m_A$ and $\beta = (1-\gamma^{-2})^{1/2}$ are the Lorentz factors for the boost, while $E'_{\rm max} = \gamma(E+\beta p)$ and $E'_{\rm min} = \gamma(E-\beta p)$ refer to the maximum and minimum energy, respectively. When $a$ is produced as monochromatic, namely $dN/dE' = \delta(E - E_\star)$, then the above expression reduces to the typical box spectrum $dN/dE = 1/(2\beta\gamma p_\star)$ for $\gamma\left(E_\star - \beta p_\star\right) \le E \le \gamma\left(E_\star + \beta p_\star\right)$. Fig. 6.7 illustrates the electron energy spectrum $dN/dE_e$ as a function of the initial electron energy $E_e$ for DM particles with mass $m_\chi$ = 250 MeV. The blue line refers to the muon channel, while the magenta curve represents the pion channel. The electron channel, being a vertical line at $m_\chi$ = 250 MeV, is not shown. We remind the reader that our $\pi^+\pi^-$ case is intended as one possible representative case of DM annihilations into light quarks. A more detailed analysis would require using the spectra computed in Ref. [460], that consider the annihilation into light quark pairs and the subsequent production of many light hadronic resonances besides pions. This approach would imply choosing a specific model for the annihilation diagram (e.g. via a light mediator with arbitrary but defined couplings to quarks).

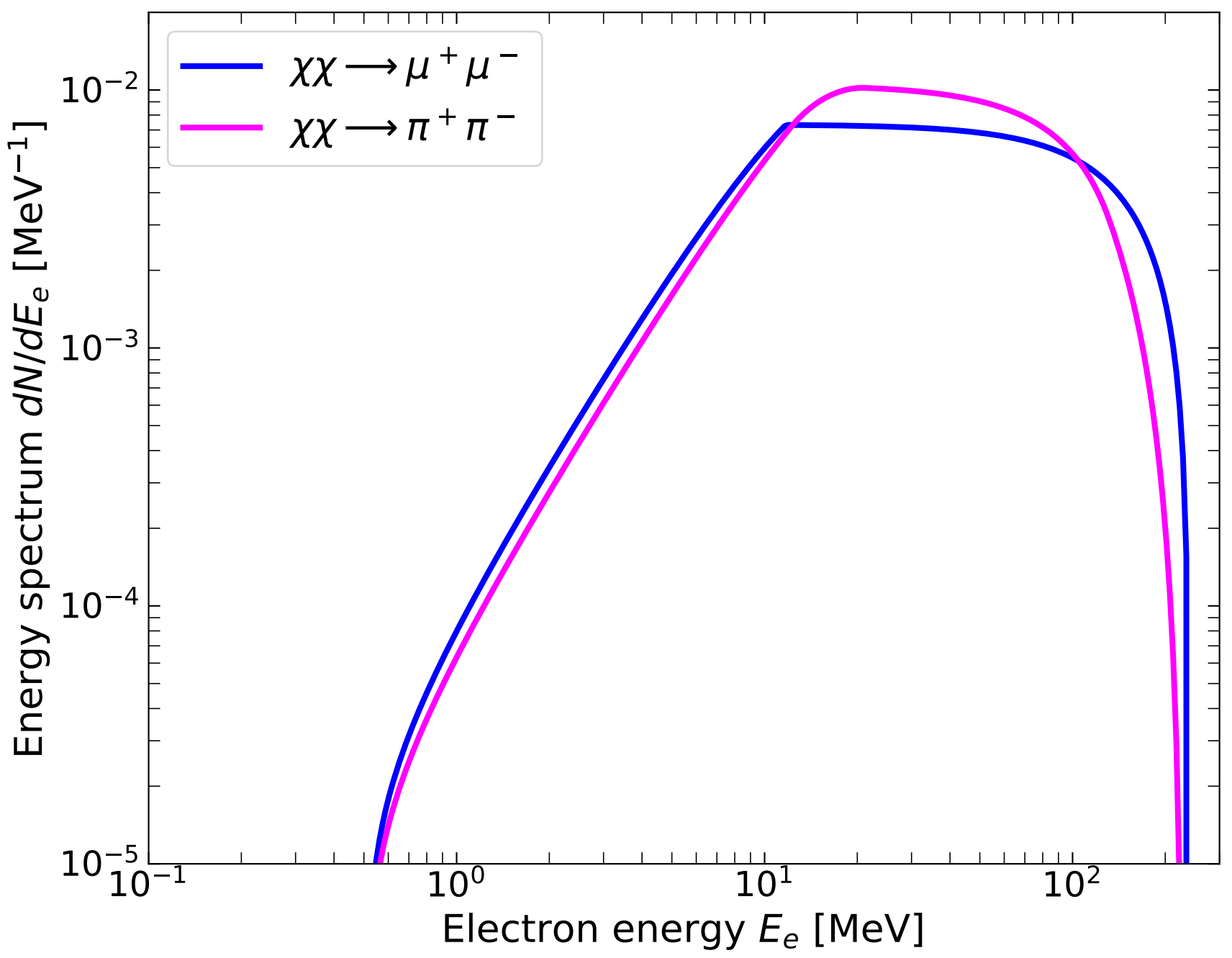

**Fig. 6.7.** Energy spectrum as a function of the initial electron energy for the muon (blue line) and pion (magenta) channels. The mass of the DM particles is set to $m_\chi$ = 250 MeV.

**Energy losses.** Besides the injection term, the second component to consider while computing the spectral number density of Eq. (6.29) is the total energy loss function

$$b_{\rm tot}(E,\vec{x}) \equiv -\frac{{\rm d}E}{{\rm d}t} = b_{\rm Coul+ioniz} + b_{\rm brem} + b_{\rm syn} + b_{\rm ICS}\,, \tag{6.33}$$

which takes into account all the energy loss processes experienced by the $e^{\pm}$ in the local galactic environment where they are injected. The energy losses due to Coulomb interactions and ionization on neutral matter are given by

$$b^{\rm neut}(E,\vec{x}) = \frac{9}{4} c\,\sigma_T\, m_e \sum_i n_i Z_i \left(\log\frac{E}{m_e} + \frac{2}{3}\log\frac{m_e}{\Delta E_i}\right), \tag{6.34}$$

where $\sigma_T \simeq 6.65\,10^{-29}$ m$^2$ is the Thomson cross-section and $n_i$ represents the number density per unit volume of a gas species $i$ with atomic number $Z_i$. The average excitation energy $\Delta E_i$ is worth 15 eV for hydrogen and 41.5 eV for helium. In the case of ionized matter, the loss coefficient reads

$$b^{\rm ion}(E,\vec{x}) = \frac{3}{4} c\,\sigma_T\, m_e n_e^{\rm pl} Z_i \left(\log\frac{E}{m_e} + 2\log\frac{m_e}{E_{\rm pl}}\right), \tag{6.35}$$

where $n_e^{\rm pl}$ represents the electron number density of the plasma. The characteristic plasma energy is $E_{\rm pl} = \sqrt{4\pi n_e r_e^3 m_e/\alpha}$, with $r_e \simeq 2.8\,10^{-15}$ being the classical electron radius. The degree of ionization $\alpha = n_{\rm ion}/(n_{\rm ion} + n_{\rm n})$ is the fraction of ionized matter, with $n_{\rm ion}$ being the ion density and $n_{\rm n}$ the density of neutral particles. Hence, the total energy losses for ionization and Coulomb interactions are given by

$$b_{\rm Coul\,+\,ioniz} = b^{\rm neut} + b^{\rm ion}\,. \tag{6.36}$$

Note that both $b^{\rm neut}$ and $b^{\rm ion}$ are ruled by the constant term in the brackets, therefore this type of energy losses has a very weak (logarithmic) dependence on the input $e^{\pm}$ energy.

Regarding the bremsstrahlung process, it holds

$$b_{\rm brem}(E,\vec{x}) = c \sum_i n_i(\vec{x}) \int_0^E {\rm d}E_\gamma E_\gamma \frac{{\rm d}\sigma_i}{{\rm d}E_\gamma}\,, \tag{6.37}$$

where $E_\gamma$ is the energy of the bremsstrahlung photon and ${\rm d}\sigma_i/{\rm d}E_\gamma$ represents the differential cross-section, which depends on the properties of the gas. For a fully ionised gas, the losses are

$$b^{\rm ion}(E,\vec{x}) = \frac{3}{2\pi}\alpha_e\,\sigma_T\, n_i Z_i (Z_i+1)\left(\log\frac{2E_e}{m_e} - \frac{1}{3}\right) E\,, \tag{6.38}$$

where $\alpha_e \simeq 1/137$ is the fine-structure constant. In the case of neutral matter, it holds

$$b^{\rm neut}(E,\vec{x}) = \frac{3}{8\pi}\alpha_e\,\sigma_T\, n_i \left(\frac{4}{3}\phi_1^i + \phi_2^i\right) E_e\,, \tag{6.39}$$

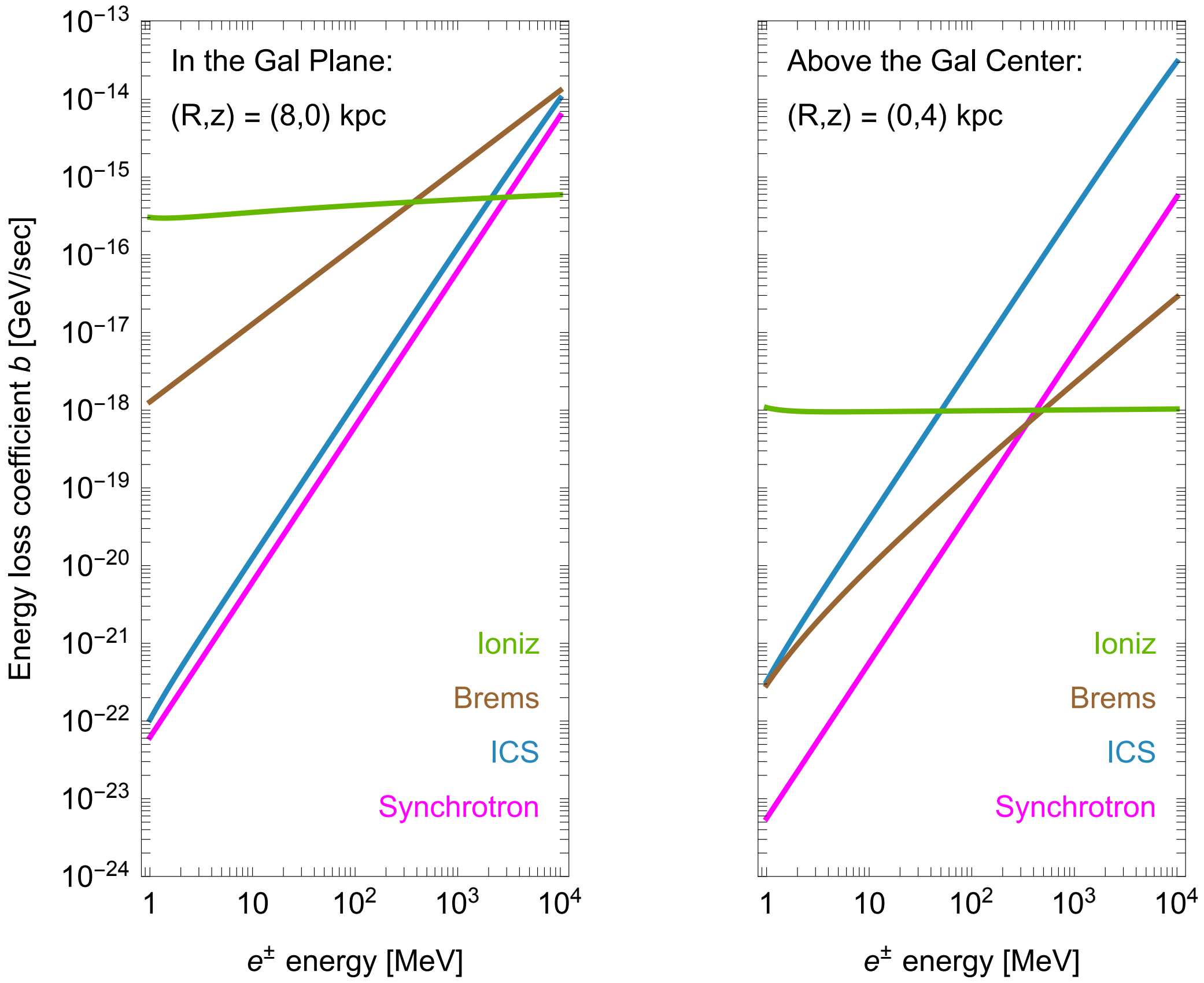

**Fig. 6.8.** Energy losses as a function of the electrons and positrons input energy for two different locations: on the Galactic Plane, near the Sun (left panel) and 4 kpc above the GC (right panel). Different colours refer to different loss mechanisms: ionization (green), bremsstrahlung (brown), ICS (blue), synchrotron (magenta).

where $\phi^i_{1,2}$ are the scattering functions and they are essentially constant for relativistic $e^{\pm}$. Note that both $b^{\rm ion}$ and $b^{\rm neut}$ scale linearly with $E_e$ at leading order. The synchrotron energy losses are given by

$$b_{\rm syn}(E,\vec{x}) = \frac{4\,c\,\sigma_T}{3m_e^2}\frac{B^2}{8\pi}E^2\,. \tag{6.40}$$

The synchrotron losses are proportional to the square of the initial $e^{\pm}$ energy. Regarding the magnetic field, we adopt the MF1 configuration of Ref. [461] (based on Ref. [462]), according to which

$$B = B_0\,\exp\left(-\frac{r-R_0}{R_B} - \frac{z}{z_B}\right)\,, \tag{6.41}$$

with $B_0 = 4.78\ \mu$G, $R_B = 10$ kpc and $z_B = 2$ kpc. The impact of the choice of galactic magnetic field by using different models will be discussed in Chapter 7.

Last but not least, the ICS energy loss rate can be described as

$$b_{\rm ICS}(E,\vec{x}) = \iint {\rm d}\epsilon\,{\rm d}E_\gamma\,(E_\gamma-\epsilon)\,\frac{{\rm d}N}{{\rm d}t\,{\rm d}\epsilon\,{\rm d}E_\gamma}\,, \tag{6.42}$$

where $\frac{\mathrm{d}N}{\mathrm{d}t\,\mathrm{d}\epsilon\,\mathrm{d}E_\gamma}$ represents the scattering rate on photons with initial energy $\epsilon$ and final energy $E_\gamma$, while the factor $\left(E_\gamma - \epsilon\right)$ stands for the energy that the $e^\pm$ lose in the ICS process. It is convenient to change variable

$$E_\gamma \Longrightarrow q = \frac{\tilde{E}_\gamma}{\Gamma\left(1-\tilde{E}_\gamma\right)} \tag{6.43}$$

with $\tilde{E}_\gamma = E_\gamma/\gamma m_e$, $\gamma = E/m_e$, $\Gamma = 4\epsilon\gamma/m_e$. Thus, Eq. (6.42) becomes

$$b_{\mathrm{ICS}}\left(E,\vec{x}\right) = \int_0^\infty \mathrm{d}\epsilon \int_{1/4\gamma^2}^1 \mathrm{d}q\, \frac{\mathrm{d}E_\gamma}{\mathrm{d}q} \left(E_\gamma(q) - \epsilon\right) \frac{\mathrm{d}N}{\mathrm{d}t\,\mathrm{d}\epsilon\,\mathrm{d}q}\,. \tag{6.44}$$

When $\Gamma << 1$, the scattering occurs in the Thomson regime, which represents the classical limit of the more general Klein-Nishina regime. In our case, we deal with relativistic particles, thus it is necessary to adopt the Klein-Nishina formalism. In this regime, the scattering rate can be expressed as

$$\frac{\mathrm{d}N}{\mathrm{d}t\,\mathrm{d}\epsilon\,\mathrm{d}q} = \frac{3\sigma_T\, c}{4\gamma^2}\frac{n_\gamma}{\epsilon} f_{\mathrm{ICS}} \tag{6.45}$$

where

$$f_{\mathrm{ICS}} = 1 + 2q\left(\ln q - q + \frac{1}{2}\right) - \frac{1-q}{2}\frac{(\Gamma q)^2}{1+\Gamma q}\,, \tag{6.46}$$

and $n_\gamma$ represents the density of the radiation field in the energy range $\mathrm{d}\epsilon$, sometimes also indicated as $\mathrm{d}n_\gamma/\mathrm{d}\epsilon$ in the literature. Plugging Eq. (6.45) into Eq. (6.44), we get

$$b_{\mathrm{ICS}}\left(E,\vec{x}\right) = 3\sigma_T \int_0^\infty \mathrm{d}\epsilon\,\epsilon \int_{1/4\gamma^2}^1 \mathrm{d}q\, n_\gamma \frac{\left(4\gamma^2 - \Gamma\right)q - 1}{(1+\Gamma q)^3} f_{\mathrm{ICS}}(q)\,. \tag{6.47}$$

For low electron energies, the leading contribution in Eq. (6.47) is $4\gamma^2 q$, implying that ICS losses are essentially proportional to $E^2$, in analogy with synchrotron losses. This is even more evident in the Thomson limit, where ICS losses reduce to

$$b_{\mathrm{ICS}}\left(E,\vec{x}\right) = \frac{4c\,\sigma_T}{3\,m_e^2}E^2 \int_0^\infty \mathrm{d}\epsilon\,\epsilon\, n_\gamma(\epsilon,\vec{x})\,, \tag{6.48}$$

where $u_\gamma = \int_0^\infty \mathrm{d}\epsilon\,\epsilon\, n_\gamma(\epsilon,\vec{x})$ is the energy density of the photon bath. Fig 6.8 illustrates the energy loss function as a function of the input $e^\pm$ energy. The trends discussed above are easily discernible: the ionization curve (green) is quite constant, the bremsstrahlung component (brown) has a linear behaviour in $E$, while the synchrotron (magenta) and ICS (blue) losses scale with $E^2$. The left panel refers to a location on the Galactic Plane, in the vicinity of the Sun. In this region and for the energy range of interest, the $e^\pm$ lose most of their energy via ionization and bremsstrahlung. This is not surprising since there is a high density of interstellar atomic and molecular gas in the galactic plane, therefore interactions with gas are favoured. The right panel refers to a location well outside the Galactic Plane, on the vertical line above the GC. For $e^\pm$ energies above approximately 40 MeV, the ICS emission is the dominant loss mechanism. Therefore, lines of sight that avoid the Galactic Plane, at high latitudes, are preferred for our purposes. It is important to keep in mind that these quantities are subject to astrophysical uncertainties, stemming for instance from the uncertainties in the gas distribution, in the density of the photon bath and in the intensity of the galactic magnetic field. We will discuss the impact of such uncertainties on our signal in Chapter 7.

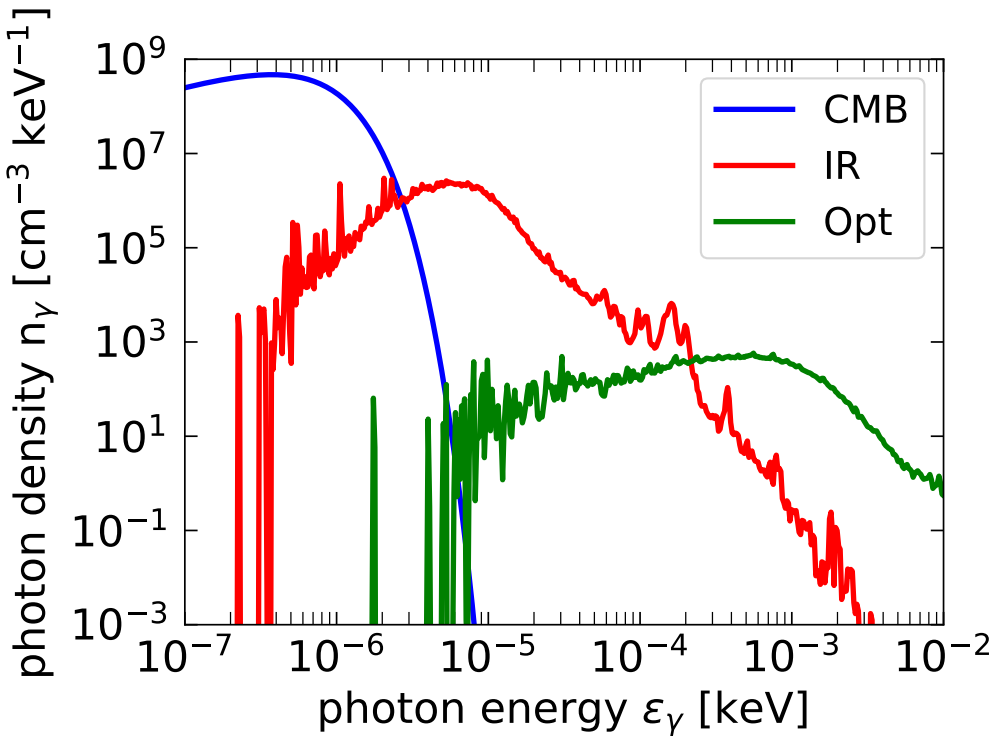

**Fig. 6.9.** Photon number density as a function of the initial photon energy for CMB (blue), infrared from dust (red) and starlight (green).

| Type | $\epsilon$ (eV) | $E_e$ (GeV) | $E_\gamma$ (keV) |
|---|---|---|---|
| CMB | $10^{-4}$ | 5 | 40 |
| IR | $10^{-2}$ | 0.5 | 40 |
| Opt | 10 | 0.05 | 400 |

**Table 6.1.** Estimates of the photon energy shift due to the inverse Compton scattering of sub-GeV electrons. The columns contain, in order of appearance: the type of radiation field, the initial energy of the photon, the initial energy of the electron, the final energy of the photon.

**Interstellar radiation fields.** In our analysis we take into considerations three kinds of photon fields: CMB, infrared light from dust (IR) and starlight (Opt). The CMB energy density is well characterised by the isotropic black-body radiation

$$n_{\rm CMB} = \frac{\epsilon^2}{\pi^2 (\hbar c)^3} \frac{1}{e^{\epsilon/k_B T} - 1}\,, \tag{6.49}$$

while for the IR and Opt components we adopt the interstellar radiation field (ISRF) maps extracted from the GALPROP code [463], as in Refs. [464, 367, 461]. Fig. 6.9 illustrates the photon number density as a function of the initial energy for the three radiation fields under consideration. The final energy $E_\gamma$ of the photon after the ICS reads

$$E_\gamma = \gamma^2 \left(1 + \frac{v}{c}\cos\alpha\right)^2 \epsilon\,, \tag{6.50}$$

where $\alpha$ is the angle of incidence between the initial directions of the electron and photon, while $\epsilon$ is the energy of the radiation field before the scattering. Head-on collisions deliver the maximum energy to the photon, corresponding to

$$E_{\gamma,\max} = \gamma^2 \left(1 + \frac{v}{c}\right)^2 \epsilon \approx 4\gamma^2 \epsilon\,. \tag{6.51}$$

An estimate of the final energy of the photons after the scattering with GeV and MeV electrons is provided in Table 6.1. The first column indicates the type of radiation field, the second column specifies its initial energy, while the third and the fourth columns show the initial energy of the electron and the final energy of the photon, respectively. One can observe that after the ICS, these low-energy radiation fields become hard X rays and soft $\gamma$ rays. Notably, these energies fall in the energy range probed the INTEGRAL data, as will be illustrated in Chapter 7.

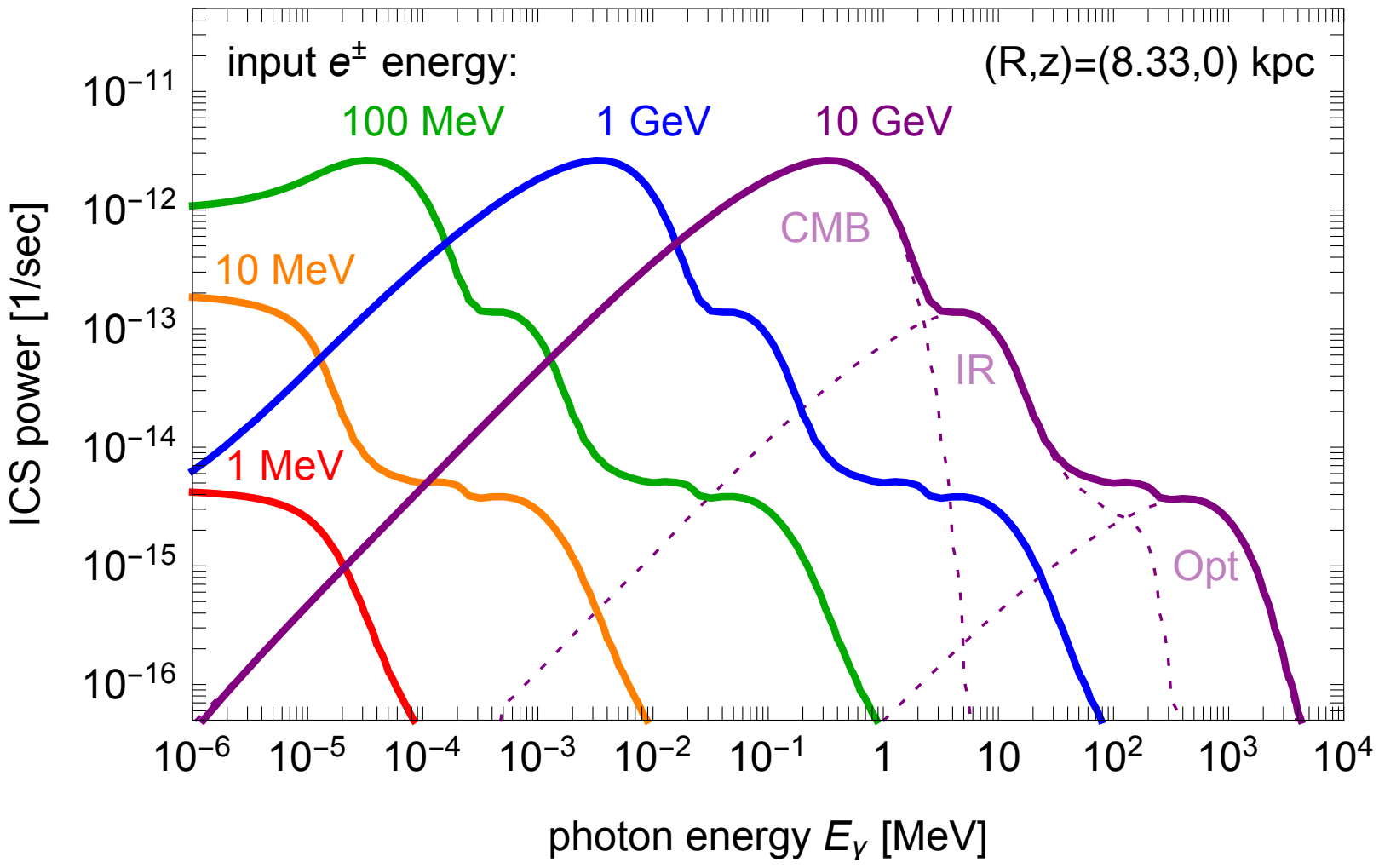

**Fig. 6.10.** Total inverse Compton scattering differential power as a function of the emitted photon energy, at the location of the Sun. Different colours refer to different input $e^{\pm}$ energies, listed near each curve. In the 10 GeV case, the contributions from the three components of the photon bath (dashed lines) are displayed.

### 6.3.2 Inverse Compton scattering power

The differential power emitted into photons due to the ICS radiative processes is defined as

$$\mathcal{P}_{\rm ICS}\left(E_\gamma, E_e, \vec{x}\right) = \int \mathrm{d}\epsilon\, E_\gamma \left(E_\gamma - \epsilon\right) \frac{\mathrm{d}N}{\mathrm{d}t\mathrm{d}\epsilon\mathrm{d}E_\gamma}\,, \tag{6.52}$$

which is equivalent to the computation of the ICS energy losses, except that no integral on $E_\gamma$ (or likewise $q$) is performed. Thus, we can adopt the same change of variable to get

$$\mathcal{P}^i_{\rm ICS}(E_\gamma, E_e, \vec{x}) = c\, E_\gamma \int \mathrm{d}\epsilon \frac{3\sigma_T}{4\epsilon\,\gamma^2}\, n_\gamma(\epsilon)\, f_{\rm ICS}\left(E_\gamma, E_e\right)\,. \tag{6.53}$$

The total differential power is given by $\mathcal{P}_{\rm ICS} = \sum_i \mathcal{P}^i_{\rm ICS}$, where $i$ runs over the different contributions to the photon bath. The power is illustrated in Fig. 6.10 as a function of the final photon energy in a location corresponding to the solar system. The different colours denote different input energies of the electrons and positrons. The wiggles in the total differential power are directly related to the behaviour of the photon density distributions. This is apparent by looking at the purple curve, corresponding to $E_{e^\pm} = 10$ GeV, where the contributions from the individual radiation fields are emphasised (dashed lines).

### 6.3.3 Inverse Compton scattering flux

The emissivity is obtained as a convolution of the density of the emitting medium with the power that it radiates:

$$j(E_\gamma, \vec{x}) = 2\int_{m_e}^{M_\chi} \mathrm{d}E_e\, \mathcal{P}_{\mathrm{ICS}}(E_\gamma, E_e, \vec{x})\, n_e(E_e, \vec{x})\ , \tag{6.54}$$

where the factor 2 takes into account that annihilation events of DM particles give rise to an equal populations of electrons and positrons. The differential flux of the ICS photons that we measured on Earth can be written in terms of the emissivity $j(E_\gamma, \vec{x})$ of all cells located along the line of sight at position $\vec{x}$:

$$\frac{\mathrm{d}\Phi_{\mathrm{ICS}\gamma}}{\mathrm{d}E_\gamma\,\mathrm{d}\Omega} = \frac{1}{E_\gamma}\int_{\mathrm{l.o.s.}} \mathrm{d}s\, \frac{j(E_\gamma, \vec{x})}{4\pi}\ . \tag{6.55}$$

Note that the spherical symmetry of the system around the centre of the Milky Way is broken by the distribution of the ambient light, which mostly lies in the galactic disk.

The final step to compute the full spectrum of photons from annihilating DM consists of integrating the contributions in Eqs. (6.9) and (6.55) over the selected region of observation:

$$\frac{\mathrm{d}\Phi_{\mathrm{Tot},\gamma}}{\mathrm{d}E_\gamma} = \int_{b_{\mathrm{min}}}^{b_{\mathrm{max}}}\int_{\ell_{\mathrm{min}}}^{\ell_{\mathrm{max}}} \mathrm{d}b\,\mathrm{d}\ell\,\cos b\left(\frac{\mathrm{d}\Phi_{\mathrm{prompt},\gamma}}{\mathrm{d}E_\gamma\,\mathrm{d}\Omega} + \frac{\mathrm{d}\Phi_{\mathrm{ICS},\gamma}}{\mathrm{d}E_\gamma\,\mathrm{d}\Omega}\right)\ . \tag{6.56}$$

The constraints on DM particles can be obtained by comparing the total flux with the data in the same region, as will be explained in more detail in Chapter 7. Fig. 6.11 illustrates two examples of the energy flux as a function of the final photon energy $E_\gamma$. The left panel refers to DM particles with a mass of 150 MeV and $\sigma v = 3 \times 10^{26}$ cm$^3$/s, annihilating into $\mu^+\mu^-$. The region of observation is a rectangular area around the Galactic Plane, corresponding to $|b| < 15°$ and $|l| < 30°$. The blue curves refer to the prompt components: FSR (dashed) and radiative decay (dash-dotted). The dotted lines refer to the contribution from ICS on the different components of the ambient bath: infrared (red) and starlight (green), while the CMB does not appear in the figure because its contribution is subdominant in this energy range. For illustrative purposes, the INTEGRAL data (red bars) and the FERMI-LAT measurements (blue crosses) are displayed. We note that the total DM flux (black line) does not reach the FERMI energy window, but it produces a signal in X rays that can be constrained by the INTEGRAL telescope. Also, the prompt contributions stay well below the data and they are several order of magnitude lower with respect to the ICS component. Therefore, including the ICS contribution on the ambient radiation fields is essential, being the leading component to the total flux in the energy range relevant for INTEGRAL, and thus it will lead to stronger constraints, as we will discuss in Chapter 7. On the right panel, we have the case of DM particles with a lower mass, $m_\chi$ = 10 MeV, which annihilate into electron-positron pairs. This scenario is constrained only using the FSR, because in this case the ICS becomes important at energies much lower than the range probed by the INTEGRAL data. This evidence will have an impact in the final constraints. Note that the FERMI-LAT measurements are shown here just for reference, as they are not the focus of this analysis. Also, we employed a different data set for the INTEGRAL measurements with respect to the one shown in Fig. 6.11 (which is meant to be illustrative of the points raised by our analysis), as will be illustrated in Chapter 7.

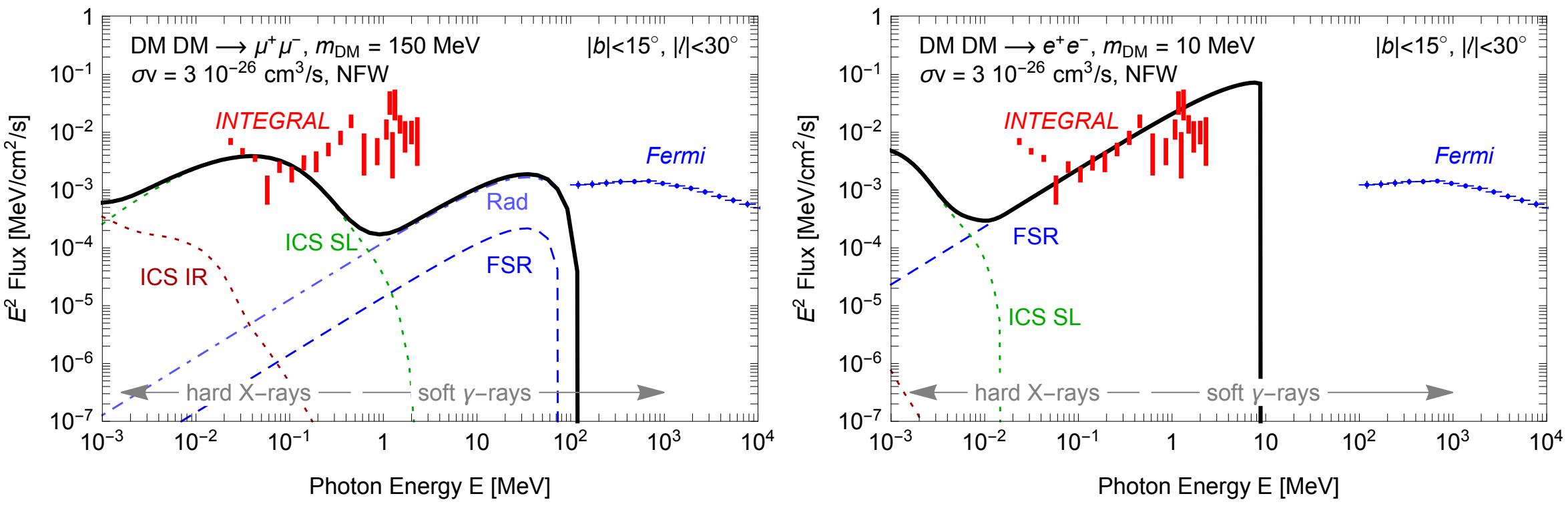

**Fig. 6.11.** *Left:* Photon spectrum produced by DM particles with mass $m_\chi$ = 150 MeV and cross-section $\sigma v = 3 \times 10^{26}$ cm$^3$/s, annihilating into $\mu^+\mu^-$. We show the total flux in black and the individual components in colour: final state radiation (FSR, blue dashed), radiative decay (Rad, blue dash-dotted), inverse Compton scattering over starlight (ICS SL, green dotted) and over infrared light (IR, brown dotted). The INTEGRAL and FERMI data are displayed with red bars and blue crosses, respectively. *Right:* The same as the left panel, for a 10 MeV DM particle annihilating into $e^+e^-$.

CHAPTER 7

# INTEGRAL X-RAY CONSTRAINTS ON SUB-GEV DARK MATTER

## 7.1 INTEGRAL data

The INTEGRAL space telescope is a mission of the European Space Agency to observe hard X rays and soft $\gamma$ rays. One of the instruments aboard is the spectrometer SPI, whose measurements are employed in this analysis. The data were collected in the period 2003-2009[1], corresponding to a total exposure of about $10^8$ seconds and cover the energy range from 20 keV up to a few MeV. They are provided in two forms:

- as an energy spectrum of the total diffuse flux in a rectangular region of observation centred around the GC, namely $|b| < 15°$, $|\ell| < 30°$ (Figs. 6-7 in Ref. [465]);
- as an angular flux in latitude and longitude bins (Figs. 4-5 in Ref. [465]).

For illustrative purposes, the first data set is shown in Fig. 6.11. However, in our analysis we adopt the latter set of data, from which we mask the Galactic Plane (GP). This is based on two reasons, both of which make the GP less attractive from an analysis perspective. Firstly, the GP is bright in X rays due to numerous astrophysical sources, dust radiation and ICS emission from cosmic rays. These sources have no connection to DM and represent a significant background noise. In order to adopt a conservative approach in the derivation of the DM bounds, we do not attempt to model and subtract the galactic X-ray emission[2]. Instead, in our analysis we avoid the region of most contamination. Secondly, due to the high density of interstellar gas, the $e^{\pm}$ emissivity in the GP is dominated by scattering processes on gas, notably Coulomb interactions, ionization and bremsstrahlung. This complexity of the GP induces uncertainties in the prediction of the photon flux, which can be avoided by looking at higher latitudes. Therefore we choose to focus on relatively gas-poorer regions in which the ICS emissivity constitutes the most relevant process. Note that this approach represents a conservative choice. The data adopted in our analysis are adapted from Fig. 5 of Ref. [465]. They are divided into five energy bands: 27−49 keV, 49−90 keV, 100−200 keV, 200−600 keV and 600−1800 keV; each one including 21 bins in latitude, except the fifth band which features only 15 bins. The longitude window is $-23.1° < \ell < 23.1°$ for the first four bands and $-60° < \ell < 60°$ for the fifth one. For each DM annihilation channel, we compute the total photon flux from DM annihilation in each energy band and per each latitude bin, as discussed in Chapter 6. Fig. 7.1 displays the INTEGRAL data (coloured points) together with one example of the predicted photon flux from DM particles with a mass of 150 MeV, annihilating into $\mu^+\mu^-$ (black points). The DM flux increases approaching the central latitude bins, as expected since the DM density becomes larger towards the centre of the Milky Way. If the emission was purely of prompt origin, the angular profile would strictly follow the square of the DM density profile. However this is not the case here. The reason is that the leading contribution in the configuration under consideration is represented by the ICS component, which is moulded by the spatial behaviour related to the $e^{\pm}$ energy losses and to the density of the target radiation fields. As a result, the DM flux in the very central latitude bin, corresponding to lines of sight crossing the GP, is significantly lower compared to the neighbouring bins. This feature is a consequence of a high interstellar gas density in the GP, which favours the energy losses via interactions between gas and e$^{\pm}$, therefore suppressing the ICS emissivity. In the analysis, we remove ("mask") the three central latitude bins in order to exclude most of the signal from the GP. For the first four energy bands, the three central bins cover the interval $-3.9° < b < 3.9°$, which corresponds to masking all lines of sight passing within approximately 0.6 kpc above and below the vertical of the GC.

[1] To the best of the author's knowledge, the most recent whole-sky data (or over large patches of sky) for INTEGRAL dates back to the time interval 2003-2009.

[2] As we will explained in Sec. 7.2, we also derive the "optimistic constraints", using a template for the astrophysical component. However, we consider the conservative bounds as our reference constraints.

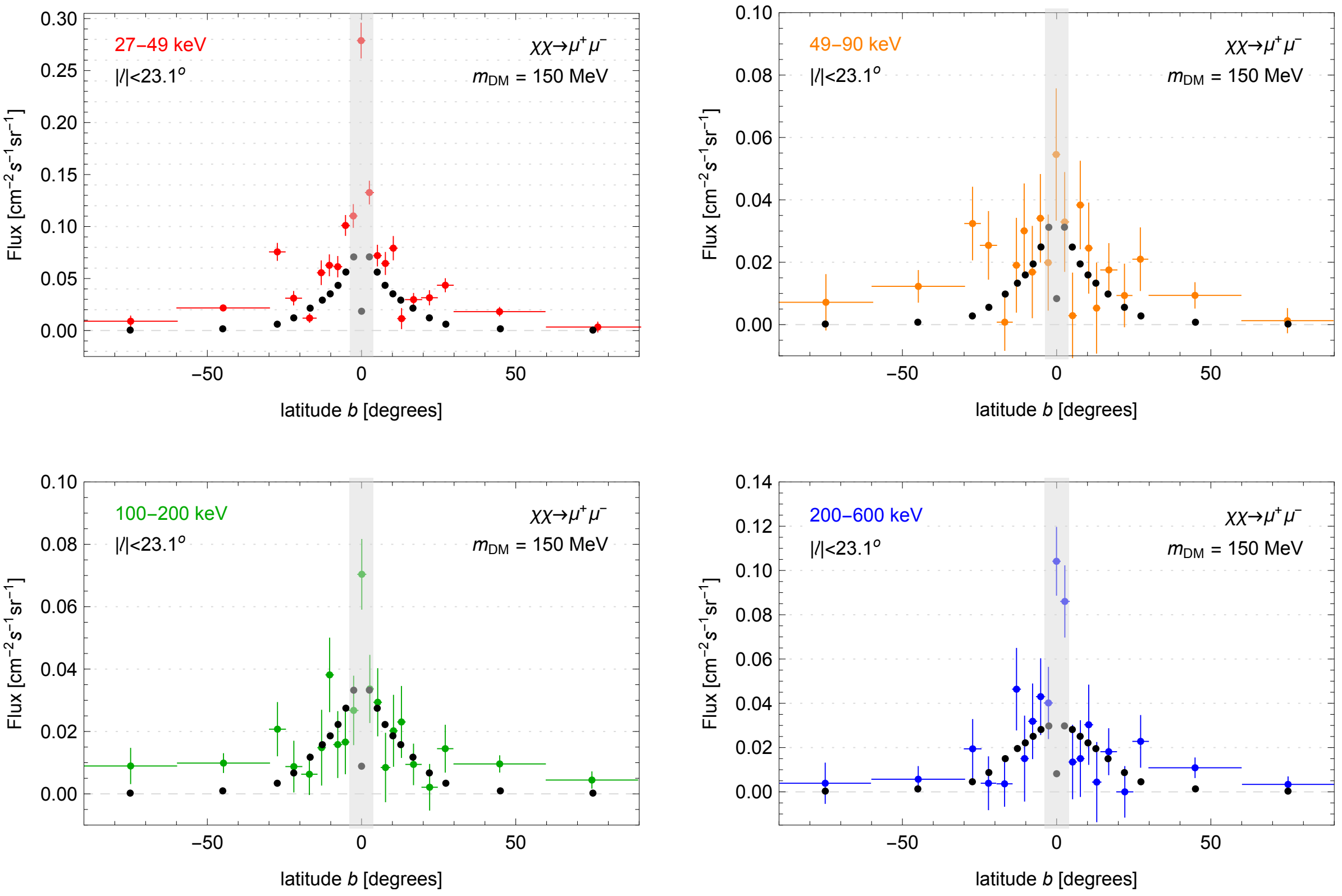

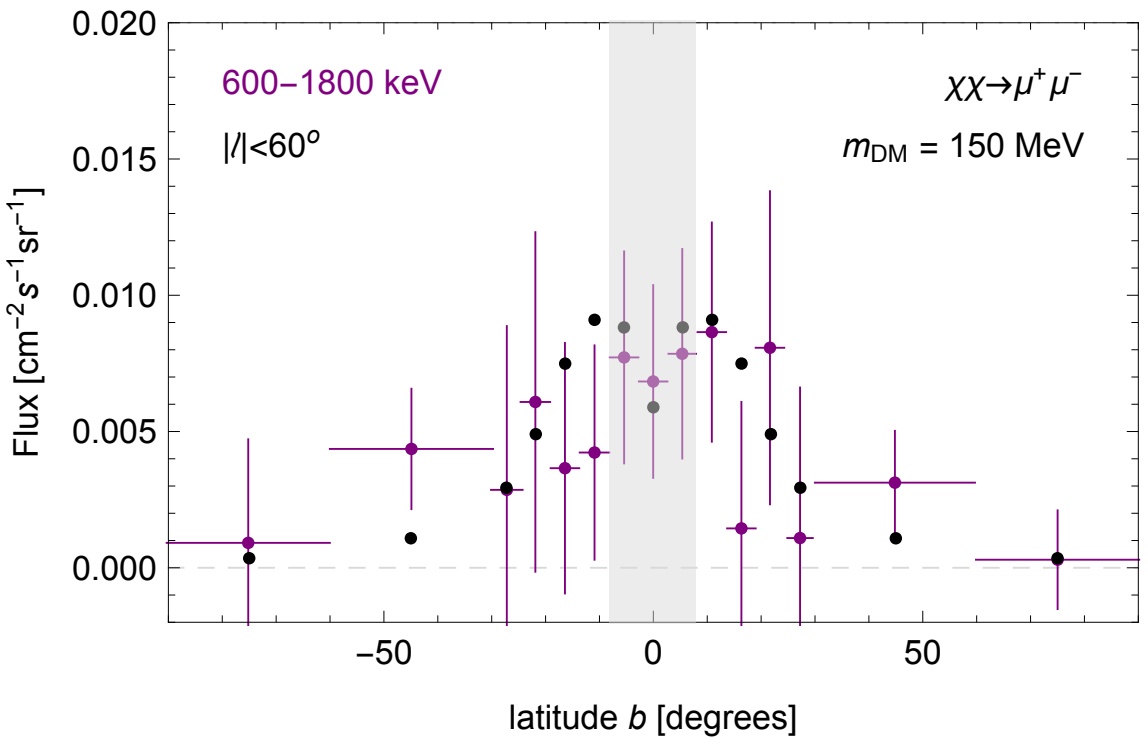

**Fig. 7.1.** The predicted dark matter angular flux for dark matter particles with mass of 150 MeV, annihilating into $\mu^+\mu^-$ (black points) compared to the INTEGRAL data (coloured points). In each panel the annihilation cross-section is set at its limit, such that it maximises the flux without exceeding the data in the energy band by more than $2\sigma$. The shaded grey regions cover the low-latitude bins that we do not use in the analysis.

## 7.2 Constraints on $\langle \sigma v \rangle$

The constraints on the DM velocity-averaged annihilation cross-section $\langle \sigma v \rangle$ are obtained in two different ways:

- conservative bounds, where no astrophysical background is included;
- optimistic bounds, where a model for the astrophysical galactic X-ray emission is included and the DM component is added on top of the background.

The two approaches are discussed below.

**Conservative constraints.** The constraints on the DM cross-section are derived from the comparison between the predicted DM flux $\Phi_{\rm DM}$ and the measured flux $\phi$, requiring that the former does not exceed the latter by more than an appropriate amount. More precisely, we define a test statistic

$$\chi^2_{\rm cons} = \sum_{\rm bands} \sum_i \frac{\left(\max\left[(\Phi_{{\rm DM},i}(\langle \sigma v \rangle) - \phi_i), 0\right]\right)^2}{\sigma_i^2}\,, \tag{7.1}$$

where the first sum runs over the five INTEGRAL energy bands and the second over the latitude bins $i \in$ {b bins} (excluding the three central ones), and $\sigma_i$ indicates the uncertainty on the $i$−th data point. The quantity $\chi^2_{\rm cons}$ corresponds to computing a global "effective" $\chi^2$ that includes only the data bins where the DM flux is higher than the measured value. This means that bins where the predicted DM flux is smaller than the observed value are considered compatible with the observations. The DM flux starts to introduce some tension only when it exceeds the data points and this tension progressively increases with the DM annihilation cross-section. We perform a raster scan for each DM mass, which is equivalent to having a number of unconstrained non-negative nuisance parameters for the background in each latitude bin. As a result, our statistic is equivalent to a $\Delta\chi^2$, distributed as a $\chi^2$ with one degree of freedom (see also Ref. [466] for a similar approach). We derive the bound on $\langle\sigma v\rangle$ by requiring $\chi^2_{\rm cons} = 4$, which corresponds to a $2\sigma$ limit. The interested reader can find a detailed discussion on similar test statistics in Ref. [467]. The first three panels in Fig. 7.2 show the constraints imposed by each energy band separately. These limits are obtained with the same $\chi^2_{\rm cons}$ criterion defined in Eq. (7.1), applied independently in each energy band. The dominant constraint comes mostly from the 49−90 keV band, although not always. The highest energy band (600−1800 keV) almost always provides the weakest bound. The relative strength of the constraints depends on various factors, including statistical fluctuations in the data and size of the errors bars in the different energy bands, but most notably from the position (in energy) of the ICS peak contribution relative to the INTEGRAL data. To get a better understanding of the origin of the constraints shown in the first three panels, the last (bottom right) panel of Fig. 7.2 shows the bounds obtained by considering the contributions from FSR, Rad and ICS separately. These bounds are computed using all the data points in all the five energy bands simultaneously. When present, the ICS constraint provides the most stringent bounds. The reason is that for large DM masses the ICS flux is significantly higher than the FSR and Rad components in the energy range probed by the INTEGRAL data, as already illustrated in the left panel of Fig. 6.11. The peak of the ICS flux shifts to lower energies as the DM mass decreases, because only lower-energy $e^\pm$ can be produced: this feature is apparent in the ICS power depicted in Fig. 6.10. As a consequence, for DM masses below approximately 30 MeV, the ICS contribution peaks at energies lower than the INTEGRAL window, therefore becoming ineffective and leaving the FSR to dominate the constraint. However, this remark affects only the electron channel, since it is the only one kinematically open at these low DM masses. Finally, the global $2\sigma$ constraints derived by applying Eq. (7.2) are displayed in Fig. 7.3. It is worth emphasising that these bounds are not just the lower envelope of the curves in the first three panels of Fig. 7.2. Indeed the overall bounds are obtained by considering the total X-ray flux (ICS + FSR + Rad) and by using all the data points, which implies combining all the five energy bands and all the angular bins (except the three central ones). Note that no horizontal line in correspondence of $\langle \sigma v \rangle = 3 \times 10^{26}$ cm/s is shown, since this is the canonical value for thermal WIMP particles. In our case, we are considering sub-GeV DM, whose production mechanism is not supposed to be the thermal freeze-out in the standard cosmological picture. In our phenomenological work we are interested in deriving the bounds in the DM parameter space and we are agnostic on the production mechanism. Figure 7.3 also shows the comparison with the existing constraints in the literature for the mass range of interest. Essig et al. [468] have derived bounds on light DM using a compilation of X-ray and soft $\gamma$-ray data from HEAO-1, INTEGRAL, COMPTEL, EGRET and FERMI-LAT. Among the annihilation channels that we study, they consider only $e^+e^-$. They do not include the ICS and Rad contributions to the photon flux and they use the INTEGRAL data in the region $|b| < 15°$, $|\ell| < 30°$ rather than the latitude bins that we employ (from which we exclude the GP). Their bound is displayed with a dash-dotted green line: it is comparable with our limits at small DM masses, becoming stronger in the mass range 5−40 MeV due to the inclusion of the COMPTEL data, then becoming weaker for $m_\chi \gtrsim 50$ MeV when the ICS emission sets in[3].

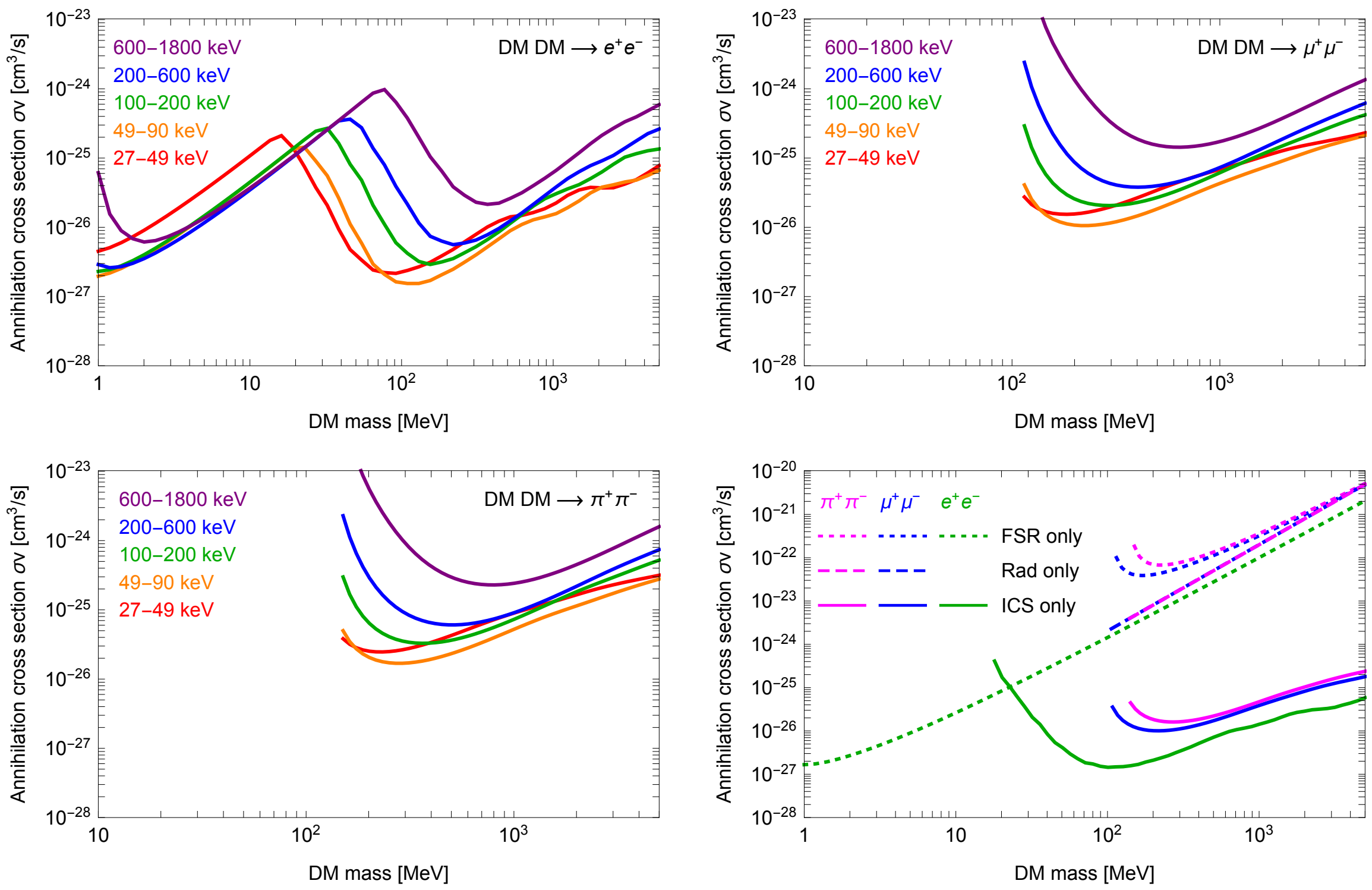

**Fig. 7.2.** *First three panels:* Conservative constraints from each INTEGRAL energy band for the three annihilation channels under consideration (in order of appearance: electron, muon and pion). *Bottom-right panel*: Comparison between the conservative bounds obtained by considering the contributions from final state radiation (FSR, dotted), radiative decay (Rad, dashed) and inverse Compton scattering (ICS, solid) separately and by combining the five energy bands. For each probe, we use the colour code specified in the legend: green for the $\chi\chi \to e^+e^-$ annihilation channel, blue for $\chi\chi \to \mu^+\mu^-$ and magenta for $\chi\chi \to \pi^+\pi^-$.

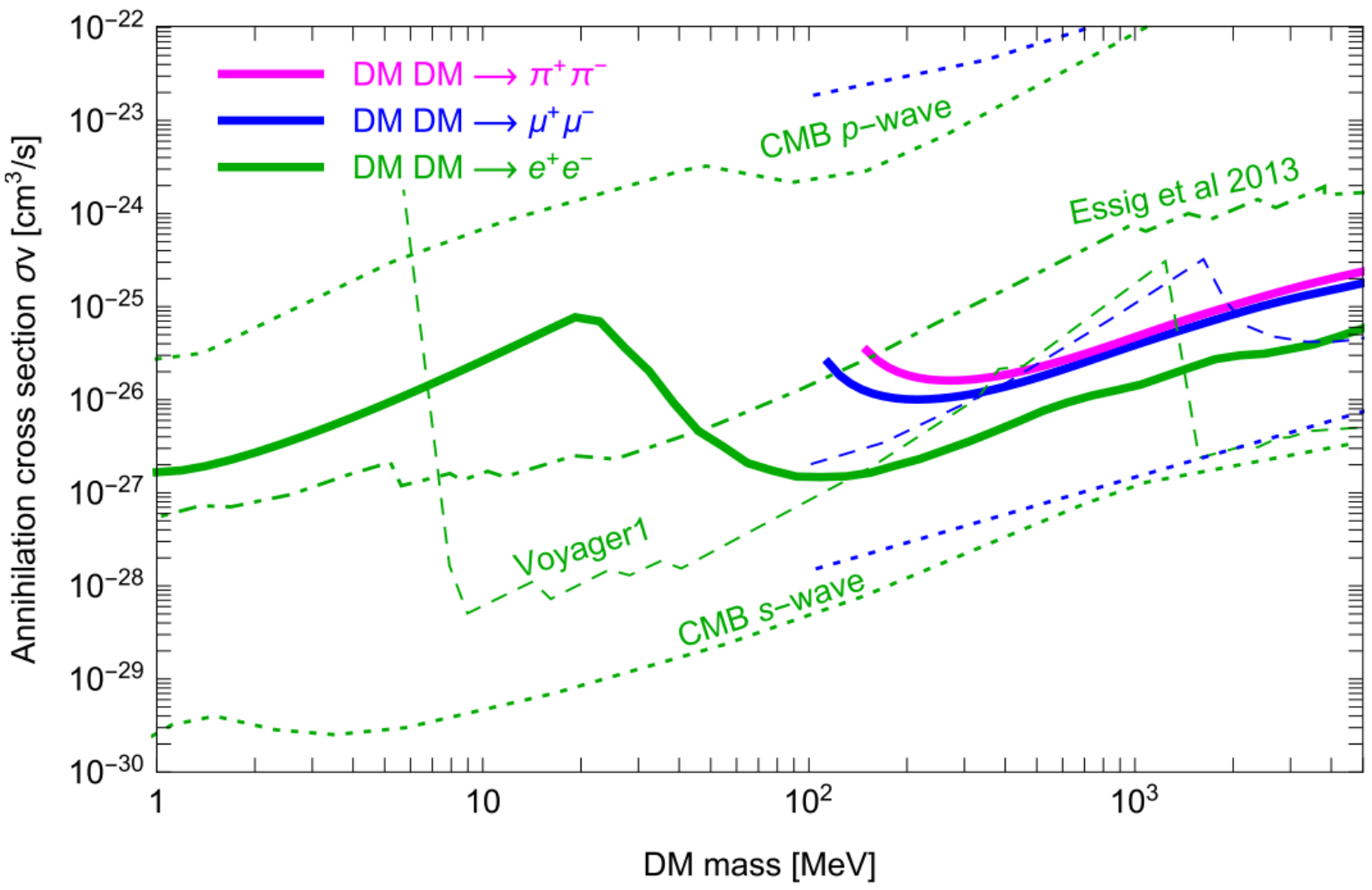

**Fig. 7.3.** Our conservative constraints on sub-GeV DM from INTEGRAL data (thick solid lines), compared to the existing bounds: from VOYAGER 1 $e^\pm$ data (dashed green and blue lines), from a compilation of X-ray data (dash-dotted green line) and from CMB assuming s−wave (dotted green and blue lines in the lower portion of the plot) or p−wave annihilation (dotted green and blue lines in the upper portion of the plot).

Boudaud el al. [470] have derived their constraints on the $e^+e^-$ and $\mu^+\mu^-$ channels using the low energy measurements of the local cosmic-ray flux, carried out by the VOYAGER 1 spacecraft outside of the heliosphere. It is worth recalling that sub-GeV cosmic rays are deflected by the solar magnetic field making them inaccessible from Earth. The only exception is represented by the VOYAGER 1 spacecraft which crossed the heliosphere in 2012, allowing to measure the flux of local cosmic rays. Boudaud et al. considered different propagation assumptions as well as two DM halo profiles (NFW and a cored halo with constant central DM density). Here we report the bounds corresponding to their model B, characterised by weak reacceleration and NFW profile. Their constraints (dashed curves) intertwine with ours over the mass range under consideration, being stronger in the mass range 7−100 MeV and weaker otherwise.

The CMB constraints derived in Ref. [471] are the most stringent across the whole mass range of interest. They are given in Ref. [471] for the $e^+e^-$ channel and in the earlier study of Ref. [472] for the $\mu^+\mu^-$ channel. However, they hold under the assumption that the DM annihilation cross-section is speed-independent ($s$-wave). Instead, if the DM cross-section is $p$-wave, namely $\langle\sigma v\rangle \propto v^2$, the CMB bounds weaken considerably. The degradation of the constraints can be explained by recalling that the CMB is sensitive to the energy injection from DM annihilations at high redshift, approximately at the time of recombination or slightly later. For $p$-wave annihilating DM, such injection was suppressed since DM was very cold (slow) back then. In the galactic halo, at present time, DM particles move faster as an effect of the gravitational collapse that formed large-scale structures, therefore they annihilate more efficiently. In other words, a large value for the annihilation cross-section at present-day is allowed as it corresponds to a much smaller value, and hence a limited effect, at the time of the CMB. The bounds obtained in our analysis as well as the other bounds shown in Fig. 7.3 are sensitive to the DM annihilation at present time. Therefore they are independent on the s−wave/p−wave assumption, under the standard hypothesis of a constant DM speed in the galactic halo, as it is usually assumed in the literature. If we introduce a radial dependence of the DM speed, the p−wave bounds are affected: we have estimated a departure from the s−wave constraints of a factor $\mathcal{O}(40\%)$, for typical assumptions on the DM speed and density profile in the Galaxy. This assessment is discussed in Section 7.3. Also note that the CMB bounds in Fig. 7.3 are rescaled by a factor $(v/v_{\rm ref})^2 = (220/100)^2$ since they are provided in the literature for $v_{\rm ref} = 100$ km/s while we consider $v \simeq 220$ km/s in the Milky Way.

[3]Laha et al. [469], in v1 on the arXiv, also present a result in agreement with Essig et al. [468], while in v2 the bound is no longer present.

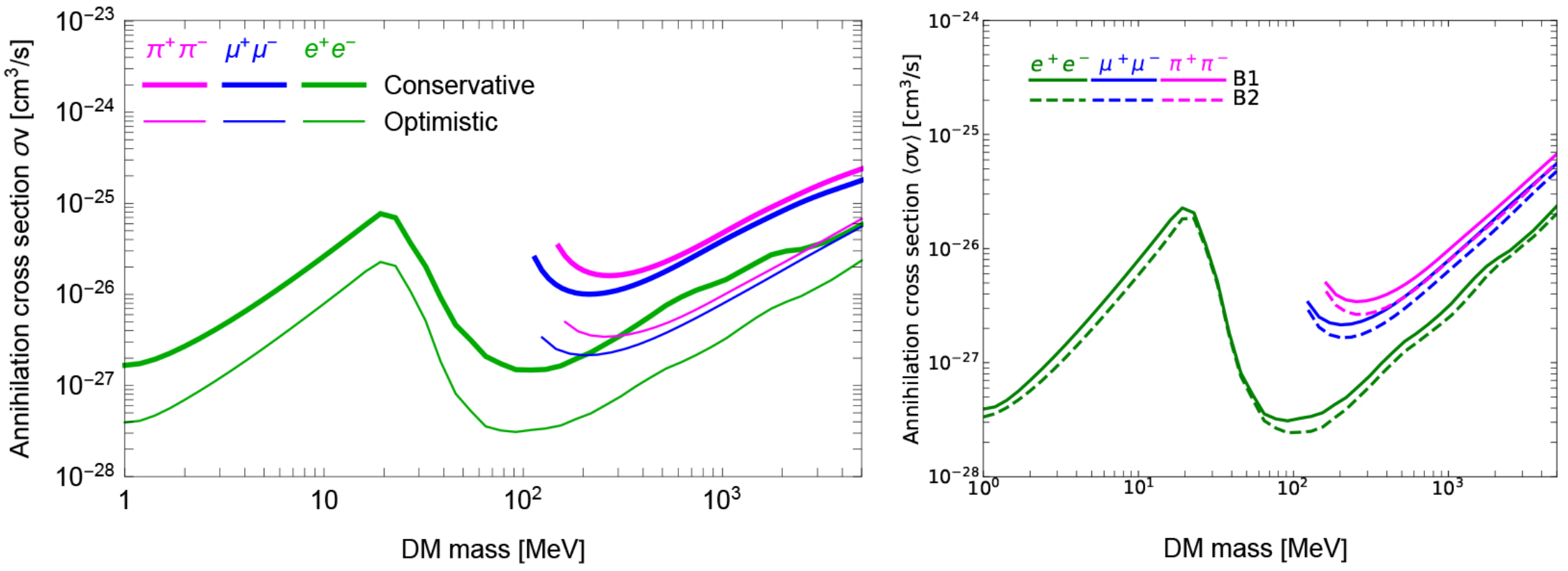

**Fig. 7.4.** *Left:* Conservative constraints (thick lines) and optimistic constraints (thin lines), derived by including the astrophysical background. *Right:* Variation of the bounds due to different normalisations of the astrophysical background. The solid lines refer to one single normalisation, while the dashed curves refer to the case with different normalisations in the five energy bands.

Lastly, the FERMI-LAT constraints as computed by the Collaboration (e.g. Ref. [473]) are not provided for DM masses below a few GeV, therefore they are not displayed.

In conclusion, if DM annihilation is a p−wave, our constraints are the most stringent at present time in the literature for DM particles with a mass in the range 150 MeV − 1.5 GeV and to the best of our knowledge, they represent the only existing constraints on $\langle\sigma v\rangle$ for sub-GeV DM annihilating into $\pi^+\pi^-$.

**Optimistic constraints.** The main source of background consists of diffusive ICS from cosmic rays of astrophysical origin. The conventional sources of primary cosmic rays are supernova remnants, pulsar winds, active galactic nuclei, quasars and the interstellar medium. Taken collectively, these astrophysical emissions constitute a background flux $\phi_B$ and the template, computed within GALPROP, has been obtained by Ref. [465] in the attempt to explain the measured flux. To derive the DM bounds, we employ the following standard procedure. We multiply the astrophysical background flux $\phi_B$ by an overall energy-independent normalisation factor $N_B$. We consider the $\chi^2$ statistic

$$\chi^2_{\rm opt} = \sum_{\rm bands}\sum_i \frac{\left[\Phi_{{\rm DM}\,i}\,(\langle\sigma v\rangle) + N_B\,\phi_B - \phi_i\right]^2}{\sigma_i^2}\,. \tag{7.2}$$

We identify the pair $\left(N_{B,0}, \langle\sigma v\rangle_0\right)$ that minimises the test statistic, corresponding to the value $\chi_0$. Then, for each DM mass we scan the values of $(N_B, \langle\sigma v\rangle)$ and we impose a constraint on $\langle\sigma v\rangle$ by requiring $\Delta\chi^2 = \chi^2 - \chi_0^2 = 4$, for any $N_B$. The left panel of Fig. 7.4 displays the bounds obtained with the two procedures discussed above: conservative and optimistic. The optimistic constraints are more stringent by about half an order of magnitude. This is expected since by including the astrophysical contribution the room for exotic physics is reduced, leading to stronger constraints. We also explore the influence of using a different normalisation of the background in each energy band. The right panel of Fig. 7.4 illustrates the constraints with one single normalisation of the background (B1, solid lines) compared to the case with a different normalisation in each band (B2, dashed lines). This assumption turned out to only mildly affect the overall bounds in the DM parameter space.

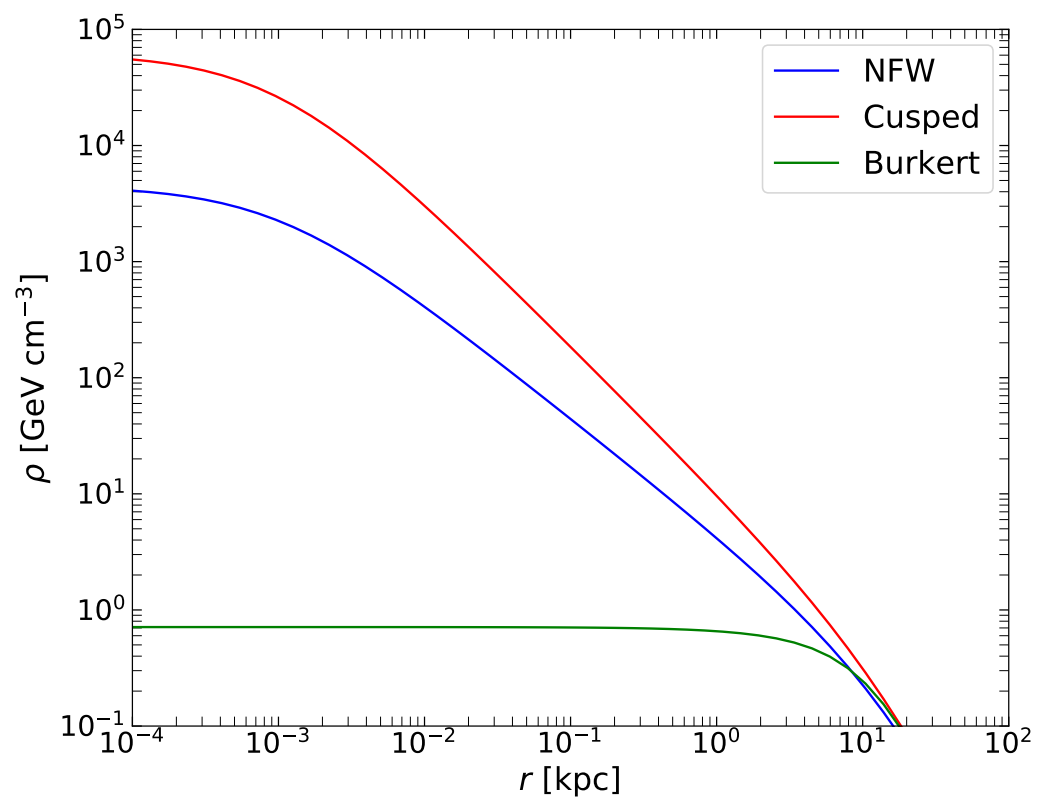

**Fig. 7.5.** Dark matter density as a function of the galactocentric distance for different halo models: NFW (blue), cusped (red), Burkert (green).

| Profile | $\rho_s$ (GeV/cm$^3$) | $r_s$ (kpc) | $\gamma$ |
|---|---|---|---|
| NFW | 0.184 | 24.42 | 1 |
| Cusped | 0.184 | 24.42 | 1.26 |
| Burkert | 0.712 | 12.67 | - |

**Table 7.1.** Normalisation $\rho_s$ and scale radius $r_s$ for different DM halo profiles.

## 7.3 Uncertainties

The DM flux is subject to different types of astrophysical uncertainties, notably the configuration of the galactic magnetic field, the density of the interstellar gas and radiation fields. The DM modelling is also a source of uncertainty, considering that the functional form of the DM halo profile is still debated. In the following we estimate the impact of these uncertainties on our constraints.

**Dark matter density profile**. The DM density distribution in the Milky Way is an uncertain quantity and the choice of the DM profile affects all the different components in the total flux: FSR, Rad and the three ICS contributions. In order to estimate its impact, we compute the constraints with three different DM halo profiles: NFW (which represents our standard choice), a cusped profile and a Burkert model. The functional form of the first two profiles can be expressed

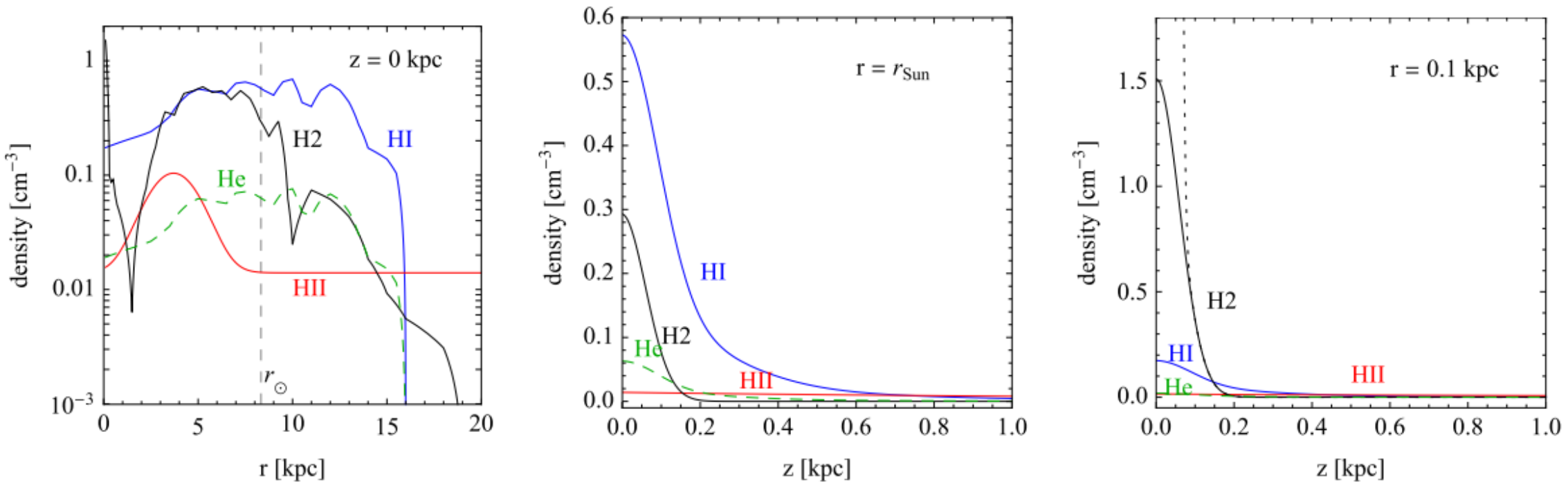

**Fig. 7.6.** *Left:* Density of the gas as a function of the distance from the GC. *Centre and Right:* Density of the gas as a function of the vertical distance from the GP in two different locations: in the position of the Sun (central panel), in a spot near the GC (right panel). Credit: [474].

as

$$\rho(r) = \frac{\rho_s}{\left(\frac{r}{r_s}\right)^{\gamma}\left(1+\frac{r}{r_s}\right)^{3-\gamma}}, \tag{7.3}$$

where the NFW profile is characterised by $\gamma = 1$, while the cusped profile is assumed to have $\gamma = 1.26$. On the other hand, the functional form of the Burkert model is

$$\rho_B = \frac{\rho_s}{\left(1+\frac{r}{r_s}\right)\left[1+\left(\frac{r}{r_s}\right)^2\right]}. \tag{7.4}$$

The values of the normalisation $\rho_s$ and scale radius $r_s$ are listed in Table 7.1. These two parameters are determined by requiring the DM density to be 0.3 GeV/cm$^3$ at the location of the Sun and the total mass of our Galaxy to be $4.7 \times 10^{11}$ M$_\odot$. The profiles taken into account are all function of the galactocentric distance $r$ and they all assume spherical symmetry. Fig. 7.5 illustrates the three DM profiles as a function of the distance from the centre of the Milky Way. The Burkert function differs significantly from the other two, being cored and several orders of magnitude lower as we approach the GC. In particular, the major difference among all the three profiles becomes greater as we move closer to the centre of our Galaxy. The top-left panel of Fig. 7.9 displays the constraints that we obtain by varying the DM profile: the cusped function provides the most stringent bounds (dotted lines), as opposed to the Burkert model which is associated with the weakest limits (dash-dotted) and the NFW bounds (solid) lie in between.

**Gas density**. The interstellar gas is mostly made up of atomic and molecular hydrogen. The latter is concentrated on the GP and can be probed indirectly by measuring the 2.6 mm line of CO. This emission line is associated to the transition of the CO molecules excited by the collisions with molecular hydrogen. On the other hand, atomic hydrogen is more spread around the galactic halo and can be traced through the 21cm line, as already explained in Chapter 4. The helium distribution is assumed to be similar to the hydrogen one, but its abundance is approximately

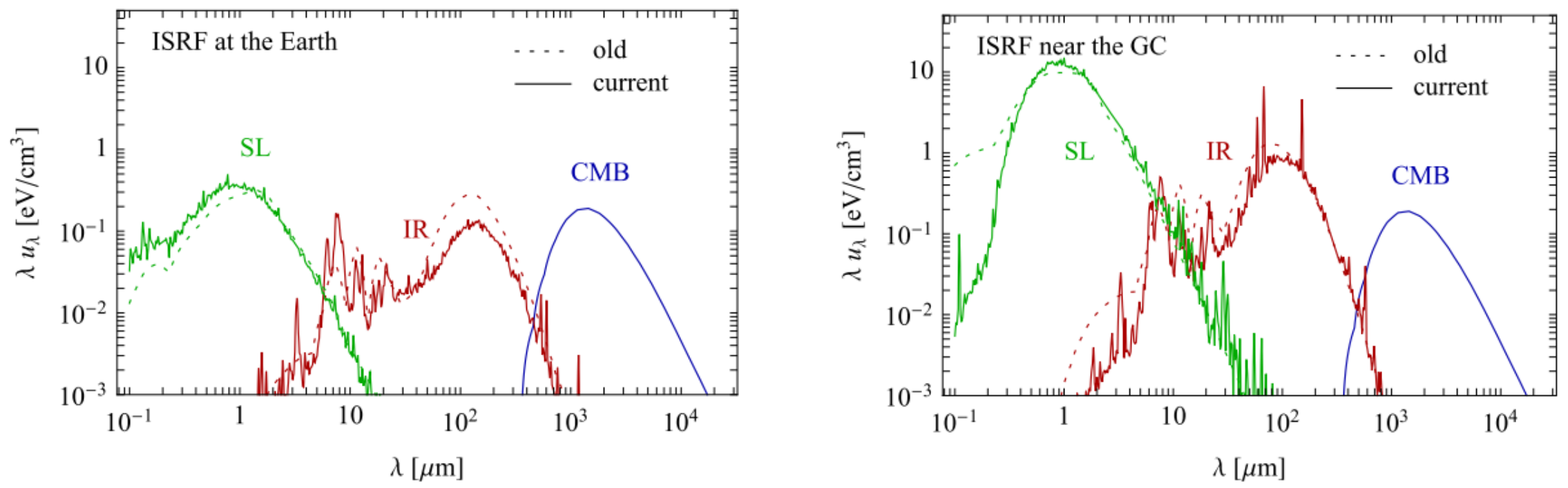

**Fig. 7.7.** Energy density as a function of the wavelength for the interstellar radiation fields in two locations: at the Earth (left) and near the GC (right). The different colours refer to the different component of the ambient bath: starlight (SL, green), infrared emission (IR, red), CMB (blue). The solid lines refer to more recent measurements, while the dashed curves refer to maps previously used in the literature. We note that the difference in the maps over a few years is not dramatic: it mostly consists of additional detected features, such as the spikes in the optical and infrared light. Credit: [461].

10 times lower [475]. The ionised hydrogen is a subdominant component near the GC, but it becomes non-negligible as we move towards the outskirts of our Galaxy. Finally, heavier elements consist of less than 1% of the total gas and they are negligible for the purpose of this analysis. Regarding our treatment of the interstellar gas, we adopt the maps of the gas density, as released by PPPC4DMID [474] (based on Ref. [476]). They are illustrated in Fig. 7.6 for different slices of the Milky Way: the left panel shows the density on the GP as a function of the galactocentric radius, while the central and right panels display the density as a function of the vertical distance from the GP at a radial distance corresponding to the Sun and in a spot near the GC, respectively.

However, the present understanding of the gas density in the Galaxy is characterised by significant uncertainties, which can affect the energy losses by Coulomb, ionisation, bremsstrahlung, and in turn modify the spectrum of the emitting $e^{\pm}$. In order to take into account this source of uncertainties, we vary the normalisation of the overall gas density in the Milky Way by a factor of two. The top-right panel in Fig. 7.9 illustrates the constraints for different assumptions on the gas density and we found that it affects only mildly our bounds.

**Interstellar radiation field density**. The ICS signal and the ICS energy losses depend on the interactions of the $e^{\pm}$ with the ambient light. Therefore, an accurate knowledge of the ISRF density and its uncertainties is necessary. Fig. 7.7 displays the energy density $u = \int d\epsilon\, \epsilon\, n_\gamma$ as a function of the wavelength for the three components in the photon bath under consideration. The left panel refers to the position of the Earth, while the right panel relates to a location near the GC. As expected, the density of infrared and optical light near the GC is higher with respect to the Earth, due to a higher astrophysical background close to the centre of the Milky Way.

| Model | $B_0$ ($\mu$G) | $r_D$ (kpc) | $z_D$ (kpc) | Ref |
|---|---|---|---|---|
| MF1 | 4.78 | 10 | 2 | [478] |
| MF2 | 5.1 | 8.5 | 1 | [479, 480] |
| MF3 | 9.5 | 30 | 4 | [481] |

**Table 7.2.** Parameters and references in the literature for the three configurations of the galactic magnetic field taken into consideration. MF1 represents our reference model.

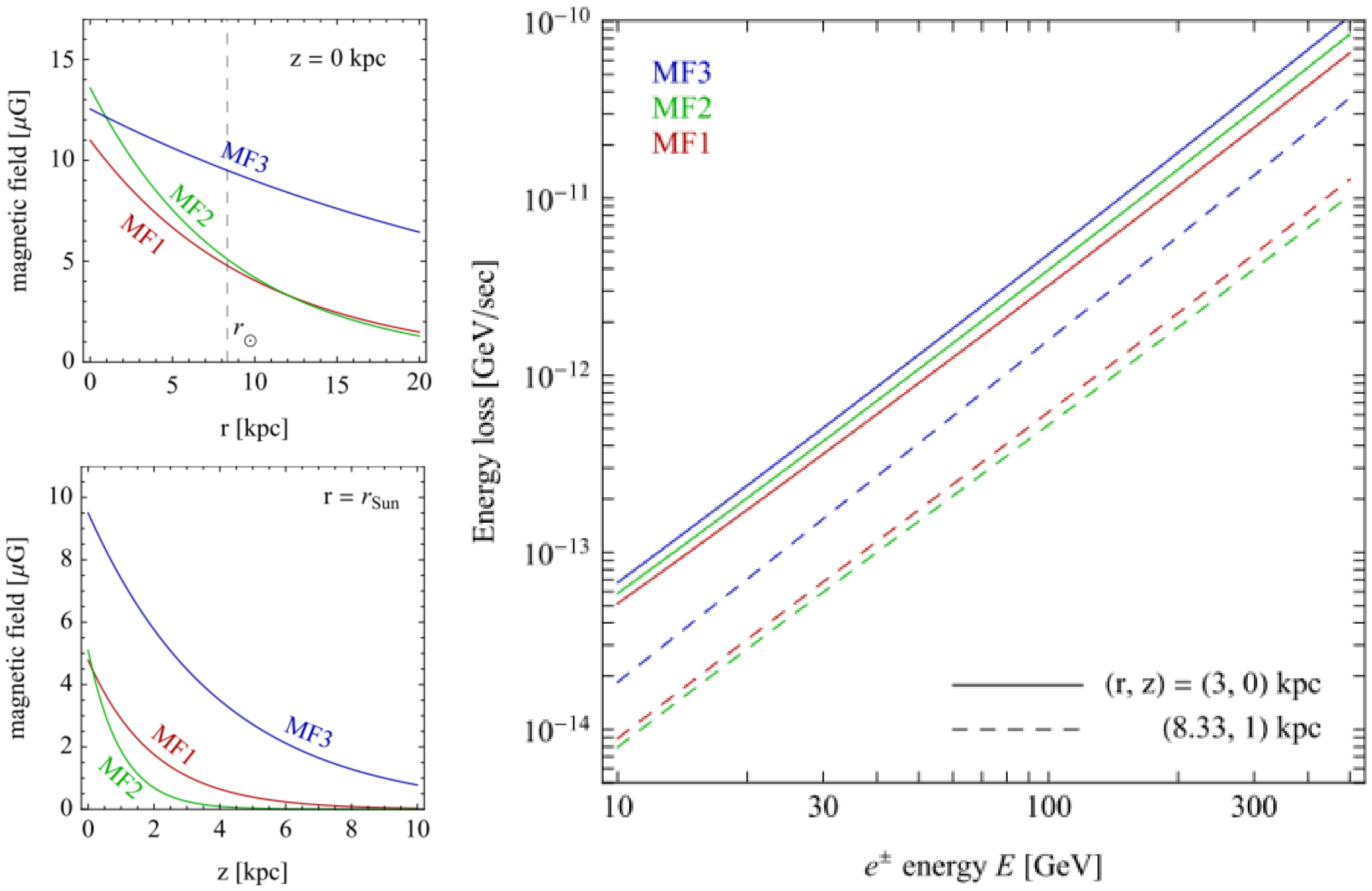

**Fig. 7.8.** *Left:* Magnetic field as a function of the galactocentric distance on the GP (top) and as a function of the vertical distance from the GP at the location of the Sun (bottom). *Right:* Energy losses as a function of the $e^{\pm}$ energy for MF1 (red), MF2 (green), MF3 (blue) models in two locations: on the GP (solid) and 1 kpc above the Sun (dashed). Credit: [461].

The solid lines refer to the GALPROP maps [463], based on the calculations performed with the FRANKIE code [477], while the dashed curves refer to maps previously used in the literature [367]. We note that more recent maps have more features (such as the spikes in the optical and infrared light) and they are less smooth. There are also some normalisation differences, order of a few. In order to take into account these uncertainties, we vary the intensity of the ISRF in the Galaxy by a factor of two. However, Fig. 7.9 (bottom-left panel) shows that the DM constraints are only slightly affected by the normalisation of the ISRF density.

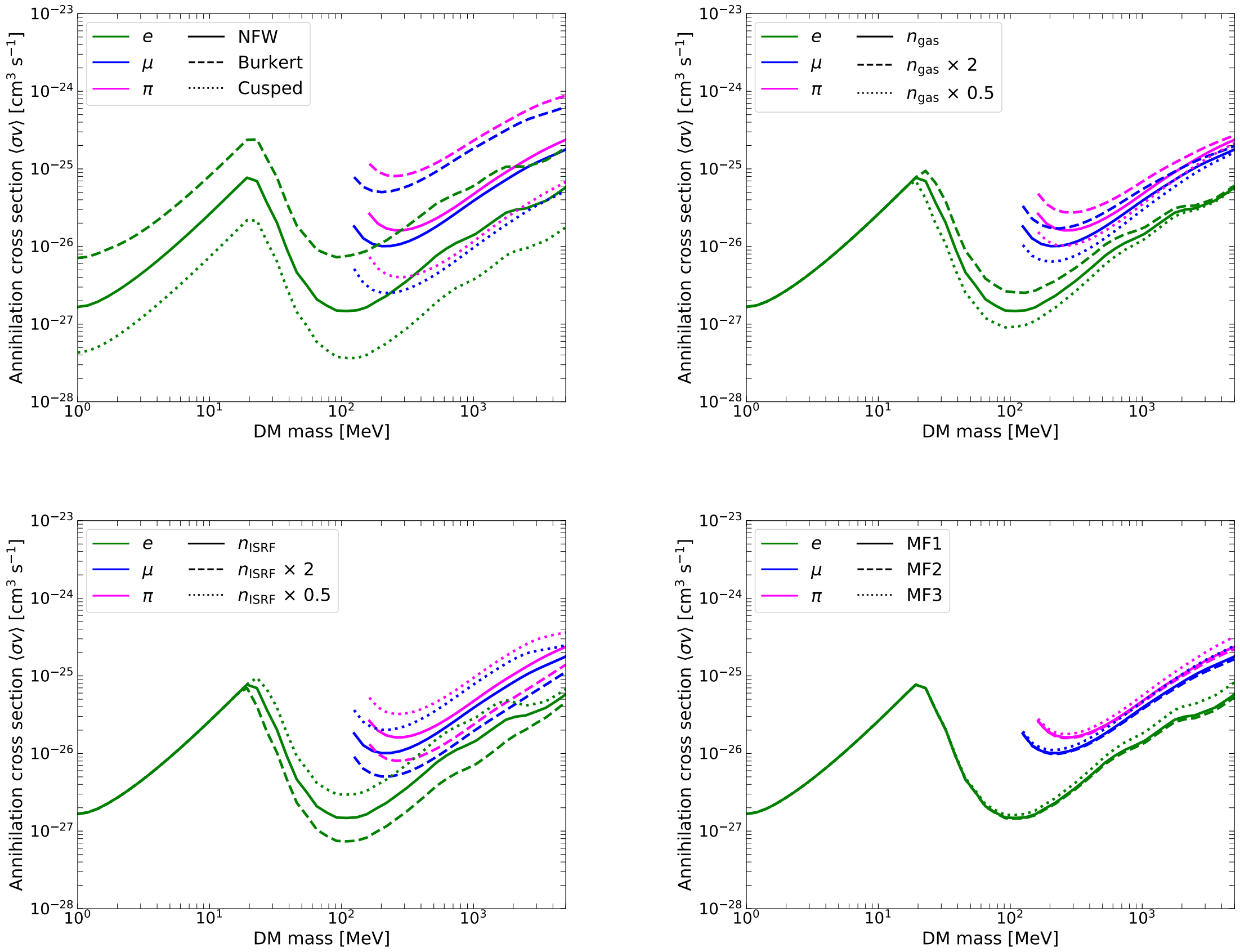

**Fig. 7.9.** Variation of the DM constraints due to the uncertainty on: DM density profile (top left), gas density (top right), interstellar radiation field density (bottom left), configuration of the galactic magnetic field (bottom right).

**Galactic magnetic field**. The energy losses by synchrotron radiation depend on the configuration of the galactic magnetic field $B$, which also carries significant uncertainties. Following Refs. [478, 480, 461], we model $B$ as a double power law in the radial and vertical distances from the GC:

$$B = B_0 \exp\left(-\frac{r - R_\odot}{R_D} - \frac{z}{z_D}\right) \tag{7.5}$$

where the set of parameters $(B_0, r_D, z_D)$ is representative of different models. Table 7.2 shows the parameters and the references in the literature for the three configurations taken into account in our analysis, denoted as MF1, MF2 and MF3. Fig. 7.8 displays the magnetic field on the GP as a function of the galactocentric distance (top-left panel) and as a function of the vertical coordinate below and above the Sun (bottom-left) for the different models. MF1 represents our reference model. MF2 has a steeper behaviour both in $r$ and $z$, and it is characterised by slightly higher values in the vicinity of the GC. Finally, MF3 is typically higher with respect to the other two configurations, especially at the location of the Sun (with the only exception of regions close to the GC). The energy losses as a function of the initial $e^{\pm}$ energy are displayed on the right panel of Fig. 7.8. The solid lines refer to a location on the GP, while the dashed curves refer to a position outside of the GP, namely 1 kpc above the Sun. We note that the energy losses are only mildly dependent on the configuration of the magnetic field and they are higher on the GP. Regarding the impact of the three configurations of the magnetic field on our constraints, Fig. 7.9 (bottom-right) illustrates the bounds that we obtain by adopting MF1 (solid lines), MF2 (dashed) and MF3 (dotted) models. The impact is essentially negligible. This is not surprising since changing the configuration of the magnetic field affects only the energy losses by synchrotron emission, which are subdominant in our regime of interest (see Fig. 6.8).

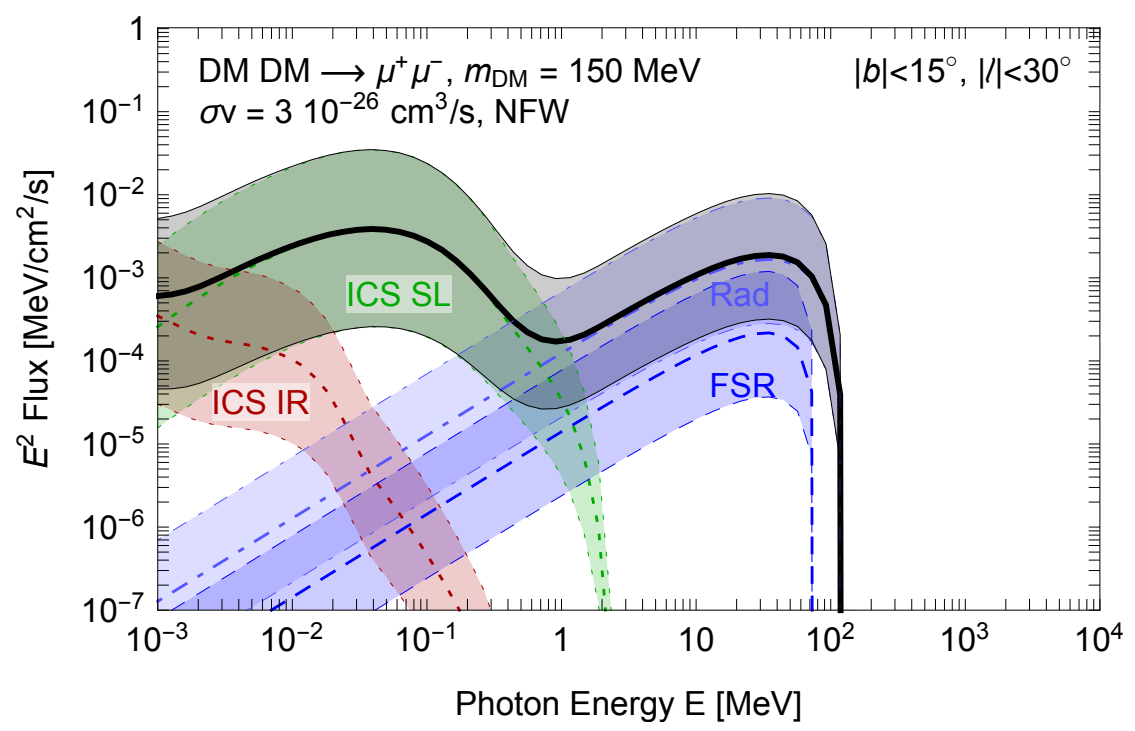

**Fig. 7.10.** *Left:* Impact of the astrophysical uncertainties on the same photon spectra of Fig. 6.11. The shaded bands correspond to the cumulative effect of the uncertainty on the DM profile, on the gas density, on the interstellar radiation field density and on the galactic magnetic field. Each coloured band corresponds to a different component of the total flux: blue shade for FSR and Rad, green shade for ICS on starlight, salmon shade for ICS on infrared light. *Right:* Variation of the bounds due to the astrophysical uncertainties.

The overall effect of these uncertainties is displayed in the left panel of Fig. 7.10, which illustrates the impact of the astrophysical uncertainties on the same photon spectra of the left panel of Fig. 6.11. The shaded bands correspond to the cumulative effect on the total signal due to the uncertainty on the DM profile, on the gas density, on the ISRF and on the galactic magnetic field. Each coloured band corresponds to a different component of the total signal: blue shade for FSR and Rad, green shade for ICS on starlight and salmon shade for ICS on infrared light[4]. Similarly, the right panel of Fig. 7.10 shows the overall impact of the astrophysical uncertainties on the DM constraints. The shaded regions denote the variation of the bounds for the three annihilation channels. The most stringent limits (dotted lines) are obtained by using a cusped profile, gas and ISRF density a factor of two larger with respect to the standard scenario (solid) and MF3 configuration for the magnetic field. The weakest bound (dashed) are determined with the Burkert profile, MF2 configuration, half of the gas and ISRF density with respect to the standard case. The two boundaries of the coloured bands represent these two extreme scenarios with all the sources of uncertainties conspiring in the same direction. As a result of the uncertainties, we can observe that the constraints span up to two orders of magnitude, with the DM profile having the greatest impact.

[4] As mentioned in Chapter 6.3.3, the ICS signal on CMB is subdominant, thus it does not appear in the figure.

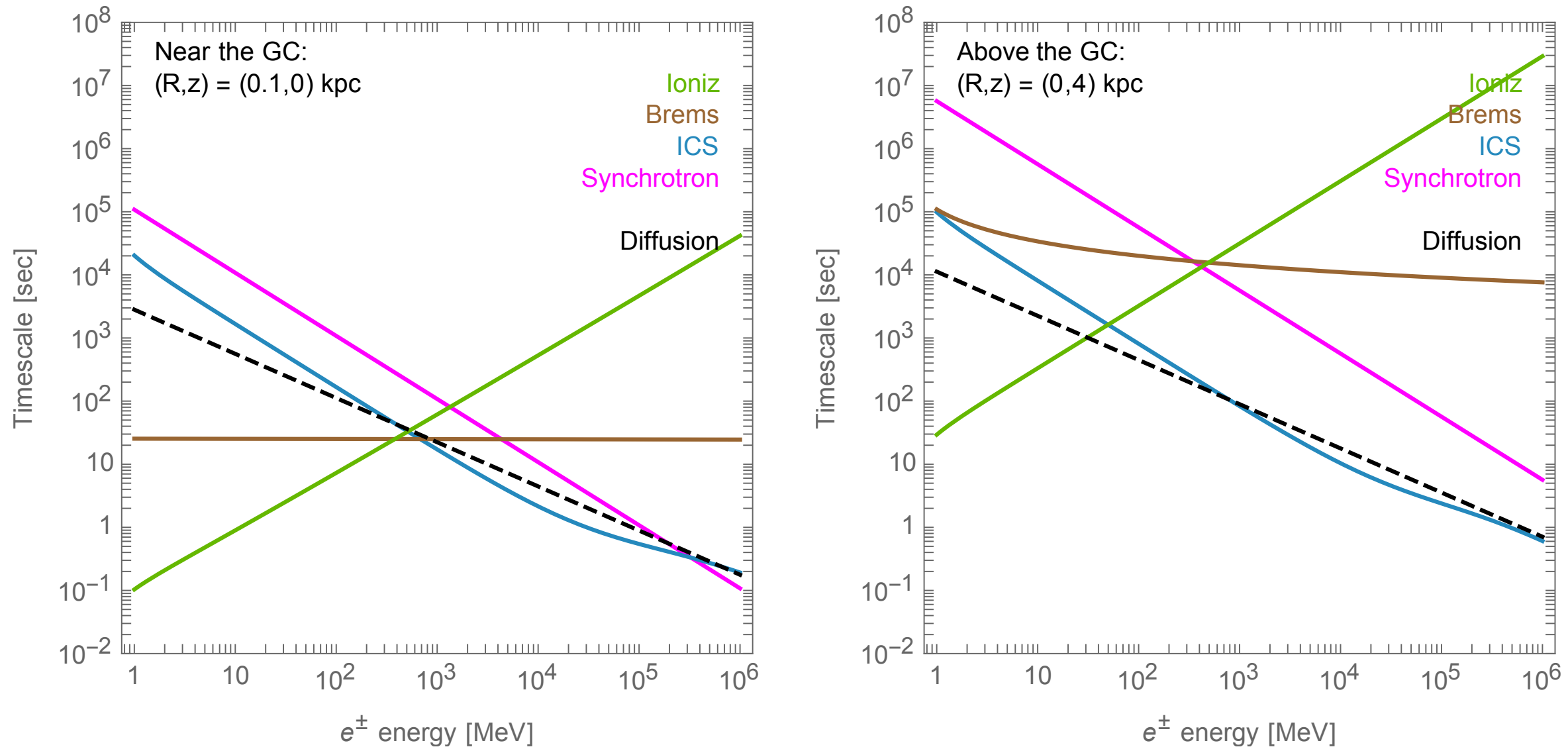

**Fig. 7.11.** Timescale of the diffusion (dashed) as a function of the $e^{\pm}$ energy compared to the energy loss processes (ionization in green, bremsstrahlung in brown, inverse Compton scattering in blue, synchrotron in magenta) in two locations: near the GC (left) and well above the GC (right).

In the following we will add some remarks concerning the on-the-spot approximation and s−wave/p−wave assumptions.

**On-the-spot approximation**. As introduced in Chapter 6, the on-the-spot approximation consists of neglecting the diffusion of electrons and positrons in the Galaxy. Physically speaking, it is equivalent to assume that the $e^{\pm}$ undergo ICS in the same location where they are produced by the DM particles. In the following we motivate this assumption. In particular, Fig. 7.11 displays the diffusion time (dashed black curve) in the Galaxy as a function of the $e^{\pm}$ energy, compared to the timescale of the energy loss processes. Near the GC (left panel), the diffusion timescale is evaluated for a length of 0.5 kpc, corresponding to a particle drifting of approximately 3 degrees, if it is located close to the GC. This value coincides with the size of the smallest angular bins that we consider. In practice, if a particle drifts less than 3 degrees, it does not "leak" into the neighbouring bin, hence diffusion is not significant. At high latitudes (right panel) the diffusion timescale is evaluated for a length of 1 kpc, since the angular bins of the INTEGRAL data are wider. Near the GC and for the sub-GeV energies of interest, the diffusion time is *longer* than the loss timescales by ionization (green line), bremsstrahlung (brown) and ICS (blue). In other words, the $e^{\pm}$ lose energy via these processes much faster than they diffuse, thus the on-the-spot assumption is a good approximation in this case. Instead, at high latitude the diffusion scale is comparable or (in a restricted range of energies) a bit shorter than the energy losses. Therefore, neglecting the diffusion leads to underestimate an effect which is the same order of magnitude of the energy losses and the on-the-spot approximation is not recommended. However, the signal is largely dominated by the regions in the vicinity of the GC and the high-latitude bins never impose significant constraints. Also note that these figures are in good agreement with other works in the literature (see e.g. Ref. [482]). In summary, while certainly intrinsically limited, we have justified the on-the-spot approximation, showing its validity in the regions and energy ranges of interest.

**s−wave/p−wave**. The dependence of our analysis on the DM velocity distribution deserves some comments. The velocity-averaged annihilation cross-section can be decomposed as

$$\langle \sigma v \rangle \simeq a + b\, v^2 \,, \tag{7.6}$$

where the parameter $a$ weights the s−wave contribution, while the parameter $b$ is associated to the p−wave term. The quantity that we constrain in our analysis is the "averaged annihilation cross-section times the velocity" *in the galactic halo at present time*, here denoted by $\langle \sigma v \rangle_{\rm now}$. If we assume a constant average velocity for the DM particles everywhere in the Galaxy, as it is typically done in this kind of analyses, our bounds are strictly independent on the s−wave or p−wave assumption for the DM annihilation cross-section. Instead, if we include a possible variation of the DM velocity in the Galaxy, the p−wave bounds will be affected. It is noteworthy that the position-dependence of the DM velocity is subject to debate. In order to estimate its impact, we can adopt the classical parameterization of Ref. [483]: $v(r) \propto r^{\beta} \rho(r)$, where $\rho$ is the usual DM density profile. The exponent is found to be $\beta \simeq 1.9$ in DM-only simulations [484] and $\beta \simeq 1.64$ in simulations including baryons [485]. By choosing the latter value and the NFW profile with the parameters of Table 7.1, we show in Fig. 7.12 (left panel) the behaviour of the DM velocity as a function of the position in the Galaxy. The right panel illustrates the quantity $\rho(r)^2 v^2$ along the line of sight at $\theta = 20°$ with respect to the direction towards the GC, which corresponds to the typical directions entering into our regions of interest. This quantity represents the relevant factor in the computation of the X-ray and $\gamma$-ray signal for the case of p−wave annihilation. The assumption of varying the DM velocity (red line) results in reducing $\rho(r)^2 v^2$ in the regions near the GC, being the DM particles slower as we approach the centre of the Milky Way, and this effect is only partly compensated by the central density peak. The overall impact in this case is quite limited: the integrated value (which affects the measured flux and, therefore, the bounds) is found to change by only 14%. For lines of sight at higher latitudes, the effect is even smaller, while closer to the GC, the value varies at most by 44%. Keep in mind that in our analysis we remove the lines of sight that approach the GC, corresponding to $\theta \leq 4°$. The variation of these values is related to one single line of sight, while the actual signal is mediated over the angular bins, resulting in a quite small overall impact. Instead, the CMB is sensitive to the "annihilation times velocity" at very high redshift, here denoted by $\langle\sigma v\rangle_{\rm past}$. It is useful to recall that the DM velocity at recombination depends on the decoupling temperature. As long as the DM particles are coupled to the plasma, their speed is $v \sim \sqrt{3T_\gamma/m_\chi}$, with $T_\gamma$ being the plasma temperature. When the decoupling occurs, the DM temperature starts decreasing and its value at redshift $z$ will be $T_\chi = T_{\rm D}\left(z/z_{\rm D}\right)^2$, where $T_{\rm D}$ and $z_{\rm D}$ represent the decoupling temperature and redshift, respectively. As a result, the DM speed will reduce to $v \sim \sqrt{3T_\chi/m_\chi}$. If the annihilation process is s−wave, then by definition $\langle\sigma v\rangle \propto a$ with $a$ being speed-independent. Therefore, $\langle\sigma v\rangle_{\rm now} = \langle\sigma v\rangle_{\rm past}$ and the CMB bounds apply directly on the same plane as the local ones. If the annihilation process is p−wave, then by definition $\langle\sigma v\rangle \propto bv^2$ and therefore the CMB bounds and the local constraints are linked by the conversion factor $(v_{\rm past}/v_{\rm now})^2$. The fact that $v_{\rm past} \ll v_{\rm now}$ explains why the CMB bounds in the p−wave case are orders of magnitude looser than the local ones, when put on the same plane. A precise comparison is difficult to draw since $v_{\rm past}$ is redshift-dependent, and therefore the conversion factor needs to be computed as an integral over $z$.

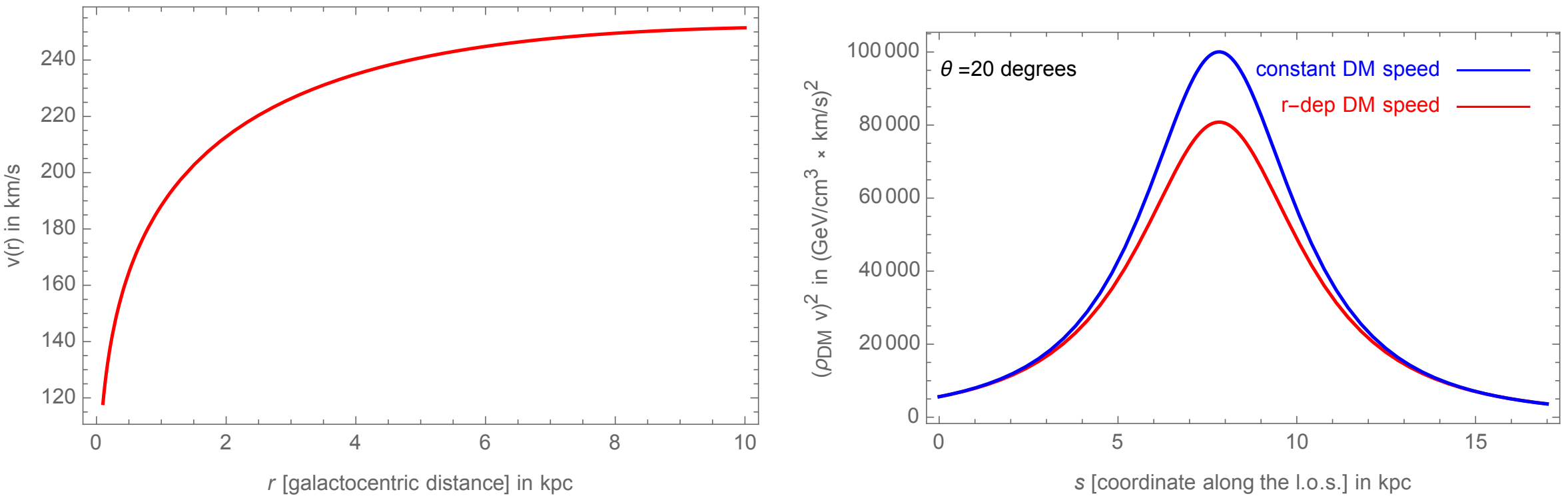

**Fig. 7.12.** *Left:* DM velocity as a function of the galactocentric distance. *Right:* The quantity $\rho_{\rm DM}(r)^2 v^2$, representing the relevant factor in the computation of the X-ray flux, as a function of the line of sight for $\theta = 20°$. The blue curve denotes the case with constant velocity in the Galaxy, while the red line indicates the effect of varying the DM velocity in case of p−wave annihilation.

## 7.4 Summary

We have derived the constraints on dark matter particles in the mass range between 1 MeV and 5 GeV, by comparing the INTEGRAL X-ray data with the prediction of the X-ray flux from annihilation events of DM particles. We have considered three annihilation channels: electrons, muons and pions. In some intervals of masses our constraints prove to be comparable with previous results derived using X−ray data and the $e^\pm$ measurements from VOYAGER 1. The CMB s−wave limits are stronger over the whole mass range, under the assumption that the DM annihilation cross-section at present-day has the same value as at the time of recombination. However, considering p−wave annihilations, the CMB bounds become much weaker along the entire mass range of interest, making our constraints the strongest to-date on the present-day cross-section for masses between 150 MeV and 1.5 GeV. The improvement of the DM bounds comes from the previously neglected flux from inverse Compton scattering. In particular, the electrons and positrons produced by annihilating dark matter can interact with the ambient photons in the Milky Way. After the scattering process, the energy of the photons is typically a few orders of magnitude lower than the mass of the dark matter particles. This allows us to probe sub-GeV dark matter using the INTEGRAL telescope, thereby deriving novel constraints and help plugging the "MeV" gap. In the near future, data from eASTROGAM (300 keV − 3 GeV) or AMEGO (200 keV − 10 GeV) will hopefully cover the "MeV gap", making it possible to directly search for sub-GeV annihilating dark matter. At the same time, upcoming full-sky data from the eROSITA X-ray telescope (0.3 − 10 keV) will be sensitive to the inverse Compton scattering emission from sub-GeV dark matter, allowing us to improve the reach of the technique we used in this analysis.

# Part III

# Dark filaments playing over the radio

# Chapter 8

## Radio signals from dark matter filaments

Galaxy clusters are thought to be connected by diffuse filaments, forming the cosmic web. The detection of these filaments could yield information on the magnetic field strength, cosmic-ray population and temperature of the inter-cluster gas. Nonetheless, the faint and large-scale nature of these bridges makes direct detection very challenging. The first robust detection of the stacked radio signal from large ($1-15$ Mpc) filaments that connect pairs of luminous red galaxies, by using multiple independent all-sky radio maps, was reported in Ref. [3]. The authors detect an average surface brightness between the clusters from synchrotron (radio) emission with a significance greater than $5\sigma$. This signal appears compatible with the non-thermal synchrotron emission from the cosmic web and provides direct evidence for one of the pillars of our understanding of structure formation in the Universe. However, at the time of writing, an astrophysical explanation does not fully account for the detected emission. A fascinating possibility that we consider is that the radio signal is produced by DM particles in the filament. Annihilating and decaying DM can produce pairs of electrons and positrons which can produce radio waves via synchrotron emission, if they move in a magnetic field. In this chapter, we briefly introduce the measurement of the detected excess, which motivates our work. Then, in Section 8.2 we present the theoretical prediction of the radio signal expected from DM particles, commenting on the various components that are necessary to compute the brightness temperature of the emission. Finally, in Section 8.3 we show that the detected excess is compatible with a DM origin, under certain circumstances. We also derive the constraints on the DM lifetime for different assumptions on the magnetic field in the filamentary structure.

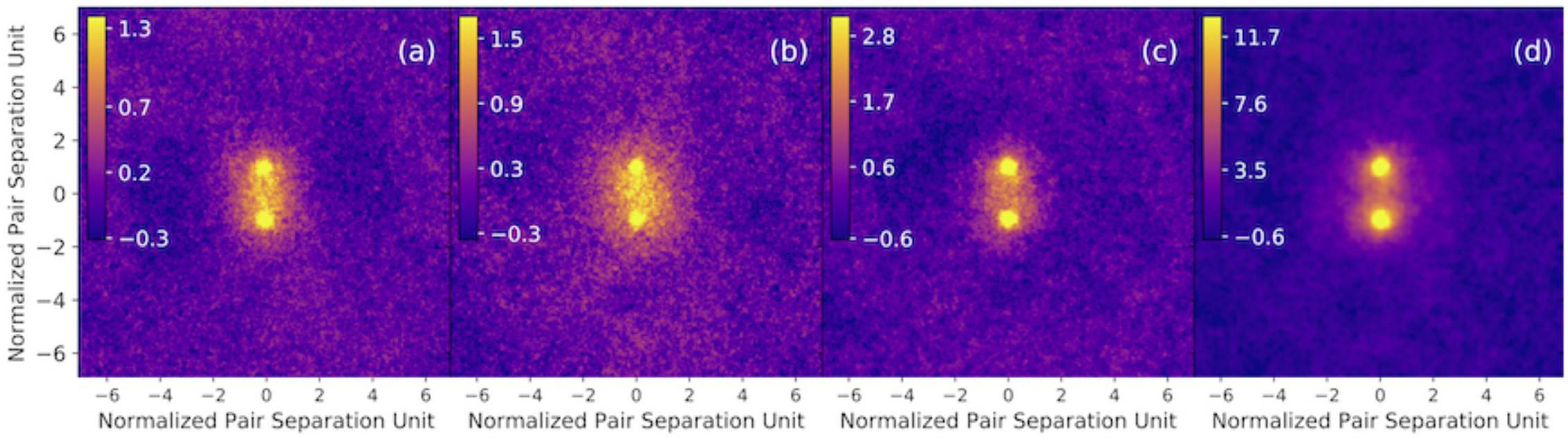

**Fig. 8.1.** Radio excess measured in nearby pairs of red luminous galaxies. From left to right the columns show the final stack images for: GLEAM 154 MHz, GLEAM 118 MHz, GLEAM 88 MHz, OVRO-LWA 73 MHz. All the colour bars have units of temperature in K.

## 8.1 Observation of the brightness temperature

One method for detecting faint diffuse filamentary emission is to use image stacking. The main advantage of stacking is that the faint emission coherently adds, while the noise and uncorrelated emission do not, leading to an increase of the signal above the noise. The drawback is that a large number $N$ of samples is required, since the noise decreases with the increase of the sample as $\sim 1/\sqrt{N}$. Such a requirement could be problematic when dealing with filaments because the location of filaments is not well known, also the number of known clusters over the whole sky is only of order of thousands. Thus a proxy for clusters with much greater numbers is necessary. Luminous red galaxies (LRGs) are acknowledged to be powerful tracers of the large-scale structure. They are massive early-type galaxies that usually reside in (or near) the centre of galaxy clusters [486, 487, 488]. Thus, pairs of LRGs that reside near each other on the sky and in physical space can act as a suitable proxy for physically nearby pairs of clusters, which may be connected by inter-cluster bridges or filaments. In this regard, for the first time Ref. [3] applies the stacking technique to multiple radio maps in order to look for synchrotron emission from filaments, and more specifically, to investigate whether a stacked filament detection can be made by stacking pairs of LRGs near each other in physical space[1].

The data from the GaLactic and Extragalactic All-sky Murchison Widefield Array (GLEAM) survey [490, 491] and from the Owens Valley Radio Observatory Long Wavelength Array (OVRO-LWA) [492] are utilised. From the GLEAM survey, three maps are considered: GLEAM Blue, GLEAM Green and GLEAM Red, whose frequencies are shown in Table 8.1 (second column). For the OVRO-LWA data they employ the 73 MHz map. The stacking results are illustrated in Fig. 8.1. The sub-plots display the final stack images in the different surveys, from left to right: GLEAM Blue (154 MHz), GLEAM Green (118 MHz), GLEAM Red (88 MHz) and OVRO-LWA (73 MHz). The axes denote the normalised pair separation between the LRGs and the colour bars provide the temperature in Kelvin. The measured surface brightness temperatures from each survey are listed in Table 8.1 (third column) and range from 1 K down to 0.1 K, depending on the frequency. Thus, we can observe that the physically related pairs of LRGs exhibit a positive average signal that increases with decreasing frequency.

[1]They employ the LRG catalogue from the Sloan Digital Sky Survey (SDSS) Data Release 7 [489].

| Survey | $\nu$ [MHz] | $\langle T_{\rm fil} \rangle$ [K] |
|---|---|---|
| GLEAM Blue | 154 | $0.10 \pm 0.04$ |
| GLEAM Green | 118 | $0.22 \pm 0.06$ |
| GLEAM Red | 88 | $0.44 \pm 0.09$ |
| OVRO-LWA | 73 | $1.1 \pm 0.2$ |

**Table 8.1.** Relevant information regarding the measurements of the radio signal from the different surveys considered. We show the frequency of observation, as well as the average brightness temperature of the filaments detected.

Ref. [3] considered numerous explanations to account for this interesting observation. In particular, possible instrumental or systematic effects as well as unsubtracted point sources have been excluded as a potential cause of the radio signal. The diffuse radio emission from galaxies or clusters seems also unlikely. A comparison with simulations of shocked intergalactic gas finds that the observed radio signal is 30−40 times higher than predicted. At the time of writing, an explanation for the detected excess from a purely astrophysical origin seems difficult. An enticing possibility is that the observed radio signal originates from DM particles in the filaments. This scenario is the focus of the next section.

## 8.2 Dark matter radio signal

An intriguing explanation that has been considered for the excess detected in the diffuse radio background is the emission of synchrotron radiation from DM annihilation and decay events (e.g. [137]). These models typically consider the bulk of the DM and the emission to be coming from halos (both in galaxies and clusters). Nevertheless, an appealing and less travelled road is to consider the contribution coming from DM filaments. In the following, we derive an estimate of the brightness temperature associated with a radio signal of DM origin. In our analysis, for definiteness, we take into consideration two annihilation channels

$$\chi\chi \to e^+e^- \tag{8.1}$$

$$\chi\chi \to b\bar{b}\,, \tag{8.2}$$

as well as two decay channels:

$$\chi \to e^+e^- \tag{8.3}$$

$$\chi \to b\bar{b}\,. \tag{8.4}$$

The electron channel is representative of a leptonic DM channel, while the $b$ quark channel is representative of a hadronic DM channel. For the conservation of energy, the annihilation channels are kinematically open whenever $m_\chi > m_i$, where the subscript denotes the channel: $i = e, b$. Instead, for the decay channels the conservation of energy requires $m_\chi > m_i/2$. Once the electrons and positrons are produced, their propagation can be described via the transport equation of Eq. (6.22), where we neglect the advection and convection terms[2]. Thus, Eq. (6.22) reduces to

$$-K(E)\,\nabla^2 n_e - \frac{\partial}{\partial E}(b_{\rm tot}\, n_e) = Q_e\,, \tag{8.5}$$

where the first term from the left represents the diffusion part, being $K(E)$ the diffusion coefficient and $n_e$ denoting the $e^\pm$ number density. The second term accounts for the energy losses and $Q_e$ is the DM source term. As already discussed in Chapter 6, the source term in the case of annihilating DM reads

$$Q_e(E) = \langle\sigma_a v\rangle \frac{\rho^2}{2m_\chi^2} \times \frac{\mathrm{d}N_e}{\mathrm{d}E}(E)\,, \tag{8.6}$$

where $\langle\sigma_a v\rangle$ is the annihilation cross-section, $\frac{\mathrm{d}N_e}{\mathrm{d}E}(E)$ is the electron energy spectrum and $\rho$ is the DM density in the filament. In case of decaying dark matter, it holds

$$Q_e(E) = \frac{\rho}{\tau_D m_\chi} \times \frac{\mathrm{d}N_e}{\mathrm{d}E}(E)\,, \tag{8.7}$$

where $\tau_D$ is the particle lifetime. The $e^\pm$ energy spectrum produced by DM particles via the $b\bar{b}$ channel is illustrated in Fig. 8.2 (left). The solid lines refer to the annihilation production, while the dashed curves denote the decay case. Different colours refer to different masses of the DM particles: $m_\chi$ = 10 GeV (green), 100 GeV (blue), 1 TeV (red). Higher values of the DM mass are associated with higher multiplicities. The electron channel, being a vertical line, is not shown. At the time of writing, filaments represent an intriguing, yet poorly known structure. Therefore, we estimate the size of the signal from DM filaments to a first order approximation, under some conservative assumptions. In particular, we assume a constant DM density and for the ICS signal over the radiation fields, we consider only CMB[3]. We model the filament as a cylinder with diameter $D = 2$ Mpc, length $L = 8$ Mpc and total mass $M = 4 \times 10^{13} M_\odot$ [493]. Under the assumption of uniform density in the filament, we have that

$$\rho = \frac{M}{(\pi R^2 L)} = 1.6 \times 10^{12} M_\odot\,\mathrm{Mpc}^{-3} \simeq 10\rho_c\,. \tag{8.8}$$

The filaments are assumed edge-on and perpendicular to the line of sight. The electron number density has been derived in three different cases: strong confinement, free escape and partial confinement. The first two cases represent two extreme situations in which the diffusion term in the transport equation can be neglected, while the third case is an intermediate scenario, where we provide an estimate of the effect associated to the diffusion of the electrons and positrons. Hereinafter we will discuss the three regimes separately.

[2] This assumption is motivated since we are considering filaments connecting clusters, and not the region inside a galaxy.

[3] Considering that the filaments are not very luminous, it is reasonable to neglect the contribution from other interstellar radiation fields. Yet this is a conservative assumption, since including additional photon components

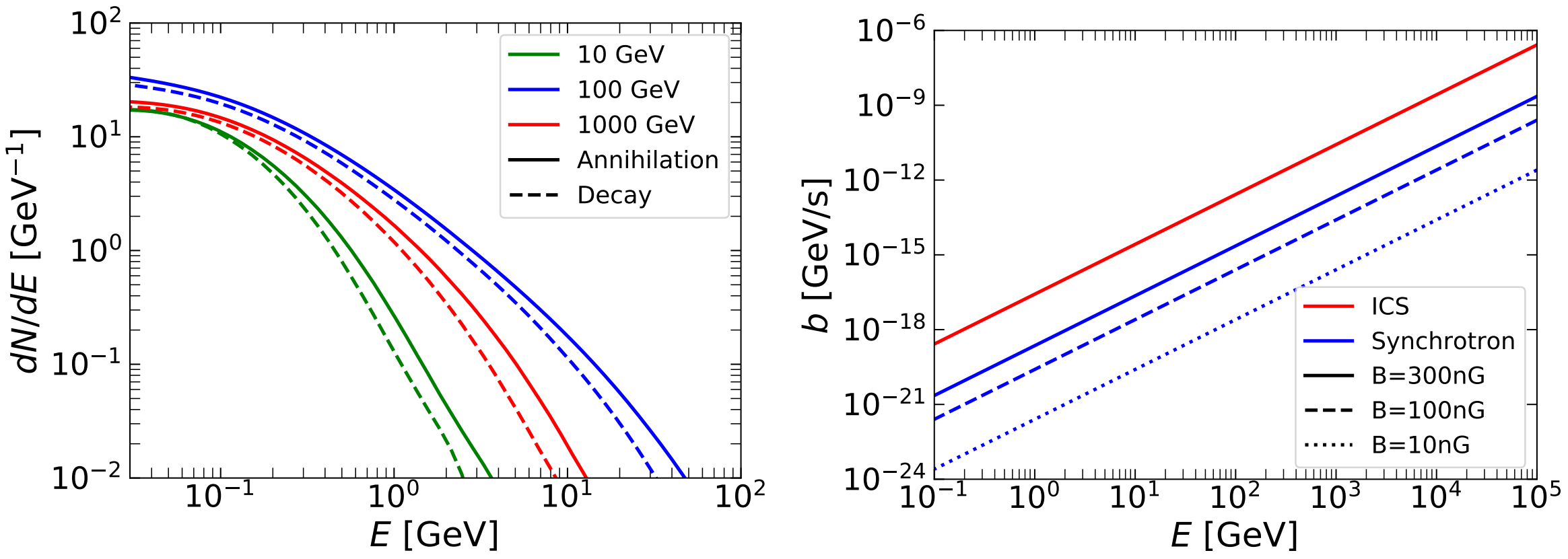

**Fig. 8.2.** *Left:* Energy spectrum for annihilating (solid lines) and decaying (dashed) DM. The production channel is $b\bar{b}$. Different colours denote different DM masses: 10 GeV (green), 100 GeV (blue), 1 TeV (red). *Right:* Energy losses as a function of the electron energy for inverse Compton scattering (red) and synchrotron emission (blue). The latter case is shown for three different values of the magnetic field: 300 nG (solid), 100 nG (dashed), 10 nG (dotted).

**Strong confinement.** This scenario applies to the case of strong turbulence with energy loss in-place. The equilibrium energy spectrum of the $e^{\pm}$ reads

$$n_e(E_e, \vec{x}) = \frac{1}{b_{\rm tot}(E_e, \vec{x})} \int_{E_e}^{m_\chi} {\rm d}\tilde{E}_e \, Q_e\left(\tilde{E}_e, \vec{x}\right) , \tag{8.9}$$

in analogy with the situation considered in Part II, notably in Eq. (6.29). The total energy loss coefficient $b_{\rm tot}$ considered in this case includes synchrotron emission and ICS on CMB:

$$b_{\rm tot} = b_{\rm sync} + b_{\rm ICS} \, . \tag{8.10}$$

Let us recall the expressions of the energy loss coefficients. For the synchrotron losses, it holds

$$b_{\rm sync} = \frac{4 \, c \, \sigma_T}{3 \, m_e^2} E^2 \, u_B \quad \text{with} \quad u_B = \frac{B^2}{2\mu_0} \, , \tag{8.11}$$

where $B$ is the magnetic field within the filament. The energy loss function for ICS on CMB in the Thomson regime reads

$$b_{\rm ICS} = \frac{4 \, c \, \sigma_T}{3 m_e^2} E^2 u_{\rm CMB} \quad \text{with} \quad u_{\rm CMB} = \int_0^\infty {\rm d}\epsilon \, \epsilon \, n_{\rm CMB} \simeq 0.260 \, {\rm eV} \quad {\rm cm}^{-3} \, , \tag{8.12}$$

where $n_{\rm CMB}$ is the number density of CMB photons, given by the black-body radiation spectrum of Eq. (6.49). The evolution of the energy losses with the electron energy is illustrated in Fig. 8.2 (right). The red line denotes the ICS energy loss, while the blue curves refer to the synchrotron losses for different values of the magnetic field: 300 nG (solid), 100 nG (dashed), 10 nG (dotted). Along the entire energy range, the synchrotron losses are subdominant.

would enhance the signal.

**Free escape.** This scenario is at the opposite extreme with respect to the previous one: if the turbulence and the magnetic confinement are negligible, the $e^{\pm}$ can freely escape. The electron number density can be approximated as

$$n_e(E) = \frac{1}{4\pi c}\int \mathrm{d}\phi\,\mathrm{d}\theta\,\sin\theta \int \mathrm{d}s\, Q_e(E) \approx \frac{D}{c}\frac{\Delta\Omega}{4\pi} Q_e(E)\,. \tag{8.13}$$

The average separation measured between two LRGs connected by a filament is found to be 82 arcmin, and as mentioned above, we adopt the diameter of the filament to be a quarter of the length. Thus, $\Delta\Omega \simeq 82\times 20\,\mathrm{arcmin}^2 \simeq 1.4\times 10^{-4}$ sr. The energy lost by the $e^{\pm}$ crossing the filament is estimated as follows. The time interval necessary to traverse the dimension $D$ at relativistic speeds is

$$\Delta t_{\mathrm{cross}} \approx \frac{D}{c} \approx 2\times 10^{14}\,s \tag{8.14}$$

and the energy loss due to ICS[4] is

$$b(E) \equiv \left|\frac{\mathrm{d}E}{\mathrm{d}t}\right| \simeq 2.7\times 10^{-17}\,\frac{\mathrm{GeV}}{\mathrm{s}}\left(\frac{E}{\mathrm{GeV}}\right)^2 . \tag{8.15}$$

Thus, the energy lost by the $e^{\pm}$ during the crossing is

$$\frac{\Delta E}{E} = \frac{b(E)\Delta t_{\mathrm{cross}}}{E} \approx 10^{-3}\left(\frac{E}{\mathrm{GeV}}\right) . \tag{8.16}$$

As a result, in the energy interval of interest (GeV range), the energy losses are negligible.

**Partial confinement.** The magnetic containment depends on the turbulent component of the magnetic field. In order to assess how effective confinement can actually be (and therefore if and how much the strong confinement regime approximation is proper), we can compare the confinement timescale with the cooling timescale and with the age of the Universe, which sets an upper limit on the same age of the filament. Regarding the confinement time, it scales as

$$\tau_{\mathrm{conf}}(E) \sim \frac{D^2}{2K(E)} . \tag{8.17}$$

For the diffusion coefficient $K(E)$, we consider a typical behaviour:

$$K(E) = K_0\left(\frac{E}{\mathrm{GeV}}\right)^{\delta} , \tag{8.18}$$

where we adopt a Kolmogorov spectrum $\delta = 1/3$. The normalisation $K_0$ is unknown, therefore we estimate it by relating its value in the filament to the one obtained in galaxies. Let's start by relating $K_0$ to the magnetic field $B$ and to its fluctuations $\delta B$ via [494]

$$K_0 \propto \left(\frac{B}{\delta B}\right)^2 B^{-\delta} . \tag{8.19}$$

[4]The synchrotron loss can be neglected, being subdominant here.

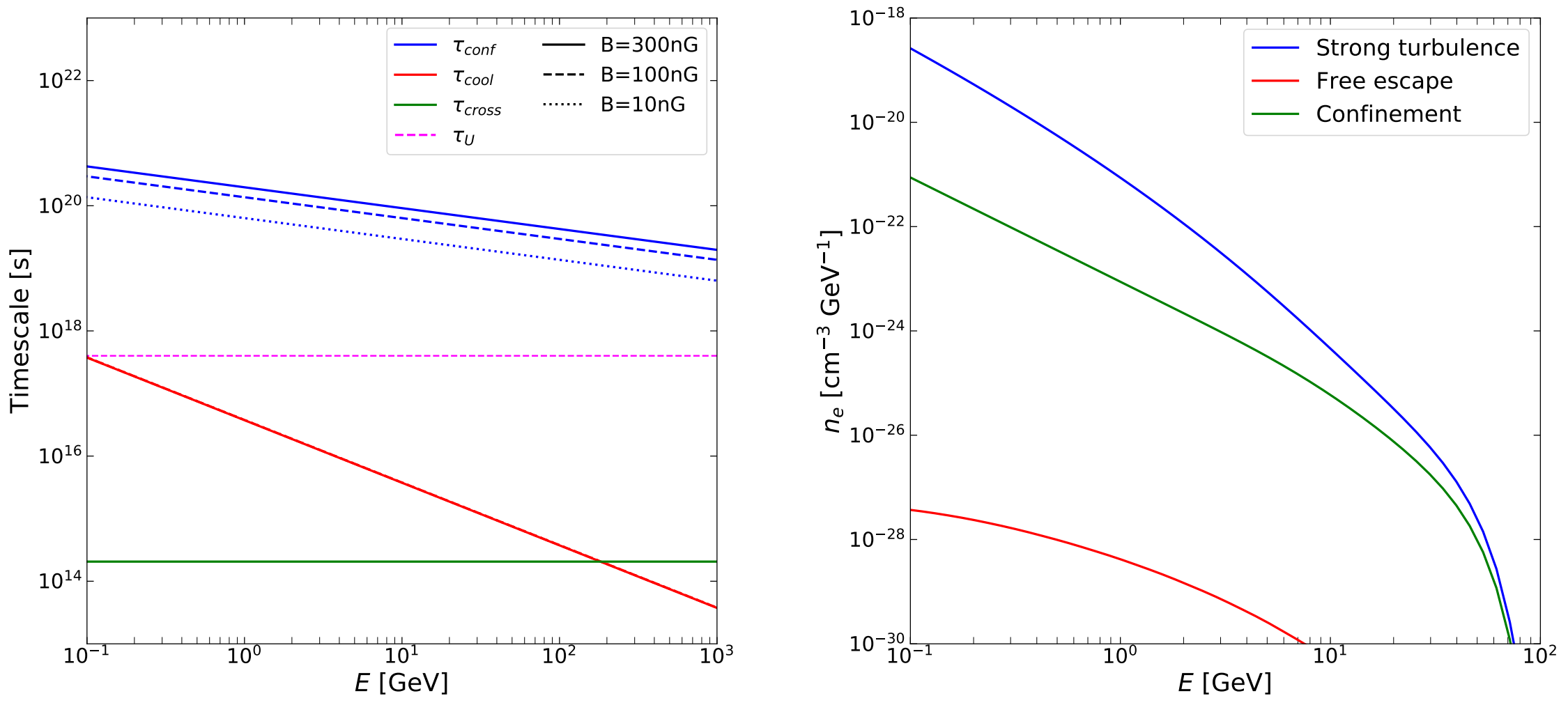

**Fig. 8.3.** *Left:* Evolution of the confinement time $\tau_{\rm conf}$ (blue), cooling time $\tau_{\rm cool}$ (red) and free-streaming time $\tau_{\rm cross}$ (green) with the electron energy. The dashed horizontal magenta line denotes the age of the Universe $\tau_U \simeq 4\times10^{17}$ s. Different choices of the magnetic field are displayed: 300 nG (solid), 100 nG (dashed), 10 nG (dotted). *Right:* Number density as a function of the electron energy for three regimes: strong turbulence (blue), free escape (red), confinement (green) with $k = 5\times10^4$. They all refer to DM particles with a mass $m_\chi = 100$ GeV, annihilating into $b\bar{b}$ with a cross-section $\langle\sigma v\rangle = 3\times10^{-26}$ cm$^3$/s. The magnetic field is assumed to be 100 nG.

We then rescale $K_0$ to the Milky Way value $K_0^{\rm MW} = 3\times10^{28}{\rm cm}^2{\rm s}^{-1}$ as follows

$$K_0^{\rm fil} \sim K_0^{\rm MW} \times k\left(\frac{B_{\rm MW}}{B}\right)^{\delta}, \tag{8.20}$$

where the coefficient $k$ reads

$$k = \left(\frac{B}{\delta B}\right)^2_{\rm fil}\left(\frac{B}{\delta B}\right)^{-2}_{\rm MW}. \tag{8.21}$$

This coefficient represents the proportionality constant between the filament and the Milky Way magnetic environments:

$$\left(\frac{B}{\delta B}\right)^2_{\rm fil} = k\left(\frac{B}{\delta B}\right)^2_{\rm MW}. \tag{8.22}$$

To estimate the depletion of the source brightness due to the imperfect containment of the electrons before they can radiate at the frequency of interest, we adopt an effective approach as in Ref. [138]. The source spectrum reduces to

$$\frac{{\rm d}N_e}{{\rm d}E} \longrightarrow \frac{{\rm d}N_e}{{\rm d}E}\exp\left(\frac{-\tau_{\rm cool}}{\tau_{\rm conf}}\right), \tag{8.23}$$

where the cooling time $\tau_{\rm cool}$ due to the ICS and synchrotron losses reads

$$\tau_{\rm cool} = \frac{E}{b_{\rm tot}(E)}. \tag{8.24}$$

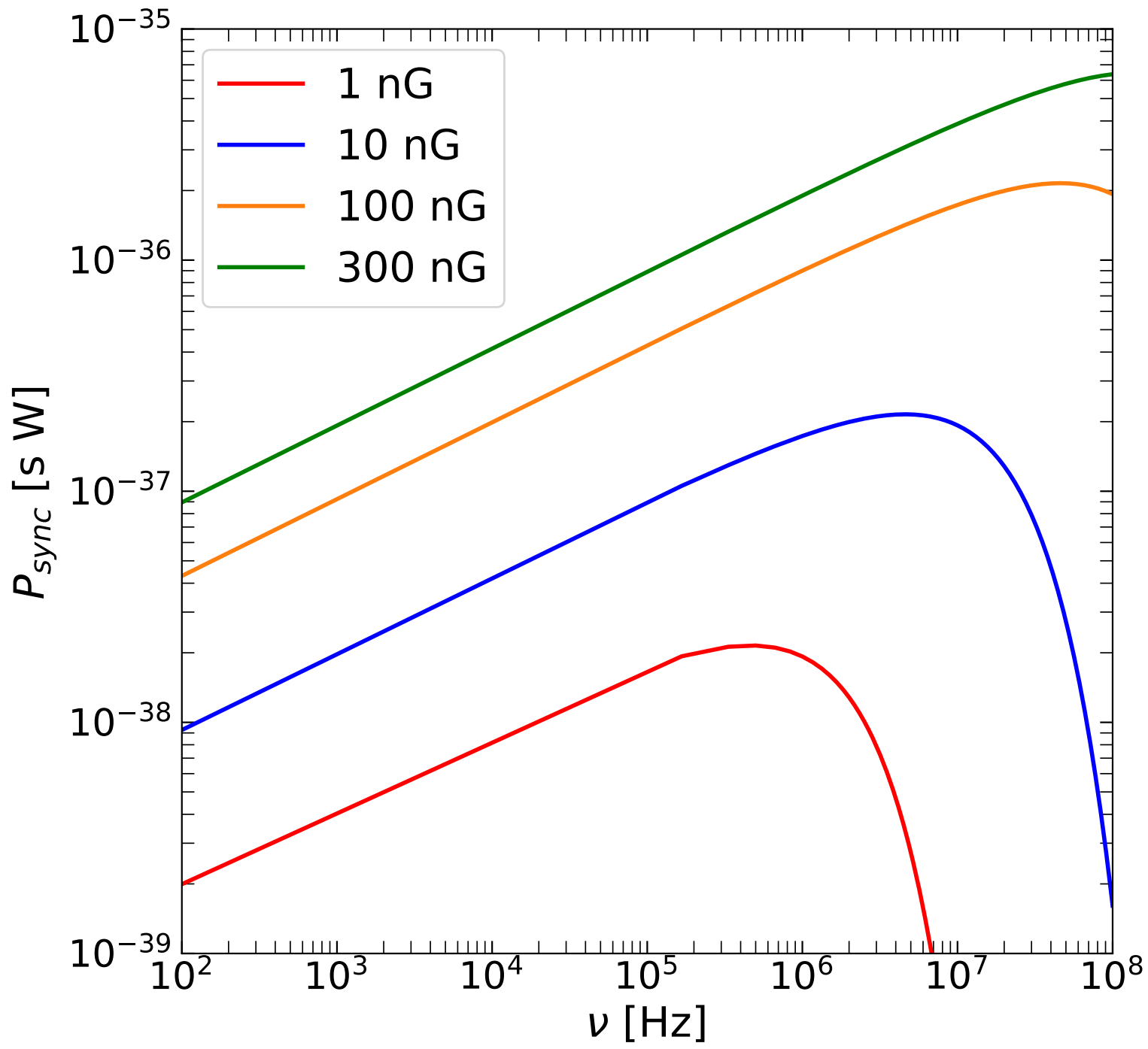

**Fig. 8.4.** Evolution of the synchrotron emitted power with the frequency, for an isotropic distribution of $e^{\pm}$ moving in a constant magnetic field. The values of the magnetic field shown are: 1 nG (red), 10 nG (blue), 100 nG (orange), 300 nG (green).

The rationale is that if the time interval spent by the $e^{\pm}$ with energy $E$ inside the filaments as a consequence of the diffusion is much greater than the time span required to loose energy, then the source term is basically unaffected. Instead, when the confinement time becomes shorter than the cooling time, the electron is removed from the energy $E$ and therefore we penalise the source term at that energy. Note that spatial diffusion does not affect energy. In addition, in our case we are assuming a constant source density, which would cancel the diffusion term in the transport equation even if K is small. Also note that $\tau_{\rm conf}$ could become smaller than the free-streaming traversal time $\tau_{\rm cross}$: in this case, we are in the regime of free escape.

Fig. 8.3 illustrates the three timescales $\tau_{\rm conf}$ (blue), $\tau_{\rm cool}$ (red) and $\tau_{\rm cross}$ (green) as a function of the electron energy for different choices of the magnetic field: 300 nG (solid), 100 nG (dashed), 10 nG (dotted). The age of the Universe $\tau_U \simeq 4 \times 10^{17}$ s is also shown as a reference timescale (dashed magenta curve). Note that the confinement time is much larger than the other timescales (including the age of the Universe). This suggests a good level of confinement in the filament, leaning in favour of the strong confinement regime, which represents our reference scenario. The three curves of cooling time corresponding to different magnetic fields are superimposed as expected, since the energy losses are ruled by the ICS process and the synchrotron term is highly subdominant. A mild dependence on $B$ appears for the confinement time. This figure assumes

$k = 1$, corresponding to

$$\left(\frac{B}{\delta B}\right)^2_{\text{fil}} = \left(\frac{B}{\delta B}\right)^2_{\text{MW}} . \tag{8.25}$$

The confinement time is typically larger than the cooling time in the energy range of interest for $k \lesssim 10^4$. The evolution of the number density with the electron energy is illustrated in Fig. 8.3 (left). The curves refer to DM particles with a mass of 100 GeV, thermal cross-section $\langle \sigma v \rangle = 3 \times 10^{-26}$ cm$^3$/s, annihilating into $b\bar{b}$ quarks. We assume a constant magnetic field $B = 100$ nG. The three regimes under consideration are displayed with different colours: strong turbulence (blue), free escape (red), confinement (green). For the latter, two cases can be discriminated depending on the value of the coefficient $k$. For $k \lesssim 10^4$, the green curve will essentially superimpose with the strong turbulence line (this case is not shown). Instead, if $k \gtrsim 10^4$, the confinement regime deviates from the strong turbulence scenario. The green line in the figure assumes $k = 5 \times 10^4$.

The second element that we need in order to compute the radio signal is the synchrotron emission power. In this regard, the power emitted at frequency $\nu$ by an isotropic distribution of $e^\pm$ in a constant magnetic field $B$ reads

$$P_{\text{sync}}(E,\nu) = \frac{\sqrt{3}\, e^3 c}{4\pi\epsilon_0 m_e c^2} B\, F(\nu/\nu_c) \, , \tag{8.26}$$

where the critical frequency $\nu_c$ is

$$\nu_c \equiv \frac{3c^2 e}{4\pi} \frac{BE^2}{\left(m_e c^2\right)^3} . \tag{8.27}$$

The synchrotron kernel $F(t)$ depends on the modified Bessel function of the second kind $K_n$ of order $n$, via

$$F(t) \equiv t \int_t^\infty \mathrm{d}z\, K_{5/3}(z) \, . \tag{8.28}$$

Fig. 8.4 illustrates the synchrotron power as a function of the frequency for different choices of the magnetic field: 1 nG (red), 10 nG (blue), 100 nG (orange), 300 nG (green). The spectrum has a broad maximum, as it is apparent for the 1 nG and 10 nG cases in the figure. The maximum of the emitted radiation corresponds to $\nu_{\text{max}} = 0.29\, \nu_c$. Considering that $\nu_c \propto B$, higher magnetic fields move the peak of the spectrum to a higher frequency. For instance, in the 300 nG curve the peak is not visible since it occurs at a frequency higher than those shown in the figure.

In analogy with the ICS emissivity discussed in Chapter 6.3.3, the synchrotron emissivity is the convolution of the synchrotron power with the electron number density:

$$j_{\text{sync}}(\nu) = 2 \int \mathrm{d}E\, P_{\text{sync}}(E,\nu)\, n_e(E) \, , \tag{8.29}$$

where the factor two takes into account that annihilation and decay events of DM particles produce an equal population of electrons and positrons. The intensity at frequency $\nu$ can be obtained by:

$$I(\nu) = \frac{1}{\Delta\Omega} \int \frac{\mathrm{d}\Omega\, \mathrm{d}s}{4\pi} j_{\text{sync}}(\nu) \tag{8.30}$$

| $\nu$ (MHz) | $\langle T_{\rm fil}\rangle$ (mK) | Model A (dec) (mK) | Model B (dec) (mK) | Model C (dec) ($\mu$K) | Model D (ann) ($\mu$K) | Model E (ann) ($\mu$K) |
|---|---|---|---|---|---|---|
| 154 | $100 \pm 40$ | 23 | 80 | 0.1 | 0.07 | 0.01 |
| 118 | $220 \pm 60$ | 110 | 226 | 0.2 | 0.13 | 0.01 |
| 88 | $440 \pm 90$ | 498 | 648 | 0.5 | 0.27 | 0.04 |
| 73 | $1100 \pm 200$ | 1175 | 1219 | 1.0 | 0.44 | 0.07 |

**Table 8.2.** Filament temperature for decaying and annihilating dark matter. The dark matter mass and lifetime refer to: Model A: decay into $e^+e^-$, $m_\chi = 5$ GeV, $\tau_D = 10^{26}$ s; Model B: decay into $e^+e^-$, $m_\chi = 8$ GeV, $\tau_D = 2\times10^{26}$ s; Model C: decay into $b\bar{b}$, $m_\chi = 1$ TeV, $\tau_D = 6\times10^{27}$ s; Model D: annihilation into $e^+e^-$, $m_\chi = 100$ GeV, $\langle\sigma v\rangle = 3\times10^{-26}$ cm$^3$ s$^{-1}$. Model E: annihilation into $b\bar{b}$, $m_\chi = 100$ GeV, $\langle\sigma v\rangle = 3\times10^{-26}$ cm$^3$ s$^{-1}$. In all cases, $B = 100$ nG. The first two columns recall the observed frequency and the average temperature of the filament, respectively.

and by applying our assumption on the geometry of the filament (edge-on), it reduces to

$$I(\nu) = \frac{D}{4\pi}\, j_{\rm sync}(\nu)\,. \tag{8.31}$$

In order to compare our predictions with the observations, we quote our results in terms of the brightness temperature, which is defined as

$$T_b = \frac{I(\nu)\, c^2}{2\, k_B\, \nu^2}\,, \tag{8.32}$$

where $k_B$ is the Boltzmann constant. Table 8.2 shows our estimate of the brightness temperature at the frequencies of the four surveys (first column), for a selection of the DM parameters and for a magnetic field $B = 100$ nG. The second column indicates the observed average brightness temperature associated with the detected filaments. Models A-C refer to decaying DM, and the lifetimes are set at the current bounds[5]. Model D and E refer to annihilating DM with mass $m_\chi = 100$ GeV and with a canonical thermal cross-section, which is close to its current bound [496, 471, 497, 498]. Since the filament is a significantly less dense structure as compared to a typical galaxy, the annihilating DM signal (which is proportional to $\rho^2$) turns out to be suppressed as compared to the decaying DM case (proportional to $\rho$) when we sit at the constraints for $\langle\sigma v\rangle$ and $\tau_D$ obtained in galaxies. For the annihilation case, we obtain brightness temperatures below the $\mu$K level. In our estimate, we assumed a homogeneous density: a density profile steeper toward the central axis of the filament or the presence of clumps inside the filament could boost the annihilation signal by a factor that depends on the specific details of the mass distribution inside the filament, which is currently mysterious. In particular, a boost factor of 100 could be feasible, making the annihilating flux at the level of few to tens of $\mu$K.

[5]The lifetime of Model A and B is within the uncertainties of Ref. [495] [207, 495]. Their bounds are displayed with the letter "E" in Fig. 8.6 and refer to the most stringent case of a NFW profile, while they are expected to reduce by approximately a factor 4 in the case of a Burkert profile.

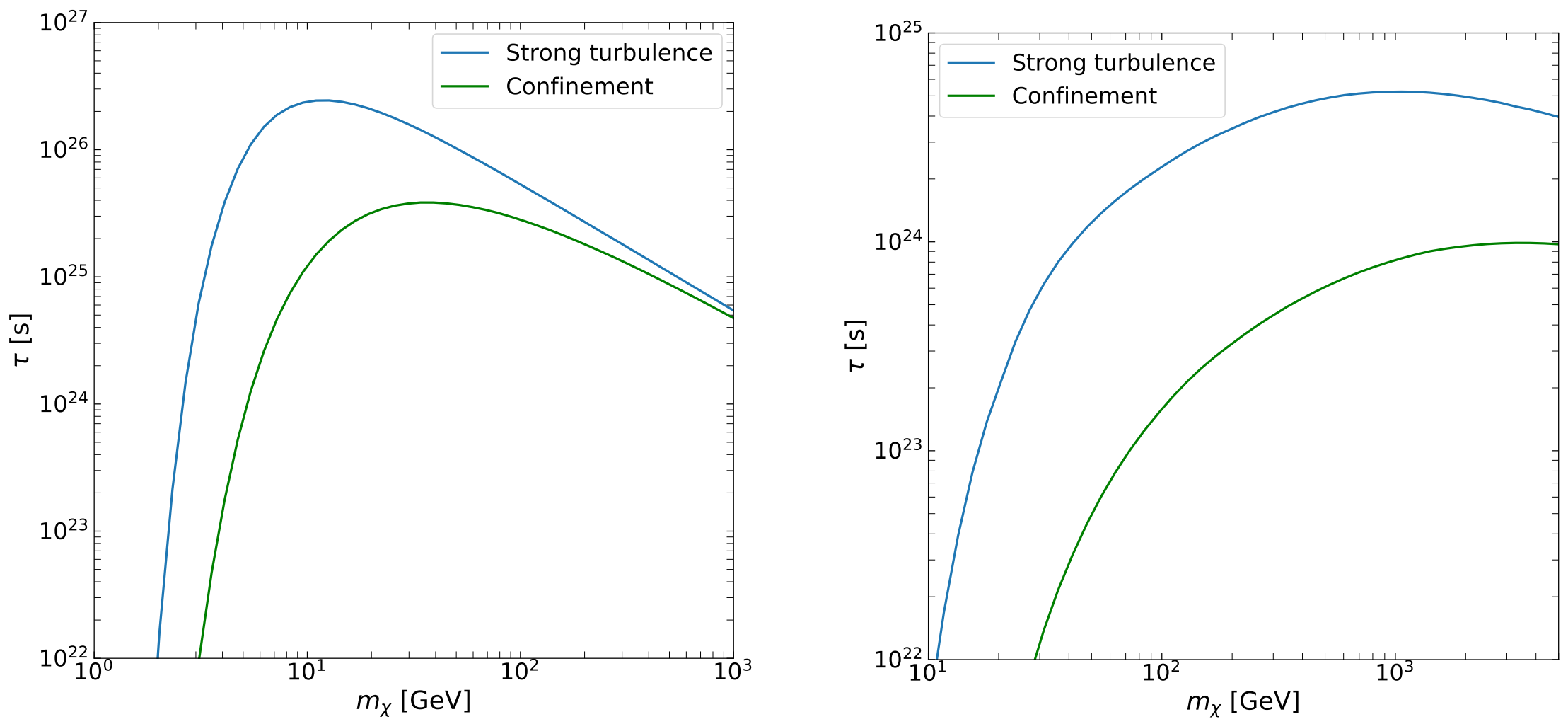

**Fig. 8.5.** Constraints on DM decay lifetime as a function of the DM mass, for the electron (left) and bottom quark channels (right). The blue line denotes the strong turbulence scenario while the green line refers to the partial confinement regime with $k = 5 \times 10^4$. The magnetic field is assumed to be $B = 100$ nG.

On the contrary, decaying DM can provide much larger brightness temperatures. For hadronic annihilation, the strongest bounds on the lifetime are around $\tau_D > 10^{28}$ s [207], while for the $e^+e^-$ channel we have $\tau_D > 2 \times 10^{25}$ s [468, 499, 500, 495] [6] . Table 8.2 shows that in the case of decaying DM, brightness temperatures from tens to more than a thousand of mK can be obtained for the leptonic channel. Notably, if the DM mass is below 10 GeV, it turns out to be possible to approach the observed emission level.

Also note that the results presented above refer to a filament mass $M = 4 \times 10^{13} M_\odot$, a magnetic field $B = 100$ nG and optimal magnetic containment of the electrons in the filament. The decaying and annihilating signals scale as $MB^n$ and $M^2B^n$, respectively, with $n < 2$, depending on the actual electron spectrum. An increase of the filament mass of an order of magnitude would directly reflect into a higher brightness temperature, while a reduction of $B$ to 10 nG would reduce the predicted temperature by a factor of at least 50.

## 8.3 Constraints on the dark matter lifetime

We can derive the constraints on the DM lifetime from the observed filament temperature. We adopt a similar approach to the one discussed in Section 7.2 for the conservative constraints from the X-ray flux. The DM bounds can be obtained by requiring that the predicted brightness temperature $\mathcal{T}_{\rm DM}$ does not exceed the measured temperature $\langle T_{\rm fil}\rangle$ by more than an appropriate value. We use the test statistic

$$\chi^2 = \sum_{i\in\{\nu\}} \frac{\left(\max\left[(\mathcal{T}_{{\rm DM},i}(\tau) - \langle T_{\rm fil}\rangle_i), 0\right]\right)^2}{\sigma_i^2}\,, \tag{8.33}$$

where the sum runs over the measured frequencies $\{\nu\} = \{73, 88, 118, 154\}$ MHz, and $\sigma_i$ denotes the uncertainty on $\langle T_{\rm fil}\rangle_i$. This test statistic is equivalent to the $\chi^2$ discussed in Chapter 7. We also follow the same procedure to derive the constraints: we perform a raster scan for each DM mass and require $\chi^2 = 4$, which corresponds to the $2\sigma$ bounds. Fig. 8.5 illustrates the constraints on the lifetime $\tau$ that we obtain for the $e^+e^-$ (left) and $b\bar{b}$ (right) channels for a magnetic field $B = 100$ nG. The strong confinement (blue) and partial confinement (green) regimes are shown, while the free escape scenario is omitted, since it does not produce a detectable signal. The partial confinement case shown in the figure corresponds to $k = 5\times10^4$, while for $k = 1$ the green curve will superimpose on the blue line. Both for the $e^+e^-$ and the $b\bar{b}$ channels, the strong confinement provides the most stringent constraints throughout the entire mass interval under consideration. Figure 8.6 illustrates the constraints on the lifetime for DM particles decaying into $e^+e^-$, for the strong turbulence scenario and different choices of the magnetic field: 10 nG (blue), 100 nG (orange), 300 nG (green). Clearly higher values of $B$ provide stronger constraints. For comparison, the current bounds in the literature are also shown (dashed and dash-dotted). The curve $A$ refers to the limits from the CMB measurements [500]. As already discussed in Chapter 7, the energy injection due to annihilating or decaying DM at high redshift would leave an imprint in the thermal history of the Universe and would alter the CMB anisotropy spectrum. The line $B$ denotes the constraints obtained in Ref. [468], using multiple data sets of the X-ray and $\gamma$-ray flux. The curve labelled $C$ represents the bounds derived in Ref. [499] by comparing the predict $\gamma$-ray signal of DM origin with the photon count distribution measured by FERMI-LAT. Also Ref. [504] employed the FERMI-LAT data. The measurements of the isotropic $\gamma$-ray background allowed them to derive pretty strong constraints for DM masses above 10 GeV (curve D). These four bounds (A−D) assume a NFW density profile. Ref. [495] (dotted, curve E) investigated Galactic DM by employing radio and microwave maps from PLANCK, assuming an Einasto profile and the configuration MF1 (discussed in Chapter 7) for the Galactic magnetic field. A precise comparison between the above-mentioned constraints is difficult to draw since they are sensitive to the assumption on the astrophysical and DM modelling (namely DM density profile and magnetic field). We note that our constraints are competitive for DM masses in the range 3−10 GeV and for magnetic fields in excess of about 130 nG.

[6] A confirmation of the EDGES observations [501] could improve the bound to $\tau_D > 3 \times 10^{26}$ s [502] for masses below 10 GeV and $\tau_D > 2 \times 10^{27}$ s for masses above [503, 504].

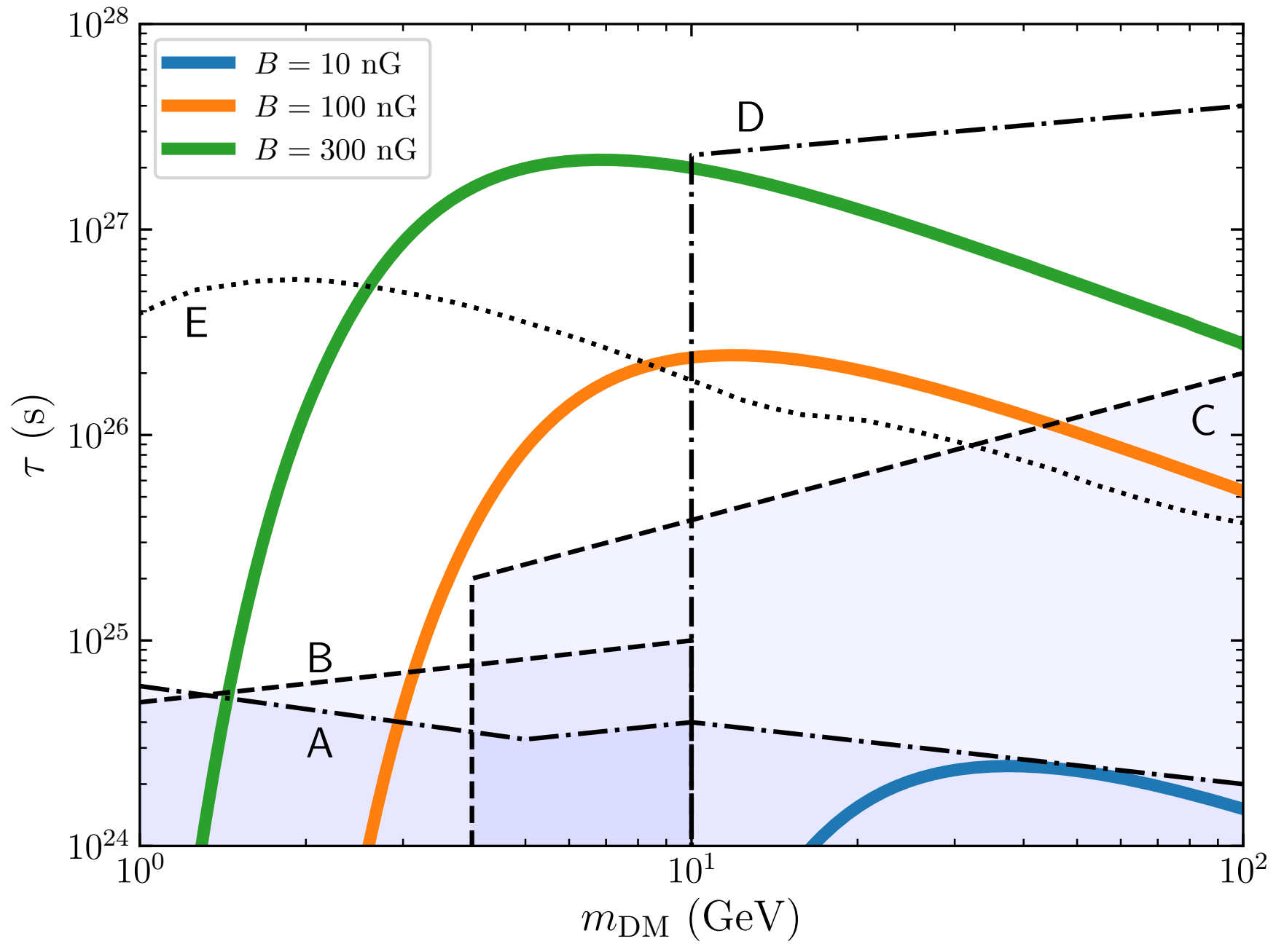

**Fig. 8.6.** Lower bounds at the 95% confidence level on the dark matter lifetime $\tau$ as a function of the dark matter mass $m_\chi$, for decay into $e^+e^-$ and for some representative values of the magnetic field $B$. Dashed, dotted and dash-dotted lines show the current bounds from [500] (curve A), [468] (B), [499] (C), [504] (D) and [495] (E).

## 8.4 Summary

A recent observation of cosmic filaments in the radio band from using a stacking technique, opens a promising avenue to get a better understanding of the physics behind these elusive bridges. The signal expressed in terms of brightness temperature covers a range between 1 K down to 0.1 K and is difficult to explain with traditional astrophysical arguments. We explore the enticing opportunity that this emission originates from DM particles in the filaments. We considered both annihilation and decay events and we focused on the electron and $b$ quark channels, as representative of leptonic and hadronic scenarios, respectively. Our estimates suggest that annihilating DM would produce filament brightness temperatures at most at the $\mu$K level. Instead, decaying DM temperatures span approximately between 10 and 1000 mK, depending on the frequency, without conflicting with current bounds on the particle lifetime. The largest temperatures, close to those observed, are obtained for a DM particle with a mass around 5−10 GeV decaying into $e^+e^-$ pairs. Hadronic decays are instead disfavoured. We also derive the $2\sigma$ constraints on the DM decay lifetime by comparing our predictions of the brightness temperature with the observed values. It turns out that our bounds are competitive for DM masses in the range 3−10 GeV and for magnetic fields above 130 nG.

# Part IV

# Conclusions and prospects

# Conclusions and prospects

More than 40 years after the revolutionary work of the pioneer Vera Rubin on the rotation curves of spiral galaxies, the fundamental essence of dark matter still remains a riddle. The most favoured explanation is that DM consists of a new elementary particle (or particles), which could annihilate or decay and indirectly produce an enormous variety of astrophysical messengers, including photons. In this thesis we examined the possibilities offered by multi-wavelength indirect detection searches. We investigated three different signals across the electromagnetic spectrum ($\gamma$ rays, X rays, radio waves) produced on a range of different scales (galactic, extragalactic, inter-cluster filaments).

In Part I we employed the cross-correlation technique, to disentangle a subdominant DM signal from the overwhelming astrophysical background. We performed the first theoretical prediction of the cross-correlation signal between the unresolved $\gamma$-ray background and the 21cm emission line produced by the spin-flip transition of neutral hydrogen atoms. Our benchmark experiment for detecting $\gamma$ rays is FERMI-LAT, and the measurements of HI will be performed by MEERKAT and the Square Kilometre Array (SKA). The constraints on DM properties that are attainable with the combination of SKA with the current FERMI-LAT statistics are found to be comparable to those accessible with other techniques utilising the unresolved components of the $\gamma$-ray background. As shown in Fig. 5.9, this combination would allow us to test a DM particle with thermal annihilation cross-section for masses up to 130 GeV. The enhanced capabilities of SKA Phase 2, combined with a future generation of $\gamma$-ray telescopes, should allow us to investigate the whole mass window for weakly interacting massive particles up to the TeV scale and potentially detect a DM signal. In future work we can include the cross-correlation signal between the HI in our Galaxy and the Galactic $\gamma$-ray flux as well as the cross-correlation between DM particles and astrophysical sources. These two additional components could, to some extent, enhance the total signal and lead to stronger constraints. Moreover, this cross-correlation channel can be extended to analyse different electromagnetic signals and astrophysical messengers, as well as different gravitational tracers of the matter distribution in the Universe. In fact, since different wavelengths are associated to different physical mechanisms of production and different DM particle masses, the advantage of the cross-correlation technique is that it allows the investigation of different DM candidates. Particularly, indirect signatures of axion-like particles (ALP) can be concealed within radio emissions that we observe on Earth. Therefore, the cross-correlation between the radio flux and a gravitational tracer can be employed to constrain the ALP lifetime and mass. This kind of analysis would be an innovative and promising way to study these DM candidates, especially in view of the new and upcoming observational data (e.g. the Dark Energy Survey for weak-lensing shear) as well as the next-generation detectors (e.g. the Square Kilometre Array for HI intensity mapping, Euclid for cosmic shear).

In Part II we focused on the X-ray flux produced by annihilating DM within our Galaxy. The main innovation brought to the DM searches in this energy range is that they can produce pairs of electrons and positrons, which can up-scatter the low-energy radiation fields in the Milky Way halo via inverse Compton scattering. By comparing our theoretical model of the total flux to the INTEGRAL data, we derived competitive bounds for particles with a mass range between 150 MeV and 1.5 GeV, as shown in Fig. 7.3. In future work, we can improve our constraints at lower DM masses of just a few MeV, by also taking into account other experiments at lower X-ray energy, where the signal from inverse Compton scattering peaks. Additionally we can investigate the X-ray flux for decaying DM particles. In fact, the formalism which we adopted to study annihilating DM, can be extended to constrain decaying DM in the Milky Way. The same technique can be employed to investigate the presence of DM subhalos, which would boost the total signal and determine constraints on the annihilation cross-section (or decay lifetime) from analysing extragalactic X-ray emission.

In Part III we have focused on an exotic signal emerging from the radio maps of the GLEAM survey. This emission seems compatible with a filament, but an explanation in terms of standard astrophysical processes seems unlikely. We provided an interpretation in terms of synchrotron radiation from DM particles emanating from inside the filaments. The observed brightness temperatures can be recovered by DM candidates with a mass in the range of 5 – 10 GeV, decaying into electron-positron pairs, as illustrated in Table 8.2. We obtained competitive constraints on the decay lifetime for DM masses between 3 GeV and 10 GeV as shown in Fig 8.6, assuming a magnetic field above 130 nG. In our analysis we made the conservative assumption of a constant DM density and ignored the impact of subhalos. Thus a non-uniform density profile, in addition to the contribution of substructures could boost the signal from annihilating DM. A more in-depth analysis of the internal structure of the filaments, as well as accounting for the cosmic ray diffusion, are necessary steps to get a deeper understanding of these cosmic bridges.

In conclusion, along our journey in the realm of DM, we have considered different signals arising on different scales and at different frequencies. We have shown that a multi-wavelength approach has the potential to provide complementary and invaluable information on the puzzling conundrum of DM. On our road to solve this mystery, there are still many stones left unturned, and many observations to be scrutinised. Thanks to our dynamic community and the prospective data from upcoming experiments, we certainly have a bright future ahead!

# Part V

# Appendices

APPENDIX A

# Bremsstrahlung emissivity

The bremsstrahlung emission from hot gas is adopted to estimate the mass of a cluster, as discussed in Chapter 1. Also, it represents a relevant source of energy losses for the sub-GeV electrons and positrons near the galactic plane, which has been considered in the derivation of the inverse Compton scattering signal in Chapter 6. We believe it is instructive to show the derivation of the thermal bremsstrahlung emissivity.

The bremsstrahlung radiation is the electromagnetic emission from a charged particle, which is decelerated by the Coulomb field produced by another charged particle. Typically it refers to an electron decelerated by an ion. This effect is called "thermal bremsstrahlung" if electrons and ions are in thermal equilibrium. The case of radiation from plasma is also known as free-free emission. Astrophysical applications of this formalism are the study of the radio emission from hydrogen-rich regions with a typical temperature of $10^4$ K, galactic bulges with a temperature around $10^7$ K, diffusion X-ray radiation emitted by hot intergalactic gas in clusters, where the temperatures are about $10^7 - 10^8$ K. The bremsstrahlung radiation is the dominant cooling mechanism in plasma with temperature above $10^7$ K. In order to derive the bremsstrahlung emissivity, let us consider an electron with charge $-e$ passing near an ion of charge $+Ze$, with impact factor $b$. The physical process and the geometry of the problem are illustrated in Fig. A.1. The power radiated by the electron is given by the Larmor's expression:

$$P = -\frac{\mathrm{d}E}{\mathrm{d}t} = \frac{2}{3}\frac{e^2}{4\pi\epsilon_0 c^3}\,|\ddot{\boldsymbol{r}}|^2\,, \tag{A.1}$$

where $\epsilon_0$ is the vacuum permittivity and the acceleration $\ddot{\boldsymbol{r}}$ is given by the Coulomb's law:

$$\boldsymbol{F} = m\ddot{\boldsymbol{r}} \simeq -\frac{Ze^2}{4\pi\epsilon_0 r^2}\,\hat{\boldsymbol{u}}_r \tag{A.2}$$

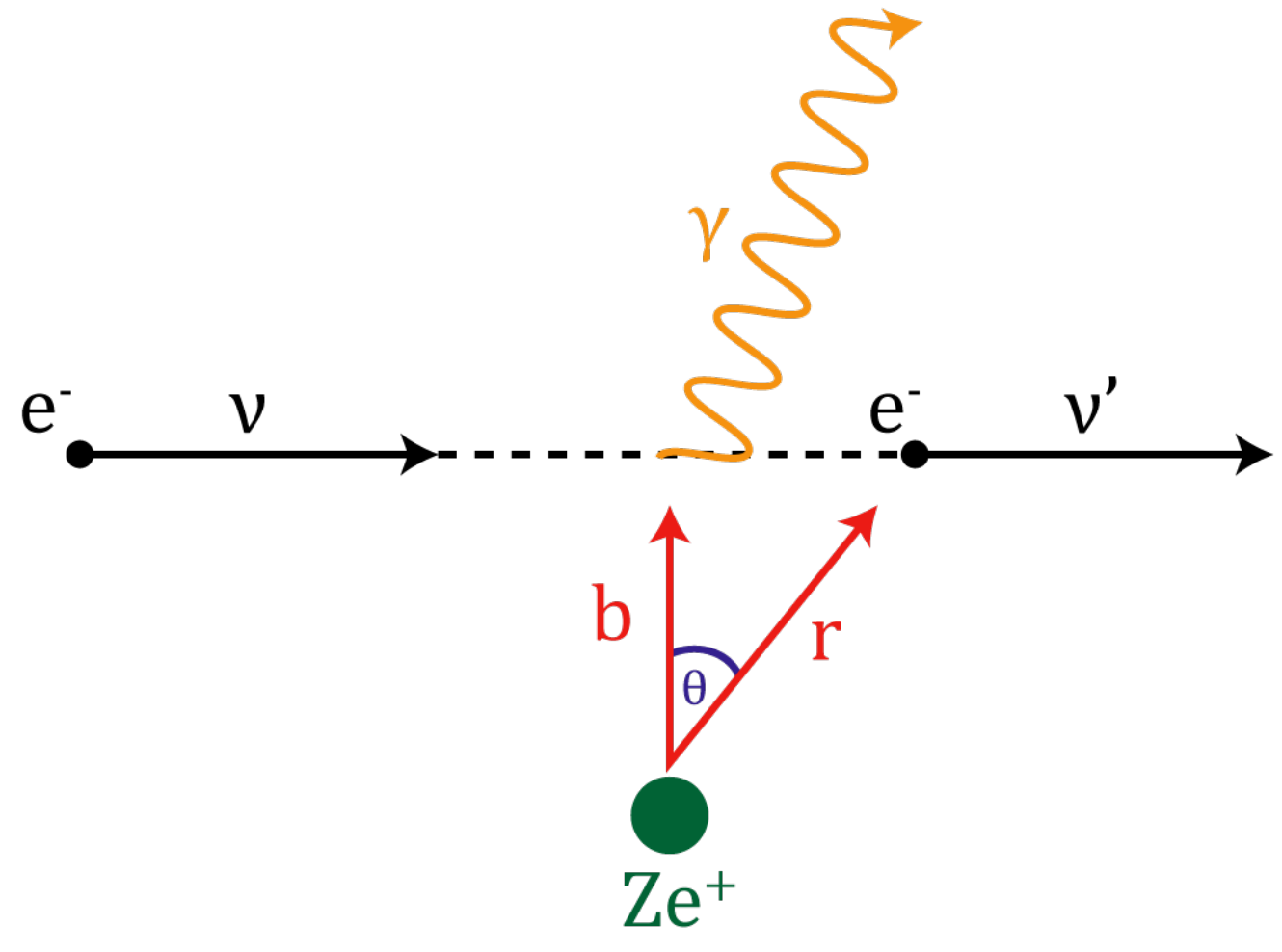

**Fig. A.1.** In the bremsstrahlung process a charged particle (usually an an electron) is decelerated by another charged particle (e.g. an ion with charge $Ze^+$). For the conservation of energy, the outcome of this deceleration is the production of radiation. $v$ and $v'$ indicate the velocity before and after the photon emission. The impact factor $b$ denotes the closest distance between the two particles, while $r$ represents the generic separation and $\theta$ is the angle between these two vectors.

with the unit vector $\hat{\boldsymbol{u}}_r$ parallel to the direction of the electromagnetic force $\boldsymbol{F}$. Thus,

$$\ddot{\boldsymbol{r}} = -\frac{Ze^2}{4\pi\epsilon_0 m_e r^2}\,\hat{\boldsymbol{u}}_r\,, \tag{A.3}$$

where $m_e$ is the mass of the deflected particle (here the electron) and $r = r(t)$ is the distance between the electron and the ion in the instant $t$. Inserting Eq. (A.3) in Eq. (A.1), we get

$$P = -\frac{\mathrm{d}E}{\mathrm{d}t} = \frac{2}{3}\frac{Z^2 e^6}{(4\pi\epsilon_0 c)^3\, m_e^2\, r^4}\,. \tag{A.4}$$

It is worth noting that $P$ is inversely proportional to $m_e^2$ and $r^4$. Therefore, the emitted bremsstrahlung power is relevant for light particles and for the closest distance, which is the impact factor $b$. The total emitted radiation can be written as an integration of the spectral density $I_\omega$ over the angular frequency $\omega$:

$$E = \int_0^\infty \mathrm{d}\omega\, I_\omega\,. \tag{A.5}$$

In order to derive the expression for $I_\omega$, recall that the Fourier transform of $\ddot{\boldsymbol{r}}(t)$ is defined as

$$\ddot{\boldsymbol{r}}(\omega) = \frac{1}{\sqrt{2\pi}}\int_{-\infty}^{\infty} \mathrm{d}t\, \exp(i\omega t)\, \ddot{\boldsymbol{r}}(t)\,. \tag{A.6}$$

The Parseval's theorem relates $\ddot{\boldsymbol{r}}(t)$ to $\ddot{\boldsymbol{r}}(\omega)$ by

$$\int_{-\infty}^{+\infty} \mathrm{d}t\,|\ddot{\boldsymbol{r}}(t)|^2 = \int_{-\infty}^{+\infty} d\omega\,|\ddot{\boldsymbol{r}}(\omega)|^2\,, \tag{A.7}$$

thus integrating Eq. (A.1) over the time and applying the Parseval's theorem, we get

$$\int_{-\infty}^{\infty} \mathrm{d}t \, \frac{\mathrm{d}E}{\mathrm{d}t} = \int_{-\infty}^{\infty} \mathrm{d}t \, \frac{e^2}{6\pi \epsilon_0 c^3} |\ddot{\boldsymbol{r}}(t)|^2 = \int_{-\infty}^{\infty} \mathrm{d}\omega \, \frac{e^2}{6\pi\epsilon_0 c^3} |\ddot{\boldsymbol{r}}(\omega)|^2 \,. \tag{A.8}$$

Since the acceleration is a real function, it holds

$$\int_{0}^{\infty} \mathrm{d}\omega \, \frac{e^2}{6\pi \epsilon_0 c^3} |\ddot{\boldsymbol{r}}(\omega)|^2 = \int_{-\infty}^{0} \mathrm{d}\omega \, \frac{e^2}{6\pi \epsilon_0 c^3} |\ddot{\boldsymbol{r}}(\omega)|^2 \,. \tag{A.9}$$

By comparing Eq. (A.5) and Eq. (A.8) we get

$$\int_{0}^{\infty} \mathrm{d}\omega \, I_\omega = \int_{0}^{\infty} \mathrm{d}\omega \, \frac{e^2}{3\pi \epsilon_0 c^3} |\ddot{\boldsymbol{r}}(\omega)|^2 , \tag{A.10}$$

therefore the spectral density is given by :

$$I_\omega = \frac{e^2}{3\pi\epsilon_0 c^3} |\ddot{\boldsymbol{r}}(\omega)|^2 \,. \tag{A.11}$$

Physically speaking, the spectral density corresponds to the total energy per unit bandwidth emitted by the electron during the time interval of interaction. If we consider small angle deviations, that is to say that the path of the electron is approximately linear, the change in velocity can be obtained by integrating the component of the acceleration which is orthogonal to the path:

$$\Delta\dot{r} = \int_{-\infty}^{\infty} \mathrm{d}t \, \ddot{r}_\perp(t) \,. \tag{A.12}$$

The normal acceleration is given by

$$\ddot{r}_\perp(t) = |\ddot{\boldsymbol{r}}(t)| \cos\theta(t) , \tag{A.13}$$

where $\theta(t)$ is the angle between $b$ and $r$ in the instant $t$, as outlined in Fig. A.1.

$$\begin{cases} r^2 & = & b^2 + \dot{r}^2 \, t^2 \\ b & = & r \, \cos\theta \,. \end{cases} \tag{A.14}$$

Thus,

$$\cos\theta = \frac{b}{\sqrt{b^2 + \dot{r}^2 \, t^2}} \tag{A.15}$$

and we obtain

$$\ddot{r}_\perp(t) = \frac{Ze^2}{4\pi \epsilon_0 \, m_e} \frac{b}{\left(b^2 + \dot{r}^2 \, t^2\right)^{3/2}} \,. \tag{A.16}$$

Therefore, the change in velocity takes the simple form

$$\Delta\dot{r} = \frac{2Ze^2}{4\pi \epsilon_0 \, m_e \, b \, \dot{r}} \,. \tag{A.17}$$

The period of time during which the interaction between the two charged particles takes place is known as collision time and it can be defined as

$$\tau = \frac{b}{\dot{r}} \,. \tag{A.18}$$

Considering that

$$\begin{cases} \ddot{r} \approx \dfrac{\Delta \dot{r}}{\sqrt{2\pi}} & \omega\tau \ll 1 \\ \ddot{r} \approx 0 & \omega\tau \gg 1 \end{cases}$$

the radiated energy per unit frequency for the collision of a single electron is

$$I_\omega = \frac{1}{24\pi^4 \epsilon_0^3 c^3} \frac{Z^2 e^6}{m_e^2 b^2 \dot{r}^2}. \tag{A.19}$$

Now, let us consider a region with a large number of electrons and ions with number density $n_e$ and $n_i$, respectively. For the sake of simplicity, we assume that all the electrons have the same speed $v$. The number of collisions that each electron undergoes in an infinitesimal surface between $b$ and $b + \mathrm{d}b$ is $2\pi n_i b \,\mathrm{d}b$. In a unit time, the flux of electrons that collide on an ion is $n_e v$. Thus, the overall number of collisions per unit volume is $2\pi n_e n_i v b \,\mathrm{d}b$ and the total emissivity (power per unit of volume and frequency) is

$$\begin{aligned} k_\nu &= 2\pi v n_e n_i \int_{b_{\min}}^{b_{\max}} \mathrm{d}b \, I_\omega b \\ &= \frac{2\pi v n_e n_i}{24\pi^4 \epsilon_0^3 c^3} \frac{Z^2 e^6}{m_e^2 v^2} \int_{b_{\min}}^{b_{\max}} \frac{\mathrm{d}b}{b} \\ &= \frac{n_e n_i Z^2 e^6}{12\pi^3 \epsilon_0 c^3 m_e^2 v} \ln \frac{b_{\max}}{b_{\min}}. \end{aligned} \tag{A.20}$$

Since the contribution of the emitted power becomes negligible for $\omega\tau \gg 1$, we can take $b_{\max} \equiv \dfrac{v}{\omega}$. Depending on the energetic regime under consideration, $b_{\min}$ can be selected with two possible approaches. A classical treatment is valid until the electron kinetic energy is higher than the potential energy, namely if $\frac{1}{2} m_e v^2 \ll Z^2 Ry$, where $Ry = \dfrac{m_e e^4}{2\hbar^2}$ is the Rydberg energy for an atom of hydrogen. This implies that the straight-line assumption is valid until $\Delta v \sim v$. Therefore, the condition to derive $b_{\min}^{(1)}$ in the classical regime is

$$\frac{2 Z e^2}{4\pi \epsilon_0 m_e v b_{\min}^{(1)}} = v \quad \Longrightarrow \quad b_{\min}^{(1)} = \frac{Z e^2}{2\pi \epsilon_0 m_e v^2}. \tag{A.21}$$

As for the quantum regime, we can derive the limit from the Heisenberg's uncertainty principle

$$\Delta x \Delta p \sim \hbar \tag{A.22}$$

with $\Delta x \sim b$ and $\Delta p \sim m_e v$, therefore

$$b_{\min}^{(2)} = \frac{\hbar}{m_e v}. \tag{A.23}$$

A simple prescription to determine which one is the more appropriate boundary condition is to set $b_{\min} = \max\left(b_{\min}^{(1)}, b_{\min}^{(2)}\right)$. The emissivity in Eq. (A.20) can be expressed in terms of a correction factor $g_{\mathrm{ff}}$

$$k_\nu = \frac{n_e n_i Z^2 e^6}{12\sqrt{3}\pi^3 \epsilon_0 c^3 m_e^2 v} g_{\mathrm{ff}}, \tag{A.24}$$

where the so-called Gaunt factor $g_{\rm ff}$ evolves with the electron speed and with the frequency of the bremsstrahlung emission

$$g_{\rm ff}(v,\omega) = \frac{\sqrt{3}}{\pi} \ln \frac{b_{\rm max}}{b_{\rm min}} . \tag{A.25}$$

In the regime $h\nu \ll kT$, the Gaunt factor evolves with the logarithm of the frequency

$$g_{\rm ff}(v,\omega) = \frac{\sqrt{3}}{\pi} \ln \frac{kT}{h\nu} . \tag{A.26}$$

This description can be applied to the interesting case of thermal bremsstrahlung. The term "thermal" denotes that the particles are in thermal equilibrium, namely their speed distribution is a Maxwellian function. The probability that an electron has speed in the range $\mathrm{d}^3v$ is

$$\mathrm{d}P \propto \exp\left(\frac{E}{kT}\right) \mathrm{d}^3v = \exp\left(-\frac{m_e v^2}{2kT}\right) \mathrm{d}^3v . \tag{A.27}$$

Recalling that $\mathrm{d}^3v = 4\pi v^2 \mathrm{d}v$, the probability distribution reads

$$\mathrm{d}P \propto v^2 \exp\left(-\frac{m_{\rm e} v^2}{kT}\right) \mathrm{d}v , \tag{A.28}$$

thus the total emissivity can be obtained by averaging the single-speed expression of Eq. (A.24) over the Maxwellian distribution:

$$k_\nu = \frac{\displaystyle\int_{v_{\rm min}}^{\infty} \mathrm{d}v \, k_\nu \, v^2 \exp\left(-\frac{m_e v^2}{2kT}\right)}{\displaystyle\int_0^{\infty} \mathrm{d}v \, v^2 \exp\left(-\frac{m_e v^2}{2kT}\right)} . \tag{A.29}$$

The lower bound $v_{\rm min}$ on the speed is set under the physical condition that the kinetic energy has to be at least $h\nu$ to produce a photon with frequency $\nu$

$$h\nu \leq \frac{1}{2} m_e v^2 . \tag{A.30}$$

This lower cut-off is known as photon discreteness effect and it implies $v_{\rm min} \equiv \sqrt{\frac{2h\nu}{m_e}}$. Thus, the total spectral emissivity for a thermal bremsstrahlung radiation is

$$k_\nu = \left(\frac{\pi}{6}\right)^2 \frac{Z^2 e^6}{3\pi^2 m_e^2 c^3 \epsilon_0^2} \left(\frac{m_e}{kT}\right)^2 n_e n_i \exp\left(-\frac{m_{\rm e} v^2}{kT}\right) \overline{g}_{\rm ff} , \tag{A.31}$$

where $\overline{g}_{\rm ff}$ is the velocity-averaged Gaunt factor and it is of order unity (see [505] and [506] for a thorough discussion).

APPENDIX B

# FORMALISM OF THE FOURIER POWER SPECTRUM

In Part I we constrain the velocity-averaged annihilation cross-section of DM particles in the GeV−TeV range, by applying the cross-correlation technique. This is a statistical method to estimate the correlation between two generic observables. The key estimator is the two-point correlation function in real space, or its equivalent in $k$-space: the Fourier power spectrum. In this chapter we derive the general expression of these two quantities, following the treatment of Refs. [507, 329].

The two-point correlation function (2PCF) of a generic density field $f_i(\boldsymbol{x})$ is defined as

$$\xi_{ij}(\boldsymbol{x},\boldsymbol{y}) = \langle f_i(\boldsymbol{x}) f_j(\boldsymbol{y})\rangle \,, \tag{B.1}$$

where the angle brackets indicate the ensemble average. The 3D power spectrum (PS) is the Fourier transform of the 2PCF:

$$\langle \tilde{f}_i(\chi,\boldsymbol{k}) \tilde{f}_j(\chi',\boldsymbol{k}')\rangle = (2\pi)^3\, \delta^3\left(\boldsymbol{k}-\boldsymbol{k}'\right) P_{ij}\left(\boldsymbol{k},\chi,\chi'\right) . \tag{B.2}$$

Let us write $f(\boldsymbol{x})$ as a sum of independent seeds $a$ and take the mass $M$ as the characteristic parameter[1] for these seeds:

$$f(\boldsymbol{x}) = \sum_a f(M_a,\boldsymbol{x}-\boldsymbol{x}_a) = \int \mathrm{d}M \int \mathrm{d}^3x' \sum_a \delta^3\left(\boldsymbol{x}'-\boldsymbol{x}_a\right) \delta(M-M_a) f\left(M,\boldsymbol{x}-\boldsymbol{x}'\right) . \tag{B.3}$$

The seed density reads

$$\frac{\mathrm{d}n}{\mathrm{d}M} = \left\langle \sum_a \delta^3(\boldsymbol{x}-\boldsymbol{x}_a)\, \delta(M-M_a)\right\rangle \tag{B.4}$$

[1]The mass is a suitable parameter when dealing with gravitational tracers like DM halos. In the case of astrophysical sources, a better characteristic parameter is represented by the luminosity.

The 2PCF can be expressed as

$$\xi_{ij}(\boldsymbol{x},\boldsymbol{y}) = \int \mathrm{d}M_1\,\mathrm{d}M_2\,\mathrm{d}^3x_1\,\mathrm{d}^3x_2 \left\langle \sum_a \delta^3(\boldsymbol{x}_1-\boldsymbol{x}_a)\delta(M_1-M)\sum_b \delta^3(\boldsymbol{x}_2-\boldsymbol{x}_b)\delta(M_2-M_b)\right\rangle \times \\ \times f_1(M_1,\boldsymbol{x}-\boldsymbol{x}_1)f_2(M_2,\boldsymbol{y}-\boldsymbol{x}_2)\,. \tag{B.5}$$

If we consider a seed $a$ with mass $M_1$ located at the position $\boldsymbol{x}_1$ and a seed $b$ characterised by a mass $M_2$ at the position $\boldsymbol{x}_2$, their correlation reads

$$\left\langle \sum_a \delta^3(\boldsymbol{x}_1-\boldsymbol{x}_a)\delta(M_1-M)\sum_b \delta^3(\boldsymbol{x}_2-\boldsymbol{x}_b)\delta(M_2-M_b)\right\rangle = \tag{B.6}$$

$$= \left\langle \sum_a \delta^3(\boldsymbol{x}_1-\boldsymbol{x}_a)\delta(M-M_a)\delta^3(\boldsymbol{x}_2-\boldsymbol{x}_a)\delta(M'-M_a)\right\rangle + \\ + \left\langle \sum_a \delta^3(\boldsymbol{x}_1-\boldsymbol{x}_a)\delta(M-M_a)\sum_{b\neq a} \delta^3(\boldsymbol{x}_2-\boldsymbol{x}_b)\delta(M_2-M_b)\right\rangle \tag{B.7}$$

$$= \frac{\mathrm{d}n}{\mathrm{d}M}\delta^3(\boldsymbol{x}_1-\boldsymbol{x}_2)\delta(M_1-M_2) + \frac{\mathrm{d}n}{\mathrm{d}M_1}\frac{\mathrm{d}n}{\mathrm{d}M_2}[1+\xi_s(M_1,M_2,\boldsymbol{x}_1,\boldsymbol{x}_2)]\,, \tag{B.8}$$

where $\xi_s$ is known as the seed-2PCF. Thus, Eq. (B.5) becomes

$$\xi_{ij}(\boldsymbol{x},\boldsymbol{y}) = \int \mathrm{d}M\,\mathrm{d}^3x_1\,\frac{\mathrm{d}n}{\mathrm{d}M} f_1(M,\boldsymbol{x}-\boldsymbol{x}_1)f_2(\boldsymbol{y}-\boldsymbol{x},M) + \\ + \int \mathrm{d}M_1\,\mathrm{d}M_2\,\mathrm{d}^3x_1\mathrm{d}^3x_2\,\frac{\mathrm{d}n}{\mathrm{d}M_1}\frac{\mathrm{d}n}{\mathrm{d}M_2} f_1(\boldsymbol{x}-\boldsymbol{x}_1,M_1)f(\boldsymbol{y}-\boldsymbol{x}_2,M_2)\,\xi_s(M_1,M_2,\boldsymbol{x}_1,\boldsymbol{x}_2)\,. \tag{B.9}$$

We recall that in Eq. (B.3), the density field has been expressed as a linear superposition of seeds. Therefore, when considering the mass density fluctuations, the seed-2PCF corresponds to the linear matter correlation function

$$\xi_{s,ij}\left(M_1,M_2,\boldsymbol{x}_i,\boldsymbol{x}_j\right) = \xi_{\mathrm{lin}}\left(\left|\boldsymbol{x}_i-\boldsymbol{x}_j\right|\right)\,. \tag{B.10}$$

In the case of biased tracers, such as the DM halos or the astrophysical sources, it holds

$$\xi_{s,ij}\left(M_1,M_2,\boldsymbol{x}_i,\boldsymbol{x}_j\right) \sim b_i(M_1)\,b_j(M_2)\,\xi_{\mathrm{lin}}\left(\left|\boldsymbol{x}_i-\boldsymbol{x}_j\right|\right)\,, \tag{B.11}$$

where $b_i$ denotes the bias of the observable $i$.

By performing the Fourier transform of Eq. (B.9) we get

$$P_{ij}(k) = \int \mathrm{d}M\,\frac{\mathrm{d}n}{\mathrm{d}M}\,\widetilde{f}_i^*(k|M)\,\widetilde{f}_j(k|M) + \\ + \int \mathrm{d}M_1\,\mathrm{d}M_2\,\frac{\mathrm{d}n}{\mathrm{d}M_1}\frac{\mathrm{d}n}{\mathrm{d}M_2}\,\widetilde{f}_i^*(k|M_1)\,\widetilde{f}_j(k|M_2)\,P_s(k,M_1,M_2)\,, \tag{B.12}$$

where $P_s$ corresponds to the linear matter PS in the case of the matter density fluctuations, while for biased objects it holds

$$P_s(k,M_1,M_2) = b_i(M_1)\,b_j(M_2)\,P_{\mathrm{lin}}(k)\,. \tag{B.13}$$

Thus, we obtain the decomposition of the PS into a one-halo and a two-halo contributions:

$$P_{ij}(k) = P_{ij}^{1h}(k) + P_{ij}^{2h}(k)\,, \tag{B.14}$$

where

$$P_{ij}^{1h}(k) = \int \mathrm{d}M\, \frac{\mathrm{d}n}{\mathrm{d}M}\, \widetilde{f}_i^*(k|M)\, \widetilde{f}_j(k|M) \tag{B.15}$$

$$P_{ij}^{2h}(k) = \left[\int \mathrm{d}M_1\, \frac{\mathrm{d}n}{\mathrm{d}M_1}\, b_i(M_1)\, \widetilde{f}_i^*(k|M_1)\right] \times \tag{B.16}$$

$$\times \left[\int \mathrm{d}M_2\, \frac{\mathrm{d}n}{\mathrm{d}M_2}\, b_j(M_2)\, \widetilde{f}_j(k|M_2)\right] P_{\mathrm{lin}}(k)\,.$$

APPENDIX C

# GAMMA-RAY LUMINOSITY FUNCTIONS

In Part I we compute the cross-correlation power spectrum between the unresolved $\gamma$-ray background and the neutral hydrogen distribution. Astrophysical sources represent the largest contribution to the unresolved emission and they are characterised by their $\gamma$-ray luminosity function. This chapter includes the explicit expressions of the $\gamma$-ray luminosity function (GLF) for each astrophysical source considered in our analysis: blazars (including both BL Lacertae objects and flat-spectrum radio quasars), misaligned active galactic nuclei and star-forming galaxies. The relations between their luminosity and the host dark matter halos are also reported, as given in Ref. [331].

## C.1 Blazars

We adopt the luminosity-dependent density evolution (LDDE) model of Ref. [390], according to which the GLF of blazars can be parameterized as

$$\phi_\gamma(L,z,\Gamma) = \phi(L,\Gamma) \times e(z,L)\,. \tag{C.1}$$

At redshift $z = 0$, the GLF is parameterized as a broken power law in luminosity and it follows a Gaussian distribution in photon spectral index

$$\phi(L) = \frac{A}{\ln(10)}\left(\frac{L}{\text{erg s}^{-1}}\right)^{-1}\left[\left(\frac{L}{L_\star}\right)^{\gamma_1} + \left(\frac{L}{L_\star}\right)^{\gamma_2}\right]^{-1} \exp\left[-\frac{(\Gamma-\mu(L))^2}{2\sigma^2}\right], \tag{C.2}$$

where the mean spectral index $\mu$ exhibits a weak (logarithmic) dependence on $L$:

$$\mu(L) = \mu^\star + \beta\left[\log\frac{L}{\text{erg s}^{-1}} - 46\right]\,. \tag{C.3}$$

| | $A$ [Mpc$^{-3}$ erg$^{-1}$ s] | $L_\star$ [erg s$^{-1}$] | $\gamma_1$ | $\gamma_2$ | $p_1$ | $p_2$ | $z_\star$ | $\beta$ |
|---|---|---|---|---|---|---|---|---|
| BL Lacs | $9.20\cdot10^{-11}$ | $2.43\cdot10^{48}$ | 1.12 | 3.71 | 4.50 | −12.88 | 1.67 | $4.46\cdot10^{-2}$ |
| FSRQ | $3.06\cdot10^{-9}$ | $0.84\cdot10^{48}$ | 0.21 | 1.58 | 7.35 | −6.51 | 1.47 | 0.21 |

**Table C.1.** Fixed parameters in the $\gamma$-ray luminosity functions of BL Lacertae objects and flat-spectrum radio quasars.

In the following, we assume $\Gamma = \mu^\star = 2.11$, thus we can neglect the exponential factor. The redshift dependence is encoded in the function $e(z, L)$. In the LDDE model, it holds

$$e(z,L) = \left[\left(\frac{1+z}{1+z_c(L)}\right)^{-p_1} + \left(\frac{1+z}{1+z_c(L)}\right)^{-p_2}\right]^{-1}, \tag{C.4}$$

where $z_c = z_\star \left(L/10^{48}\mathrm{erg/s}\right)^\beta$. BL Lacs and FSRQs exhibit the same functional form for the GLF, but they differ for the values of the parameters, as displayed in Table C.1. The relation between the luminosity of blazars and the mass of the host DM halo reads

$$M(L) = 10^{13} M_\odot \left(\frac{M_\star}{10^{8.8}(1+z)^{1.4}}\right)^{0.645}, \tag{C.5}$$

where

$$M_\star = 10^9 \left(\frac{L}{10^{48}\mathrm{erg/s}}\right)^{0.36}. \tag{C.6}$$

## C.2 Misaligned active galactic nuclei

The radio luminosity function (RLF) of mAGN is well known, unlike the GLF. The two luminosity functions are related via

$$\phi_\gamma(L, z) = k\,\rho_r(L_r, z)\,\frac{\mathrm{d}\log L_r}{\mathrm{d}\log L}, \tag{C.7}$$

thus the relation between the radio and $\gamma$ luminosities is necessary. The constant $k$ is tuned to reproduce the numbers of mAGN observed by the $\gamma$-ray detector. Typically, radio observations are related to the total radiation emitted by AGN. However, the flux from the central core has been individually detected in some rare cases at low radio frequencies. Core and total RLF can be related with the same reasoning of Eq. (C.7):

$$\rho_{r,\mathrm{core}}(L_{r,\mathrm{core}}, z) = \rho_{r,\mathrm{tot}}(L_{r,\mathrm{tot}}, z)\,\frac{\mathrm{d}\log L_{r,\mathrm{tot}}}{\mathrm{d}\log L_{r,\mathrm{core}}}. \tag{C.8}$$

Ref. [392] provides the RLF, defined as the number of radio sources per unit of co-moving volume and per logarithmic[1] of luminosity:

$$\rho_r(L_r,z) = \rho_l(L_r,z) + \rho_h(L_r,z), \tag{C.9}$$

[1]Base-10 logarithm.

where

$$\begin{cases} \rho_l = \rho_{l\star}\left(\dfrac{L_r}{L_{l\star}}\right)^{-\beta_l} \exp\left(-\dfrac{L_r}{L_{l\star}}\right)(1+z)^{k_l} & \text{for} \quad z < z_{l\star} \\ \rho_l = \rho_{l\star}\left(\dfrac{L_r}{L_{l\star}}\right)^{-\beta_l} \exp\left(-\dfrac{L_r}{L_{l\star}}\right)(1+z_{l\star})^{k_l} & \text{for} \quad z \geq z_{l\star} \end{cases} \tag{C.10}$$

and

$$\rho_h = \rho_{h\star}\left(\frac{L_r}{L_{h\star}}\right)^{-\beta_h} \exp\left(-\frac{L_{h\star}}{L}\right) f_h(z)\,. \tag{C.11}$$

The function $f_h$ takes the exponential form

$$f_h(z) = \exp\left\{-\frac{1}{2}\left(\frac{z - z_{h\star}}{z_{h0}}\right)^2\right\}. \tag{C.12}$$

The values of the parameters are: $\rho_{l\star} = 10^{-7.523}$ Mpc$^{-3}$, $\beta_l = 0.586$, $L_{l\star} = 10^{26.48}$ W/Hz, $k_l = 3.48$, $z_{l\star} = 0.710$, $\rho_{h\star} = 10^{-6.757}$ Mpc$^{-3}$, $\beta_h = 2.42$, $\log L_{h\star} = 27.39$ W/Hz and $z_{h\star} = 2.03$. For $z < z_{h\star}$ we use $z_{h0} = 0.568$, while for $z \geq z_{h\star}$ we adopt $z_{h0} = 0.956$. Ref. [388] derived the correlation between the core radio and the $\gamma-$ray luminosities, while Ref. [508] provides the relation between the core and total luminosities:

$$\log L = 2 + 1.008 \log L_{r,\text{core}} \tag{C.13}$$

$$\log L_{r,\text{core}}^{5\text{ GHz}} = 4.2 + 0.77 \log L_{r,\text{tot}}^{1.4\text{ GHz}}\,. \tag{C.14}$$

The reference radio frequency in Eq. (C.9) is 151 MHz. Thus, we shift all the luminosities to 151 MHz, assuming a power-law scaling [509]

$$\frac{L_r}{\nu} \propto \nu^{-\alpha_r}\,, \tag{C.15}$$

where $\alpha_r = 0.80$ for the total radio emission. The comoving volume element used in Ref. [392] is

$$\frac{\mathrm{d}^2 V_W}{\mathrm{d}z\,\mathrm{d}\Omega} = \frac{c^3 z^2 (2+z)^2}{4 H_{0,W}^3 (1+z)^3}\,, \tag{C.16}$$

where $H_{0,W} = 50$ MHz/s/Mpc. The comoving volume in the standard $\Lambda$CDM cosmology reads

$$\frac{\mathrm{d}^2 V}{\mathrm{d}z\,\mathrm{d}\Omega} = \frac{c\, d_L^2(z)}{H_0 (1+z)^2 \sqrt{(1-\Omega_\Lambda - \Omega_m)(1+z)^2 + (1+z)^3 \Omega_m + \Omega_\Lambda}}\,. \tag{C.17}$$

Thus, we need to take into account the conversion factor

$$\eta = \frac{\mathrm{d}^2 V_w/\mathrm{d}z\mathrm{d}\Omega}{\mathrm{d}^2 V/\mathrm{d}z\mathrm{d}\Omega}\,. \tag{C.18}$$

Recalling that $\rho_r$ is in unit of $\log_{10} L_r$, the GLF can be derived from the RLF via

$$\phi_\gamma(L,z) = \frac{k\,\eta}{(1+z)^{2-\Gamma}} \frac{1}{\ln(10)\, L_{\text{tot}}^{151\text{ MHz}}} \frac{\mathrm{d}L_{\text{tot}}^{151\text{ MHz}}}{\mathrm{d}L} \rho_r(L_{\text{tot}}^{151\text{ MHz}}(L))\,, \tag{C.19}$$

where $k = 3.05$ [388], the spectral index is $\Gamma = 2.37$ and the factor $(1+z)^{2-\Gamma}$ is the so-called K-correction, which addresses the redshift variation between observed and emitted energies.

The mass-to-luminosity relation for mAGN reads

$$M(L) = 10^{13}\, M_\odot \left(\frac{M_\star}{10^{8.8}\,(1+z)^{1.4}}\right)^{0.645}, \tag{C.20}$$

with

$$M_\star = 4.6\cdot 10^9 \left(\frac{L}{10^8 \mathrm{erg/s}}\right)^{0.16}. \tag{C.21}$$

## C.3 Star-forming galaxies

We adopt the model of Ref. [393] for the infrared luminosity function of SFG. They used the data of the Herschel Space Observatory, which identified three separate sub-classes of SFG: quiescent spiral galaxies, starburst galaxies and SFG hosting a concealed or low-luminosity AGN. The total IR luminosity function is the sum of these three contributions:

$$\phi_{\mathrm{IR}} = \phi_{\mathrm{spiral}} + \phi_{\mathrm{starburst}} + \phi_{\mathrm{SF\text{-}AGN}}\,. \tag{C.22}$$

Each component exhibits the following functional behaviour

$$\phi_i = \phi_{0,i}(z)\left(\frac{L_{\mathrm{IR}}}{L_{0,i}}\right)^{1-\gamma_i} \exp\left(-\frac{1}{2\sigma_i^2}\right)\log_{10}^2\left(1+\frac{L_{\mathrm{IR}}}{L_{0,i}}\right), \tag{C.23}$$

where $i$ = {spiral, starburst, SF-AGN} and $L_{0,i}$ takes the form

$$L_{0,i} = \begin{cases} L_{\star,i}\left(\dfrac{1+z}{1.15}\right)^{k_{Li}} & \text{for } z \le 1.1 \\ L_{\star,i}\left(\dfrac{2.1}{1.15}\right)^{k_{Li}} & \text{for } z > 1.1 \end{cases} \tag{C.24}$$

For the spiral component, we use

$$\phi_{0,\,\mathrm{spiral}} = \begin{cases} \phi_{\star,\mathrm{sp}}\left(\dfrac{1+z}{1.15}\right)^{k_{R1,\mathrm{sp}}} & \text{for } z \le 0.53 \\ \phi_{\star,\mathrm{sp}}\left(\dfrac{1.53}{1.15}\right)^{k_{R1,\mathrm{sp}}}\left(\dfrac{1+z}{1.53}\right)^{k_{R2,\mathrm{sp}}} & \text{for } z > 0.53 \end{cases} \tag{C.25}$$

The starburst and SF-AGN components have the same functional form for $\phi_0$:

$$\phi_{0,j} = \begin{cases} \phi_{\star,j}\left(\dfrac{1+z}{1.15}\right)^{k_{R1,j}} & \text{for } z \le 1.1 \\ \phi_{\star,j}\left(\dfrac{2.1}{1.15}\right)^{k_{R1,j}}\left(\dfrac{1+z}{2.1}\right)^{k_{R2,j}} & \text{for } z > 1.1 \end{cases} \tag{C.26}$$

with $j$={starburst, SF-AGN}. All the parameters are listed in Table C.2. As shown in Ref. [393], the different classes evolve differently and independently. The spiral galaxy population is the leading contribution at low redshift, $0 \leq z \lesssim 0.5$. At higher redshift, the SF-AGN population takes over, becoming the dominant contribution in the global IR luminosity function. The starbust galaxy population is always subdominant with the largest contribution at $z \sim 1-2$.

| | $\gamma$ | $\sigma$ | $\log_{10}(L_\star/L_\odot)$ | $\log_{10}(\phi_\star/\text{Mpc}^{-3})$ | $k_L$ | $k_{R1}$ | $k_{R2}$ |
|---|---|---|---|---|---|---|---|
| spiral | 1.0 | 0.50 | 9.78 | −2.12 | 4.49 | −0.54 | −7.13 |
| starburst | 1.0 | 0.35 | 11.17 | −4.46 | 1.96 | 3.79 | −1.06 |
| SF-AGN | 1.2 | 0.40 | 10.80 | −3.20 | 3.17 | 0.67 | 3.17 |

**Table C.2.** Parameters entering the infrared luminosity function for the three galaxy populations under consideration: spiral, starbursts, star-forming galaxies hosting an AGN.

Ref. [510] parameterised a scaling relation between the $\gamma$-ray luminosity, defined between 0.1 GeV and 100 GeV, and the IR luminosity, defined between 8 $\mu$m and 1000 $\mu$m:

$$\log_{10}\left(\frac{L_{0.1-100\text{GeV}}}{\text{erg s}^{-1}}\right) = \alpha_{\text{IR}} \log_{10}\left(\frac{L_{8-1000\mu\text{m}}}{10^{10} L_\odot}\right) + \beta_{\text{IR}} , \tag{C.27}$$

where $\alpha_{\text{IR}} = 1.09$ and $\beta_{\text{IR}} = 39.19$.

The GLF can be derived using Eqs. (C.22) and (C.27) via

$$\phi_\gamma\left(L_\gamma, z\right) = \phi_{\text{IR}} \frac{\text{d}\log_{10}(L_{\text{IR}})}{\text{d}\log_{10}\left(L_\gamma\right)} . \tag{C.28}$$

Finally, the mass-to-luminosity relation for SFG reads

$$M(L) = \frac{10^{12} M_\odot}{(1+z)^{1.61}} \left(\frac{L}{6.8 \cdot 10^{39}\ \text{erg/s}}\right)^{0.92} . \tag{C.29}$$

APPENDIX D

# DERIVATION OF THE ANGULAR POWER SPECTRUM

In Part I we derived the cross-correlation signal between the intensity fluctuations of the unresolved $\gamma$-ray background and the fluctuations in the brightness temperature of the 21cm line emitted by neutral hydrogen. The angular power spectrum is the statistical tool that we employed to correlate the two fluctuation fields. In the following, we derive the expression of the angular power spectrum in terms of the Fourier power spectrum.

Let us consider an intensity field $I_g$, associated to a generic source $g$ and a given direction $\boldsymbol{n}$:

$$I_g(\boldsymbol{n}) = \int \mathrm{d}\chi\, g(\chi, \boldsymbol{n}) \widetilde{W}(\chi)\,, \tag{D.1}$$

where $\chi = c/H(z)$ is the comoving distance. The intensity fluctuations are defined as

$$\delta I_g(\boldsymbol{n}) = I_g(\boldsymbol{n}) - \langle I_g \rangle\,, \tag{D.2}$$

where $\langle I_g \rangle$ is the average intensity. We can expand the fluctuation field in spherical harmonics

$$\delta I_g(\hat{\boldsymbol{n}}) = \langle I_g \rangle \sum_{\ell m} a_{\ell m} Y_{\ell m}(\boldsymbol{n})\,, \tag{D.3}$$

where $\ell$ is a non-negative integer and $m$ represents an integer such that $|m| \leq \ell$. The coefficients of the expansion can be written as

$$a_{\ell m} = \frac{1}{\langle I_g \rangle} \int \mathrm{d}\boldsymbol{n}\, \delta I_g(\boldsymbol{n}) Y^*_{\ell m}(\boldsymbol{n}) \tag{D.4}$$

$$= \frac{1}{\langle I_g \rangle} \int \mathrm{d}\boldsymbol{n} Y^*_{\ell m}(\boldsymbol{n}) \left[ I_g(\boldsymbol{n}) - \langle I_g \rangle \right] \tag{D.5}$$

$$= \frac{1}{\langle I_g \rangle} \int \mathrm{d}\boldsymbol{n} Y^*_{\ell m}(\boldsymbol{n}) \int \mathrm{d}\chi \left[ g(\chi, \boldsymbol{n}) \widetilde{W}(\chi) - \langle g \rangle(\chi) \widetilde{W}(\chi) \right] \tag{D.6}$$

$$= \frac{1}{\langle I_g \rangle} \int \mathrm{d}\boldsymbol{n} Y^*_{\ell m}(\boldsymbol{n}) \int \mathrm{d}\chi \frac{g(\chi, \boldsymbol{n}) - \langle g \rangle}{\langle g \rangle} \langle g \rangle \widetilde{W}(\chi)\,, \tag{D.7}$$

where $\langle g\rangle = \langle g\rangle(\chi)$. We can define the fluctuation field $f_g$ and the normalised window function $W$ as

$$f_g(\chi,\boldsymbol{n}) = \frac{g(\chi,\boldsymbol{n}) - \langle g\rangle}{\langle g\rangle}\,, \tag{D.8}$$

$$W(\chi) = \langle g\rangle\,\widetilde{W}(\chi)\,. \tag{D.9}$$

Thus, Eq. (D.7) becomes

$$a_{\ell m} = \frac{1}{\langle I_g\rangle}\int \mathrm{d}\boldsymbol{n} Y^*_{\ell m}(\boldsymbol{n})\int \mathrm{d}\chi\, f_g(\chi,\boldsymbol{n})W(\chi)\,. \tag{D.10}$$

The Fourier transform of $f_g(\chi,\boldsymbol{n})$ reads

$$f_g(\chi,\boldsymbol{n}) = \int \frac{\mathrm{d}^3k}{(2\pi)^3}\tilde{f}_g(k)e^{i\boldsymbol{k}\cdot\boldsymbol{r}} \tag{D.11}$$

and we can use the Rayleigh equation to write a plane wave in terms of spherical harmonics $Y_{\ell m}$ and spherical Bessel functions $j_l$ via

$$e^{i\boldsymbol{k}\cdot\boldsymbol{r}} = 4\pi\sum_{\ell'=0}^{\infty}\sum_{m=-\ell}^{l} i^{\ell'} j_{\ell'}(kr)Y^*_{\ell' m'}(\hat{k})Y_{\ell' m'}(\boldsymbol{n})\,, \tag{D.12}$$

where $r = \chi\boldsymbol{n}$. By plugging Eqs. (D.11) and (D.12) into Eq. (D.10) we get

$$a_{\ell m} = \frac{1}{\langle I_g\rangle}\int \mathrm{d}\boldsymbol{n}\int \mathrm{d}\chi W(\chi)\int \frac{\mathrm{d}^3k}{(2\pi)^3}4\pi\sum_{\ell'=0}^{\infty}\sum_{m=-\ell}^{l} i^{\ell'} j_{\ell'}(k\chi)Y^*_{\ell' m'}(\hat{k})Y_{\ell' m'}(\boldsymbol{n})\tilde{f}_g(k)Y^*_{\ell m}(\boldsymbol{n})\,. \tag{D.13}$$

The spherical harmonics are orthonormal functions, thus it holds

$$\int \mathrm{d}\boldsymbol{n}\, Y^*_{\ell m}(\boldsymbol{n})Y^*_{\ell' m'}(\boldsymbol{n}) = \delta_{\ell\ell'}\delta_{mm'}\,. \tag{D.14}$$

As a consequence, Eq. (D.13) reduces to

$$a_{\ell m} = \frac{1}{\langle I_g\rangle}\int \mathrm{d}\chi W(\chi)\int \frac{\mathrm{d}^3k}{2\pi^2}\tilde{f}_g(k)\sum_{\ell'=0}^{\infty}\sum_{m=-\ell}^{l} i^{\ell'} j_{\ell'}(k\chi)Y^*_{\ell' m'}(\hat{k})\delta_{\ell\ell'}\delta_{mm'} \tag{D.15}$$

$$= \frac{1}{\langle I_g\rangle}\int \mathrm{d}\chi W(\chi)\int \frac{\mathrm{d}^3k}{2\pi^2}\tilde{f}_g(k)i^{\ell} j_{\ell}(k\chi)Y^*_{\ell m}(\hat{k})\,. \tag{D.16}$$

The angular power spectrum is defined as

$$C_\ell^{ij} = \frac{1}{2\ell+1}\left\langle \sum_m a^{(i)}_{\ell m}a^{*(j)}_{\ell m}\right\rangle\,, \tag{D.17}$$

where $i,j$ refer to the two signals that we want to cross-correlate and $i = j$ applies to the auto-correlation case. Plugging Eq. (D.16) into the definition of the $C_\ell^{ij}$, we obtain

$$C_\ell^{ij} = \frac{1}{2\ell+1}\frac{1}{\langle I_i\rangle\langle I_j\rangle}\sum_m\int \mathrm{d}\chi W(\chi)\int \mathrm{d}\chi' W(\chi')\int \frac{\mathrm{d}^3k}{2\pi^2}i^{\ell} j_{\ell}(k\chi)Y^*_{\ell m}(\hat{k})$$
$$\int \frac{\mathrm{d}^3k'}{2\pi^2}i^{-\ell} j_{\ell}(k'\chi)Y_{\ell m}(\hat{k}')\left\langle \tilde{f}_i(k)\tilde{f}^*_j(k)\right\rangle\,, \tag{D.18}$$

We recall that

$$\left\langle \widetilde{f}_i(k)\widetilde{f}_j^*(k)\right\rangle = (2\pi)^3\,\delta_D^3(k-k')P_{ij}(k,\chi,\chi')\,. \tag{D.19}$$

Thus, Eq. (D.18) becomes

$$C_\ell^{ij} = \frac{1}{2\ell+1}\frac{1}{\langle I_i\rangle\langle I_j\rangle}\int \mathrm{d}\chi\, W(\chi)\int \mathrm{d}\chi'\, W(\chi')\int \frac{\mathrm{d}^3k}{2\pi^2}j_\ell(k\chi)Y_{\ell m}^*(\hat{k})$$

$$\int \frac{\mathrm{d}^3k'}{2\pi^2}j_\ell(k'\chi)Y_{\ell m}(\hat{k}')(2\pi)^3\,\delta_D^3(k-k')P_{ij}(k,\chi,\chi') \tag{D.20}$$

$$= \frac{1}{2\ell+1}\frac{1}{\langle I_i\rangle\langle I_j\rangle}\frac{2}{\pi}\sum_m\int \mathrm{d}\chi W(\chi)\int \mathrm{d}\chi' W(\chi')\int \mathrm{d}^3k\, j_\ell(k\chi)j_\ell(k'\chi')P_{ij}(k,\chi,\chi')\sum_m Y_{\ell m}^*(\hat{k})Y_{\ell m}(\hat{k})\,. \tag{D.21}$$

The Unsöld's theorem [511] states that

$$\sum_{m=-\ell}^{\ell} Y_{\ell m}^*(\hat{k})Y_{\ell m}(\hat{k}) = \frac{2\ell+1}{4\pi}\,, \tag{D.22}$$

which leads to

$$C_\ell^{ij} = \frac{1}{2\ell+1}\frac{1}{\langle I_i\rangle\langle I_j\rangle}\frac{2}{\pi}\int \mathrm{d}\chi W(\chi)\int \mathrm{d}\chi' W(\chi')\int \mathrm{d}k\, 4\pi k^2 j_\ell(k\chi)j_\ell(k\chi')P_{ij}(k,\chi,\chi')\frac{2\ell+1}{4\pi} \tag{D.23}$$

$$= \frac{1}{\langle I_i\rangle\langle I_j\rangle}\frac{2}{\pi}\int \mathrm{d}\chi W(\chi)\int \mathrm{d}\chi' W(\chi')\int \mathrm{d}k\, k^2 j_\ell(k\chi)j_\ell(k\chi')P_{ij}(k,\chi,\chi')\,. \tag{D.24}$$

In our analysis, we are interested in high multipoles, thus we can apply the Limber approximation [446, 447, 448] by adopting the following relation, valid for large multipoles[1]

$$\lim_{\ell\gg 1} j_\ell(x) = \sqrt{\frac{\pi}{2\ell+1}}\,\delta_D\left(\ell+\frac{1}{2}-x\right)\,. \tag{D.25}$$

Plugging Eq. (D.24) into Eq. (D.25) we get

$$C_\ell^{ij} = \frac{1}{\langle I_i\rangle\langle I_j\rangle}\frac{2}{\pi}\int \mathrm{d}\chi W(\chi)\int \mathrm{d}\chi' W(\chi')\int \mathrm{d}k\, k^2 P_{ij}(k,\chi,\chi')\frac{\pi}{2\ell+1}\delta_D\left(\ell+\frac{1}{2}-k\chi\right)\delta_D\left(\ell+\frac{1}{2}-k\chi'\right) \tag{D.26}$$

$$= \frac{1}{\langle I_i\rangle\langle I_j\rangle}\int \mathrm{d}\chi W(\chi)\int \mathrm{d}\chi' W(\chi')\int \mathrm{d}k\, k^2 P_{ij}(k,\chi,\chi')\frac{1}{\ell}\delta_D\left(\ell-k\chi\right)\delta_D\left(\ell-k\chi'\right)\,, \tag{D.27}$$

where in the last equation we have used the fact that $\ell \gg 1$. The delta function obeys

$$\delta\left(g(x)\right) = \sum_i \frac{\delta(x-x_{\star,i})}{|g'(x_{\star,i})|}\,, \tag{D.28}$$

[1]Eq. (D.25) can be derived from the expression obtained in Ref. [512].

where $x_{\star,i}$ corresponds to the zeros of $g(x)$, namely $g(x_{\star,i}) = 0$. In the first delta function, we take

$$\begin{cases} g(\chi'_\star) = \ell - k\chi'_\star = 0 \Longrightarrow \chi'_\star = \dfrac{\ell}{k} \\ |g'(\chi')| = |-k| = k\,, \end{cases} \tag{D.29}$$

while for the second delta function it holds

$$\begin{cases} g(k_\star) = \ell - k_\star\chi = 0 \Longrightarrow k_\star = \dfrac{\ell}{\chi} \\ |g'(k)| = |-\chi| = \chi\,. \end{cases} \tag{D.30}$$

Thus, Eq. (D.27) becomes

$$C_\ell^{ij} = \frac{1}{\langle I_i\rangle\langle I_j\rangle}\int \mathrm{d}\chi W(\chi)\int \mathrm{d}\chi' W(\chi')\int \mathrm{d}k\, k^2 P_{ij}(k,\chi,\chi')\frac{1}{\ell}\frac{1}{k\chi}\delta_D\left(\chi' - \frac{\ell}{k}\right)\delta_D\left(k - \frac{\ell}{\chi}\right). \tag{D.31}$$

Using the delta function, the factor which depends on $k$, $\ell$ and $\chi$ reduces to

$$k^2\frac{1}{\ell}\frac{1}{k\chi} = k\frac{1}{\ell\chi} = \frac{\ell}{\chi}\frac{1}{\ell\chi} = \frac{1}{\chi^2}\,. \tag{D.32}$$

Therefore, the angular power spectrum has the following convenient expression:

$$C_\ell^{ij} = \frac{1}{\langle I_i\rangle\langle I_j\rangle}\int \frac{\mathrm{d}\chi}{\chi^2} W_i(\chi)W_j(\chi)P_{ij}\left(k = \frac{\ell}{\chi}\right), \tag{D.33}$$

which we adopt in Chapter 5.

APPENDIX E

# WINDOW FUNCTION FOR ANNIHILATING DARK MATTER

The window function is one of the key elements to compute the angular power spectrum since it describes the redshift evolution of the observable under consideration. Hereinafter, we derive the window function for annihilating DM, following the method of Ref. [513].

The $\gamma$-ray intensity flux can be written as

$$I_\gamma = \int \mathrm{d}\chi\, \delta^2(\chi, \hat{\boldsymbol{n}}\chi)\, W(E_s, \chi)\,, \tag{E.1}$$

where $\delta = \rho_{\rm DM}/\overline{\rho}_{\rm DM}$ is the DM overdensity, $E_s = (1+z)E_\gamma$ represents the photon energy at the source and $E_\gamma$ denotes the observed photon energy. The intensity is generally expressed as [514]

$$E_\gamma I_\gamma = \frac{c}{4\pi} \int \mathrm{d}z\, \frac{P_\gamma\left((1+z)E_\gamma, z, \chi\hat{\boldsymbol{n}}\right)}{H(z)(1+z)^4} \exp\left(-\tau\left[(1+z)E_\gamma,\, z\right]\right)\,, \tag{E.2}$$

where $P_\gamma$ denotes the photon emissivity, namely the photon energy per unit of volume, time and energy range. It holds

$$P_\gamma(E_\gamma, z, \chi\hat{\boldsymbol{n}}) = E_\gamma \frac{\mathrm{d}N_\gamma}{\mathrm{d}E_\gamma} \frac{\langle\sigma v\rangle}{2} \left[\frac{\rho(z, \chi\hat{\boldsymbol{n}})}{m_{\rm DM}}\right]^2\,, \tag{E.3}$$

where $\mathrm{d}N_\gamma/\mathrm{d}E_\gamma$ is the energy spectrum for annihilating DM. We recall that

$$\rho_{\rm DM}(z, \chi\hat{\boldsymbol{n}}) = \overline{\rho}_{\rm DM}\, \delta(z, \chi\hat{\boldsymbol{n}}) = \Omega_{\rm DM}\, \rho_{c,0}\, (1+z)^3\, \delta(z, \chi\hat{\boldsymbol{n}})\,. \tag{E.4}$$

By replacing Eq. (E.3) in Eq. (E.2) we get

$$I_\gamma(\hat{\boldsymbol{n}},E_\gamma) = \frac{1}{4\pi}\int \mathrm{d}\chi \frac{\mathrm{d}N_\gamma}{\mathrm{d}E_\gamma}[E_\gamma(1+z)]\frac{\langle\sigma v\rangle}{2}(1+z)\left[\frac{\Omega_{\rm DM}\,\rho_c\,(1+z)^3\,\delta(\chi,\hat{\boldsymbol{n}}\chi)}{m_{\rm DM}}\right]^2 \tag{E.5}$$

$$\times\frac{1}{(1+z)^4}\exp\left(-\tau\left[(1+z)E_\gamma, z\right]\right)$$

$$= \frac{1}{4\pi}\int \mathrm{d}\chi \frac{\mathrm{d}N_\gamma}{\mathrm{d}E_\gamma}[E_\gamma(1+z)]\frac{\langle\sigma v\rangle}{2}\left(\frac{\Omega_{\rm DM}\rho_{c,0}}{m_{\rm DM}}\right)^2(1+z)^3\exp\left(-\tau\left[(1+z)E_\gamma, z\right]\right), \tag{E.6}$$

where in the first equality we have used the fact that $\mathrm{d}\chi = c/H(z)$. By comparing Eq. (E.1) with Eq. (E.6), we derive

$$W(E_\gamma, z) = \frac{1}{4\pi}\frac{\langle\sigma v\rangle}{2}\Delta^2(z)\left(\frac{\Omega_{\rm DM}\rho_{c,0}}{m_{\rm DM}}\right)^2(1+z)^3\frac{\mathrm{d}N_\gamma}{\mathrm{d}E_\gamma}[E_\gamma(1+z)]\exp\left(-\tau\left[(1+z)E_\gamma, z\right]\right), \tag{E.7}$$

where we include the clumping factor $\Delta^2(z)$ in Eq. (3.61) in order to take into account the substructures within the DM halos under study.